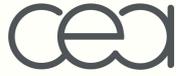
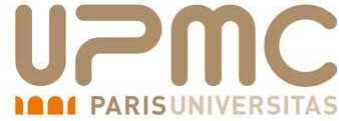

UNIVERSITÉ PIERRE ET MARIE CURIE - PARIS VI

et

INSTITUT DE PHYSIQUE THÉORIQUE - CEA/SACLAY

École Doctorale de Physique de la Région Parisienne - ED 107

### Thèse de doctorat
Spécialité: **Physique Théorique**

---

## Exact Large Deviations of the Current in the Asymmetric Simple Exclusion Process with Open Boundaries

---

présentée par Alexandre Lazarescu

pour obtenir le grade de Docteur de l'Université Pierre et Marie Curie

Thèse préparée sous la direction de Kirone Mallick

Soutenue le 25 septembre 2013 devant le jury composé de:

| | |
|---|---|
| Richard Blythe | Rapporteur |
| Alain Comtet | Examinateur |
| Martin Evans | Examinateur |
| Kirone Mallick | Directeur de thèse |
| Vincent Pasquier | Membre invité |
| Éric Ragoucy | Rapporteur |
| Frédéric Van Wijland | Examinateur |

*Pour Jean-Daniel,
qui a su m'enseigner
que la Mathématique
est une langue vivante.*

# Contents

# Remerciements

De nombreuses personnes ont contribué à rendre les trois années qu'ont duré ma thèse aussi intéressantes qu'agréables, et c'est ici que je les en remercie.

Tout d'abord, un grand merci à Richard Blythe et à Éric Ragoucy, qui ont accepté d'être rapporteurs de cette thèse et qui ont mené leur tâche à bien avec une extrême minutie, y sacrifiant une partie de leurs vacances d'été. Je remercie également tous les examinateurs qui se sont joint à eux pour évaluer mon travail : Alain Comtet, Martin Evans, Vincent Pasquier et Frédéric Van Wijland. Ce fut un réel plaisir de vous avoir dans mon jury de thèse, et vos corrections et commentaires au sujet de ce manuscrit m'ont été précieux.

Je remercie mon directeur de thèse, Kirone Mallick, pour m'avoir initié à un sujet non seulement passionnant, mais aussi empreint d'une certaine beauté esthétique, dont l'appréciation dépend certes des goûts, mais qui correspond parfaitement aux miens. Je te remercie également pour ta gentillesse, ton soutien et ta disponibilité, dont j'ai parfois abusé. J'ai énormément appris de nos interactions durant ces trois années, et j'espère que l'avenir en apportera beaucoup d'autres.

Pouvoir travailler à l'Institut de Physique Théorique a été une grande chance et un vrai bonheur pour moi. J'aimerais en remercier tous les membres pour avoir contribué à rendre ce lieu si agréable qu'on en oublie parfois à quel point il peut être pénible d'y arriver ou d'en repartir (et j'en profite pour ne pas remercier la RATP, et en particulier l'équipe qui est chargée du bon fonctionnement du RER B, si tant est que cette équipe existe, car rien n'est moins sûr). Un grand merci à Henri Orland et à Michel Bauer, qui ont, durant mon séjour, dirigé le labo de mains de maîtres. Merci à Olivier Golinelli et à Stéphane Nonnenmacher, toujours prêts à nous aider dans nos problèmes, fussent-ils matériels, humains ou existentiels. Merci à toute l'équipe administrative : Anne Capdepon, Sylvie Zaffanella, Catherine Cataldi, Laure Sauboy, Émilie Quéré, Morgane Moulin, Loic Brevas et Emmanuelle de Laborderie, ainsi qu'à l'équipe informatique : Philippe Caresmel, Patrick Berthelot, Pascale Beurtey et Laurent Sengmanivanh, pour leur grande gentillesse et leur extrême efficacité.

Je remercie également les permanents du labo, avec qui j'ai pu avoir de nombreuses discussions, d'ordre scientifique ou non, et tout particulièrement Vincent Pasquier, avec qui j'ai eu le plaisir collaborer durant ma dernière année de thèse, navigant entre des concepts algébriques d'une simplicité et d'une élégance rare, et certains des calculs les plus abominables que j'ai pu rencontrer jusqu'ici (je remercie d'ailleurs aussi Leo August Pochhammer, qui nous aura causé bien du souci, mais nous ne pouvons au fond nous en prendre qu'à nous-mêmes). J'ai aussi eu l'occasion et la joie de rencontrer certains des visiteurs du labo, comme Sylvain Prolhac (mon "grand frère de thèse"), Raphaël Chetrite (qui m'avait précédemment fait profiter d'un tas d'exercices de physique des milieux continus plus passionnants les uns que les autres), et Pavel Krapivsky (contre qui j'ai pu me mesurer au tennis de table, ce dont je garde un souvenir cuisant).

Cette thèse a aussi été l'occasion de visiter d'autres labos, à Paris ou ailleurs, et de discuter avec d'autres chercheurs de mon domaine, qui m'ont à chaque fois accueilli à bras ouverts. Je remercie Julien Tailleur (à qui je dois certaines des données numériques utilisées dans ce manuscrit) et Frédéric Van Wijland, du laboratoire Matière et Systèmes Complexes, à l'Université Paris Diderot, à qui j'ai rendu de nombreuses visites. Merci également à Hugo Touchette, Rosemary Harris et aux autres membres de la School of Mathematical Sciences, pour ces quelques jours très agréables à Queen Mary, University of London. Merci à Fabian Essler pour son accueil durant la journée que j'ai passée au Rudolf Peierls Centre for Theoretical Physics à Oxford. Merci à Martin Evans, Richard Blythe et à tous ceux que j'ai pu croiser à la School of Physics and Astronomy de l'University of Edinburgh, dont je garde un excellent souvenir, ainsi qu'à la météo écossaise pour m'avoir presque entièrement épargné. Je ne peux pas en dire autant de la météo belge, mais je remercie vivement Christian Maes, ainsi que tous les membres de l'Instituut voor Theoretische Fysica à K.U. Leuven, pour leur accueil et leur sympathie, non seulement lors du bref séjour que j'y fis il y a quelques mois, mais surtout depuis mon arrivée ici en tant que post-doc (j'y travaille en effet depuis quelques semaines, et j'y suis déjà chez moi). Merci enfin à Mieke Gorissen et Carlo Vanderzande, avec qui j'ai pu interagir lors d'une collaboration des plus fructueuses.

Revenons un moment à l'IPhT, car la liste est loin d'être finie. Comment pourrais-je en effet oublier la joyeuse ribambelle de thésards dont je fis partie, et sans qui les déjeuners aux restaurants du CEA auraient été bien différents. Trois années de thèse, ça fait cinq générations de thésards, voire un peu plus, c'est à dire un paquet de monde, et je risque fort d'en oublier quelques-uns, mais je suis sûr qu'ils me le pardonneront.

Commençons donc par ceux qui étaient déjà là à mon arrivée, et en particulier par les deux infortunés dont j'ai envahi le bureau sans autre forme de procès : Hélène G., dont émanait parfois d'étranges miaulements, et Jean-Marie, à qui aucun déterminant ne résiste. Il y avait aussi Jérôme D. et son optimisme démesuré, Clément G. et ses techniques de domptage de bureaucrates, Clément R., Emeline (qui commença ce qu'il échut à Piotr d'achever, mais nous y reviendrons), Gaëtan, Michaël (qui n'était plus thésard, mais Bon), Enrico, et puis Bruno, Roberto, Francesco et Nicolas.

Puis vient le tour de mes contemporains. Romain, d'abord : tu as été un élément absolument essentiel à ma méthode de travail (qui consistait principalement à aller dans ton bureau et à t'empêcher de travailler jusqu'à ce qu'une idée me vienne), et je t'en serai éternellement redevable, surtout si on compte le temps que tu y as perdu. Piotr, ensuite : mon petit Pitou, sans toi, je n'aurais jamais eu la joie d'entrer dans mon bureau et de découvrir mon ventilateur ligoté à ma chaise avec du fil de fer, ou la boule de ma souris (ainsi que celle de ma souris de rechange) me regardant du haut d'une armoire, ou toutes les touches de mon clavier interverties, ou mon parapluie rempli de confettis, ou encore bien d'autres taquineries de ce genre, qui n'ont cessé de m'émerveiller par leur variété et leur surprenante ingénuosité, sans oublier tous les petits mots doux et gentils que tu as pu avoir à mon égard, toujours pertinent, toujours subtil, un bonheur chaque fois renouvelé. Et puis Julien, le spécialiste de la blague au beurre (c'est plus gras, mais c'est meilleur) : une telle capacité à stimuler l'imagination, je ne la retrouverai chez personne d'autre, et c'est bien dommage. Enfin, il y avait Andrea et Stefano, que j'ai un peu moins eu l'occasion de côtoyer, ainsi que Letícia, dont le séjour à l'IPhT fut plus bref, mais non moins intense.

Il reste la relève : Alexander, d'abord, qui combla une partie du gouffre laissé par mes premiers co-bureaux, et Hélène D., qui ne combla l'autre que par intermittence, à notre grand regret. Et puis, en vrac, Thiago, Éric, Rémi, Thomas Ep., Thomas Ey., Katya (victime collatérale de mes errances dans le bureau de Romain), Antoine, Benoît, Hanna, Jerôme, Gaëlle, Christophe, Yunfeng, ainsi que Jean-Philippe, Pierre et Piotr W., que l'on aurait aimé garder plus longtemps, sans oublier les visiteurs du "K. E. fan club" : Hendrik, Ervand, Andreas, Matthias et Sebastian. Je n'ai aucune crainte à laisser le labo entre vos mains, vous en prendrez soin comme vous avez su prendre soin de moi (allant même jusqu'à me nourrir pour me sauver de l'hypoglycémie : ce gâteau au chocolat d'avant ma soutenance était un vrai régal).

Et n'oublions pas nos grands frères les post-docs : Richard, Guillaume, Laura, Tridib, Juan, et bien d'autres.

C'est grâce à vous tous que j'ai passé d'aussi bonnes années à l'IPhT, et je vous en remercie.

J'ai également pu croiser quelques thésards en dehors du labo. Je remercie tous les participants, intervenants et organisateurs de l'école d'été *Emergent Order in Biology* qui eut lieu en juillet 2012 à l'Institut d'Études scientifiques de Cargèse, et dont je suis revenu avec quelques amis et beaucoup de coups de soleil.

Je remercie aussi tous ceux qui ne m'ont pas mis à la porte à grands coups de pieds lors de mes très occasionnels (lire : fréquents) séjours invasifs au 24, rue Lhomond : Swann, Manon, Sebastian et Tommaso ; Romain, Sébastien, Alexander, Éric et Thomas ; et puis Michele, Tristan et tous ceux ayant pris part au séminaire de la jeunesse condensée, à qui, en plus d'imposer ma présence, j'ai aussi retiré le pain au chocolat de la bouche.

Je suis aussi très reconnaissant à tous mes amis, thésards ou non, d'avoir exploré avec moi le paysage culinaire parisien tout au long de cette thèse. Romain (troisième mention, mais tu les mérites bien), Mandy, Wahb et Cédric en ont été des acteurs importants : entre les quelques restaus de haut vol que nous avons testé ensemble, et vos productions personnelles tout aussi excellentes (pizzas, burgers, sushis, cocktails, fondues, et autres mexi-fat), ainsi que quelques voyages aux États-Unis pour briser la monotonie, il sera difficile de me passer de vous durant les années qui viennent (la solution étant, bien entendu, de ne pas me passer de vous : préparez vos sofas, j'arrive) ! Merci à Andy pour toute ces balades nocturnes le long des quais de Seine, merci à Inès, Phil et Manon pour m'avoir aidé à supporter la chaleur canarienne, merci à Magic, Sophie, Agnès, L.-P., Antoine, Bichette, Gaby, Ziane, Cyntia, Clément, et tous les autres anciens de l'É.N.S de Lyon, et merci à tous mes amis d'enfance, de L.L.G., et d'ailleurs (qui ne m'en voudront pas de ne pas tous les nommer : ils se reconnaîtront, dans le cas bien improbable où cette thèse leur tomberait entre les mains), pour toutes les soirées passées ensemble et tous les repas que nous avons partagés.

Merci, enfin, à mes parents, Papa et Maman (que d'autres que moi appellent parfois Razvan et Carmen, sûrement pour éviter d'éventuelles confusions). Est-ce une banalité que de dire que je ne serais pas là sans vous ? C'est bien plus qu'une simple question de causalité : vous avez toujours, et en toute circonstance, été là pour moi, vous m'avez toujours soutenu, toujours supporté, vous avez toujours fait tout votre possible pour rendre ma vie aussi plaisante qu'elle puisse l'être, et personne n'aurait pu le faire mieux que vous. Pour tout ça, et encore infiniment plus, merci.

Je finirai en remerciant quelques entités que je ne connais pas personnellement, mais qui ont contribué, à leur manière, à mon équilibre physique et psychologique durant ces trois éprouvantes années. Merci, donc, à Gearbox, Lipton, Écusson (surtout le rosé), AMC, Tyrrells, Stephen Fry, Milka, et au Palenque.

<div style="text-align: right;">
Leuven, le 19 novembre 2013,<br>
Alexandre Lazarescu.
</div>

# Informal Introduction

Consider a statistical system. Let's say, to keep things simple, a system composed of a large number of particles: they could be atoms in a box, spins on a lattice, grains of sand in a dune, cells in an organ, insects in a colony, people in a theatre, or pretty much anything. Those particles have individual characteristics (masses, charges, sizes, behavioural patterns, etc.). They might interact with one another according to some rules, and with their environment (through its geometry, the presence of external forces, reservoirs of particles, heat baths, etc.). All that can, in principle, be put in equations, using a set of laws, and a gigantic number of degrees of freedom (the position and state of each and every one of the particles), which contain all the information that there is about the state of the system at any given time. Getting that information explicitly, by solving that enormous system of equations, is of course extremely hard, and often completely futile.

Now, if we look at that system from further away, we might describe it through a reasonably small number of global quantities: pressure, temperature, density, magnetisation, elasticity, and so on. We might also find, empirically, that those quantities obey certain laws (for instance an equation of state for a gas). These macroscopic quantities and laws are, naturally, a consequence of the microscopic behaviour of the system: each of them is a massively averaged combination of all the individual parameters associated to all the particles. A great deal of information is lost through this averaging, as almost all the initial degrees of freedom are summed out and disappear. However, the result is not only much simpler than the exact microscopic description of the system, but also perfectly sufficient to describe whatever, in the system's behaviour, is relevant to us. Moreover, we sometimes discover that most of the details pertaining to the particles (such as their shapes, masses, charges, etc.) have little or no influence on the global behaviour of the system, and that it is in fact universal (with respect to those details), which is a good thing to know, and would not have been known from a purely microscopic analysis.

The role of a statistical physicist, in that context, is twofold: firstly, one must, starting from what one knows of the microscopic rules of the system, weed out all the superfluous information, obtain its macroscopic behaviour, and draw any conclusion that one might from the result. Secondly, one should, when possible, determine what microscopic elements are relevant to the macroscopic behaviour of the system, and why it is so. That second task is somewhat trickier to define than the first, but much more important: the knowledge one might draw from it would apply to a large class of systems rather than a single specific one. In short, solving one system is nice, but what we're really after is universality, because we don't want to have to solve every system by itself, it's much easier to do them all at once. The best quality in a scientist, as we all know, is laziness.

Let us now get a bit more specific. There are two kinds of statistical systems: those that are at equilibrium, and those that aren't.

The former are systems which, if let to evolve freely, go to a state where everything is at rest macroscopically (i.e. not considering the thermal fluctuations at the microscopic level). This will be the case generally for systems that are isolated and devoid of active elements, so that nothing can drive them out of staticity. The signature of this inactivity is detailed balance, which is to say the absence of probability fluxes between the microscopic configurations of the system. If that condition is met, then the system is governed by the Gibbs-Boltzmann law, which gives the probability of observing any configuration, explicitly, in terms of its energy. Of course, this doesn't make equilibrium statistical physics trivial: finding the probabilities of the microstates is usually only the first step, and the rest can be just as hard as anything else, but at least that first step took care of itself.

For all the other cases, nothing can be said *a priori*, and one would have to solve the entire dynamics of the system to get at whichever quantities one might be interested in. Among those systems that do not reach equilibrium, the simplest ones, and the ones which will be at the core of this study, are those that reach a steady state (usually abbreviated as NESS, standing for Non-Equilibrium Steady State), where all observables are independent of time. The absence of detailed balance then manifests itself through macroscopic currents (of particles, for instance), which are the result of the microscopic currents that exist between the microstates of the system. Think for instance of a tube connecting two infinite reservoirs of particles, with either a difference of potential between the reservoirs, or a field in the tube driving the particles to one side rather than the other. In both cases, once the system has reached its steady state, we can observe a current of particles from one reservoir to the other through the tube. That current, being the signature of non-equilibrium, is of particular interest to us. We'll get back to that picture in a moment.

In the quest for a substitute to the Gibbs-Boltzmann law, a lot of effort has been put in the study of so-called 'large deviation functions', which are logarithms of probabilities of observables in the limit of some large quantity (usually the size of the system or the time that has passed). They can be seen as a more precise version of the Gaussian approximation obtained from the central limit theorem, that holds information not only on the probability of likely events, but of extremely rare ones too. Analysing those functions, as well as the microscopic pathways associated to those rare events, is an important challenge to statistical physicists. One could think, for example, of a chemical reaction in a complex environment, and involving few reactants. What would interest us there is not that, most of the time, nothing happens, but exactly how often something does, and through which succession of events it does best (so that we can, for instance, try to favour that optimal pathway, or inhibit it, depending on whether we want the reaction to happen or not). Finding efficient algorithms to produce rare events in simulations, for this type of problems or others, has been a popular topic in the past decade or so, and a lot remains to be discovered and understood.

There are many more ways one could go about studying non-equilibrium systems and large deviations. One that has had a lot of success in the past is to choose a toy model that is simple enough to be mathematically tractable, yet complex enough to be physically relevant, then solve as much of it as possible, and pray for something interesting or useful to emerge along the way. The model of choice for us here is the 'asymmetric simple exclusion process' (or ASEP for short). In this model, particles jump stochastically from site to site on a one-dimensional lattice of finite size. Reservoirs of particles at fixed densities are connected to each end of the system, and particles may enter or leave the system only at those ends. The model is asymmetric in that particles jump preferentially to the right, which mimics the presence of a driving field in the bulk of the system. Finally, they interact with one another through simple hard core repulsion, accounted for by the exclusion constraint that no more than one particle may be on a given site at a given time. Because of the biased jumps, and the possibly unequilibrated reservoirs, particles flow from left to right. As we mentioned before, this current of particles is the signature of non-equilibrium and we'd like to know everything we can about it.

The ASEP has many qualities which make it perfectly suited to that endeavour, the first of which being that it is integrable, which means that one might be able, as indeed we were, to obtain exact analytic results. It is also well motivated physically, as it was first invented to study the motion of biological objects, and is still used, along with its numerous variants, to model real systems. Moreover, it can be mapped onto or related to a large number of other toy models from statistical physics or condensed matter, or even some interesting mathematical objects. For all these reasons, it is one of the most extensively studied models in non-equilibrium statistical physics. Despite that, the ASEP still holds a few secrets, and I did my best during my three years as a doctoral student to uncover one or two of them.

The outline of this thesis is as follows.

In chapter I, we define all the concepts we will need that have to do with large deviations, first in general, then in the context of Markov processes in continuous time, along with a few simple results. In particular, we define what is called the 's-ensemble', which is a statistical ensemble for Markov processes where the current is seen as a free parameter.

In chapter II, we get acquainted with the asymmetric simple exclusion process. In the first part of the chapter, we give the definition of the model, do a very brief overview of existing variants and relations to other models, and then look at what we can learn from simulations and mean-field calculations. In the second part, we present two known results that we will need later on, namely the so-called 'matrix Ansatz' invented by Derrida, Evans, Hakim and Pasquier to describe the steady state distribution and mean current of the open ASEP, and the Bethe Ansatz solution found by Prolhac and Mallick for the fluctuations of the current in the periodic ASEP. This will also allow us to rederive those results using our own notations, to avoid confusion in the following calculations.

In chapter III, we present our own results, which were published in [1], [2] and [3]. We first define our 'perturbative matrix Ansatz', as a generalisation of the original matrix Ansatz, and show how it allows to access the cumulants of the current order by order. We then present our main result, which is an exact expression for the complete generating function of the cumulants of the current, correct for any finite size of the system and any values of its parameters. Finally, we show how it was possible to *guess* that expression from calculating only the first few cumulants explicitly from our Ansatz. At that point, our result is therefore, strictly speaking, a conjecture.

In chapter IV, we analyse our result in detail, by first taking the limit of large sizes and extracting the asymptotic behaviour of our expression for the cumulants, obtaining a different result for each phase in the ASEP's diagram. We then look at what the corresponding behaviours are for the large deviation function of the current in each phase. After that, we look at the limits of an extremely large or low current imposed to the system, and gather all the information we can on the steady state of the system under those conditions. Finally, show how all this put together, along with results from a simple hydrodynamic description of the system, allows to express a conjecture for the phase diagram of the open ASEP in the s-ensemble.

In chapter V, we come back to our generalised matrix Ansatz, and generalise it yet a bit further, by introducing two free parameters in it. We show how this new and improved version relates to Baxter's Q-operator and to the algebraic Bethe Ansatz. We also show how it leads to the functional Bethe Ansatz for the open ASEP, without any restriction on its parameters. We use this result to finally prove the expression obtained in chapter III.

Before getting started, I should say a few things about this here manuscript.

I have tried, to the best of my abilities, to make the contents of this work as self-contained as possible. I was aided in that by the fact that the object of my study, i.e. the large deviations of the current in the steady state of the open ASEP, is reasonably simple to define, and requires very few sophisticated concepts or mathematical tools, if any. Most of those are presented in chapter I, so that readers with no deep knowledge of non-equilibrium statistical physics might find there everything they need to understand the rest. As for the part treating of integrability and Bethe Ansatz (i.e. chapter V and a bit from chapter II), no previous knowledge is required of the reader. The actual calculation may get somewhat involved from time to time, but I hope to have managed to present them in such a way that they can be followed without too much of an effort.

The layout of this work is more or less logical, but not necessarily linear. Chapter I contains general definitions and classic results, and can be skipped by a reader already familiar with non-equilibrium statistical physics and large deviations. The first part of chapter II only gives well known facts and results on the ASEP, and should be of interest in particular to those who have never heard of it. It is, however, very brief, as I didn't intend to do an extensive review of the field, but I have tried, at least, to give a few relevant references, so that the curious reader might know where to look. The second part of chapter II should not be skipped, since most of what is in chapters III and V is built upon it. Chapter III is in fact slightly obsolete, as a large part of it is a special case of what is done in chapter V, but since it represents the work done in two and a half years out of three, I thought I'd include it anyway. That being said, the reader can perfectly go directly from chapter II to chapter V and then back to chapter IV. Finally, chapter IV can be read at any point after chapter II, since it only relies on the final result from chapter III (or V), which is reminded at the beginning, and not on any detail of the calculations.

# CHAPTER I

# A Short Introduction to Large Deviations

In this first introductory chapter, we present a few useful concepts and results related to large deviations and Markov processes. We start by introducing the reader to large deviations using the simplest of all examples, the loaded coin toss. We then give a general definition of large deviation functions, and show how they are related to generating functions of cumulants. In a second section, we focus on temporal large deviations for Markov processes in continuous time (which we first define), by looking at probabilities of time-additive quantities. By considering a special case corresponding to the entropy production, we derive the Gallavotti-Cohen symmetry and define the so-called 's-ensemble'.

There are many more things to be said about large deviations than we do here, as we consider only what will be specifically useful to us later on. For a thorough review of this topic, one can refer to [4] and [5].

## I.1 Large numbers and large deviations

Large deviations can be understood as a more precise version of the central limit theorem, which is itself an extension of the law of large numbers. Let's first look at that through a simple example.

### I.1.1 A simple example - tossing a loaded coin

Let us consider a loaded coin, which has a probability $p$ to land on heads, and a probability $1-p$ to land on tails, with $p$ between 0 and 1. We toss that coin a very large number of times $N$, and we record the outcome of each coin toss (which we label $i$, going from 1 to $N$), as $X_i = 1$ for heads and $X_i = 0$ for tails.

We assume the tosses to be independent of each other, so that the probability of any sequence $\{X_i\}$ is simply the product of the probabilities of each individual toss:

$$\mathrm{P}(\{X_i\}) = \prod_{i=1}^{N} p^{X_i}(1-p)^{(1-X_i)}. \tag{I.1}$$

Now, let's say that we're not interested in the full sequence of outcomes, but only on the total number of heads, or, equivalently, the rate of occurrence of heads, $r = \frac{1}{N}\sum X_i$, which is between 0 and 1. The probability of any $r$ (which is here only defined if $Nr$ is an integer) is the sum of the probabilities of all sequences $\{X_i\}$ which have the same $r$. Luckily, the probability of each of those is the same, so we only need to count how many there are, which is the number of ways to place $Nr$ ones in a sequence of size $N$. This gives us an entropic coefficient $\frac{N!}{[Nr]![N(1-r)]!}$. The probability of $r$, knowing $N$, is therefore:

$$P_N(r) = \frac{N!}{[Nr]![N(1-r)]!} p^{Nr}(1-p)^{N(1-r)}. \tag{I.2}$$

What the central limit theorem tells us about this probability is that when $N$ gets large, $P_N(r)$ becomes peaked around the mean value of $r$ (which is $p$), and that $r$ is distributed around that value as a Gaussian with variance $\frac{p(1-p)}{N}$. That is to say:

$$P_N(r) \sim P_N^{Gauss}(r) = P_N(p)\ e^{-N\frac{(r-p)^2}{2p(1-p)}} \tag{I.3}$$

where $\sim$ means that the difference of the two functions goes to 0 when $N$ goes to infinity. This gives us a good approximation for values very close to $r = p$ (typically at a distance of order $N^{-1/2}$), which accounts for the most probable fluctuations of $r$, but what if we need an estimate of the probability of a completely different $r$ ? As we can see on fig.-I.1, the relative accuracy of the Gaussian approximation decays with $N$ for any $r$ that isn't $p$, which makes it a rather poor approximation whenever rare events are important.

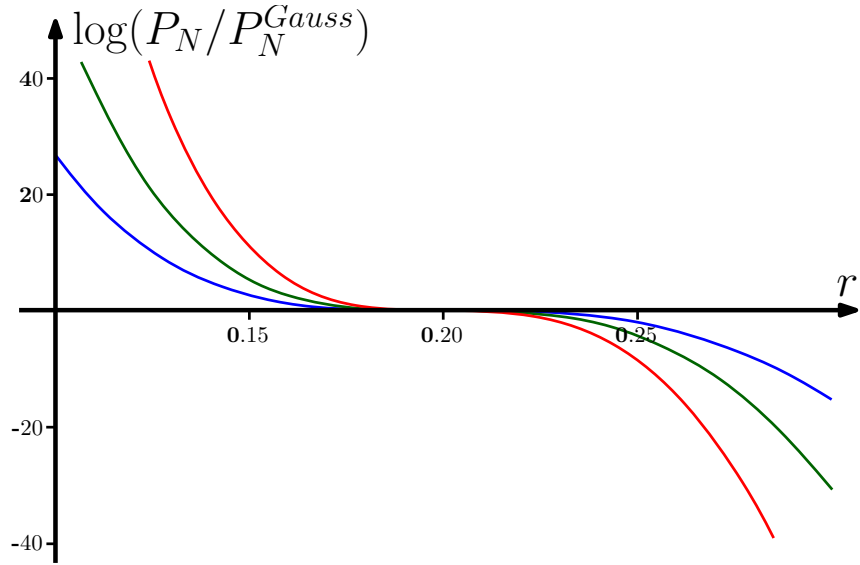

Figure I.1: Relative accuracy of the Gaussian approximation for $P_N(r)$, in logarithmic scale, for $p = 0.2$ and $N = 5000$ (blue), 10000 (green), 20000 (red).

This is where large deviations come in. We can get a much better approximation, unsurprisingly, by looking at the asymptotic behaviour of $P_N(r)$ for every $r$, and not only around $r = p$. We can calculate a function $g(r)$, called the 'large deviation function' of $r$, such that:

$$\boxed{P_N(r) \sim e^{-Ng(r)}} \tag{I.4}$$

where the ∼ now means that the ratio of the two functions is of order 1 with respect to $N$. The function $g(r)$ can be calculated by extracting the term of order $N$ in $\log(P_N(r))$ (using Stirling's approximation), and a straightforward calculation yields:

$$g(r) = r\log(r) + (1-r)\log(1-r) - r\log(p) - (1-r)\log(1-p). \tag{I.5}$$

Naturally, taking the quadratic part of $g(r)$ around $r = p$ gives the gaussian rate $\frac{(r-p)^2}{2p(1-p)}$.

If we now plot the relative accuracy of this approximation (fig.-I.2), we see that it quickly converges to a function $f(r)$ of order 1 (which is here $\sqrt{2\pi r(1-r)}$). That function comes from the term of order 1 in Stirling's approximation, it's the prefactor in eq.(I.4), and it is not important to us (we're always looking at logarithms of probabilities which scale with $N$, so the constant term is irrelevant).

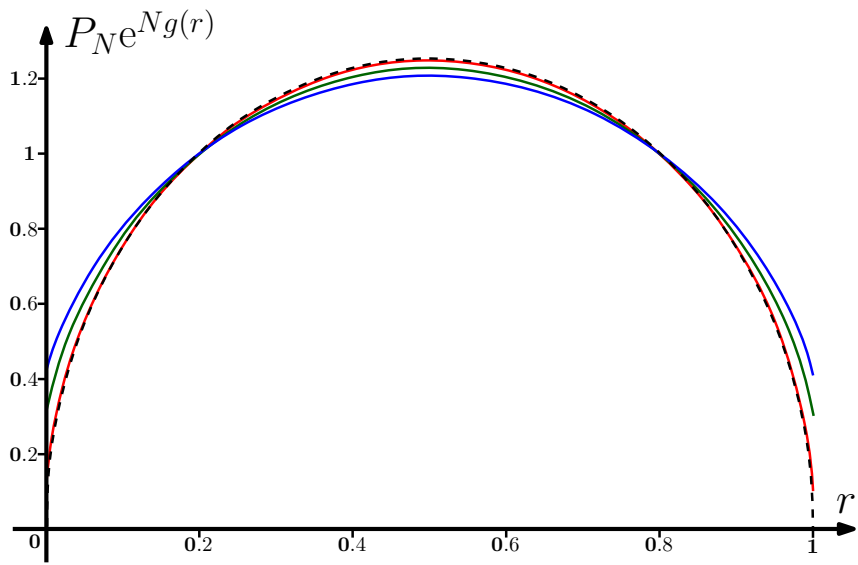

Figure I.2: Relative accuracy of the large deviations approximation for $P_N(r)$, in logarithmic scale, for $p = 0.2$ and $N = 5$ (blue), 10 (green), 100 (red), and the function $f(r) = \sqrt{2\pi r(1-r)}$ (dashed).

In some situations, however, that prefactor can become extremely important, in particular when it has poles, or any kind of singularities [6]. We won't have to worry about that in our case.

### I.1.2 General definition of large deviations

Now for a more general definition of large deviation functions: consider a system defined by a size $N$, and an observable $a$ intensive in $N$, which has a probability distribution $P_N(a)$ for each $N$. It is said that $a$ obeys a large deviations principle of rate $g(a)$ if the limit

$$g(a) = \lim_{N\to\infty}\left[-\frac{\log(P_N(a))}{N}\right] \tag{I.6}$$

is defined and finite for every $a$. Simply put, $g(a)$ is the rate of exponential decay of $\mathrm{P}_N(a)$ with respect to $N$. Its minimum is the most probable value of $a$. We then write:

$$\boxed{\mathrm{P}_N(a) \approx \mathrm{e}^{-Ng(a)}} \tag{I.7}$$

where the $\approx$ signifies precisely what is written in eq.(I.6).

It may happen that the correct scaling of $\log(\mathrm{P}_N(a))$ be $N^\alpha$, with $\alpha \neq 1$, instead of $N$. In that case, the large deviations of $a$ are said to be anomalous.

Note that $N$ is not necessarily an actual size or a number of elements, but can very well be a time span, or any variable that can be taken to infinity (although it is actually almost always time or size).

Also note that $a$ doesn't have to be a scalar observable. It can be a function (in which case $g[a]$ is a large deviation functional), or any mathematical object for which a probability can be defined in the system under consideration.

In the case where $a$ is more complex than just one scalar variable, one can easily obtain the large deviation function of a subset $a_1$ of $a$ from that of $a$, through the extremely useful *contraction principle*. What we mean by 'subset of $a$' is a collection of variables included in $a$, and independent of the other variables in $a$. The contraction principle tells us that, if $a = a_1, a_2$, and if $a$ obeys a large deviation principle with rate $g$, then the large deviation function $g_1$ of $a_1$ is equal to $g$ taken at the best possible value of $a_2$ for that $a_1$ (all other values of $a_2$ give a sub-dominant contribution). This is easily shown using a saddle-point approximation:

$$\mathrm{P}_N(a_1) = \int \mathrm{P}_N(a_1, a_2)\mathrm{d}a_2 = \int \mathrm{e}^{-Ng(a1,a2)}\mathrm{d}a_2 \sim \mathrm{e}^{-N \min_{a_2}[g(a1,a2)]} \tag{I.8}$$

so that

$$\boxed{g_1(a_1) = \min_{a_2}[g(a1, a2)].} \tag{I.9}$$

### I.1.3 Large deviation function and cumulants

Instead of considering the probability distribution of an observable, we may just as well consider its moments, which are the expectation values of that observable to integer powers. The collection of those moments bear the same information as the probability distribution, so it is natural to wonder what the consequence of a large deviation principle is on them.

Let us therefore consider the same quantities as in the previous section, and define the moments of $a$:

$$m_k^{(N)} = \langle a^k \rangle_N = \int \mathrm{P}_N(a) a^k \mathrm{d}a \tag{I.10}$$

($N$ being put in subscript under the mean value simply to signify dependence on $N$), as well as the exponential generating function [7] of those moments:

$$m_N(\mu) = \sum m_k^{(N)} \frac{(N\mu)^k}{k!} = \int \mathrm{P}_N(a) \mathrm{e}^{\mu N a} \mathrm{d}a = \langle \mathrm{e}^{\mu N a} \rangle_N \tag{I.11}$$

where the argument of the exponential is taken as $\mu N a$ for convenience in future calculations (one could of course include the factor $N$ in $\mu$).

Finally, let us define the cumulants $E_k^{(N)}$ of $a$ through their generating function:

$$E_N(\mu) = \sum_{k=1}^{\infty} E_k^{(N)} \frac{\mu^k}{k!} = \frac{1}{N} \log(m_N(\mu)) \tag{I.12}$$

so that:

$$\boxed{m_N(\mu) = e^{NE_N(\mu)} = \langle e^{\mu N a} \rangle_N = \int P_N(a) e^{\mu N a} da.} \tag{I.13}$$

We see here that $m_N(\mu)$ (and therefore $E_N(\mu)$) and $P_N(a)$ do indeed hold the same information: they are merely Laplace transforms of one another.

We can now see what happens if we replace $P_N(a)$ by its limit under the large deviation principle: $P_N(a) \to f(a) e^{-Ng(a)}$ (keeping the prefactor $f(a)$ for now, just in case). We get, in the large $N$ limit:

$$e^{NE_N(\mu)} \to e^{NE(\mu)} = \int f(a) e^{-N(g(a) - \mu a)} da. \tag{I.14}$$

It is here that the regularity of $f(a)$ becomes important. If $f(a)$ has no poles, then the integral in eq.(I.14) can be evaluated through a saddle point approximation by deforming its integration path from the real domain of definition of $a$ to one going through the saddle point of $g(a) - \mu a$, which yields:

$$E(\mu) = \max_a [\mu a - g(a)] \tag{I.15}$$

or equivalently

$$\boxed{E(\mu) = \mu a^\star - g(a^\star) \ , \ \frac{d}{da} g(a^\star) = \mu} \tag{I.16}$$

where $a^\star$ is the value of $a$ at which the maximum in eq.(I.15) is attained. That is to say that $E$ and $g$ are Legendre transforms of one another. In this case, $f(a)$ merely gives a prefactor $f(a^\star)$ in front of $e^{NE(\mu)}$. The inverse transformation formula is then:

$$g(a) = \min_\mu [\mu a - E(\mu)] \tag{I.17}$$

or

$$\boxed{g(a) = \mu^\star a - E(\mu^\star) \ , \ \frac{d}{d\mu} E(\mu^\star) = a} \tag{I.18}$$

where $\mu^\star$ is the value of $\mu$ at which the maximum in eq.(I.18) is attained.

In the case where $f(a)$ has poles between the initial integration path in eq.(I.14) and the saddle point of $g(a) - \mu a$, contour integrals around those poles survive the approximation and the value of $a^\star$ might become that of one of those poles instead of the position of the saddle point [6].

## I.2 Temporal large deviations for Markov processes in continuous time

Let us now apply those considerations to a specific context: that of a Markov process [8], for which we look at large deviations with respect to time. See also [9] for a more detailed review of this topic, and [10] for many examples of models that fall under this category.

### I.2.1 Definition and steady state

First things first, let us define what a Markov process is.

We start with discrete time, which is simpler. Consider a set of configurations $\{\mathcal{C}\}$, and a probability vector $|P(t)\rangle$ defined for discrete times $\{t = k\ \delta t\}_{k \in \mathbb{N}}$, which contains the probabilities $\mathrm{P}(\mathcal{C}; t)$ of being in state $\mathcal{C}$ at time $t$. A Markov process (or Markov chain) on $\{\mathcal{C}\}$ is a process for which the probability of being in a certain configuration at a certain time depends only on the state of the system at the previous time, which is to say that $\mathrm{P}(\mathcal{C}; t)$ is a linear combination of all the $\mathrm{P}(\mathcal{C}'; t - \delta t)$ but doesn't depend on anything that has happened before that time: the system has no memory of its past. We can therefore write:

$$\mathrm{P}(\mathcal{C}; t) = \sum_{\mathcal{C}'} W(\mathcal{C}, \mathcal{C}') \mathrm{P}(\mathcal{C}'; t - \delta t). \tag{I.19}$$

The weights $W(\mathcal{C}, \mathcal{C}')$ are the probabilities for the system to find itself at $\mathcal{C}$ knowing that it was previously at $\mathcal{C}'$, i.e. the probabilities to 'transit' from $\mathcal{C}'$ to $\mathcal{C}$. They must of course be positive, and sum to 1 (with respect to $\mathcal{C}$), which is usually written:

$$W(\mathcal{C}', \mathcal{C}') = 1 - \sum_{\mathcal{C} \neq \mathcal{C}'} W(\mathcal{C}, \mathcal{C}'). \tag{I.20}$$

These probabilities can in principle depend on time, but we will assume that they do not. For an actual physical process, most of them will be 0, because the transitions usually happen only between configurations that are close enough according to some criterion. For a system with moving particles, for instance, a transition is usually the result of only one particle moving by one step, so that all $W(\mathcal{C}, \mathcal{C}')$ where $\mathcal{C}$ and $\mathcal{C}'$ are not one step of a particle away from each other will be 0.

One can put these equations in vectorial form, by defining the discrete time Markov matrix $W$ containing the transition probabilities:

$$W = \sum_{\mathcal{C}, \mathcal{C}'} W(\mathcal{C}, \mathcal{C}') |\mathcal{C}\rangle \langle \mathcal{C}'|. \tag{I.21}$$

Equation (I.19) becomes

$$\boxed{|P_t\rangle = W |P_{t-\delta t}\rangle} \tag{I.22}$$

and (I.20) becomes

$$\boxed{\langle 1|W = \langle 1|} \tag{I.23}$$

where $\langle 1|$ is the vector with all entries equal to 1.

We can now define that same process in continuous time, by taking the limit $\delta t \to 0$. The first thing to do is to rescale the transition probabilities with respect to time. It makes physical sense to consider that those probabilities are proportional to the time step, rather than constant: if the system has some small probability to make a transition during a small time $\delta t$, it has twice the opportunities, and therefore twice the probability, to make that same transition in twice the time. We write this as $W(\mathcal{C}, \mathcal{C}') = \delta t\, w(\mathcal{C}, \mathcal{C}')$ for $\mathcal{C} \neq \mathcal{C}'$, and define the continuous time Markov matrix as:

$$M = \sum_{\mathcal{C},\mathcal{C}'} w(\mathcal{C},\mathcal{C}')|\mathcal{C}\rangle\langle\mathcal{C}'| \quad \text{with} \quad w(\mathcal{C}',\mathcal{C}') = -\sum_{\mathcal{C}\neq\mathcal{C}'} w(\mathcal{C},\mathcal{C}') \tag{I.24}$$

(the 1 from the right side of eq.(I.20) has been taken out of the matrix).

Equation (I.22) can then be rewritten as

$$|P_t\rangle - |P_{t-\delta t}\rangle = \delta t M |P_{t-\delta t}\rangle \tag{I.25}$$

and (I.23) as

$$\boxed{\langle 1|M = 0.} \tag{I.26}$$

Now the continuous time limit can be taken unambiguously in equation (I.25):

$$\boxed{\frac{d}{dt}|P_t\rangle = M|P_t\rangle} \tag{I.27}$$

and obtain what is called the 'master equation'.

The solution of this equation is, formally:

$$\boxed{|P_t\rangle = e^{tM}|P_0\rangle} \tag{I.28}$$

where $|P_0\rangle$ is the initial probability distribution.

We know, from the Perron-Frobenius theorem [11], that, unless this Markov process is reducible (if it is, for instance, the sum of two independent Markov processes on two different sets of configurations), the matrix $M$ has exactly one eigenvalue equal to 0 (of which the left eigenvector is $\langle 1|$), and all the other eigenvalues have a strictly negative real part. If we write the right eigenvector associated to the eigenvalue 0 as $|P^\star\rangle$, and decompose $e^{tM}$ on the eigenbasis of $M$ (the other eigenvalues being written as $-\lambda_i$ with $\text{Re}(\lambda_i) > 0$, and the corresponding eigenvectors as $|\psi_i\rangle$ and $\langle\psi_i|$), we get:

$$|P_t\rangle = \left(|P^\star\rangle\langle 1| + \sum e^{-t\lambda_i}|\psi_i\rangle\langle\psi_i|\right)|P_0\rangle = |P^\star\rangle + \sum e^{-t\lambda_i}|\psi_i\rangle\langle\psi_i|P_0\rangle \tag{I.29}$$

(we recall that, since $|P_0\rangle$ is a probability distribution, it sums to 1, i.e.$\langle 1|P_0\rangle = 1$).

This last equation means that for a large enough time, the system will always converge to its *steady state* $|P^\star\rangle$, regardless of $|P_0\rangle$:

$$\boxed{\lim_{t\to\infty}|P_t\rangle = |P^\star\rangle \quad \text{with} \quad M|P^\star\rangle = 0.} \tag{I.30}$$

For shorter times, one can observe a transient behaviour on top of that, given by $\sum e^{-t\lambda_i}|\psi_i\rangle\langle\psi_i|P_0\rangle$, and which does depend on the initial condition. The relaxation rate to the steady state

is then given by the slowliest decaying term in this sum, which is to say the $\lambda_i$ with the smallest real part (the real parts of the eigenvalues are responsible for the decay of the transient regime, whereas the complex parts are responsible for oscillations).

Instead of considering the evolution of a probability distribution on $\{\mathcal{C}\}$, one could look at a single realisation of the process, which is to say a trajectory $\mathcal{C}(t)$ in time (called a 'history'). The simplest way to do that is once again to start from the discrete time version. Let us write $r(\mathcal{C}) = -w(\mathcal{C},\mathcal{C})$. If the system is at configuration $\mathcal{C}$ at time $t$, it has a probability $1 - \delta t\, r(\mathcal{C})$ to stay there, and a probability $\delta t\, r(\mathcal{C})$ to escape to another configuration. The probability for the system to still be at $\mathcal{C}$ at time $t$ without having jumped, knowing that it was there at time 0, is simply the probability not to jump at any time step between 0 and $t$, so that the probability $\delta \mathrm{P}_\mathcal{C}(t)$ to actually jump at $t$ (i.e. the waiting time distribution) is given by:

$$\delta \mathrm{P}_\mathcal{C}(t) = \delta t\, r(\mathcal{C})\bigl(1 - \delta t\, r(\mathcal{C})\bigr)^{t/\delta t}. \tag{I.31}$$

In the continuous time limit, this waiting time distribution becomes a probability density $\mathrm{P}_\mathcal{C}(t)$ which has an exponential distribution with rate $r(\mathcal{C})$ (so that the Markov chain becomes a Poisson process):

$$\mathrm{P}_\mathcal{C}(t) = r(\mathcal{C})\mathrm{e}^{-t\, r(\mathcal{C})} \tag{I.32}$$

The probability rate, after a jump, to go to one particular configuration $\mathcal{C}'$ is given by $w(\mathcal{C}',\mathcal{C})/r(\mathcal{C})$ (i.e. the corresponding transition rate, rescaled as a probability). A history is therefore a succession of discrete jumps, separated by exponentially distributed waiting periods. Given an initial time $t_0$, a final time $t_N$, a set of jumping times $\{t_i\}_{i:1..N-1}$, and a set of configurations $\{\mathcal{C}_i\}_{i:1..N}$, the probability density of the history where the system is at $\mathcal{C}_i$ between times $t_{i-1}$ and $t_i$ is given by:

$$\begin{aligned}\mathrm{P}\bigl(\{t_i\};\{\mathcal{C}_i\}\bigr) =& \mathrm{P}_{\mathcal{C}_1}(t_1 - t_0)\frac{w(\mathcal{C}_2,\mathcal{C}_1)}{r(\mathcal{C}_1)}\mathrm{P}_{\mathcal{C}_2}(t_2 - t_1)\frac{w(\mathcal{C}_3,\mathcal{C}_2)}{r(\mathcal{C}_2)}\ldots\\ & \mathrm{P}_{\mathcal{C}_{N-1}}(t_{N-1} - t_{N-2})\frac{w(\mathcal{C}_N,\mathcal{C}_{N-1})}{r(\mathcal{C}_{N-1})}\ \mathrm{e}^{-(t_N-t_{N-1})r(\mathcal{C}_N)}\\ =& \mathrm{e}^{-(t_1-t_0)r(\mathcal{C}_1)}w(\mathcal{C}_2,\mathcal{C}_1)\ \mathrm{e}^{-(t_2-t_1)r(\mathcal{C}_2)}w(\mathcal{C}_3,\mathcal{C}_2)\ldots\\ & \mathrm{e}^{-(t_{N-1}-t_{N-2})r(\mathcal{C}_{N-1})}w(\mathcal{C}_N,\mathcal{C}_{N-1})\mathrm{e}^{-(t_N-t_{N-1})r(\mathcal{C}_N)}\\ =& \boxed{\prod_{i=1}^{N-1}\mathrm{e}^{-(t_i-t_{i-1})r(\mathcal{C}_i)}w(\mathcal{C}_{i+1},\mathcal{C}_i)\ \mathrm{e}^{-(t_N-t_{N-1})r(\mathcal{C}_N)}}.\end{aligned} \tag{I.33}$$

Note that the last exponential is without a coefficient $r(\mathcal{C}_N)$ because the last event to be considered is not that of a jump at $t_N$, but that of no jump before $t_N$. Also note that all the other coefficients $r(\mathcal{C}_i)$ compensate between the waiting probabilities and the transition probabilities.

### I.2.2 Additive observable

We can now define time-dependent observables on those histories and look at their large deviations in the long time limit. Let us consider an observable $A_t$ defined as a functional of a history $\mathcal{C}(t)$:
$$A_t = \mathrm{F}[\mathcal{C}(t)]. \tag{I.34}$$

For the sake of simplicity, we are only interested in observables that are additive in time, meaning that if a history $\mathcal{C}(t)$ is the concatenation of two shorter ones $\mathcal{C}_1(t)$ and $\mathcal{C}_2(t)$, the functional $F$ distributes over them:
$$\mathrm{F}[\mathcal{C}_1(t) \oplus \mathcal{C}_2(t)] = \mathrm{F}[\mathcal{C}_1(t)] + \mathrm{F}[\mathcal{C}_2(t)]. \tag{I.35}$$

This forces $F$ to be local in time (independent of time correlations). We also assume $F$ to be time-invariant, i.e. independent on the value of the initial time $t_0$ in $\mathcal{C}(t)$.

These constraints allow us to find the most general of such functionals explicitly. Consider first a history without any transitions: $\mathcal{C}(t) = \mathcal{C}_1$. In this case, time additivity can be used to show that $F[\mathcal{C}(t)]$ is proportional to the duration of the process $(t_1 - t_0)$. The proportionality coefficient may depend on $\mathcal{C}_1$, and we will call it $V(\mathcal{C}_1)$. Consider now a history with one transition: the system is in $\mathcal{C}_1$ between times $t_0$ and $t_1$, and in $\mathcal{C}_2$ between $t_1$ and $t_2$. By cutting out, using additivity, the portion of history before $t_1 - \varepsilon$ and that after $t_1 + \varepsilon$, with $\varepsilon$ going to 0, one is left with just the transition, which has a contribution to $F$ that depends only on $\mathcal{C}_1$ and $\mathcal{C}_2$. We will call it $U(\mathcal{C}_2, \mathcal{C}_1)$.

Putting those pieces together, and considering that any history can be decomposed into portions containing at most one transition, we can finally write:
$$\boxed{\mathrm{F}[\mathcal{C}(t)] = \int_{t_0}^{t_N} V\bigl(\mathcal{C}(t)\bigr) dt + \sum_{i=1}^{N} U(\mathcal{C}_i, \mathcal{C}_{i-1})} \tag{I.36}$$

which is expressed schematically on fig.-I.3.

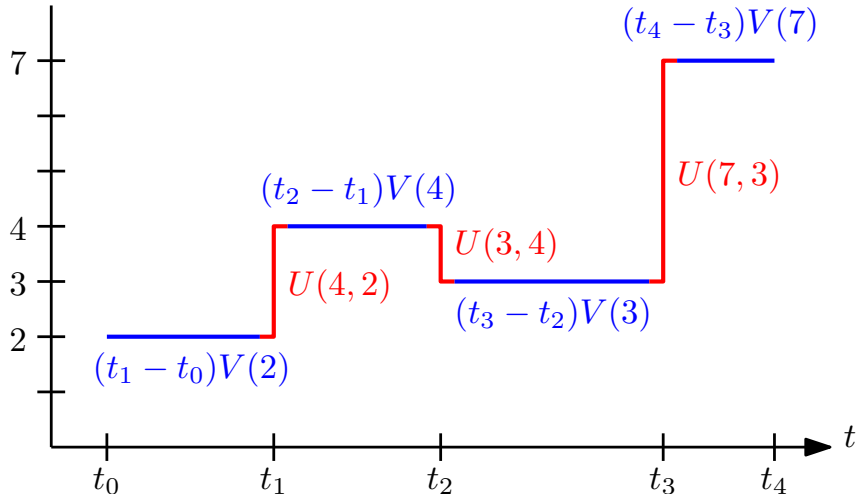

Figure I.3: Functional $F$ over a schematised history. Each straight portion contributes a simple term to the whole function. Waiting periods (in blue) give a contribution that is extensive in time, and depends only on one configuration. Transitions (in red) give a term that depends on the two configurations involved.

The part containing $V$ is the time integral of a purely spatial observable, and cannot contain any information on currents, i.e. on the non-equilibrium nature of the process.

The function $U$, however, can. Each $U(\mathcal{C}_i, \mathcal{C}_{i-1})$ can be seen as a counter for the transition between $\mathcal{C}_{i-1}$ and $\mathcal{C}_i$: every time it is used in the evolution of the system, the value of $A_t$ increases by one quantum of $U$. Whether each value of $U$ is taken as an independent variable, or given a precise value, determines what information is monitored regarding the way those transitions are used, but in many cases, one thing that makes a crucial difference on the resulting behaviour of $A_t$ is whether the system is at equilibrium or not (and, reversely, the behaviour of this observable is a good indication of the system being in or out of equilibrium [12]), as we will see in the next section.

But first, let us derive a simple and extremely useful result on the cumulants of this observable. The generating function $E_t(\nu)$ of the cumulants of $a_t = \frac{1}{t} A_t$ (the intensive version of $A_t$) can be expressed as:

$$e^{tE_t(\nu)} = \langle e^{t\nu a_t} \rangle = \int e^{\nu F[\mathcal{C}(t)]} P[\mathcal{C}(t)] \mathcal{D}[\mathcal{C}(t)] \tag{I.37}$$

where $\mathcal{D}[\mathcal{C}(t)]$ is the measure associated with histories (this can be simply defined in the discrete time case, and then taken as a formal limit for small $\delta t$).

What produces the weights $P[\mathcal{C}(t)]$, as we have seen, is the object $e^{tM}$ applied to our initial condition. A simple way to produce weights equal to $e^{\nu F[\mathcal{C}(t)]} P[\mathcal{C}(t)]$ instead is to consider a deformed Markov matrix $M_\nu$ defined as:

$$\boxed{M_\nu = \sum_{\mathcal{C},\mathcal{C}'} e^{\nu U(\mathcal{C},\mathcal{C}')} w(\mathcal{C},\mathcal{C}') |\mathcal{C}\rangle\langle\mathcal{C}'| + \nu \sum_{\mathcal{C}} V(\mathcal{C}) |\mathcal{C}\rangle\langle\mathcal{C}|} \tag{I.38}$$

(where we take $U(\mathcal{C},\mathcal{C}) = 0$).

This comes from the observation that, by writing $e^{\nu F[\mathcal{C}(t)]} P[\mathcal{C}(t)]$ in the same form as in equation (I.33), one can rearrange the terms to get:

$$e^{\nu F[\mathcal{C}(t)]} P[\mathcal{C}(t)] = \prod_{i=1}^{N-1} e^{-(t_i - t_{i-1})(r(\mathcal{C}_i) - \nu V(\mathcal{C}_i))} w(\mathcal{C}_{i+1}, \mathcal{C}_i) e^{\nu U(\mathcal{C}_{i+1}, \mathcal{C}_i)} \ e^{-(t_N - t_{N-1})(r(\mathcal{C}_N) - \nu V(\mathcal{C}_N))}. \tag{I.39}$$

By replacing $M$ by $M_\nu$ in (I.28), we get:

$$|P_\nu(t)\rangle = e^{tM_\nu} |P_0\rangle = \sum_{\mathcal{C}_N} \int e^{\nu F[\mathcal{C}(t)]} P[\mathcal{C}(t)] \mathcal{D}[\mathcal{C}(t)] \ |\mathcal{C}_N\rangle \tag{I.40}$$

where, as intended, the probabilities of histories have received an extra $e^{\nu F[\mathcal{C}(t)]}$ factor, and we can finally sum over the final configuration $\mathcal{C}_N$ and write:

$$\boxed{e^{tE_t(\nu)} = \langle 1 | P_\nu(t) \rangle = \langle 1 | e^{tM_\nu} | P_0 \rangle.} \tag{I.41}$$

Moreover, as long as $\nu$, $V$ and $U$ are real, the Perron-Frobenius theorem applies, so that the largest eigenvalue $\Lambda_\nu$ of $M_\nu$ is non-degenerate. If we write the corresponding eigenvectors as $|P_\nu\rangle$ and $\langle \tilde{P}_\nu|$, we get, for large times:

$$\boxed{e^{tM_\nu} \approx e^{t\Lambda_\nu} |P_\nu\rangle \langle \tilde{P}_\nu|.} \tag{I.42}$$

By combining equations (I.41) and (I.42) for $t$ large, we can identify $E(\nu) = E_\infty(\nu)$ with $\Lambda_\nu$, so that:

$$\boxed{e^{tE(\nu)} \approx \langle 1|e^{tM_\nu}|P_0\rangle \sim e^{t\Lambda_\nu}\langle 1|P_\nu\rangle\langle \tilde{P}_\nu|P_0\rangle.} \tag{I.43}$$

The *generating function of the cumulants* of any additive observable is therefore equal to the *largest eigenvalue of the associated deformed Markov matrix $M_\nu$*. This is a classic result from the Donsker-Varadhan theory of temporal large deviations [13–17]. Notice that this is a property of the deformed matrix, and not of the initial or final configurations: regardless of those, the long time behaviour of the generating function of the cumulants will be the same. One can easily make sense of this: if the duration of the process is large enough, then the system only takes a small time at first to reach its steady state, and gets out of it very near the end. The rest of the evolution can be considered to be around the steady state, whatever the initial and final distributions are, and the part of $A_t$ which is extensive in time comes only from there. The initial and final distributions only give a time-independent term $\langle \tilde{P}_f|P_\nu\rangle\langle \tilde{P}_\nu|P_0\rangle$, which is akin to the function $f(a)$ in eq.(I.14).

The eigenvectors $|P_\nu\rangle$ and $\langle \tilde{P}_\nu|$ are also useful in the context of large deviations. From equation (I.40) for $t$ large, by regrouping all the histories for which $F[\mathcal{C}(t)] = tf$ and the final configuration is $\mathcal{C}_N$, we get:

$$e^{tE(\nu)}|P_\nu\rangle \approx |P_\nu(t)\rangle = \sum_{\mathcal{C}_N} |\mathcal{C}_N\rangle \int e^{\nu tf} P(f \ \& \ \mathcal{C}_N) df. \tag{I.44}$$

We can then write $P(f \ \& \ \mathcal{C}_N)$ as $P(\mathcal{C}_N|f)P(f)$ (where $P(A|B)$ is the probability of $A$, knowing $B$), and invoke the large deviations principle $P(f) \approx e^{-tg(f)}$:

$$e^{tE(\nu)}|P_\nu\rangle \approx \sum_{\mathcal{C}_N} |\mathcal{C}_N\rangle \int P(\mathcal{C}_N|f) e^{t(\nu f - g(f))} df. \tag{I.45}$$

Finally, just as in eq.(I.14), a saddle-point approximation on $e^{t(\nu f - g(f))}$ yields $e^{tE(\nu)}$ and fixes the value of $f$ to $\frac{d}{d\nu}E(\nu)$. Injecting this in (I.45), we get:

$$\boxed{|P_\nu\rangle = \sum_{\mathcal{C}_N} P\Big(\mathcal{C}_N\Big|f = \frac{d}{d\nu}E(\nu)\Big)|\mathcal{C}_N\rangle.} \tag{I.46}$$

This tells us that the vector $|P_\nu\rangle$ is in fact the probability vector of the final configuration, knowing that the value of $f$ through the evolution of the system was $\frac{d}{d\nu}E(\nu)$. A similar calculation on the left eigenvector $\langle \tilde{P}_\nu|$ shows it to be the probability vector of what the initial configuration was, knowing that the value of $f$ was $\frac{d}{d\nu}E(\nu)$.

We could also look at the probability of a configuration $\mathcal{C}$ at any point during the evolution of the system. To do that, we just need to apply an observable corresponding to that information (which would be $|\mathcal{C}\rangle\langle\mathcal{C}|$) at the correct time. In this case, we need both an initial distribution $|P_0\rangle$ and a final distribution $\langle P_N|$, which will not matter in the end. Let us note $\mathcal{C}_\alpha$ the configuration at time $\alpha t$, with $\alpha < 1$. We get:

$$P(\mathcal{C}_\alpha = \mathcal{C}|\nu) = \langle P_N|e^{(1-\alpha)tM_\nu}|\mathcal{C}\rangle\langle\mathcal{C}|e^{\alpha tM_\nu}|P_0\rangle = \int e^{\nu tf} P(f \ \& \ \mathcal{C}_\alpha = \mathcal{C}) df. \tag{I.47}$$

The rightmost part of this equation gives, as before (through a saddle-point approximation), the corresponding probability conditioned on $f = \frac{d}{d\nu}E(\nu)$. The middle part can be separated in two scalar products, which we recognise as the two probabilities we obtained just before, relating to the initial or final configurations. All in all, we find that:

$$P\left(\mathcal{C}_\alpha = \mathcal{C}\Big|f = \frac{d}{d\nu}E(\nu)\right) = P\left(\mathcal{C}_N = \mathcal{C}\Big|f_1 = \frac{d}{d\nu}E(\nu)\right) P\left(\mathcal{C}_0 = \mathcal{C}\Big|f_2 = \frac{d}{d\nu}E(\nu)\right) \quad (\text{I.48})$$

where $f_1$ is the mean value of $F$ before time $\alpha t$ and $f_2$ is the men value of $F$ after time $\alpha t$. This last result tells us that, for the histories contributing most to those probabilities, the mean value of $F$ is the same for the whole evolution time as it is for the evolution only up to or from time $\alpha t$. Note that this probability distribution does not depend on $\alpha$, so that it is stationary: the probability of a given configuration, knowing $\nu$, is constant in time. All this is valid at the large $t$ limit for a finite, non-zero $\alpha$ (it is not valid, for instance, if $\alpha t$ is finite).

We can go even further than that, and divide the evolution time of the system into a large number of smaller steps. For each of those steps, we find (through the same calculations) that the mean value of $F$ is the same and equal to $\frac{d}{d\nu}E(\nu)$. From this, we conclude that $f = \frac{d}{d\nu}E(\nu)$ is not only the mean value of $F$ for the whole evolution of the system, but also the instantaneous value of $F$ at every time.

Note that all this strongly relies on $F$ being time-additive. Indeed, without that assumption, we could not write the weighted evolution of the system in the exponential form (I.40), and isolating a configuration at a given time (as we did in eq.(I.47)) would not result in two uncorrelated scalar products.

### I.2.3 Entropy - Gallavotti-Cohen symmetry and s-ensemble

We now consider a very specific observable: the entropy production $S_t$, defined as the logarithm of the ratio of the probability of a history $\mathcal{C}(t)$, and that of the same history reversed in time $\mathcal{C}^R(t)$:

$$S_t[\mathcal{C}(t)] = \log\left(\frac{P[\mathcal{C}(t)]}{P[\mathcal{C}^R(t)]}\right) \quad (\text{I.49})$$

where $\mathcal{C}^R(t)$ is obtained by reversing the order of all the events in $\mathcal{C}(t)$: all the times $t_i$ become $t_0 + t_N - t_i$, and the order of the configurations $\mathcal{C}_i$ is reversed. This is well defined only if for any allowed transition, the opposite transition is allowed as well (it won't be the case, for instance, for the totally asymmetric simple exclusion process, or TASEP, which we will introduce in the next chapter).

Considering equation (I.33) for both histories, one can see that all the exponentials appear in both terms, and therefore compensate. We are left with:

$$\boxed{S_t[\mathcal{C}(t)] = \log\left(\prod_{i=1}^{N} \frac{w(\mathcal{C}_i, \mathcal{C}_{i-1})}{w(\mathcal{C}_{i-1}, \mathcal{C}_i)}\right).} \quad (\text{I.50})$$

We recognise a special case of time additive observable, where:

$$V(\mathcal{C}) = 0 \quad , \quad U(\mathcal{C}, \mathcal{C}') = \log\big(w(\mathcal{C}', \mathcal{C})\big) - \log\big(w(\mathcal{C}, \mathcal{C}')\big) \quad (\text{I.51})$$

(we don't need to fix $U(\mathcal{C},\mathcal{C})$ to 0, as it is already verified).

The corresponding deformed Markov matrix is:

$$M_\nu = \sum_{\mathcal{C},\mathcal{C}'} w(\mathcal{C},\mathcal{C}')^{1+\nu} w(\mathcal{C}',\mathcal{C})^{-\nu} |\mathcal{C}\rangle\langle\mathcal{C}'|. \tag{I.52}$$

We notice that it has a peculiar symmetry: if we replace $\nu$ by $-1-\nu$, the exponents in $M_\nu$ are swapped, which is the same as exchanging $\mathcal{C}$ and $\mathcal{C}'$. The new matrix is therefore merely the transpose of the old one:

$$M_\nu = {}^t M_{-1-\nu}. \tag{I.53}$$

This has interesting consequences on its eigensystem: all the eigenvalues are symmetric with respect to $\nu \leftrightarrow -1-\nu$, and the associated right and left eigenvectors are exchanged. This is the famous 'Gallavotti-Cohen symmetry' [18–20].

In particular, the generating function of the cumulants of the intensive entropy production $s$, and the conditional probabilities that we have defined earlier, verify:

$$E(\nu) = E(-1-\nu) \quad , \quad P_\nu(\mathcal{C}) = \tilde{P}_{-1-\nu}(\mathcal{C}). \tag{I.54}$$

Now, the large deviation function $g(s)$ of $s = \frac{1}{t}S_t$ (the entropy production rate), as we recall, verifies

$$g(s) = \min_\nu[\nu s - E(\nu)] = \min_\nu[(-1-\nu)(-s) - E(-1-\nu)] - s = g(-s) - s \tag{I.55}$$

so that

$$g(s) - g(-s) = -s \tag{I.56}$$

or, in other words,

$$\mathrm{P}(-s) = \mathrm{e}^{-ts}\mathrm{P}(s). \tag{I.57}$$

This last equation is called the 'fluctuation theorem'. It was first observed by Evans, Cohen and Morriss in [21], then proven by Evans and Searles in [22], and later led to Gallavotti and Cohen's formulation of their symmetry. The theorem means that a negative entropy production rate is much less probable than its positive counterpart, but not impossible. This does not, as it might seem, contradict the second law of thermodynamics, which is expressed only for the expectation value of $s$. It in fact validates it, since it implies that the mean value of $s$ must be positive.

Let us now go back to the principal eigenvectors $|P_\nu\rangle$ and $\langle\tilde{P}_\nu|$ of $M_\nu$. As we said earlier, the entries of those vectors correspond, respectively, to probabilities of the final configuration, or of the initial configuration, conditioned on the observed value of $s$:

$$P_\nu(\mathcal{C}) = \mathrm{P}\left(\mathcal{C}_f = \mathcal{C} \,\Big|\, s = \frac{\mathrm{d}}{\mathrm{d}\nu}E(\nu)\right) \tag{I.58}$$

$$\tilde{P}_\nu(\mathcal{C}) = \mathrm{P}\left(\mathcal{C}_i = \mathcal{C} \,\Big|\, s = \frac{\mathrm{d}}{\mathrm{d}\nu}E(\nu)\right). \tag{I.59}$$

If we multiply the two, as we did in (I.48), we get the probability of observing a configuration, anywhere in time (but far away from the initial and final times), conditioned on $s$:

$$\boxed{P_\nu(\mathcal{C})\tilde{P}_\nu(\mathcal{C}) = \mathrm{P}\Big(\mathcal{C} \;\Big|\; s = \frac{\mathrm{d}}{\mathrm{d}\nu}E(\nu)\Big).} \tag{I.60}$$

Those quantities are very different from one another, and the most probable states according to those three distributions are in general not the same. We will be mostly interested in eq.(I.60), for reasons that will become apparent later, in section III.1.2.

The statistical ensemble defined by those probabilities, where $\nu$ is considered as a parameter (a kind of temperature, as it is conjugate to the entropy), is sometimes called the 's-ensemble' [23]. It is rather natural to consider this ensemble: since entropy production plays an important part in the system being out of equilibrium, being able to control its value and look at how the system responds may provide us with useful information of which we might be able to make sense. More on that in chapter IV.

Finally, let us examine the case of an equilibrium system. The detailed balance condition, which defines equilibrium, tells us that the steady state distribution is such that:

$$\boxed{P_{eq}(\mathcal{C})w(\mathcal{C}',\mathcal{C}) = P_{eq}(\mathcal{C}')w(\mathcal{C},\mathcal{C}')} \tag{I.61}$$

for any two configurations $\mathcal{C}$ and $\mathcal{C}'$.

Putting this in (I.52), we get:

$$M_\nu = \sum_{\mathcal{C},\mathcal{C}'} w(\mathcal{C},\mathcal{C}') \left(\frac{P_{eq}(\mathcal{C})}{P_{eq}(\mathcal{C}')}\right)^\nu |\mathcal{C}\rangle\langle\mathcal{C}'| \tag{I.62}$$

which is to say that $M_\nu$ is actually similar to $M$ through a diagonal matrix $D$, containing the equilibrium probabilities, to the power $\nu$:

$$M_\nu = D^\nu \; M \; D^{-\nu} \tag{I.63}$$

with

$$D = \sum_\mathcal{C} P_{eq}(\mathcal{C})|\mathcal{C}\rangle\langle\mathcal{C}|, \tag{I.64}$$

so that the eigenvalues $M_\nu$ and $M$ are the same. We then have:

$$\boxed{E(\nu) = 0 \quad \text{and} \quad \mathrm{P}(s) = \delta(s).} \tag{I.65}$$

There is therefore no entropy production whatsoever in the case of an equilibrium system. This, as in eq.(I.43), is a property of the transition rates, and not of the initial or final distributions. There can still be a conservative exchange of entropy between the initial and final configurations of a history (if their equilibrium probabilities are not equal), which is accounted for by the dominant eigenvectors $D^\nu|P_{eq}\rangle$ and $\langle 1|D^{-\nu}$ of $M_\nu$.

# CHAPTER II

# The Asymmetric Simple Exclusion Process : phenomenology and a few exact results

This second chapter serves to introduce the reader to the asymmetric simple exclusion process and to present a few known results which we will need later in order to construct our own. We start by defining the model, as well as a few variants, and some of the many other models and problems to which it is connected. We then take a first peek at its behaviour through a couple of simulations, and learn what we can from a mean field approach. The second part of this chapter consists of two important exact results along with the methods used to obtain them. The first is the matrix Ansatz solution for the steady state of the open ASEP, found by Derrida, Evans, Hakim and Pasquier in [24], and which we have used as a starting point to build our own matrix Ansatz for the distribution of the s-ensemble [2, 3] (see chapter III). We also present a calculation of the mean current based on this solution, which was done by Sasamoto in [25], using q-deformed algebraic relations that will appear in all our results. The second is the Bethe Ansatz solution for the full generating function of the cumulants of the current in the periodic ASEP, found by Prolhac and Mallick [26–29], and which we will need in chapter V to give the Bethe Ansatz proof of the formulae that we guess in chapter III for the open ASEP.

## II.1 Phenomenology of the open ASEP

In order to first get a bit of intuition about the ASEP rather than diving headlong into pages and pages of algebraic equations (which will be the next thing we do), let us first take a simple phenomenological look at the system's behaviour, and do a few easy mean field calculations. Of course, we will first need to give its definition. We will also do a very short review of the many uses and interests of the ASEP, and give a few references to make up for its brevity.

### II.1.1 Definition of the model and variants

Without further ado, let us define the object of all our attention: the open asymmetric simple exclusion process.

Consider a one-dimensional lattice with $L$ sites (or a row of $L$ boxes), numbered from

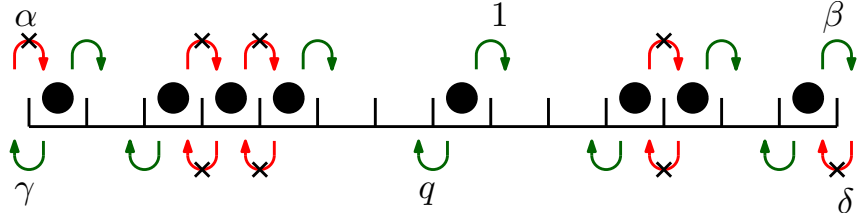

Figure II.1: Dynamical rules for the ASEP with open boundaries. The rate of forward jumps has been normalised to 1. Backward jumps occur with rate $q < 1$. All other parameters are arbitrary. The jumps shown in green are allowed by the exclusion constraint. Those shown in red and crossed out are forbidden.

1 to $L$. Each site can be empty, or carry one particle. Those particles jump stochastically from site to site, with a rate $p$ if the jump is to the right, from site $i$ to site $i+1$ (and which will be set to $p=1$ by choosing the rate of forward jumps as a time scale), and a rate $q < 1$ if the jump is to the left, from site $i$ to site $i-1$. The jumping rate is larger to the right than to the left in order to mimic the action of a field driving the particles in the bulk of the system. Each end of the system is connected to a reservoir of particles, so that they may enter the system at site 1 with rate $\alpha$ or at site $L$ with rate $\delta$, and leave it from site 1 with rate $\gamma$ or from site $L$ with rate $\beta$. Those rates allow us to define the effective densities of the two reservoirs. In all of these operations, the only constraint that must be obeyed is that of exclusion, which is to say that there cannot be more than one particle on a given site at a given time, so that a particle cannot jump to a site that is already occupied. These rules are represented schematically on fig.-II.1.

Configurations of the system are written as strings of 0's and 1's, where 0 indicates an empty site, and 1 an occupied site.

The Markov matrix governing that process is a sum of local jump operators, each carrying the rates of jumps over one of the bonds in the system:

$$M = m_0 + \sum_{i=1}^{L-1} M_i + m_L \quad \text{(II.1)}$$

with

$$m_0 = \begin{bmatrix} -\alpha & \gamma \\ \alpha & -\gamma \end{bmatrix} , \quad M_i = \begin{bmatrix} 0 & 0 & 0 & 0 \\ 0 & -q & 1 & 0 \\ 0 & q & -1 & 0 \\ 0 & 0 & 0 & 0 \end{bmatrix} , \quad m_L = \begin{bmatrix} -\delta & \beta \\ \delta & -\beta \end{bmatrix}. \quad \text{(II.2)}$$

It is implied here that $m_0$ acts as written on site 1 (and is represented in basis $\{0, 1\}$ for the occupancy of the first site), and as the identity on all the other sites. Likewise, $m_L$ acts as written on site $L$, and $M_i$ on sites $i$ and $i+1$ (and is represented in basis $\{00, 01, 10, 11\}$ for the occupancy of those two sites). Each of the non-diagonal entries represents a transition between two configurations that are one particle jump away from each other.

As we mentioned earlier, because of the asymmetry of the jumps, the particles flow to the right, and that current is an important observable, deeply connected to the non-equilibrium nature of the system. It is a special case of time-additive observable (as presented in section I.2.2) where $V = 0$ and $U$ is taken as 1 for the transitions where a particle jumps to the right, and $-1$ for those where one jumps to the left. We will see in section III.1.2 that it is strongly related to the entropy production (they are in fact equal, up to a constant), and we will be referring to probabilities conditioned on the current as the s-ensemble as well. All the results presented in this thesis are related to the characterisation of that current.

**Variants of the ASEP**

There are a few simpler cases that one can consider. The first is to force the particles to jump only to the right, by taking $q = \gamma = \delta = 0$. In this case, the model is called the totally asymmetric simple exclusion process (or TASEP), and we will often use it in our calculations, as its behaviour is identical to that of the ASEP for all intents and purposes, but much easier to deal with.

The second is the opposite limit, where the jumps are as probable to the left as they are to the right: $q = 1$. This one is called, unsurprisingly, the symmetric simple exclusion process (or SSEP). It is an example of what is called 'boundary-driven diffusive systems' (as opposed to the ASEP, which is bulk-driven by the asymmetry, and is therefore not diffusive). It behaves *very* differently from the ASEP, and we will almost never talk about it, but we will give a few references in a moment.

Somewhere in between the SSEP and the ASEP is the weakly asymmetric simple exclusion process (or WASEP), where the asymmetry $1 - q$ is taken to scale with the size of the system as $L^{-1}$. This is done in order to make the integral of the field in the bulk, which is of order $L(1-q)$, comparable with the difference of chemical potential between the reservoirs, which is a constant with respect to $L$. The ASEP and the WASEP correspond to two different ways to take the large $L$ limit in the system: in the ASEP, no rescaling is done to the driving field, so that the large size limit corresponds to a system of increasing length, with the lattice spacing remaining constant, which is relevant to model a system which is really discrete (think for instance of ribosomes on a long string of mRNA, or any other example of discrete biological transport). In the WASEP, on the contrary, the field is rescaled as $L^{-1}$, so that the large size limit corresponds to a system of fixed length, with a smaller and smaller lattice spacing, going to a continuous system when $L$ reaches infinity, perhaps describing something like waves of density in a fluid in a tube. We will be using the WASEP at one point, in section IV.2.1, because, amazingly, taking the weak asymmetry to infinity after having taken the continuous limit gives correct results for the ASEP in some regimes.

One can also consider different geometries for the model. There is for instance the ASEP with periodic boundary conditions, i.e. on a ring (fig.-II.2-b). In this case, there are no reservoirs, and the number of particles in the system is conserved. This makes it somewhat easier to deal with: the steady-state distribution is uniform, and the coordinate version of the Bethe Ansatz can be used to solve it, as we will see in section II.2.2 (but, as we will also see, this doesn't mean that solving it is in any way trivial).

The ASEP can be considered on an infinite lattice instead (c.f. lower part of fig.-II.2-d). In this case, there is in general no steady state (for generic initial conditions), and the observable of choice is instead the large time behaviour of the transient regime, which lasts forever.

Finally, one can put more than one type of particles in the system, and consider the multispecies ASEP (fig.-II.2-c). The exchange rates must then be defined between any two different species of particles. The simplest case to consider (and the most tractable one) is that where the types of particles are numbered, from 0 (for holes) to $K$ (for the 'fastest' particles), and where a particle of type $k$ sees all lower types $k' < k$ as holes, which is to say that the rates of exchange of two particles of types $k_1 < k_2$ are 1 for $k_2 k_1 \to k_1 k_2$ and $q$ for $k_1 k_2 \to k_2 k_1$ (those rates are represented on fig.-II.2-c, where different species of particles bear different colours, and are numbered by their rank).

**Brief overview of the ASEP's family tree**

We mentioned biological transport earlier for a good reason: the first definition of an ASEP-like model was made in 1968 in [30, 31] precisely in order to study the dynamics of ribosomes on mRNA (fig.-II.2-a). It is still used today in that context, usually after a few modifications to make it slightly more realistic, such as making the particle reservoirs finite [32] or even shared between several systems [33], changing the jumping rates from site to site [34], changing the jumping cycle by adding an inactive state for particles [35], allowing them to attach or detach in the middle of the chain [36], and so on. These are only a few recent examples, but a thorough review can be found in [37].

It has also been noticed that the ASEP is strongly related to the XXZ spin chain with spin $\frac{1}{2}$ [38] (the Markov matrix of the ASEP and the Hamiltonian of the spin chain are related through a simple matrix similarity). This fact goes deeper than a simple mapping between two systems: the XXZ spin chain is well known and well studied because it has the mathematical property of being 'integrable', meaning that it can be solved using the Bethe Ansatz [39], and that we can expect exact analytical results from it [40]. Thanks to this, many results have been obtained for the ASEP by adapting the Bethe Ansatz to its formalism [26–29, 41–45] (as section II.2.2 demonstrates), and even results that have been found by other means (such as those presented in II.2.1, for instance, or any of our own results) are in fact consequences of that property (as we shall see in chapter V, where we will also express our results in the formalism of the open XXZ chain). The downside of this, one could say, is that those methods are only transposable to other integrable systems, but the undeniable upside is that we might have access to very precise results, which could lead to discovering universal features of non-equilibrium systems.

Another model related to the ASEP is that of random surface growth [46] (fig.-II.2-d). In that model, a wall, made of square blocks (with corners pointing up and down), grows by a Tetris-like procedure where blocks can fall in valleys at rate 1, or lift off of peaks at rate $q$. The relation to the ASEP is rather obvious if one replaces upward slopes (when reading from left to right) by holes, and downward slopes by particles (adding a block means replacing 'down up' by 'up down', i.e. 10 by 01, and removing one is the opposite operation). In this context, the situation that is usually considered is that of the infinite ASEP, with a simple initial condition, such as a given mean density to the right of site 0 and another one to the left (the simplest one being all 1s to the left and all 0s to the right [47, 48], as represented by the dashed line in fig.-II.2-d), although more general ones can be considered [49]. One of the most interesting quantities, just as for finite size, is the total current of particles that went over the bond at the centre of the system, which is equal to the number of blocks that have been added above that site, i.e. to the height of the surface (represented in green in fig.-II.2-d; each green block corresponds to one of the particles that crossed to the right of the system). After a first breakthrough by Johansson in [47], the fluctuations of that height were conjectured [50] and then proven [51] to be related to the famous Tracy-Widom distributions governing the eigenvalues of random matrices [52, 53]. Some even more complex quantities have been studied, such as the general n-point correlations of the height [54]. Moreover, there is in a whole class of systems, the KPZ universality class (named after Kardar, Parisi and Zhang, authors of the seminal article that started it all [46]), that are governed by the same laws, such as the directed random polymer [55, 56], or the delta-Bose gas [57]. Experimental evidence of the relevance of that model has been obtained only very recently in a liquid crystal undergoing a phase transition [58]. This subject has generated many more works than could be summed up here, and the reader can find more information in reviews such as [59–62].

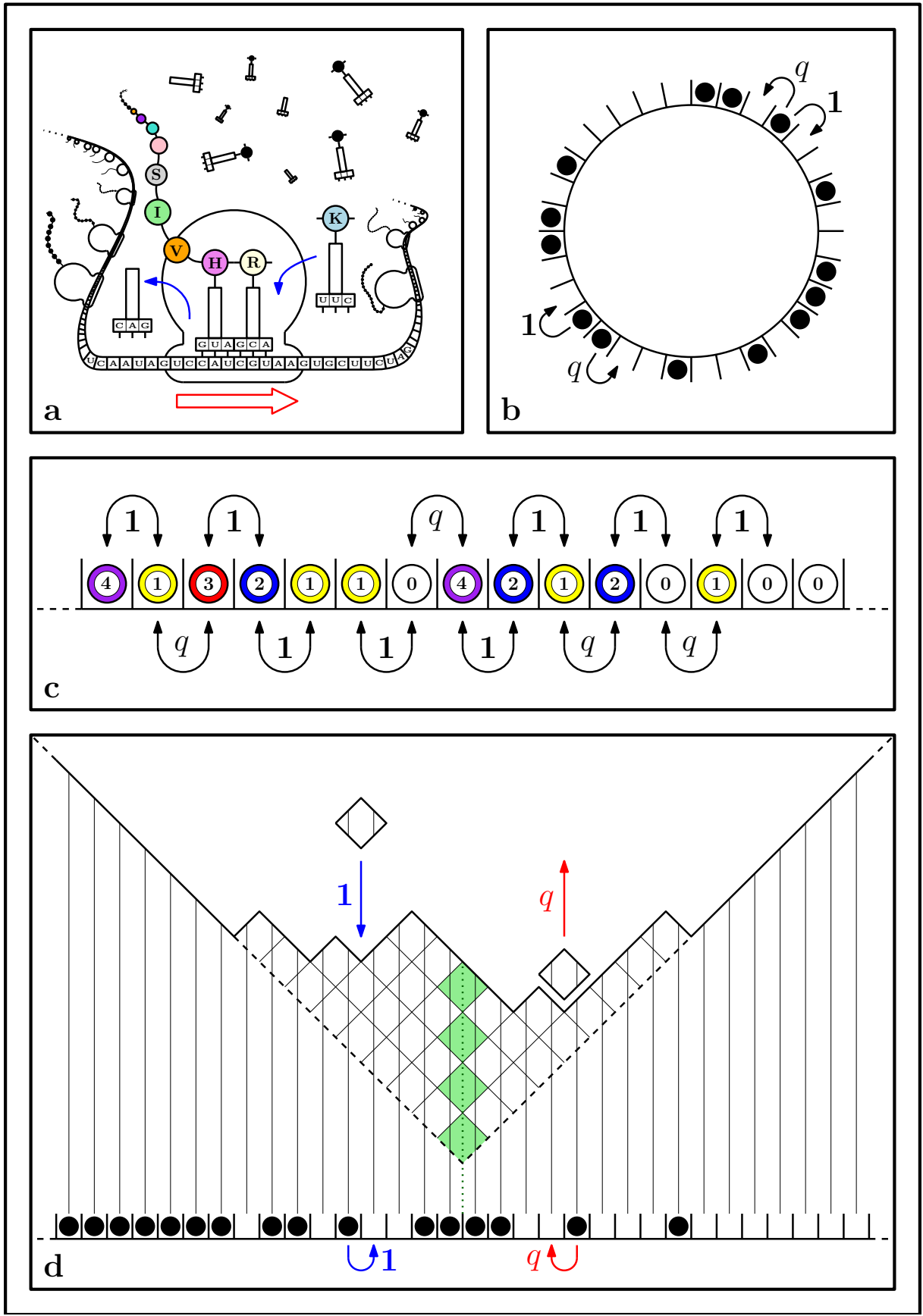

Figure II.2: The ASEP's family album: a) Ribosomes on mRNA, b) Periodic ASEP, c) Multispecies ASEP, d) Surface growth.

The ASEP can be related to many more models and mathematical objects, such as chains of quantum dots [63], alternating sign matrices [64, 65] (through its connection to the XXZ chain), continued fractions [66], Brownian excursions [67–69], Askey-Wilson polynomials [25, 70, 71], and a large family of combinatorial objects which all have a connection to Catalan numbers [72].

**Earlier results**

All these interesting connections notwithstanding, the ASEP is a very popular model in itself [73–75] (it has even been referred to as the Ising model of non-equilibrium systems [76]), and has been the subject of a tremendous number of works.

The SSEP, for one, has established itself as an archetype of diffusive systems with interactions, for which many universal results have been found, such as the cumulants of the current in a periodic system [77] or an open one [78]. Those results all have to do with the so-called 'macroscopic fluctuation theory' (or MFT) [79–82], developed to deal with the fluctuations of diffusive systems through a hydrodynamic approach [83]. As for results more specific to the SSEP or the WASEP, the large deviation functional of the density profiles was expressed in [84], leading to the joint large deviations functional for the current and the density [85] which we will be using in section IV.2.1. The cumulants of the current for the open SSEP were found in [86], and were observed to depend on a single variable and not on the two boundary densities independently. This lead to the discovery of a surprising symmetry connecting the non-equilibrium SSEP (with different reservoir densities) to a system at equilibrium [87–89]. The full cumulants of the current for the periodic WASEP were found in [90]. In that case, as was found in [91] and further analysed in [92], the system undergoes a phase transition in the s-ensemble where, for a low enough current, the optimal density profiles become time-dependent (we will come back to this in section IV.4.7. A similar transition can be found for the activity (the sum of jumps, regardless of direction) in the SSEP [93]. See [75] for a review of some of these results.

The periodic ASEP, with its fixed number of particles and its trivial steady state (all the configurations are equally probable, as long as they have the correct number of particles), has mostly been studied for the fluctuations of the current. The full generating function of those was found for the TASEP in 1998 [94, 95], and although the second cumulant for the ASEP was found prior to that [96], the complete generating function was only obtained more than 10 years later [26–29]. Some other results were obtained for the periodic TASEP, such as the gap (i.e. the characteristic time of the transient regime) [97, 98], and, very recently, the whole distribution of the spectrum of the Markov matrix [99]. The s-ensemble was also investigated, for the limit of very large currents, and the probabilities of the configurations were found to be those of a Dyson-Gaudin gas (the discrete analogue of a Coulomb gas) [100]. We will come back to that last observation in section IV.3.3.

The open ASEP is richer than the periodic case, but much harder to deal with. The structure of the steady state itself is quite intricate: it was first found in [101] for the TASEP thanks to some surprising recurrence relations between the weights of the configurations for successive sizes. It was then generalised to the ASEP by expressing those relations in algebraic form [24], giving birth to the 'matrix Ansatz', which we will present in section II.2.1. Depending on the values of the two reservoir densities, the system can find itself in three different phases, which was discovered for the TASEP in [102] as an interesting feature of non-equilibrium systems (since, for equilibrium systems with short range interactions, transitions cannot be induced by boundaries). This phase diagram was refined in [103] where sub-phases were found with different correlation lengths. Those results were extended to the ASEP in [25, 104] (part of which we present in section II.2.1). The 2-point correlation function [105] and then the complete n-point function [67] were calculated for the TASEP, for some values of the boundary densities, and the same was later done for the ASEP in [71, 106]. Most of these results rely on the matrix Ansatz, and a review of those results and methods can be found in [107]. See also [74] for a review of various results for the steady state of the open ASEP.

Other properties of the steady state were analysed, such as the static density, current and activity distributions [108, 109], the large deviation function of the density profiles [110], or the reverse bias regime (where the boundaries impose a current floving to the left) [111, 112]. A hydrodynamic description, named 'domain wall theory' (or DWT) [83, 113–116] where states of the system are approximated by regions of constant density separated by discontinuities called shocks, was proposed to describe the large scale dynamics of the system, even in the transient regime, but no full equivalent of the MFT has yet been devised.

One of the main reasons why the open ASEP is more difficult to study than its circular sibling is that the Bethe Ansatz cannot be used as easily in this case. The coordinate version of the Ansatz (where the particles are treated as plane waves, as presented in section II.2.2) relies on the number of particles being fixed, and breaks down in the open case. Variants of the coordinate Bethe Ansatz were used successfully to build excited eigenstates of the system for some special cases of the boundary parameters [41, 42, 45], and, in conjunction with numerical analysis, to find the relaxation speed of the system (i.e. the gap of the Markov matrix) [43, 44, 117], as well as the asymptotic large deviation function of the current inside of the Gaussian phases [118] (see section IV.2.1), but it was not known whether it could be applied to the open ASEP in general. In chapter V, we show how to construct the functional Bethe Ansatz, which allows, in principle, to access the whole spectrum of the Markov matrix (deformed with respect to the current). Using that, we can find the complete generating function of the cumulants of the current, of which only the second was known, for the TASEP [119]. The eigenvectors, however, are still out of reach as far as we know.

Many variants of the ASEP have also been studied. A matrix Ansatz, akin to the one we mentioned before, was found for the steady state of the periodic multispecies ASEP [120, 121]. The case of a single defect particle was analysed, for itself [122–124] or used as a way to mark the position of a shock [125, 126]. Different updates procedures were considered and compared for the discrete time case [127, 128]. A system with two interacting chains was studied in [129]. The ASEP was also considered with entry and exit of particles in the bulk of the system [130], disordered [131] or smoothly varying [132] jumping rates, a single slow bond [133,134], repulsive nearest-neighbour interactions [135], or on a two-dimensional grid [136].

Finally, on the numerical front, the ASEP has been used to develop and test numerical algorithms aimed at producing and analysing rare events, such as a variant of the 'density matrix renormalisation group' (DMRG) algorithm [137, 138], numerical implementations of the MFT [139–141], and the so-called 'cloning algorithm' [142–145].

## II.1.2 Simulation of the system and observations

Now that we have defined our system, and before doing any calculation, let us simply watch it evolve through a few simulations.

To make things simpler, we consider the TASEP, so that we only have two parameters $\alpha$ and $\beta$ that we can vary (we will keep the size $L$ of the system fixed). We may also notice that the system is symmetric under the exchange of particles with holes, of $\alpha$ with $\beta$ and of left with right: the holes enter the system at the right boundary with rate $\beta$, jump to the left with rate 1, and exit at the left boundary with rate $\alpha$. This means that we can consider only cases where $\alpha \leq \beta$, and deduce the rest through that symmetry.

On the next page are represented three examples that are characteristic of the various behaviours that the system may show. In each case, the system size is taken to be $L = 200$, and the initial condition is a step profile centred around the 100th site, with a mean density $\alpha$ to the left and $1 - \beta$ to the right (since the system goes, after a certain time, to a steady state that does not depend on the initial condition, we are free to choose that which suits best to what we want to observe). For each situation, we represent, on the left, the time evolution of a single realisation of the system, with respect to the site $i$ and the time $t$ (the grey scale gives the local mean density, averaged over a few time steps); at the centre we give that same time evolution averaged over a large number of realisations (the colour scale gives the local mean density); finally, at the right, we show a plot of the final density profile, which is close to the density in the steady state of the system. Note that the time scale for the temporal plots was chosen arbitrarily, but is the same for all three examples.

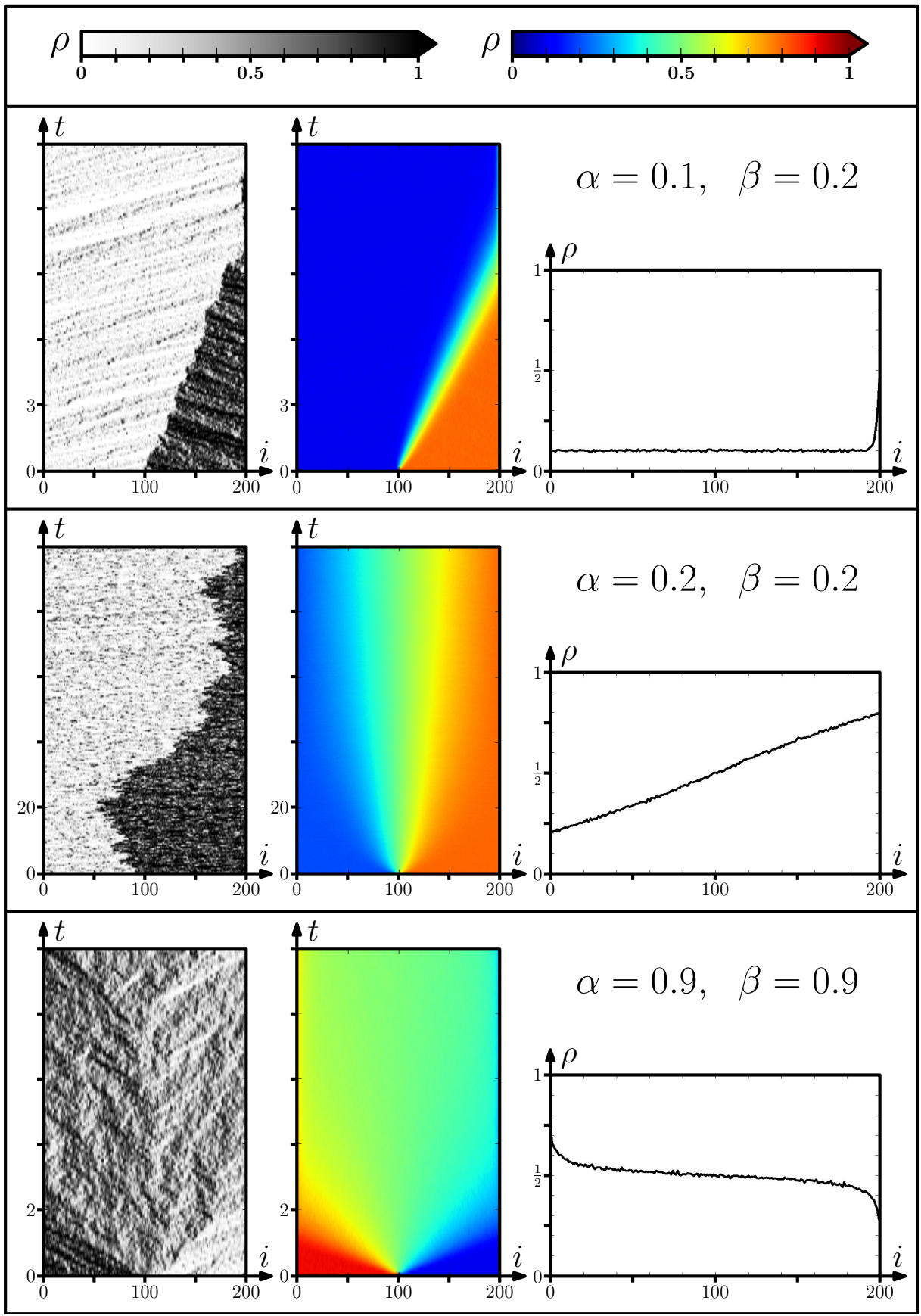

What we observe is this: in the first case, $\alpha$ is very low, and $\beta$ is slightly higher, which is to say that more holes enter the system from the right than particles from the left. That can be seen on the first plot, where the bunch of particles at the right of the system decreases in size every time a hole is added, and grows when a particle is added. We clearly see that with time, it tends to shrink at a fairly constant rate, which is to say that the system empties itself into the right reservoir. The steady state has a low density almost everywhere, except very close to the right boundary, where the low exit rate creates a blockage. We notice that the value of that density seems to be $\alpha$. The symmetric situation (with $\beta$ lower than $\alpha$) would have produced a profile with a high density (close to $1 - \beta$) everywhere except near the left boundary.

In the second case, $\alpha$ is still very low, but now equal to $\beta$. The entry rate of holes is equal to that of particles, so that the boundary between the low and high density zones in the system now performs a random walk (to the left each time a particle is added, and to the right each time one is removed), which is not biased any more, as is visible on the leftmost plot. Depending on the actual value of $\alpha = \beta$, that boundary is more or less localised: if that value is very small, the jumps in the bulk happen so much faster than the addition or removal of particles that we get a very clear domain wall between a region with a density close to 0 and one close to 1. After a long time, that domain wall will have randomly walked everywhere in the system, meaning that the steady state is a superposition of step profiles with the same probability for the jump to be at every site, so that the averaged mean density is actually linear with respect to the position (as seen on the plot at the right). On the middle plot, the parabolic behaviour of the iso-density lines (which is due to the diffusion of the domain wall) is clearly visible.

In the third case, the situation is entirely opposite: $\alpha$ and $\beta$ are both very high (and are equal, but this time it makes no difference). We see that the initial downward step profile rapidly fans out and converges to a profile where the density is close to $\frac{1}{2}$ everywhere, but with large fluctuations everywhere, and rather wide boundary zones where it becomes higher (at the left) or lower (at the right). Little can be said on this case from the simulation alone, other than the fact that a density of $\frac{1}{2}$ seems optimal for the system to carry a large current: were it lower, there would be fewer particles, so less current as well, but a higher density would cause the particles to jam more and move more slowly.

For other values of $\alpha$ or $\beta$, we would obtain one of those three behaviours: the first one if $\alpha$ is small, and smaller than $\beta$ (and the symmetric case if $\beta$ is smaller than $\alpha$); the second one if $\alpha$ is small, and equal to $\beta$; and the third one if $\alpha$ and $\beta$ are both large enough.

We are grateful to J. Tailleur for providing us with the numerical data for this section.

### II.1.3 Mean-field calculations

Let us now clarify these observations through a few simple mean field calculations. These were first carried out in [102] and [101], and will provide us with the phase diagram of the system.

As we recall, the master equation reads:

$$\frac{d}{dt}|P_t\rangle = M|P_t\rangle \tag{II.3}$$

where $M$ is the Markov matrix of the open ASEP:

$$M = m_0 + \sum_{i=1}^{L-1} M_i + m_L \tag{II.4}$$

with

$$m_0 = \begin{bmatrix} -\alpha & \gamma \\ \alpha & -\gamma \end{bmatrix}, \ M_i = \begin{bmatrix} 0 & 0 & 0 & 0 \\ 0 & -q & 1 & 0 \\ 0 & q & -1 & 0 \\ 0 & 0 & 0 & 0 \end{bmatrix}, \ m_L = \begin{bmatrix} -\delta & \beta \\ \delta & -\beta \end{bmatrix}. \tag{II.5}$$

We shall write the configurations of the system as $\mathcal{C} = \{n_i\}_{i:1..L}$, where $n_i \in \{0,1\}$ is the occupancy of site $i$. If we trace equation (II.3) over all $n_j$'s except for one at site $i$ which is taken to be 1 (which means projecting it onto $\langle 1 | \delta_{n_i,1}\rangle$), we get an equation for the time evolution of the mean density at that site:

$$\frac{d}{dt}\langle n_i \rangle = \langle 1 | \delta_{n_i,1} M | P_t \rangle. \tag{II.6}$$

For any site $i$, only matrices $M_{i-1}$ and $M_i$ from (II.5) contribute to that equation (all the others are not affected by $\delta_{n_i,1}$ and disappear under the action of $\langle 1 |$, since their columns sum to 0). We get, for each site:

$$\frac{d}{dt}\langle n_1 \rangle = \alpha \langle (1 - n_1) \rangle - \gamma \langle n_1 \rangle$$
$$- \langle n_1(1 - n_2) \rangle + q\langle n_2(1 - n_1) \rangle, \tag{II.7}$$

$$\frac{d}{dt}\langle n_i \rangle = \langle n_{i-1}(1 - n_i) \rangle - q\langle (1 - n_{i-1})n_i \rangle$$
$$- \langle n_i(1 - n_{i+1}) \rangle + q\langle n_{i+1}(1 - n_i) \rangle, \tag{II.8}$$

$$\frac{d}{dt}\langle n_L \rangle = \langle n_{L-1}(1 - n_L) \rangle - q\langle (1 - n_{L-1})n_L \rangle$$
$$- \beta \langle n_L \rangle + \delta \langle (1 - n_L) \rangle \tag{II.9}$$

which we can rewrite as

$$\boxed{\frac{d}{dt}\langle n_i \rangle = J_{i-1} - J_i} \tag{II.10}$$

with

$$J_0 = \alpha \langle (1 - n_1) \rangle - \gamma \langle n_1 \rangle, \tag{II.11}$$
$$J_i = \langle n_{i-1}(1 - n_i) \rangle - q\langle (1 - n_{i-1})n_i \rangle, \tag{II.12}$$
$$J_L = \beta \langle n_L \rangle - \delta \langle (1 - n_L) \rangle. \tag{II.13}$$

We recognise each $J_i$ as the mean current flowing from site $i$ to site $i+1$. Equation (II.10) is simply the continuity equation for the mean density at site $i$: its variation is equal to the current coming from the left minus the current leaving to the right. The steady state is then obtained by taking the time derivative equal to 0 in (II.10), meaning that all the $J_i$'s have to be equal, which is to say that the current must be conserved throughout the system.

At this point, we have made no approximation yet. However, since the currents depend on the local 2-point correlations $\langle n_{i-1} n_i \rangle$, (II.10) is not a closed system of equations on the $\langle n_i \rangle$s. To solve it from there, we would have to write equations on the 2-point correlations, which would involve the 3-point correlations, and so on. This is where we need to make the mean field approximation: $\langle n_{i-1} n_i \rangle \sim \langle n_{i-1} \rangle \langle n_i \rangle$. By writing $\rho_i = \langle n_i \rangle$ for simplicity, we get:

$$J = \alpha(1 - \rho_1) - \gamma \rho_1 \tag{II.14}$$
$$= \rho_{i-1}(1 - \rho_i) - q \rho_i (1 - \rho_{i-1}) \tag{II.15}$$
$$= \beta \rho_L - \delta(1 - \rho_L). \tag{II.16}$$

The second of these equations gives a recursion relation between $\rho_i$ and $\rho_{i-1}$, which can then be used $L-1$ times to express $\rho_L$ as a function of $\rho_1$. Then, the first and last equations can be used to get a single equation on $J$, which fixes its value as a function of all the parameters of the system ($L$, $q$ and the four boundary rates). These calculations can be found in [101] for the TASEP. We are only interested in the large size limit, so we will use a method similar to that of [102].

Before going further, let us make one remark: from what we saw on the simulations in the previous section, we expect that, at least for some values of the parameters, the density be constant in a large portion of the system, possibly all the way up to one of the boundaries. If this is the case, at the left boundary, for instance, and we write that density $\rho_a$, then from eq.(II.15) inside of that constant density domain, we get $J = (1-q)\rho_a(1-\rho_a)$, and eq.(II.14) becomes:

$$\alpha(1 - \rho_a) - \gamma \rho_a = (1-q)\rho_a(1 - \rho_a). \tag{II.17}$$

Writing $\rho_a = \frac{1}{1+a}$, for convenience in future calculations, we find:

$$\boxed{a = \frac{1}{2\alpha}\Big[(1 - q - \alpha + \gamma) + \sqrt{(1 - q - \alpha + \gamma)^2 + 4\alpha\gamma}\Big].} \tag{II.18}$$

Doing the same at the left boundary, we get a density $\rho_b = \frac{b}{1+b}$ with:

$$\boxed{b = \frac{1}{2\beta}\Big[(1 - q - \beta + \delta) + \sqrt{(1 - q - \beta + \delta)^2 + 4\beta\delta}\Big].} \tag{II.19}$$

Those two densities $\rho_a$ and $\rho_b$ both depend only on the parameters of their respective boundary, and can be considered as the effective densities of the reservoirs to which the system is connected.

Let us now take $L$ to infinity. We want to take a continuous limit for $\rho$, which will be easier to deal with, and useful for later. Let us therefore write:

$$\rho_i = \rho(x_i) \quad \text{with} \quad x_i = \frac{i - 1/2}{L}. \tag{II.20}$$

We then expand eq.(II.15) around $x = \frac{i-1}{L}$ (which is halfway between $x_{i-1}$ and $x_i$), using $\rho_{i-1} \sim \rho(x) - \frac{1}{2L}\nabla\rho(x)$ and $\rho_i \sim \rho(x) + \frac{1}{2L}\nabla\rho(x)$, up to order $L^{-1}$, obtaining:

$$\boxed{J = (1-q)\rho(1-\rho) - \frac{1+q}{2L}\nabla\rho} \tag{II.21}$$

We will assume that the correct boundary conditions to take on $\rho$ are $\rho(0) = \rho_a$ and $\rho(1) = \rho_b$.

Looking at equation (II.21), we see that the sign of $\nabla\rho$ depends on the difference between $J$ and $(1-q)\rho(1-\rho)$. We can first argue that $J$ cannot be larger than $\frac{1-q}{4}$, which is the maximal value taken by $(1-q)\rho(1-\rho)$ (or, for that matter, smaller than 0), otherwise $|\nabla\rho|$ would be larger than some constant of order $L$, and $\rho$ would diverge. This means that there is a density $\rho_c \leq 1/2$ such that $J = (1-q)\rho_c(1-\rho_c)$. We then have (fig.-II.3):

$$\nabla\rho < 0 \quad \text{for} \quad \rho < \rho_c, \tag{II.22}$$
$$\nabla\rho > 0 \quad \text{for} \quad \rho_c < \rho < (1-\rho_c), \tag{II.23}$$
$$\nabla\rho < 0 \quad \text{for} \quad (1-\rho_c) < \rho, \tag{II.24}$$

which is to say that $\rho$ gets away from $\rho_c$ and closer to $(1-\rho_c)$.

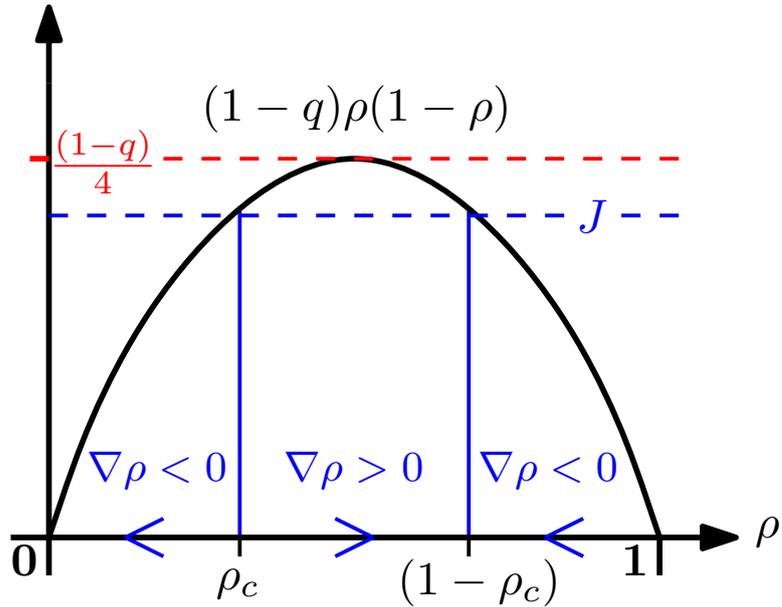

Figure II.3: Variations of $\rho$ depending on its position with respect to $\rho_c$ and $(1-\rho_c)$. For $x$ increasing, $(1-\rho_c)$ is an attractive fixed point, and $\rho_c$ is repulsive.

Considering these relations, and the boundary conditions $\rho_a$ and $\rho_b$, we can now easily determine $J$.

First of all, if $\rho_a > \frac{1}{2}$ and $\rho_b < \frac{1}{2}$, and since $\rho$ cannot decrease inside of the region $[\rho_c, 1-\rho_c]$, we must have $\rho_c = \frac{1}{2}$ (which reduces that region to a single point). We therefore have $J = \frac{1-q}{4}$ throughout this phase, which is called the 'maximal current phase' (MC), due to the fact that $J$ takes its highest possible value. In that phase, the density profile is constant and equal to $\rho_c = \frac{1}{2}$ except near the boundaries.

For the other cases, we note that $\nabla\rho$ is of order $L$ for a finite distance between $\rho$ and either $\rho_c$ or $1-\rho_c$, which means that the variations of $\rho$, when it isn't constant at one of these two values, are extremely steep. It is therefore equal to either one of these values for the most part. Let's say it is equal to $\rho_c$ at some point. From that point, going to lower values of $x$, the profile can never leave $\rho_c$, so that we must have $\rho_a = \rho_c$. As for $\rho_b$, it cannot be higher than $1 - \rho_c$. Conversely, if $\rho = 1 - \rho_c$ at some point, then the profile cannot leave $1 - \rho_c$ for larger $x$'s, and we must have $\rho_b = 1 - \rho_c$, with $\rho_a$ no lower than $\rho_c$. All those possible profiles are summarised on fig.-II.4.

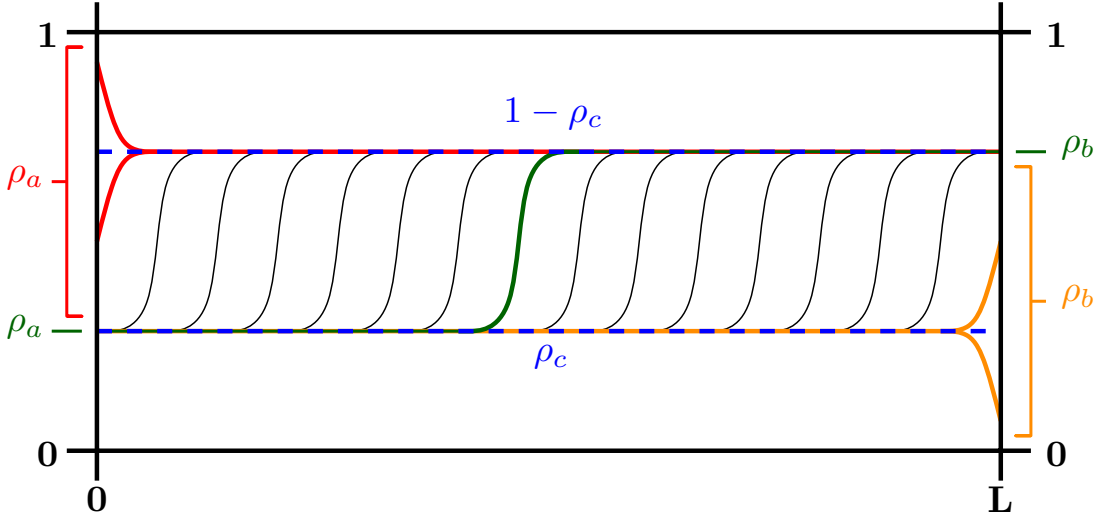

Figure II.4: Possible density profiles for a given $\rho_c$. All the profiles with $\rho_a$ in the red region converge to $\rho_b = 1 - \rho_c$. All the profiles with $\rho_b$ in the orange region come from $\rho_a = \rho_c$.

This allows us to identify two more phases. When $\rho_a < 1 - \rho_b$ and $\rho_a < \frac{1}{2}$, we have $J = (1-q)\rho_a(1-\rho_a)$ and $\rho = \rho_a$ except near the right boundary; this is called the 'low density phase' (LD). When $1 - \rho_b < \rho_a$ and $1 - \rho_b < \frac{1}{2}$, we have $J = (1-q)\rho_b(1-\rho_b)$ and $\rho = \rho_b$ except near the left boundary; this is called the 'high density phase' (HD), and is identical to the LD phase through a left↔right and particle↔hole symmetry. There is also the special case where $\rho_a = 1 - \rho_b$ and $\rho_a < \frac{1}{2}$, in which the profile goes from $\rho_c$ to $1 - \rho_c$ very fast around some $x$ which can be anywhere in the system (and the mean density, being the sum of all these possible profiles, is linear between $\rho_c$ and $1 - \rho_c$). That type of rapid transition between two constant regions being called a shock, this region is called the 'shock line' (SL).

We can finally draw the phase diagram of the system (fig.-II.5). The transitions between the MC phase and the HD and LD phases are continuous in both the current and the density profiles. The transition over the SL, however, is discontinuous in the profiles (the mean density goes from $\rho_c$ to $1 - \rho_c$), but still continuous in the current.

The only thing that remains to be seen is how the profiles decay near the boundaries. To do that, let's write $\rho = \frac{1}{2} + u$ and $\rho_c = \frac{1}{2} + u_c$. Equation (II.21) becomes:

$$\nabla u = 2L\frac{1-q}{1+q}\left(u_c^2 - u^2\right). \tag{II.25}$$

If $u_c = 0$ (i.e. if we're in the MC phase), we get $\frac{1}{u} = 2L\frac{1-q}{1+q}x + K$ (where $K$ is

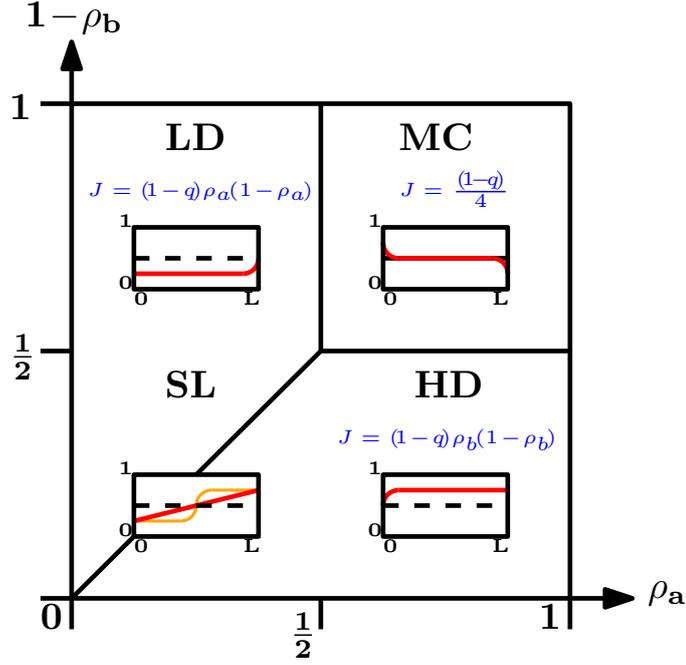

Figure II.5: Phase diagram of the open ASEP. The values of the mean current in each phase are given in blue, and the mean density profiles are represented in the insets.

an integration constant), so that $u \sim \frac{1}{L(x-x_0)}$, and the decay is algebraic around both boundaries.

If $u_c \neq 0$, we get $\frac{1}{2u_c} \log\left(\frac{u_c+u}{u_c-u}\right) = 2L\frac{1-q}{1+q}x + K$, so that $u \sim \pm u_c + A\mathrm{e}^{x/x_0}$, and the decay is exponential around one of the boundaries, or around the shock, with a characteristic length $x_0 \sim \frac{1}{L}$ (which gives the width of the boundary layers or of the shock).

## II.2   A few exact results for the ASEP

In this section, we present two important exact results for the ASEP, which we will need later in presenting our own. To be consistent with the following chapters, we will do so using notations that might differ from those of the articles in which they originally appeared.

### II.2.1   Matrix Ansatz for the open ASEP - steady state and mean current

The first of these important results is the matrix Ansatz solution for the steady state of the open ASEP, devised by Derrida, Evans, Hakim and Pasquier in [24], using the recursion relations found by Derrida, Domany and Mukamel in [101] for the TASEP.

Let us first recall the Markov matrix of the open ASEP:

$$M = m_0 + \sum_{i=1}^{L-1} M_i + m_L \tag{II.26}$$

with

$$m_0 = \begin{bmatrix} -\alpha & \gamma \\ \alpha & -\gamma \end{bmatrix} \ , \ M_i = \begin{bmatrix} 0 & 0 & 0 & 0 \\ 0 & -q & 1 & 0 \\ 0 & q & -1 & 0 \\ 0 & 0 & 0 & 0 \end{bmatrix} \ , \ m_L = \begin{bmatrix} -\delta & \beta \\ \delta & -\beta \end{bmatrix}. \quad \text{(II.27)}$$

The steady state $|P^\star\rangle$ is the vector that verifies $M|P^\star\rangle = 0$.

It was noticed in [101] that for the open TASEP, the unnormalised weights of configurations for systems of successive sizes obey a certain recursion relation. The simplest case to consider is $\alpha = \beta = 1$. In that case, if one considers the weights $p^\star_L(\mathcal{C})$ of the steady state for a system of size $L$, normalised so that the smallest weight, for a given size, be 1, then those weights obey:

$$p^\star_L(A10B) = p^\star_{L-1}(A1B) + p^\star_{L-1}(A0B) \quad \text{(II.28)}$$

where $A$ and $B$ are any strings of 0s and 1s of total size $L - 2$. The configurations to which that relation cannot be applied, which are of the form $(0^k 1^{L-k})$, all have a weight of 1. Using those two observations, one can get the probability of any configuration in less than $2^{L-1}$ operations.

In order to make that result more efficient, it was recast in algebraic form in [24], where it was realised that if two matrices $D$ and $E$ could be found, such that $DE = D + E$, then a product of $L$ of those matrices would verify the same recursion relations as the $p^\star_L$'s, where $E$ would play the role of 0 and $D$ that of 1:

$$ADEB = ADB + AEB \quad \text{(II.29)}$$

where $A$ and $B$ are any products of $D$s and $E$s. We can now turn those matrix products into scalars by projecting them between two well-chosen vectors $\langle\!\langle W\|$ and $\|V\rangle\!\rangle$ (where we use the double bracket notation for vectors from the space on which $D$ and $E$ act, which we shall call the 'auxiliary space'), such that $\langle\!\langle W\|E^k D^{L-k}\|V\rangle\!\rangle = 1$ for any $k$ and $L$.

Following this, the un-normalised probabilities, for a system with six sites, of configurations 101101, 110001 and 011010, are given by, respectively, $\langle\!\langle W\|DEDDED\|V\rangle\!\rangle$, $\langle\!\langle W\|DDEEED\|V\rangle\!\rangle$ and $\langle\!\langle W\|EDDEDE\|V\rangle\!\rangle$.

In general, the probability of any configuration $\mathcal{C} = \{n_i\}_{i:1..L}$ takes the form:

$$\boxed{P^\star(\mathcal{C}) = \frac{1}{Z_L} \langle\!\langle W\| \prod_{i=1}^{L} (n_i D + (1 - n_i) E) \|V\rangle\!\rangle} \quad \text{(II.30)}$$

with the normalisation factor equal to

$$Z_L = \langle\!\langle W\|(D+E)^L\|V\rangle\!\rangle \quad \text{(II.31)}$$

so that $\sum_{\mathcal{C}} P^\star(\mathcal{C}) = 1$.

The algebraic relations between matrices $D$ and $E$ can be generalised to the ASEP, and the relations that the boundary vectors have to verify can be obtained as well. They

are:

$$DE - q\,ED = (1-q)(D+E), \quad (\text{II.32})$$

$$\langle\!\langle W \| (\alpha E - \gamma D) = (1-q)\langle\!\langle W \|, \quad (\text{II.33})$$

$$(\beta D - \delta E) \| V \rangle\!\rangle = (1-q) \| V \rangle\!\rangle. \quad (\text{II.34})$$

We will now give a simple proof of this Ansatz.

**Proof of the matrix Ansatz**

Let us consider the two-dimensional vectors $X$ and $\hat{X}$ as:

$$X = \begin{bmatrix} E \\ D \end{bmatrix} \quad , \quad \hat{X} = (1-q)\begin{bmatrix} 1 \\ -1 \end{bmatrix} \quad (\text{II.35})$$

where the indices of those vectors correspond to occupancies of 0 and 1 (they are vectors in the physical space, the entries of which are matrices in the auxiliary space). Note that if one chooses a different normalisation for $D$ and $E$, such as dividing them by $(1-q)$ for instance (which is relevant when taking the $q \to 1$ limit), equations (II.32) to (II.34), as well as the expression of $\hat{X}$, must be modified accordingly, as in [107].

This allows us to rewrite eq.(II.30) as:

$$|P^\star\rangle = \frac{1}{Z_L} \langle\!\langle W \| \prod_{i=1}^{L} X^{(i)} \| V \rangle\!\rangle \quad (\text{II.36})$$

where the superscript $(i)$ serves to mark the site to which each $X$ corresponds, and the product between the $X$s is seen as a tensor product in the physical space, and as a matrix product in the auxiliary space. In all that follows, we will write products from the perspective of the auxiliary space: $DE$, for instance, is a matrix product in the auxiliary space, although it is a tensor product in configuration space, since it corresponds to two sites. Applying $M$ to $|P^\star\rangle$, on the other hand, is matrix product in configuration space, but a tensor product in the auxiliary space, and so, whenever both products appear in a calculation, we will note the product in configuration space using $\cdot$, as in eq.(II.37).

Let us now see how each of the individual $M_i$'s act on this object. Let us start with the bulk matrices. Since the action of $M_i$ involves only sites $i$ and $i+1$, we only have to consider a small part $X^{(i)} X^{(i+1)}$ of the whole product:

$$M_i \cdot X^{(i)} X^{(i+1)} = \begin{bmatrix} 0 & 0 & 0 & 0 \\ 0 & -q & 1 & 0 \\ 0 & q & -1 & 0 \\ 0 & 0 & 0 & 0 \end{bmatrix} \cdot \begin{bmatrix} EE \\ ED \\ DE \\ DD \end{bmatrix} = \begin{bmatrix} 0 \\ DE - q\,ED \\ q\,ED - DE \\ 0 \end{bmatrix} = (1-q)\begin{bmatrix} 0 \\ D+E \\ -D-E \\ 0 \end{bmatrix}$$

$$= \boxed{\hat{X}^{(i)} X^{(i+1)} - X^{(i)} \hat{X}^{(i+1)}} \quad (\text{II.37})$$

where we used eq.(II.32) to get the third equality. Writing this in terms of $X$ and $\hat{X}$ makes it easy to take the sum of this equation over $i$ (excluding the boundary matrices): each $\hat{X}^{(i)}$ with $i$ from 2 to $L-1$ appears twice, once from the action of $M_{i-1}$ and once from that of $M_i$, but with opposite signs, so that they cancel out. The only terms that remain are one with $\hat{X}^{(1)}$, and one with $-\hat{X}^{(L)}$, so that:

$$\sum_{i=1}^{L-1} M_i |P^\star\rangle = \frac{1}{Z_L} \langle\!\langle W \| \hat{X}^{(1)} \prod_{i=2}^{L} X^{(i)} \| V \rangle\!\rangle - \frac{1}{Z_L} \langle\!\langle W \| \prod_{i=1}^{L-1} X^{(i)} \hat{X}^{(L)} \| V \rangle\!\rangle. \tag{II.38}$$

We now only need to show that each of those two terms are taken care of by one of the boundary matrices. First the left boundary:

$$m_0 \cdot \langle\!\langle W \| X^{(1)} = \begin{bmatrix} -\alpha & \gamma \\ \alpha & -\gamma \end{bmatrix} \cdot \begin{bmatrix} \langle\!\langle W \| E \\ \langle\!\langle W \| D \end{bmatrix} = \begin{bmatrix} \langle\!\langle W \| (\gamma D - \alpha E) \\ \langle\!\langle W \| (\alpha E - \gamma D) \end{bmatrix} = (1-q) \begin{bmatrix} -\langle\!\langle W \| \\ \langle\!\langle W \| \end{bmatrix}$$

$$= \boxed{-\langle\!\langle W \| \hat{X}^{(1)}} \tag{II.39}$$

where we used eq.(II.33) to get the third equality. This cancels the first term in (II.38). Now the right boundary:

$$m_L \cdot X^{(L)} \| V \rangle\!\rangle = \begin{bmatrix} -\delta & \beta \\ \delta & -\beta \end{bmatrix} \cdot \begin{bmatrix} E \| V \rangle\!\rangle \\ D \| V \rangle\!\rangle \end{bmatrix} = \begin{bmatrix} (\beta D - \delta E) \| V \rangle\!\rangle \\ (\delta E - \beta D) \| V \rangle\!\rangle \end{bmatrix} = (1-q) \begin{bmatrix} \| V \rangle\!\rangle \\ -\| V \rangle\!\rangle \end{bmatrix}$$

$$= \boxed{\hat{X}^{(L)} \| V \rangle\!\rangle} \tag{II.40}$$

where we used eq.(II.34). This cancels the second term in (II.38), and concludes the proof that $M | P^\star \rangle = 0$.

**Average current**

We can put this Ansatz to good use right away by calculating the average current in the steady state of the ASEP. This calculation was done by Sasamoto in [25], while the simpler equivalent for the TASEP can be found in [24].

The average current flowing between sites $i$ and $i+1$ is the expectation value of the observable $\delta_{n_i,1}\delta_{n_{i+1},0} - q\, \delta_{n_i,0}\delta_{n_{i+1},1}$, which is equal to 1 if $n_i = 1$ and $n_{i+1} = 0$ (i.e. if a forward jump is possible), and to $-q$ if $n_i = 0$ and $n_{i+1} = 1$ (i.e. if a backward jump is possible).

We then have:

$$\langle J_i \rangle = \langle 1 | \delta_{n_i,1}\delta_{n_{i+1},0} - q\, \delta_{n_i,0}\delta_{n_{i+1},1} | P^\star \rangle \tag{II.41}$$

$$= \frac{1}{Z_L} \langle\!\langle W \| (D+E)^i (DE - q\, ED)(D+E)^{L-i-2} \| V \rangle\!\rangle \tag{II.42}$$

$$= \frac{1}{Z_L} \langle\!\langle W \| (D+E)^i (1-q)(D+E)(D+E)^{L-i-2} \| V \rangle\!\rangle, \tag{II.43}$$

$$J = (1-q) \frac{1}{Z_L} \langle\!\langle W \| (D+E)^{L-1} \| V \rangle\!\rangle \tag{II.44}$$

where we get from line 2 to line 3 using eq.(II.32). We recognise $Z_{L-1}$ in this last equation, so that we have:

$$\boxed{J = (1-q) \frac{Z_{L-1}}{Z_L}} \tag{II.45}$$

and we are left having only to calculate $Z_L$.

In order to do this, we need an explicit representation for $D$, $E$, and the boundary vectors. The simplest way to get one is to define two more matrices $d$ and $e$ by:

$$d = \sum_{n=1}^{\infty}(1-q^n)\|n-1\rangle\!\rangle\langle\!\langle n\| = \begin{bmatrix} 0 & (1-q) & 0 & 0 & \cdots \\ 0 & 0 & (1-q^2) & 0 & \\ 0 & 0 & 0 & (1-q^3) & \\ 0 & 0 & 0 & 0 & \\ \vdots & & & & \ddots \end{bmatrix} \quad (\text{II.46})$$

and

$$e = \sum_{n=0}^{\infty}\|n+1\rangle\!\rangle\langle\!\langle n\| = \begin{bmatrix} 0 & 0 & 0 & 0 & \cdots \\ 1 & 0 & 0 & 0 \\ 0 & 1 & 0 & 0 \\ 0 & 0 & 1 & 0 \\ \vdots & & & & \ddots \end{bmatrix} \quad (\text{II.47})$$

where the vectors $\|n\rangle\!\rangle$ and $\langle\!\langle n\|$ form an orthonormal basis of the auxiliary space, which is of infinite dimension.

Those two matrices satisfy:

$$\boxed{de - q\,ed = (1-q)} \quad (\text{II.48})$$

which is the algebra of a q-deformed harmonic oscillator [146], of which $d$ is the annihilation operator, and $e$ the creation operator.

If we now take $D = 1 + d$ and $E = 1 + e$, they do verify eq.(II.32). There are several other representations that we could have chosen, but this is the simplest one where $D$ and $E$ don't depend on the boundary parameters. As for why they need to be infinite-dimensional, one may refer to [24].

Using this and equs.(II.33) and (II.34), we can find the correct boundary vectors. Let us define:

$$\langle\!\langle W\| = \sum_{n=0}^{\infty} W_n \langle\!\langle n\|, \quad (\text{II.49})$$

$$\|V\rangle\!\rangle = \sum_{n=0}^{\infty} V_n \|n\rangle\!\rangle, \quad (\text{II.50})$$

and write (II.33) and (II.34) in terms of the $W_n$'s and $V_n$'s:

$$\alpha W_{n+1} + (\alpha - \gamma - 1 + q)W_n - \gamma(1-q^n)W_{n-1} = 0, \quad (\text{II.51})$$

$$\beta(1-q^{n+1})V_{n+1} + (\beta - \delta - 1 + q)V_n - \delta V_{n-1} = 0. \quad (\text{II.52})$$

At this point, there are a number of mathematical objects that we need to define, and that will serve us on many occasions in the future.

First of all, we need to replace the boundary parameters $\alpha$, $\beta$, $\gamma$ and $\delta$ by four other parameters that will be better suited to what we need to do. Let us therefore write:

$$a = \frac{1}{2\alpha}\left[(1-q-\alpha+\gamma) + \sqrt{(1-q-\alpha+\gamma)^2 + 4\alpha\gamma}\right], \tag{II.53}$$

$$\tilde{a} = \frac{1}{2\alpha}\left[(1-q-\alpha+\gamma) - \sqrt{(1-q-\alpha+\gamma)^2 + 4\alpha\gamma}\right], \tag{II.54}$$

$$b = \frac{1}{2\beta}\left[(1-q-\beta+\delta) + \sqrt{(1-q-\beta+\delta)^2 + 4\beta\delta}\right], \tag{II.55}$$

$$\tilde{b} = \frac{1}{2\beta}\left[(1-q-\beta+\delta) - \sqrt{(1-q-\beta+\delta)^2 + 4\beta\delta}\right], \tag{II.56}$$

and reversely:

$$\alpha = \frac{(1-q)}{(1+a)(1+\tilde{a})}, \tag{II.57}$$

$$\gamma = -\frac{a\tilde{a}(1-q)}{(1+a)(1+\tilde{a})}, \tag{II.58}$$

$$\delta = -\frac{b\tilde{b}(1-q)}{(1+b)(1+\tilde{b})}, \tag{II.59}$$

$$\beta = \frac{(1-q)}{(1+b)(1+\tilde{b})}. \tag{II.60}$$

The quantities $a$ and $b$ are the same as were defined earlier in (II.18) and (II.19) in relation to the boundary conditions of the system in the large $L$ limit. This is, of course, not a coincidence.

We can now rewrite (II.51) and (II.52) as:

$$\boxed{W_{n+1} - (a + \tilde{a})W_n + a\tilde{a}(1-q^n)W_{n-1} = 0,} \tag{II.61}$$

$$\boxed{(1-q^{n+1})V_{n+1} - (b + \tilde{b})V_n + b\tilde{b}V_{n-1} = 0.} \tag{II.62}$$

Now, let us introduce a few q-deformed functions [147]. The parameter $q$ involved is the same as the rate of backward jumps in the ASEP. Since we will never use any other parameter to define these functions, we will omit it in our notations.

This is the q-Pochhammer symbol of order $n$:

$$(x)_n = \prod_{k=0}^{n-1}(1-q^k x) \tag{II.63}$$

which can also be defined for $n = \infty$, if $q < 1$ (which it was defined to be):

$$\boxed{(x)_\infty = \prod_{k=0}^{\infty}(1-q^k x).} \tag{II.64}$$

If we have a product of several infinite q-Pochhammer symbols, we may write them in the same symbol using commas, as $(x)_\infty(y)_\infty = (x,y)_\infty$, or $(x)_\infty(y)_\infty(z)_\infty = (x,y,z)_\infty$, etc.

This last function verifies:

$$(x)_\infty = (1-x)(qx)_\infty \tag{II.65}$$

which might come in handy. For instance, consider this simple equation:

$$(1-z)\sum_{n=0}^\infty \frac{z^n}{(q)_n} = \sum_{n=0}^\infty \frac{q^n z^n}{(q)_n}. \tag{II.66}$$

This tells us that the series expansion of the inverse of $(z)_\infty$ is:

$$\frac{1}{(z)_\infty} = \sum_{n=0}^\infty \frac{z^n}{(q)_n}. \tag{II.67}$$

This is called the q-exponential, from the fact that the limit $q \to 1$ gives a regular exponential (after rescaling $z$ to $z(1-q)$).

We can also expand the product of two q-exponentials:

$$\frac{1}{(xz, yz)_\infty} = \sum_n H_n(x,y) \frac{z^n}{(q)_n} \tag{II.68}$$

where one can find $H_n$ to be equal to

$$H_n(x,y) = \sum_{k=0}^n \frac{(q)_n}{(q)_k (q)_{n-k}} x^k y^{n-k}. \tag{II.69}$$

$H_n$ is called the q-Hermite polynomial of order $n$ (because the limit $q \to 1$ gives the regular Hermite polynomial of order $n$). It is usually defined using only one variable, which corresponds to the fact that we have $H_n(x,y) = y^n H_n(x/y, 1) = x^n H_n(1, y/x) = (xy)^{n/2} H_n(\sqrt{x/y}, \sqrt{y/x})$, but we will keep writing it with two variables for convenience.

By multiplying eq.(II.68) by $(1-xz)(1-yz)$, we find a recursion relation on the $H_n$'s:

$$H_{n+1}(x,y) - (x+y)H_n(x,y) + xy(1-q^n)H_{n-1}(x,y) = 0. \tag{II.70}$$

Another useful identity on the q-Hermite polynomials is the q-Mehler formula [148]:

$$\sum_{n=0}^\infty H_n(x,y) H_n(z,t) \frac{\lambda^n}{(q)_n} = \frac{(xyzt\lambda^2)_\infty}{(xz\lambda, xt\lambda, yz\lambda, yt\lambda)_\infty}. \tag{II.71}$$

Let it be noted that this formula is in principle only valid when all the arguments of the q-Pochhammer symbols on the right are smaller than 1 in module (note that this doesn't constrain the arguments of the functions $H_n$ directly, but only their products with $\lambda$).

We can now finally get back to the calculation of $Z_L$.

Comparing (II.70) to (II.61) and (II.62), we find that:

$$\boxed{W_n = H_n(a, \tilde{a}) \quad , \quad V_n = \frac{H_n(b, \tilde{b})}{(q)_n}.} \tag{II.72}$$

Now, considering that $Z_L = \langle\!\langle W \| (2 + d + e)^L \| V \rangle\!\rangle$, a good way to proceed would be to find the eigenvectors of $d + e$, and expand $\|V\rangle\!\rangle$ on that basis. Let us define:

$$\|z\rangle\!\rangle = \sum_{n=0}^{\infty} \frac{H_n(z, \frac{1}{z})}{(q)_n} \|n\rangle\!\rangle \quad , \quad \langle\!\langle z\| = \sum_{n=0}^{\infty} H_n(z, \frac{1}{z}) \langle\!\langle n\|. \tag{II.73}$$

Applying $d + e$ on those and using eq.(II.70) yields:

$$\boxed{(d + e)\|z\rangle\!\rangle = (z + 1/z)\|z\rangle\!\rangle \quad , \quad \langle\!\langle z\|(d + e) = \langle\!\langle z\|(z + 1/z).} \tag{II.74}$$

Luckily, there is a reasonably simple closure identity involving that basis, which is given by [147]:

$$1 = \frac{(q)_\infty}{2} \oint_{c_1} \frac{dz}{i2\pi z} (z^2, z^{-2})_\infty \|z\rangle\!\rangle \langle\!\langle z\| \tag{II.75}$$

where $c_1$ is the complex unit circle.

Injecting this relation right next to $\|V\rangle\!\rangle$ in $Z_L$, and using (II.74), gives:

$$Z_L = \frac{(q)_\infty}{2} \oint_{c_1} \frac{dz}{i2\pi z} (z^2, z^{-2})_\infty (2 + z + z^{-1})^L \langle\!\langle W\|z\rangle\!\rangle \langle\!\langle z\|V\rangle\!\rangle. \tag{II.76}$$

Finally, using the q-Mehler formula (II.71) on $\langle\!\langle W\|z\rangle\!\rangle$ and $\langle\!\langle z\|V\rangle\!\rangle$ gives:

$$\langle\!\langle W\|z\rangle\!\rangle = \frac{(a\tilde{a})_\infty}{(az, a/z, \tilde{a}z, \tilde{a}/z)_\infty} \quad , \quad \langle\!\langle z\|V\rangle\!\rangle = \frac{(b\tilde{b})_\infty}{(bz, b/z, \tilde{b}z, \tilde{b}/z)_\infty}. \tag{II.77}$$

This is rigorous only for $|a| < 1$, $|\tilde{a}| < 1$, $|b| < 1$ and $|\tilde{b}| < 1$. These conditions are always verified for $\tilde{a}$ and $\tilde{b}$, but not for $a$ and $b$. However, we can extend this result to all values of $a$ and $b$ by analytic continuation.

Combining all those relations, and getting rid of a constant term $(q, a\tilde{a}, b\tilde{b})_\infty$, we can write the definitive form of $Z_L$:

$$\boxed{\boxed{Z_L = \frac{1}{2} \oint_S \frac{dz}{2i\pi} \frac{F(z)}{z}}} \tag{II.78}$$

with

$$\boxed{F(z) = \frac{(1 + z)^L (1 + z^{-1})^L (z^2, z^{-2})_\infty}{(az, a/z, \tilde{a}z, \tilde{a}/z, bz, b/z, \tilde{b}z, \tilde{b}/z)_\infty}.} \tag{II.79}$$

For $a < 1$ and $b < 1$, the domain of integration is the unit circle, which is to say that we take the residues at all the poles of $F$ that are in $S = \{0; aq^k, \tilde{a}q^k, bq^k, \tilde{b}q^k\}_{k \in \mathbb{N}}$, and not at the other ones (which are their inverses). Since $Z_L$ is analytic in all the parameters, the poles of $F$ at which we have to take a residue are always the same, even if one of them leaves the unit circle. For that reason, the integral in (II.78) must be done around $S$, and not on the unit circle.

The case of the TASEP is much simpler, as always. If $q = \gamma = \delta = 0$, $F$ reduces to:

$$\boxed{F(z) = \frac{(1+z)^L(1+z^{-1})^L(1-z^2)(1-z^{-2})}{(1-az)(1-\frac{a}{z})(1-bz)(1-\frac{b}{z})}} \qquad \text{(II.80)}$$

with $S = \{0, a, b\}$ and $a = \frac{1-\alpha}{\alpha}$, $b = \frac{1-\beta}{\beta}$.

We can use this expression of $Z_L$, either for the ASEP or for the TASEP, to find the phase diagram of the current $J = (1-q)\frac{Z_{L-1}}{Z_L}$ in a more rigorous fashion.

If $a < 1$ and $b < 1$, the contour integral in (II.78) can be done on the unit circle. Because of the term $(1+z)^L(1+z^{-1})^L$, $F(z)$ has a saddle point at $z = 1$ for $L$ large. $Z_L$ therefore behaves as $4^L$, and we get $J = \frac{1-q}{4}$. We recognise the maximal current (MC) phase.

If $a > 1$ and $a > b$, the contour integral is dominated by the largest pole on the real axis, which is $a$ (see section IV.1.1 for more details on that). $Z_L$ then behaves as $(1+a)^L(1+a^{-1})^L$, and we get $J = (1-q)\frac{a}{(1+a)^2}$. Remembering that we have $\rho_a = \frac{1}{1+a}$, we finally get $J = (1-q)\rho_a(1-\rho_a)$. This is the low density (LD) phase.

If $b > 1$ and $b > a$, we find the same with $b$ instead of $a$, i.e. with $1 - \rho_b$ instead of $\rho_a$, and we get the high density (HD) phase.

Moreover, the conditions $a > 1$ or $b > 1$ correspond to $\rho_a < \frac{1}{2}$ and $1 - \rho_b < \frac{1}{2}$, and $a = b$ corresponds to $\rho_a = 1 - \rho_b$, so that the boundaries of the three phases we find are indeed the same as in the previous section.

Let us finally note that, using the matrix Ansatz, we can also calculate the 2-point correlations of the density [67, 105, 106], and find them to be vanishing entirely for $L \to \infty$ (algebraically in the MC phase, and exponentially everywhere else), which validates the mean field approach of the previous section. However, the behaviour of the profiles near the boundaries in the MC phase is not correct in the mean field approximation (it should behave as $1/\sqrt{L(x-x_0)}$ rather than $1/L(x-x_0)$).

## II.2.2 Coordinate Bethe Ansatz for the periodic ASEP

Here, we present the method used by Prolhac and Mallick in [26–29] to obtain the complete generating function of the cumulants of the current in the periodic ASEP, using the coordinate (and then the functional) Bethe Ansatz.

Instead of being connected to reservoirs through boundary matrices, the periodic ASEP has one extra bulk matrix that connects site $L$ to site $1$. Because of that, the number of particles $N$ is now conserved.

We will here be interested in the total current going over all of the bonds of the system. From what we saw in section I.2.2, the deformed Markov matrix for that observable is:

$$M_\mu = \sum_{i=1}^{L} M_i(\mu) \qquad \text{(II.81)}$$

with

$$M_i(\mu) = \begin{bmatrix} 0 & 0 & 0 & 0 \\ 0 & -q & e^\mu & 0 \\ 0 & qe^{-\mu} & -1 & 0 \\ 0 & 0 & 0 & 0 \end{bmatrix} \qquad (\text{II}.82)$$

(this is the case where $V = 0$ and $U = 1$ for forward jumps and $-1$ for backward jumps), and the cumulant generating fuction $E(\mu)$ is its largest eigenvalue.

**Coordinate Bethe Ansatz**

The reasoning leading to the coordinate Bethe Ansatz goes, schematically, like this: in the case of non-interacting particles (i.e. without the exclusion constraint), the eigenstates of the Markov matrix would be plane waves, with some quantised fugacities (due to the periodicity of the chain). In our case, the particles are interacting, but only if they are next to one-another, so that they still behave like plane waves if they don't touch. When they do collide, they might exchange their fugacities, but not modify their values entirely (as this would also modify the corresponding eigenvalue of the Markov matrix), so the natural Ansatz to make is a superposition of plane waves, with some coefficients depending on which particle has which fugacity. Those coefficients, and the fugacities, are unknown quantities to be determined. Thanks to the integrability of the chain, this actually works.

According to the coordinate Bethe Ansatz [149, 150], the eigenvectors of $M_\mu$ can be written as:

$$\boxed{|\psi\rangle = \sum_{\{x_i\}} \left[ \sum_{\sigma \in S_N} A_\sigma \prod z_{\sigma(i)}^{x_i} \right] |\{x_i\}\rangle} \qquad (\text{II}.83)$$

where, unlike the notation we have used until now, configurations are designated by the ordered positions $x_i$ of all the particles. The coefficient of a given configuration $\{x_i\}$ is, in this Ansatz, a sum of plane waves $\prod z_{\sigma(i)}^{x_i}$ with fugacities $z_j$. Each element in that sum corresponds to one distribution of the fugacities among the particles (designated by a permutation $\sigma$ of the indices of the $z$'s, belonging to $S_N$, the group of permutations of size $N$), and bears a coefficient $A_\sigma$ that doesn't depend on the positions $\{x_i\}$. Using $M_\mu$ on this expression, while assuming that it is indeed one of its eigenvectors ($M_\mu|\psi\rangle = E|\psi\rangle$), gives us a system of coupled equations involving the $z_i$'s, the $A_\sigma$'s, and $E$, that we then have to solve.

Let us first look at a configuration $\{x_i\}$ where all the particles are far away from each other (let's assume that there are few enough particles for that to be possible). The configurations leading to that one after applying $M_\mu$ to $|\psi\rangle$ are those where one of the particles is at $x_i - 1$ or $x_i + 1$, and is about to jump. Each forward jump corresponds to a term $\frac{e^\mu}{z_j}$ in $E$, and each backward jump to a term $qe^{-\mu}z_j$. We must also take into account the escape rate $-1 - q$ for each particle. All in all, we find that the eigenvalue $E$ is given by:

$$E = \sum_{j=1}^{N} \left( \frac{e^\mu}{z_j} + qe^{-\mu} z_j - 1 - q \right). \qquad (\text{II}.84)$$

Let us now consider a configuration where two particles are next to each another, at positions $x_i$ and $x_i + 1$, and the others are far away. In the previous case, no two particles were close enough to exchange their fugacities, and the action of $M_\mu$ was independent of the $A_\sigma$'s. In this case, it will involve couples of $\sigma$'s that differ by one transposition $\tau_{i,j}$ (corresponding to the exchange of two fugacities $z_i$ and $z_j$). When we apply $M_\mu$ to the configurations leading to the one we chose, all the jumps made by the isolated particles have the same contribution to $E$ as previously, and we don't have to worry about them. The only part of the weight of our configuration in $|\psi\rangle$ that we have to consider is $A_\sigma z_{\sigma(i)}^{x_i} z_{\sigma(i+1)}^{x_i+1} + A_{\sigma \circ \tau_{i,i+1}} z_{\sigma(i+1)}^{x_i} z_{\sigma(i)}^{x_i+1}$, for a given $\sigma$ (the rest of the fugacities can be factored out), where $\tau_{i,i+1}$ is the transposition of $i$ with $i+1$, and $\circ$ is the composition operation. The only difference with the previous situation is that two of the four configurations that leads to this one through a jump are forbidden, because they would have the two particles standing on the same site. Since $E$ is the same anyway, it must be that the contribution they would have had to $E$ is 0.

That contribution is $qe^{-\mu} A_\sigma z_{\sigma(i)}^{x_i+1} z_{\sigma(i+1)}^{x_i+1} + e^\mu A_{\sigma \circ \tau_{i,i+1}} z_{\sigma(i+1)}^{x_i} z_{\sigma(i)}^{x_i}$ from the particle with fugacity $z_{\sigma(i)}$, and $e^\mu A_\sigma z_{\sigma(i)}^{x_i} z_{\sigma(i+1)}^{x_i} + qe^{-\mu} A_{\sigma \circ \tau_{i,i+1}} z_{\sigma(i+1)}^{x_i+1} z_{\sigma(i)}^{x_i+1}$ from the one with fugacity $z_{\sigma(i+1)}$, without forgetting the escape rate $(-1-q)\left(A_\sigma z_{\sigma(i)}^{x_i} z_{\sigma(i+1)}^{x_i+1} + A_{\sigma \circ \tau_{i,i+1}} z_{\sigma(i+1)}^{x_i} z_{\sigma(i)}^{x_i+1}\right)$. Their sum must be 0, which gives:

$$\frac{A_{\sigma \circ \tau_{i,i+1}}}{A_\sigma} = -\frac{e^\mu - (1+q) z_{\sigma(i+1)} + qe^{-\mu} z_{\sigma(i)} z_{\sigma(i+1)}}{e^\mu - (1+q) z_{\sigma(i)} + qe^{-\mu} z_{\sigma(i)} z_{\sigma(i+1)}}. \tag{II.85}$$

A nice property of integrability [40, 151] is that the three-body scattering matrix factorises in terms of the two-body scattering matrix, which is to say that the same calculation with three particles gives merely the composition of two of those relations, so that we don't have to consider any more complicated situations.

Let us now look at the consequence of periodicity. Since there is no preferred site from which to start numbering the particles, we could have written $\{x_2, ..., x_L, x_1 + L\}$ instead of $\{x_1, x_2, ..., x_L\}$ for any configuration. This must make no difference on $|\psi\rangle$. It does, however, change $\sigma$ into $\sigma \circ t^+$ (where $t^+$ is the circular permutation which raises each number by 1 and sends $N$ to 1), and $z_{\sigma(1)}^{x_1}$ into $z_{\sigma(1)}^{x_1+L}$. These two changes must therefore compensate, so that:

$$\frac{A_{\sigma \circ t^+}}{A_\sigma} = z_{\sigma(1)}^{-L}. \tag{II.86}$$

Before going further, we should make a simple change of variables on the $z_i$'s, which will make all future calculations much easier. Let us define, for each $z_i$, a new quantity $y_i$ such that:

$$y = -\frac{1 - e^{-\mu} z}{1 - qe^{-\mu} z} \quad , \quad z = e^\mu \frac{1+y}{1+qy}. \tag{II.87}$$

Equation (II.85) becomes:

$$\boxed{\frac{A_{\sigma \circ \tau_{i,i+1}}}{A_\sigma} = \frac{y_{\sigma(i+1)} - q y_{\sigma(i)}}{q y_{\sigma(i+1)} - y_{\sigma(i)}}} \tag{II.88}$$

and the eigenvalue $E$ is given by:

$$E = (1-q) \sum_{i=1}^{N} \left( \frac{1}{1+y_i} - \frac{1}{1+qy_i} \right). \tag{II.89}$$

Now, since what we want is $E$, and not $|\psi\rangle$, we had better find a way to get rid of the $A_\sigma$'s in eq.(II.88). In order to do that, we need to reconstruct permutation $t^+$ as a product of $N-1$ transposition: $t^+ = \tau_{1,2} \circ \tau_{2,3} \circ \cdots \circ \tau_{N-2,N-1} \circ \tau_{N-1,N}$. Considering that $\frac{A_{\sigma \circ \tau_{1,2}}}{A_\sigma} \frac{A_{\sigma \circ \tau_{1,2} \circ \tau_{2,3}}}{A_{\sigma \circ \tau_{1,2}}} = \frac{A_{\sigma \circ \tau_{1,2} \circ \tau_{2,3}}}{A_\sigma}$, and adding one ratio at a time, one can get $\frac{A_{\sigma \circ t^+}}{A_\sigma}$ as a product of $N-1$ equations of the form (II.85), where, in each of them, $\sigma(1)$ is exchanged with one of the other indices. Since we have the choice of $\sigma(1)$ from the start, what we finally get is a system of $N$ coupled equations, one for each $z_i$:

$$e^{L\mu} \left( \frac{1+y_i}{1+qy_i} \right)^L = -\prod_{j=1}^{N} \frac{y_i - qy_j}{qy_i - y_j}. \tag{II.90}$$

These are called the Bethe equations. The left side of the equation comes from eq.(II.86) written in terms of $y_i$, and the right side comes from the product of all the ratios from (II.85), and one more for $i = j$ to complete the product (which accounts for the minus sign).

**Functional Bethe Ansatz**

It is now time to switch to the functional Bethe Ansatz (which is not, in fact, another form of the Bethe Ansatz, but merely another way to write the Bethe equations).

Consider a polynomial $Q$, defined as:

$$Q(x) = Q_0 \prod_{i=1}^{N} (1 - x \, y_i) \tag{II.91}$$

where $Q_0$ is a constant. The roots of $Q$ are the inverses of the $y_i$'s from before.

Let us also define $h(y)$ as:

$$h(y) = \frac{(1+y)^L}{y^N}. \tag{II.92}$$

In terms of those two functions, the Bethe equations (II.90) become:

$$e^{L\mu} \frac{h(y_i)}{h(qy_i)} = -\frac{Q(q/y_i)}{Q(1/qy_i)} \tag{II.93}$$

for all $y_i$'s. Rearranging this equation, we find that there must be a polynomial $T(y)$ with roots at all the $y_i$'s, such that:

$$T(y)Q(1/y) = h(y)Q(1/qy) + e^{-L\mu} h(qy)Q(q/y) \tag{II.94}$$

($T(y)$ is in fact a polynomial in $y$ and $y^{-1}$, but we could always multiply the whole equation by a large enough power of $y$).

We can also express $E$ in terms of $Q$ in two different ways, as:

$$E = (1-q)\frac{d}{dy}\log\left(\frac{Q(q/y)}{Q(1/y)}\right)\bigg|_{y=-1} = (1-1/q)\frac{d}{dy}\log\left(\frac{Q(1/qy)}{Q(1/y)}\right)\bigg|_{y=-1/q} \quad \text{(II.95)}$$

and then, using eq.(II.94), in terms of $T$:

$$E = (1-q)\frac{d}{dy}\log\left(\frac{T(y)}{h(qy)}\right)\bigg|_{y=-1} = (1-1/q)\frac{d}{dy}\log\left(\frac{T(y)}{h(y)}\right)\bigg|_{y=-1/q} \quad \text{(II.96)}$$

or

$$E = -N - (L-N)q + (1-q)\frac{d}{dy}\log(T(y))\big|_{y=-1} \quad \text{(II.97)}$$

$$= -(L-N) - Nq + (1-1/q)\frac{d}{dy}\log(T(y))\big|_{y=-1/q}. \quad \text{(II.98)}$$

(the fact that there are two versions of each of those equations comes from the particle↔hole symmetry of the system).

Through a procedure detailed in [27] and [152], we can find a polynomial $P(y)$, defined as:

$$P(y) = \prod_{i=1}^{L-N}(1 - y/q\tilde{y}_i) \quad \text{(II.99)}$$

which is solution of the same polynomial equation as $Q$, but with $1/y$ being replaced by $qy$:

$$T(y)P(qy) = h(qy)P(y) + e^{-L\mu}h(y)P(q^2y). \quad \text{(II.100)}$$

This polynomial $P$ is in fact the one we would have gotten instead of $Q$ if we had chosen to track the holes instead of the particles in the coordinate Bethe Ansatz. It is also, due to the particle↔hole symmetry, the polynomial for the particles for a system with $L-N$ particles instead of $N$.

We can combine equations (II.94) and (II.100) to get rid of $T$, and obtain:

$$P(y)Q(1/y) = h(y) + e^{-L\mu}P(qy)Q(q/y). \quad \text{(II.101)}$$

Considering the lowest or the highest power of $y$ in this equation, we get two expressions for $Q_0$, in terms of the $y_i$'s or of the $\tilde{y}_i$'s (but we will only use the first one):

$$Q_0 = \frac{\prod_{i=1}^{N}(-1/y_i)}{(1 - e^{-L\mu}q^N)} = \frac{\prod_{i=1}^{L-N}(-q\tilde{y}_i)}{(1 - e^{-L\mu}q^{L-N})}. \quad \text{(II.102)}$$

Let us now define a function $W(y)$ as:

$$W(y) = -\log\left(\frac{P(y)Q(1/y)}{e^{-L\mu}P(qy)Q(q/y)}\right) \quad \text{(II.103)}$$

and a constant $B$ as:
$$B = -\frac{e^{L\mu}}{Q_0}. \tag{II.104}$$

Let us also define a convolution kernel $K$, as:
$$K(z,\tilde{z}) = \sum_{k=1}^{\infty} \frac{q^k}{1-q^k}\left((z/\tilde{z})^k + (z/\tilde{z})^{-k}\right) \tag{II.105}$$

and the associated convolution operator $X$:
$$X[f](z) = \oint_{c_1} \frac{d\tilde{z}}{\imath 2\pi \tilde{z}} f(\tilde{z}) K(z,\tilde{z}). \tag{II.106}$$

Using those, one can find that $-\log\bigl(P(qy)Q(q/y)/Q_0\bigr) = X[W](y)$. It can be done simply by separating all the monomials from $P$ and $Q/Q_0$ in the logarithms, and expanding them. Each positive or negative power $\pm k$ of $y$ has the same coefficient in both sides, up to a factor $q^k$ in the left side of the equation, and $(1-q^k)$ in $W$. The convolution then serves to replace that $(1-q^k)$ by $q^k$.

All in all, we can finally rewrite eq.(II.101) in terms of only one unknown function $W$:
$$\boxed{W(z) = -\ln\left(1 - Bh(z)e^{X[W](z)}\right)} \tag{II.107}$$

which is a logarithmic equivalent of a Fredholm integral equation of the second kind [153]. This gives us, in principle, $W(z)$ in terms of $Bh(z)$. The next step is to use the equations we found earlier to express both $E$ and $\mu$ in terms of $B$, which will give us $E$ in terms of $\mu$ implicitly.

Let us remark here that, until now, we never had to specify which eigenvector of $M_\mu$ it was that we were considering. Everything we have done so far is therefore valid for any eigenvalue $E$ which can be obtained by Bethe Ansatz. For the next step, however, we need to input information specific to the steady state.

We know that in that state, all configurations are equiprobable (this is easily checked), so that all the $z_i$'s must be 1, i.e. all the $y_i$'s are equal to 0. This tells us that:
$$\lim_{\mu \to 0} B = 0 \tag{II.108}$$

and that, for $\mu$ small enough, all the $y_i$'s are inside the unit circle. We can also show that the opposite goes for the $\tilde{y}_i$'s: they are infinite for $\mu = 0$ and they stay outside of the unit circle for a small $\mu$.

The two equations that we will need to use to get $E(\mu)$ and $\mu$ in terms of $B$ are the left part of eq.(II.95), and eq.(II.103) for $y=0$, which we rewrite here:
$$E(\mu) = (1-q)\frac{d}{dy}\log\left(\frac{Q(q/y)}{Q(1/y)}\right)\bigg|_{y=-1} \quad , \quad L\mu = -W(0) \tag{II.109}$$

As we see, $E(\mu)$ is not written in terms of $W$, but only of $Q$. This is where the behaviour of the $y_i$'s and $\tilde{y}_i$'s becomes useful. Because of their respective positions on the complex plane, and if we take $z$ to be on the unit circle, each factor $\log(1 - y_i/z)$ (coming from $Q$) in $W$ can be expanded as a series in $1/z$, and each factor $\log(1 - z/q\tilde{y}_i)$ (coming from $P$) as a series in $z$ without any pole inside of the unit circle. Simply put, $P$ is holomorphic inside of the unit circle. This assures us that a contour integral over the unit circle will only pick up contributions from $Q$, and not from $P$.

All that being said, we can finally write $E(\mu)$ and $\mu$ in terms of contour integrals, as:

$$\boxed{L\mu = -\oint_{c_1} \frac{dz}{i2\pi z} W(z) = -\sum_{k=1}^{\infty} C_k \frac{B^k}{k}} \tag{II.110}$$

and

$$\boxed{E(\mu) = -(1-q)\oint_{c_1} \frac{dz}{i2\pi(1+z)^2} W(z) = -(1-q)\sum_{k=1}^{\infty} D_k \frac{B^k}{k}.} \tag{II.111}$$

The coefficients $C_k$ and $D_k$ are what we get after expanding $W$ in powers of $B$ in this expression (we recall that $B \sim 0$ for $\mu$ small), and can be written in terms of contour integrals over combinations of $h$ and $K$.

The factor $L$ in eq.(II.110) comes from the fact that we counted the current over all the bonds of the system. To get the equivalent formulae for the current over only one bond, we just have to take that $L$ away (or to replace $\mu$ by $\mu/L$).

For the simpler case of the TASEP, we have $q = 0$, so $K(z, z') = 0$, and $W = -\ln\big(1 - Bh(z)\big)$. In this case, $W$ is easily expanded in $B$, and we find:

$$C_k = \oint_{c_1} \frac{dz}{i2\pi z} h^k(z) = \binom{kL}{kN} \tag{II.112}$$

and

$$D_k = \oint_{c_1} \frac{dz}{i2\pi(1+z)^2} h^k(z) = \binom{kL-2}{kN-1} \tag{II.113}$$

where $\binom{X}{Y} = \frac{X!}{Y!(X-Y)!}$ is a binomial coefficient.

Those are the same results as were found by Derrida and Lebowitz in [94].

Now that we have the explicit expression of those coefficients, we can find the cumulants of the current by inverting eq.(II.110) to find $B$ in terms of $\mu$, and then inject the result into eq.(II.111), in order to get $E(\mu)$ explicitly as an exponential generating series in $\mu$ (the coefficients $E_k$ of which are precisely the cumulants of the current). This doesn't give a closed formula for any $E_k$, but they can be calculated order by order, in a finite number of calculation steps for a given $k$. The first few of those are given by:

$$\begin{aligned}
E_1 &= (1-q)L\frac{D_1}{C_1}, \\
E_2 &= (1-q)L^2\frac{D_1 C_2 - D_2 C_1}{2C_1^3}, \\
E_3 &= (1-q)L^3\frac{3D_1 C_2^2 - 2D_1 C_1 C_3 - 3D_2 C_1 C_2 + 2D_3 C_1^2}{6C_1^5}.
\end{aligned} \tag{II.114}$$

Because of the complexity of eqs.(II.110) and (II.111), it is not possible, in practice, to find the large deviation function of the cumulants of the current explicitly for a finite $L$. It can be done in the large $L$ limit, however. The result for the case of where $L = 2N$, along with the calculations involved, can be found in sections IV.1.4, IV.2.3 and part of IV.4, which treat of the open ASEP at the HD-MC transition (we will see that this its behaviour on that transition is exactly that of a half-filled periodic ASEP).

# CHAPTER III

## Perturbative Matrix Ansatz and Fluctuations of the Current

Now that we have reviewed all the tools and pre-existing results that we might need, we can finally get to the core of this thesis, namely the generalisation of the matrix Ansatz [24] to the calculation of the exact generating function of the cumulants of the current in the open ASEP. We will refer to this new method as 'perturbative matrix Anatz', for reasons that will become clear in a few pages.

This chapter gives a logical (rather than chronological) account of the results, as opposed to the order in which they were published. That is to say, we will present the method before the results, whereas while we were exploring the subject in search of a solution, we managed to make our way to the final formulae [1,2] long before we had a neat idea of how the proof worked. Needless to say that a fair amount of guesswork entered into play, and those formulae were published as conjectures, first for the simpler case of the TASEP [1], then for the general ASEP [2]. In both cases, the results were backed by exact numerical calculations on systems of small size and by DMRG calculations from our collaborators M. Gorissen and C. Vanderzande. All this will be presented in more detail further down in section III.4.

The layout of this chapter is the following:

- In section III.1, we apply what we saw in chapter I and give a few results related to the Gallavotti-Cohen symmetry, which are well-known, but will be useful later on.

- The perturbative matrix Ansatz method is presented in part section III.2, the contents of which are pretty much the same as part 4 of [3].

- Section III.3 contains the formulae for the generating function of the cumulants of the current in the open ASEP and its simpler expression for the open TASEP.

- Section III.4 only contains calculation details and can be skipped entirely: we obtain the first few cumulants, which allows us to infer the form of the complete solution, and take a look at the numerical verifications that were carried out to validate our result.

It should be noted that, although, at this point, the formulae in section III.3 are still, strictly speaking, conjectured, we take care of that issue in chapter V (which can be read instead of section III.2), where we further generalise our perturbative matrix Anatz and relate it to the algebraic Bethe Ansatz [39] and to Baxter's Q-operator [40].

## III.1 Current-counting Markov matrix and Gallavotti-Cohen symmetry

We first define the most general current-counting Markov matrix for the open ASEP. After a quick derivation of the Gallavotti-Cohen symmetry, which relies on the relation between the macroscopic current of particles and the microscopic production of entropy in the system, we reduce the deformed Markov matrix to a simpler form.

### III.1.1 Current-counting Markov matrix

Let us first quickly recall the jumping rules of the open ASEP:

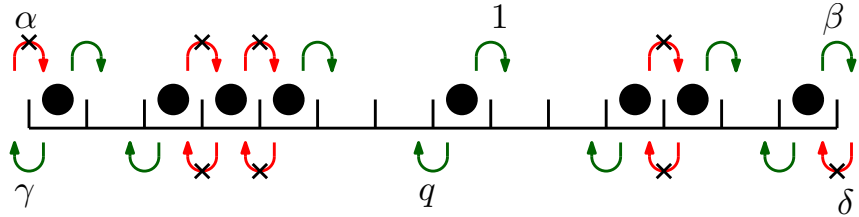

Figure III.1: Dynamical rules for the ASEP with open boundaries. The rate of forward jumps has been normalised to 1. Backward jumps occur with rate $q < 1$. All other parameters are arbitrary. The jumps shown in green are allowed by the exclusion constraint. Those shown in red and crossed out are forbidden.

Just as for the periodic case in the last section of chapter II, the observable we're interested in is the particle current, which is time-additive, so that the results from chapter I apply. In principle, we could count the currents over each of the bonds independently, with different fugacities $e^{\mu_i}$ between each neighbouring sites $i$ and $i+1$ (where $i$ goes from 0 to $L$, site 0 being the left reservoir, and site $L+1$ the right reservoir). The most general deformed Markov matrix for the current is therefore:

$$M_{\{\mu_i\}} = m_0(\mu_0) + \sum_{i=1}^{L-1} M_i(\mu_i) + m_L(\mu_l) \tag{III.1}$$

with

$$m_0(\mu_0) = \begin{bmatrix} -\alpha & \gamma e^{-\mu_0} \\ \alpha e^{\mu_0} & -\gamma \end{bmatrix} \;,\; M_i(\mu_i) = \begin{bmatrix} 0 & 0 & 0 & 0 \\ 0 & -q & e^{\mu_i} & 0 \\ 0 & qe^{-\mu_i} & -1 & 0 \\ 0 & 0 & 0 & 0 \end{bmatrix} \;,\; m_L(\mu_L) = \begin{bmatrix} -\delta & \beta e^{\mu_L} \\ \delta e^{-\mu_L} & -\beta \end{bmatrix} \tag{III.2}$$

(where, as before, it is implied that $m_0$ acts as written on site 0 in the basis $\{0,1\}$ and as the identity on the other sites, and the same goes for $m_L$ on site $L$; similarly, $M_i$ is expressed by its action on sites $i$ and $i+1$ in the basis $\{00, 01, 10, 11\}$ and acts as the identity on the rest of the system).

The largest eigenvalue $E(\{\mu_i\})$ of that matrix is the joint generating function of the cumulants of all the local currents, and the left and right eigenvectors carry the probabilities of configurations conditioned on the values of the integrated currents going to or coming from the steady state (as explained in chapter I).

It would stand to reason that, since the system is one-dimensional, the current should be conserved from one reservoir to the other, and that the bond over which it is measured should make no difference at all in the long time limit. This is easy enough to prove, and is done in the next section.

### III.1.2  Gallavotti-Cohen symmetry

In this section, we prove that two current-counting Markov matrices $M_{\{\mu_i\}}$ and $M_{\{\mu'_i\}}$ are similar (and therefore have the same eigenvalues) as long as $\sum_{i=0}^{L} \mu_i = \sum_{i=0}^{L} \mu'_i$, which is to say that the current and its fluctuations only depend on $\mu = \sum_{i=0}^{L} \mu_i$, regardless of how the fugacities are distributed: the currents through each of the bonds are all exactly equivalent. This is a well known result, which can be found in [18], and was also used in [86].

Let us consider the diagonal matrix $R_i(\lambda_i)$ (with $1 \leq i \leq L$) which multiplies by $e^{\lambda_i}$ all configurations for which site $i$ is occupied. The transformation $R_i(\lambda_i)^{-1} M_{\{\mu_i\}} R_i(\lambda_i)$ acts only on $M_{i-1}(\mu_{i-1})$ and $M_i(\mu_i)$, and a trivial calculation gives

$$R_i(\lambda_i)^{-1} M_{i-1}(\mu_{i-1}) R_i(\lambda_i) = M_{i-1}(\mu_{i-1} - \lambda_i) \tag{III.3}$$

and

$$R_i(\lambda_i)^{-1} M_i(\mu_i) R_i(\lambda_i) = M_i(\mu_i + \lambda_i). \tag{III.4}$$

One can therefore transfer any fraction $\lambda$ of $\mu = \sum_{i=0}^{L} \mu_i$ from one bond to the previous or the next one. Using this, one can go from $\{\mu_i\}$ to $\{\mu'_i\}$ step by step, or simply derive the global similarity matrix $R_{\{\mu_i\}}^{\{\mu'_i\}}$ such that

$$\boxed{M_{\{\mu'_i\}} = \left(R_{\{\mu_i\}}^{\{\mu'_i\}}\right)^{-1} M_{\{\mu_i\}} R_{\{\mu_i\}}^{\{\mu'_i\}}} \tag{III.5}$$

which one can easily find to be

$$R_{\{\mu_i\}}^{\{\mu'_i\}} = \prod_{i=1}^{L} R_i\left(\sum_{j=0}^{i-1} (\mu_j - \mu'_j)\right). \tag{III.6}$$

There is a particular set of weights $\{\mu_i\}$ defined by

$$\{\mu_0 = \nu \log\left(\frac{\alpha}{\gamma}\right), \quad \mu_i = \nu \log(1/q), \quad \mu_L = \nu \log\left(\frac{\beta}{\delta}\right)\} \tag{III.7}$$

for which $M_{\{\mu_i\}}$ becomes:

$$m_0 = \begin{bmatrix} -\alpha & \gamma^{1+\nu}\alpha^{-\nu} \\ \alpha^{1+\nu}\gamma^{-\nu} & -\gamma \end{bmatrix} \;,\; M_i = \begin{bmatrix} 0 & 0 & 0 & 0 \\ 0 & -q & q^{-\nu} & 0 \\ 0 & q^{1+\nu} & -1 & 0 \\ 0 & 0 & 0 & 0 \end{bmatrix} \;,\; m_L = \begin{bmatrix} -\delta & \beta^{1+\nu}\delta^{-\nu} \\ \delta^{1+\nu}\beta^{-\nu} & -\beta \end{bmatrix} \tag{III.8}$$

which is the deformed Markov matrix measuring the entropy production. We see immediately, as before, that

$$M_{-1-\nu} = {}^t M_\nu \tag{III.9}$$

which proves the Gallavotti-Cohen symmetry for the eigenvalues and between the left and right eigenvectors of $M_\nu$ with respect to the transformation $\nu \leftrightarrow (-1-\nu)$.

Considering that $\mu = \nu \log\left(\frac{\alpha\beta}{\gamma\delta q^{L-1}}\right)$, we also obtain the Gallavotti-Cohen symmetry related to the current, namely

$$E(\mu) = E\left(-\log\left(\frac{\alpha\beta}{\gamma\delta q^{L-1}}\right) - \mu\right) \tag{III.10}$$

which is also valid for the other eigenvalues of $M_\mu$, and the corresponding relations between the right and left eigenvectors, as well as a simple relation between the microscopic entropy production $s$, conjugate to $\nu$, and the macroscopic current $j$, conjugate to $\mu$:

$$s = j \log\left(\frac{\alpha\beta}{\gamma\delta q^{L-1}}\right). \tag{III.11}$$

The equilibrium case, where $\alpha\beta = \gamma\delta q^{L-1}$ (which is to say that the boundary chemical potentials compensate exactly the field in the bulk of the system), is somewhat pathological: $s = 0$, so that there is no entropy production, and all the odd cumulants of $j$ vanish, but not the even ones. In this case only, the current contains more information than the entropy production.

All this being said, we can now consider, without any loss of generality, the case where only the first bond (between the leftmost reservoir and the first site) is marked: $\mu_0 = \mu$, $\mu_{i\neq 0} = 0$, so that the individual jump matrices we will work with are given by

$$m_0(\mu) = \begin{bmatrix} -\alpha & \gamma e^{-\mu} \\ \alpha e^{\mu} & -\gamma \end{bmatrix} \;,\; M_i = \begin{bmatrix} 0 & 0 & 0 & 0 \\ 0 & -q & 1 & 0 \\ 0 & q & -1 & 0 \\ 0 & 0 & 0 & 0 \end{bmatrix} \;,\; m_L = \begin{bmatrix} -\delta & \beta \\ \delta & -\beta \end{bmatrix}. \tag{III.12}$$

Our goal is to find a way to get at the largest eigenvalue $E(\mu)$ of this matrix, and at its corresponding eigenvectors $\|P_\mu\rangle\rangle$ and $\langle\langle \tilde{P}_\mu\|$.

Let us make one final important remark. We have just shown that $E(\mu)$ doesn't depend on the choice of $\{\mu_i\}$. The eigenvectors, however, do (they are related to one another by multiplication with the diagonal similarity matrix expressed in (III.6)). This is easily understood if we remember that, for instance, $\|P_\mu\rangle\rangle$ gives probabilities of a *final* configuration conditioned on a value of the mean current. If we count the current on the last bond, and we want a high current, the best way for the system to do that is to do a final rush, and make all the particles leave (through that last bond) just before the time is up. The most probable final configuration will therefore be completely empty. But if we count the current on the first bond, the system had better fill itself up (through that first bond) just before the end, and the most probable final configuration is now completely full.

What doesn't depend on $\{\mu_i\}$, however, is the probability of a configuration at some point far enough from the beginning or the end of the run: the coefficients from the similarity matrix used on $\|P_\mu\rangle\!\rangle$ compensate with their inverses from $\langle\!\langle \tilde{P}_\mu\|$ (because that matrix is diagonal), so that, for any configuration $\mathcal{C}$, the probability $\tilde{P}_\mu(\mathcal{C})P_\mu(\mathcal{C})$ is independent of $\{\mu_i\}$. That is why we will only consider this quantity, and not the other two, to be physically relevant.

## III.2 Perturbative matrix Ansatz for the open ASEP

Let us now look at some of the main results of this thesis: the definition and proof of the perturbative matrix Ansatz.

This section is largely based on [3]. We first define all the necessary objects to construct our Ansatz, then we present the Ansatz itself, give its proof, and an alternative formulation which will be useful for carrying out calculations in section III.4.

### III.2.1 Definitions

We recall that we have defined, for the matrix Ansatz for the steady state of the open ASEP (section II.2.1), matrices $D$ and $E$ and vectors $\langle\!\langle W\|$ and $\|V\rangle\!\rangle$ such that:

$$
\begin{aligned}
DE - q\,ED &= (1-q)\,(D+E), \\
\langle\!\langle W\|(\alpha E - \gamma D) &= (1-q)\langle\!\langle W\|, \\
(\beta D - \delta E)\|V\rangle\!\rangle &= (1-q)\|V\rangle\!\rangle,
\end{aligned}
\tag{III.13}
$$

and we rewrote those equations in terms of $d$ and $e$ defined by:

$$
\begin{aligned}
D &= 1 + d, \\
E &= 1 + e
\end{aligned}
\tag{III.14}
$$

obtaining:

$$de - q\,ed = (1-q), \tag{III.15}$$

$$\langle\!\langle W\|\,[\alpha(1+e) - \gamma(1+d)] = (1-q)\langle\!\langle W\|, \tag{III.16}$$

$$[\beta(1+d) - \delta(1+e)]\,\|V\rangle\!\rangle = (1-q)\|V\rangle\!\rangle. \tag{III.17}$$

We also noticed that the algebra verified by matrices $d$ and $e$ is that of a q-deformed harmonic oscillator, where $e$ is the creation operator, and $d$ the annihilation operator.

We will now need another matrix $A_\mu$, defined by:

$$eA_\mu = \mathrm{e}^\mu\,A_\mu e, \tag{III.18}$$

$$A_\mu d = \mathrm{e}^\mu\,dA_\mu, \tag{III.19}$$

and two more vectors $\langle\!\langle \tilde{W} \|$ and $\| \tilde{V} \rangle\!\rangle$ such that:

$$\langle\!\langle \tilde{W} \| \, [\alpha(1-e) - \gamma(1-d)] = 0, \tag{III.20}$$

$$[\beta(1-d) - \delta(1-e)] \, \| \tilde{V} \rangle\!\rangle = 0. \tag{III.21}$$

We finally construct two transfer matrices $T_\mu$ and $U_\mu$ acting on the same space $M_\mu$ (i.e. on the configurations of the ASEP). For that, we first need to define a $2 \times 2$ matrix $X$, the entries of which are matrices $d$, $e$ or $1$, and is given, in basis $\{0,1\}$, by:

$$\boxed{X = \begin{bmatrix} 1 & e \\ d & 1 \end{bmatrix}} \tag{III.22}$$

(the sum of the rows of $X$ gives back the vector that was also called $X$ in section II.2.1).

This matrix is the building block of $U_\mu$ and $T_\mu$ (just as the vector was for the matrix Ansatz), which we define by expressing the transfer weight from a configuration $\mathcal{C}' = \{n'_i\}$ to a configuration $\mathcal{C} = \{n_i\}$ of the ASEP:

$$\boxed{U_\mu(\mathcal{C}, \mathcal{C}') = \frac{1}{Z_L} \langle\!\langle W \| A_\mu \prod_{i=1}^{L} X_{n_i, n'_i} \| V \rangle\!\rangle} \tag{III.23}$$

with $Z_L = \langle\!\langle W \| (2 + d + e)^L \| V \rangle\!\rangle$, and

$$\boxed{T_\mu(\mathcal{C}, \mathcal{C}') = \langle\!\langle \tilde{W} \| A_\mu \prod_{i=1}^{L} X_{n_i, n'_i} \| \tilde{V} \rangle\!\rangle.} \tag{III.24}$$

The construction of those matrices is similar to that of the steady state vector of the original matrix Ansatz: each weight is a product of matrices between two vectors, ordered from one boundary to the other, and each matrix depends only on the occupancies of the corresponding site. Since it is a transfer matrix, and not a vector, there is, for each site, an initial occupancy and a final occupancy. The matrix to be placed in the product is $d$ if the occupancy goes from 0 to 1, $e$ if it goes from 1 to 0, and the identity if the occupancy is unchanged, which is to say that matrices $d$ and $e$ act on the space of configurations as, respectively, creation and annihilation operators. Since it is the opposite as in their internal space (i.e. that of the q-deformed harmonic oscillator), it can also be said that $d$ transfers a particle from the oscillator to a site, and $e$ transfers a particle from a site to the oscillator.

The weights of those two matrices are entirely determined by the algebra defined above. We may note that the matrix $A_\mu$ is set between the left boundary vector and the first matrix because it is the bond between the left reservoir and the first site over which we count the current. For a general set of parameters $\{\mu_i\}$, we would have to add a matrix $A_{\mu_i}$ on each appropriate bond, i.e. between $X_{n_i, n'_i}$ and $X_{n_{i+1}, n'_{i+1}}$, in both of the

products above, so that:

$$U_{\{\mu_i\}}(\mathcal{C}, \mathcal{C}') = \frac{1}{Z_L} \langle\!\langle W \| A_{\mu_0} \prod_{i=1}^{L} X_{n_i, n_i'} A_{\mu_i} \| V \rangle\!\rangle, \tag{III.25}$$

$$T_{\{\mu_i\}}(\mathcal{C}, \mathcal{C}') = \langle\!\langle \tilde{W} \| A_{\mu_0} \prod_{i=1}^{L} X_{n_i, n_i'} A_{\mu_i} \| \tilde{V} \rangle\!\rangle. \tag{III.26}$$

### III.2.2 Statement of the perturbative matrix Ansatz

We give here the main results pertaining to those two transfer matrices $U_\mu$ and $T_\mu$. The derivation of those results will be done in the next section.

First of all, their product commutes with the deformed Markov matrix:

$$\boxed{[M_\mu, U_\mu T_\mu] = 0.} \tag{III.27}$$

Furthermore, for $\mu = 0$, i.e. for no deformation with respect to the current, $T_0$ is a projector onto the constant vector $\{1\}_i$ on both sides (i.e. a matrix with all entries equal to 1), and $U_0 T_0$ is a projector onto the principal eigenvectors of $M$, which allows to recover the original matrix Ansatz:

$$T_0 = [1]_{\mathcal{C},\mathcal{C}'} = |1\rangle\langle 1|, \tag{III.28}$$

$$U_0|1\rangle = |P^\star\rangle. \tag{III.29}$$

From this, it follows that $U_\mu T_\mu$ is a quasi-projector onto the principal eigenvectors of $M_\mu$:

$$U_\mu T_\mu \sim |P_\mu\rangle\langle \tilde{P}_\mu| + \mathcal{O}(\mu) \tag{III.30}$$

up to a multiplicative constant, so that when used repeatedly, for instance on $|P^\star\rangle$, it allows to access increasingly precise approximations in powers of $\mu$ of those vectors. Put into equations, this reads:

$$|P_\mu\rangle = \frac{1}{Z_L^{(k)}} (U_\mu T_\mu)^k |P^\star\rangle + \mathcal{O}\left(\mu^{k+1}\right), \tag{III.31}$$

$$\langle \tilde{P}_\mu| = \frac{1}{Z_L^{(k)}} \langle 1|(U_\mu T_\mu)^k + \mathcal{O}\left(\mu^{k+1}\right), \tag{III.32}$$

where $Z_L^{(k)} = \langle 1|(U_\mu T_\mu)^k|P^\star\rangle$.

Finally, by applying $M_\mu$ to (III.31), one obtains the corresponding approximation for $E(\mu)$:

$$\boxed{E(\mu) = \frac{\langle 1|M_\mu (U_\mu T_\mu)^k|P^\star\rangle}{\langle 1|(U_\mu T_\mu)^k|P^\star\rangle} + \mathcal{O}\left(\mu^{k+2}\right)} \tag{III.33}$$

which means that the ratio of matrix products on the right hand side of the equation yields the cumulants of the current up to order $k + 1$.

Those results hold for any integer $k$, so that, in essence, we have complete exact expressions for $|P_\mu\rangle$, $\langle \tilde{P}_\mu|$ and $E(\mu)$, expanded as infinite series in $\mu$, which is why we described this Ansatz as perturbative.

Let us also note that this Ansatz applies just as well to the periodic ASEP, by replacing $U_\mu T_\mu$ by a single matrix $T_\mu^{per}$ defined, if the marked bond is the one between sites $L$ and 1, as:

$$T_\mu^{per}(\mathcal{C}, \mathcal{C}') = \frac{1}{Z_L} Tr\big[A_\mu \prod_{i=1}^{L} X_{n_i, n_i'}\big] \qquad \text{(III.34)}$$

which is to say that a trace replaces the boundary vectors, consistently with the translational invariance of the system.

### III.2.3 Validation of the perturbative matrix Ansatz

This section contains the proofs of all the previous statements. We first verify the commutation relation (III.27). The technical part of the proof can be found in [3] (which is appended at the end of this manuscript). All the other results stem out from the same simple observation (III.30), and are taken care of next.

**Commutation of $U_\mu T_\mu$ and $M_\mu$**

Relation (III.27) is the main reason why our construction is of interest: since $U_\mu T_\mu$ commutes with $M_\mu$, it has the same eigenvectors, and we can use that to obtain information on $M_\mu$ without having to diagonalise it directly.

Before proving this relation, we first need to set a few notations. For convenience, we will here write $U_\mu$ and $T_\mu$ as:

$$U_\mu = \frac{1}{Z_L} \langle\!\langle W \| A_\mu \prod_{i=1}^{L} X^{(i)} \| V \rangle\!\rangle, \qquad \text{(III.35)}$$

$$T_\mu = \langle\!\langle \tilde{W} \| A_\mu \prod_{i=1}^{L} X^{(i)} \| \tilde{V} \rangle\!\rangle, \qquad \text{(III.36)}$$

with

$$X^{(i)} = \begin{bmatrix} 1 & e \\ d & 1 \end{bmatrix} \qquad \text{(III.37)}$$

as we did in section II.2.1. As previously, the product symbol in these expressions represents a matrix product in the auxiliary space, and a tensor product in configuration space. All the matrices $X^{(i)}$ are identical, but they act on different sites, so we will need to keep track of their position in the matrix products, hence the superscript.

Let us now define the 'hat' matrices $\hat{X}$ by $\hat{X}_{n,n'} = (-1)^n \frac{(1-q)}{2} X_{n,n'}$:

$$\hat{X}^{(i)} = \frac{(1-q)}{2} \begin{bmatrix} 1 & e \\ -d & -1 \end{bmatrix} = \frac{(1-q)}{2} \begin{bmatrix} 1 & 0 \\ 0 & -1 \end{bmatrix} \cdot X^{(i)} \qquad \text{(III.38)}$$

(where we note $\cdot$ the product in the 2-dimensional space corresponding to the occupancy number on one site).

The first step in proving (III.27) is to see how the matrix products react to commuting with one of the local jump matrices $M_i$. Since $M_i$ acts only on sites $i$ and $i+1$, only the part $X^{(i)}X^{(i+1)}$ of the matrix products is affected. If a fugacity is set between those sites, $M_i$ and $X^{(i)}X^{(i+1)}$ become $M_i(\mu_i)$ and $X^{(i)}A_{\mu_i}X^{(i+1)}$, and we will write the general relation for the latter, the former being simply the case $\mu_i = 0$. What we find is:

$$\boxed{[M_i(\mu_i), X^{(i)}A_{\mu_i}X^{(i+1)}] = \hat{X}^{(i)}A_{\mu_i}X^{(i+1)} - X^{(i)}A_{\mu_i}\hat{X}^{(i+1)}} \qquad \text{(III.39)}$$

and the proof of this can be found in [3] (cf. end of the manuscript) or in section V.1.1. This equation is related to (II.37), and will be used in the same way.

We may note that this relation is somewhat similar to an integration by parts, or to its discrete equivalent. $M_i(\mu_i)$ can be seen as a local differential operator (which it is in the continuous limit), and the difference (rather than the sum) of two partial differentiations, one to the right and one to the left, gives the difference of two boundary terms, one at $i$ and one at $i+1$, where matrices $X$ are replaced by $\hat{X}$. This relation also verifies Chasles' theorem, so that, by applying all of the bulk matrices $M_i$ or their deformed relatives, the boundary terms cancel one another except for the extremal ones. In other words, we have:

$$\left[\sum_{i=1}^{L-1} M_i, \prod_{i=1}^{L} X^{(i)}\right] = \hat{X}^{(1)} \prod_{i=2}^{L} X^{(i)} - \prod_{i=1}^{L-1} X^{(i)} \hat{X}^{(L)} \qquad \text{(III.40)}$$

(this is written without fugacities, but the same is true with a general set of $\mu_i$, provided matrices $A_{\mu_i}$ are put where needed in the matrix product).

These calculations, involving 'hat matrices' cancelling one another, have also appeared in [121] to validate the matrix Ansatz for the steady state of the multispecies ASEP on a ring.

This gives us two independent relations for the commutation of the bulk part of $M_\mu$ with $U_\mu$ and $T_\mu$:

$$\left[\sum_{i=1}^{L-1} M_i, U_\mu\right] = \frac{1}{Z_L} \langle\!\langle W \| A_\mu \hat{X}^{(1)} \prod_{i=2}^{L} X^{(i)} \| V \rangle\!\rangle - \frac{1}{Z_L} \langle\!\langle W \| A_\mu \prod_{i=1}^{L-1} X^{(i)} \hat{X}^{(L)} \| V \rangle\!\rangle, \qquad \text{(III.41)}$$

$$\left[\sum_{i=1}^{L-1} M_i, T_\mu\right] = \langle\!\langle \tilde{W} \| A_\mu \hat{X}^{(1)} \prod_{i=2}^{L} X^{(i)} \| \tilde{V} \rangle\!\rangle - \langle\!\langle \tilde{W} \| A_\mu \prod_{i=1}^{L-1} X^{(i)} \hat{X}^{(L)} \| \tilde{V} \rangle\!\rangle, \qquad \text{(III.42)}$$

which we may write as:

$$\left[\sum_{i=1}^{L-1} M_i, U_\mu\right] = \hat{U}_\mu^{(1)} - \hat{U}_\mu^{(L)}, \qquad \text{(III.43)}$$

$$\left[\sum_{i=1}^{L-1} M_i, T_\mu\right] = \hat{T}_\mu^{(1)} - \hat{T}_\mu^{(L)}, \qquad \text{(III.44)}$$

so that:

$$\left[\sum_{i=1}^{L-1} M_i, U_\mu T_\mu\right] = \hat{U}_\mu^{(1)} T_\mu - \hat{U}_\mu^{(L)} T_\mu + U_\mu \hat{T}_\mu^{(1)} - U_\mu \hat{T}_\mu^{(L)}. \qquad \text{(III.45)}$$

The last step is to check that

$$[m_0(\mu), U_\mu T_\mu] = -\hat{U}_\mu^{(1)} T_\mu - U_\mu \hat{T}_\mu^{(1)}, \qquad (\text{III.46})$$

$$[m_L, U_\mu T_\mu] = \hat{U}_\mu^{(L)} T_\mu + U_\mu \hat{T}_\mu^{(L)}. \qquad (\text{III.47})$$

As before, $m_0$ acts only on site 1, so that only $X^{(1)}$ is affected, and the same goes for $m_L$ and site $L$. These relations are proven in [3] (cf. end of the manuscript) and in section V.2.1.

Combining all these equations, we see that the terms from commutation in the bulk are exactly cancelled out by those from each boundary, which proves (III.27). Let us finally note that both $U_\mu$ and $T_\mu$ are necessary in (III.46) and (III.47), so that the commutation does not hold independently for the two matrices, but only for their product.

**Projectors and perturbative expansion in $\mu$**

An important remark that needs to be made at this point is that the representation we use for $d$ and $e$ in $U_\mu$ is not necessarily the same as that for $T_\mu$. Specifically, for $\mu = 0$, one particular solution to (III.15), (III.20) and (III.21) for the elements of $T_0$ is $d = e = A_0 = 1$, so that for any $\mathcal{C}'$ and $\mathcal{C}$, we have $T_0(\mathcal{C}, \mathcal{C}') = \langle\!\langle \tilde{W} \| \tilde{V} \rangle\!\rangle$ which we can set to 1. This proves (III.28). The same does not work for $U_0$, because it is not compatible with equations (III.16) and (III.17).

Furthermore, projecting $U_0$ onto $|1\rangle$ means summing over all configurations $\mathcal{C}'$ in (III.23), so that

$$\langle \mathcal{C} | U_0 | 1 \rangle = \sum_{\mathcal{C}'} U_\mu(\mathcal{C}, \mathcal{C}') = \langle\!\langle W \| A_0 \prod_{i=1}^{L} (X_{n_i, 0} + X_{n_i, 1}) \| V \rangle\!\rangle. \qquad (\text{III.48})$$

We can set $A_0$ to 1, and remark that for $n_i = 0$, we have $(X_{n_i,0} + X_{n_i,1}) = 1 + e = E$ and that for $n_i = 1$, we have $(X_{n_i,0} + X_{n_i,1}) = d + 1 = D$, so that this expression is exactly that of $P^\star(\mathcal{C})$ as given in eq.(II.30), which proves (III.29).

Using relations (III.27), (III.28) and (III.29) together, at $\mu = 0$, we get

$$[M, U_0 T_0] = 0 = \left( M | P^\star \rangle \langle 1 | \right) - \left( | P^\star \rangle \langle 1 | M \right). \qquad (\text{III.49})$$

Since we know that $\langle 1 | M = 0$ (because $M$ is a stochastic matrix), this implies that $M | P^\star \rangle = 0$, which yields the original matrix Ansatz that we saw in section II.2.1.

This alternative proof of (II.30) relies on the fact that the transfer matrix $U_\mu T_\mu$ is a projector in the limit $\mu \to 0$. It would be interesting to determine whether for other situations with matrix product states, one can generically find a transfer matrix that commutes with a deformation of the dynamics of the system and is a projector in the non-deformed limit. One could for instance look at the ASEP in discrete time with different versions of the update [128], or at the multispecies ASEP on a ring [121].

To prove (III.31) and (III.32), we use the relations derived above. Since, for $\mu = 0$, the matrix $U_0 T_0$ is the projector onto the principal eigenspace of $M$, one can write, for an infinitesimal $\mu$:

$$U_\mu T_\mu = \Lambda_\mu | P_\mu \rangle \langle \tilde{P}_\mu | + r_\mu \qquad (\text{III.50})$$

where $\Lambda_\mu \sim 1 + \mathcal{O}(\mu)$ is the largest eigenvalue of $U_\mu T_\mu$, and $r_\mu \sim \mathcal{O}(\mu)$ is the part of $U_\mu T_\mu$ that is orthogonal to its principal eigenspace (i.e. orthogonal to $|P_\mu\rangle\langle\tilde{P}_\mu|$), and has eigenvalues of order $\mu$. In other words, $U_\mu T_\mu$ is almost a projector, with an error $r_\mu$ of order $\mu$.

Since $r_\mu|P_\mu\rangle = 0$ and $\langle\tilde{P}_\mu|r_\mu = 0$, one has that:

$$(U_\mu T_\mu)^k = \Lambda_\mu^k |P_\mu\rangle\langle\tilde{P}_\mu| + r_\mu^k \tag{III.51}$$

so that the difference from the projector onto $|P_\mu\rangle\langle\tilde{P}_\mu|$ is now $r_\mu^k \sim \mathcal{O}(\mu^k)$.

Let us now remark that the parts of $|P^\star\rangle$ and of $\langle 1|$ which are not in the kernel of $r_\mu$ are of order $\mu$, so that both $r_\mu^k|P^\star\rangle$ and $\langle 1|r_\mu^k$ are of order $\mu^{k+1}$. It follows that $(U_\mu T_\mu)^k|P^\star\rangle$ is proportional to $|P_\mu\rangle$ with an error of order $\mu^{k+1}$ (and the same goes for $\langle\tilde{P}_\mu|$), which proves (III.31).

Equation (III.33) is then proven by simply applying $M_\mu$ to $(U_\mu T_\mu)^k|P^\star\rangle$:

$$\langle 1|M_\mu (U_\mu T_\mu)^k|P^\star\rangle = E(\mu)\Lambda_\mu^k \langle 1|P_\mu\rangle\langle\tilde{P}_\mu|P^\star\rangle + \langle 1|M_\mu r_\mu^k|P^\star\rangle, \tag{III.52}$$

$$\langle 1|(U_\mu T_\mu)^k|P^\star\rangle = \Lambda_\mu^k \langle 1|P_\mu\rangle\langle\tilde{P}_\mu|P^\star\rangle + \langle 1|r_\mu^k|P^\star\rangle, \tag{III.53}$$

where $\langle 1|M_\mu r_\mu^k|P^\star\rangle$ is of order $\mu^{k+2}$ because $\langle 1|M_\mu$ is of order $\mu$ and $r_\mu^k|P^\star\rangle$ is of order $\mu^{k+1}$, and $\langle 1|r_\mu^k|P^\star\rangle$ is of order $\mu^{k+2}$ for the reason given above. The ratio of those two equations is therefore equal to $E(\mu)$ up to order $\mu^{k+1}$.

### III.2.4 Formulation as a matrix product

In this section, we give a different formulation of equations (III.31) and (III.32), in terms of a matrix product akin to (II.30) rather than a product of transfer matrices. This alternative formulation is particularly useful for the explicit calculations of the cumulants of the current that we will carry out in section III.4.

The main point that needs to be made here is that, unlike the Markov matrix, which is a sum of elementary matrices, the transfer matrices $U_\mu$ and $T_\mu$ are products of the elementary matrices $X_{n,n'}$, so that a product of those transfer matrices can be seen as a tensor network, the tensors being of order four: $[X_{n,n'}]_{i,j}$, where $i$ and $j$ are the internal indices of $X$ (cf. fig.-III.2).

A consequence of this is that the object $(U_\mu T_\mu)^k|P^\star\rangle$ can be written in terms of the columns of this tensor network instead of the lines. Let us therefore define, by recursion (and denote the product between successive rows by a tensor product $\otimes$):

$$D^{(k+1)} = (1 \otimes 1 + d \otimes e) \otimes D^{(k)} + (1 \otimes d + d \otimes 1) \otimes E^{(k)}, \tag{III.54}$$

$$E^{(k+1)} = (1 \otimes 1 + e \otimes d) \otimes E^{(k)} + (e \otimes 1 + 1 \otimes e) \otimes D^{(k)}, \tag{III.55}$$

$$A_\mu^{(k+1)} = A_\mu \otimes A_\mu \otimes A_\mu^{(k)}, \tag{III.56}$$

with $D_0 = D$, $E_0 = E$ and $A_\mu^{(0)} = 1$, and

$$\|V^{(k+1)}\rangle\!\rangle = \|V\rangle\!\rangle \otimes \|\tilde{V}\rangle\!\rangle \otimes \|V^{(k)}\rangle\!\rangle, \tag{III.57}$$

$$\langle\!\langle W^{(k+1)}\| = \langle\!\langle W\| \otimes \langle\!\langle \tilde{W}\| \otimes \langle\!\langle W^{(k)}\|, \tag{III.58}$$

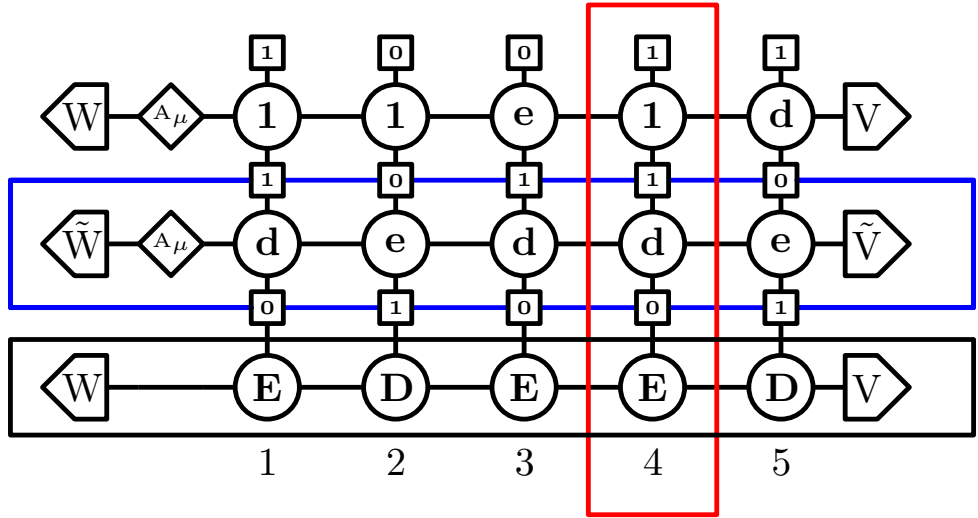

Figure III.2: One of the tensor networks that compose $U_\mu T_\mu |P^\star\rangle$ when expanded in terms of the intermediate configurations at each step of the product. Lines represent the transfer matrices $U_\mu$ and $T_\mu$, whereas columns represent $E^{(k)}$ or $D^{(k)}$. See the detailed explanation below.

with $\|V_0\rangle\rangle = \|V\rangle\rangle$ and $\langle\langle W_0\| = \langle\langle W\|$.

In this formalism, equation (III.31) becomes:

$$\langle\mathcal{C}|P_\mu\rangle = \frac{1}{Z_L^{(k)}} \langle\langle W^{(k)} \| A_\mu^{(k)} \prod_{i=1}^{L} \left(n_i D^{(k)} + (1-n_i) E^{(k)}\right) \|V^{(k)}\rangle\rangle + \mathcal{O}\left(\mu^{k+1}\right). \qquad \text{(III.59)}$$

We can do just the same with (III.32), building the matrices and vectors recursively towards the right rather than the left.

To give a simple explicit example, in fig.-III.2 is shown one of the tensor networks which compose $U_\mu T_\mu |P^\star\rangle$ for $L=5$, namely $U_\mu(\mathcal{C},\mathcal{C}')T_\mu(\mathcal{C}',\mathcal{C}'')P^\star(\mathcal{C}'')$, where we chose $\mathcal{C} = \{1,0,0,1,1\}$, $\mathcal{C}' = \{1,0,1,1,0\}$ and $\mathcal{C}'' = \{0,1,0,0,1\}$. The blue rectangle corresponds to $T_\mu(\mathcal{C}',\mathcal{C}'')$, the black one to $P^\star(\mathcal{C}'')$, and the red rectangle is one of the elements in $D^{(1)}$, namely $X_{1,1} \otimes X_{1,0} \otimes E$. Summing over the second and third indices in any column gives $E^{(1)}$ or $D^{(1)}$, depending on whether the first (upper) index is 0 or 1.

The equivalent form for the periodic case is the same, with a trace instead of the boundary vectors, and with only one tensor product per stage of the recursion, starting from $D^{(0)} = E^{(0)} = 1$, so that matrices of order $k$ in the open case correspond to matrices of order $2k+1$ in the periodic case.

While the transfer matrix formulation (III.31) is better suited to the algebraic proof of the Ansatz, this matrix product formulation (III.59) is useful for doing explicit calculations, like those of the cumulants of the current (cf. section III.4). Naturally, all the calculations done here using the transfer matrix formulation can be done with these matrix products (and were, for the most part, done that way in order to obtain the results in [1,2]), but are much more cumbersome and convoluted. Let us also note that the matrices $E^{(k)}$ and $D^{(k)}$ are related to the ones used in the matrix Ansatz solution of the multispecies periodic ASEP [121].

## III.3 Cumulants of the Current

Using the matrix product formulation of our perturbative matrix Ansatz, we can infer, from low order calculations, the general formula for the generating function of the cumulants of the current in the open ASEP. In this section, we present the final result, and what it reduces to for the simpler case of the TASEP. Everything in this section was at first conjectured from the calculations we present in the next one, and then proven through what can be found in chapter V.

We first recall a few definitions that we used in section II.2. From section II.2.1, we need:

$$a = \frac{1}{2\alpha}\Big[(1 - q - \alpha + \gamma) + \sqrt{(1 - q - \alpha + \gamma)^2 + 4\alpha\gamma}\Big], \tag{III.60}$$

$$\tilde{a} = \frac{1}{2\alpha}\Big[(1 - q - \alpha + \gamma) - \sqrt{(1 - q - \alpha + \gamma)^2 + 4\alpha\gamma}\Big], \tag{III.61}$$

$$b = \frac{1}{2\beta}\Big[(1 - q - \beta + \delta) + \sqrt{(1 - q - \beta + \delta)^2 + 4\beta\delta}\Big], \tag{III.62}$$

$$\tilde{b} = \frac{1}{2\beta}\Big[(1 - q - \beta + \delta) - \sqrt{(1 - q - \beta + \delta)^2 + 4\beta\delta}\Big], \tag{III.63}$$

and

$$\boxed{F(z) = \frac{(1+z)^L (1+z^{-1})^L (z^2, z^{-2})_\infty}{(az, a/z, \tilde{a}z, \tilde{a}/z, bz, b/z, \tilde{b}z, \tilde{b}/z)_\infty}.} \tag{III.64}$$

From section II.2.2, we redefine the convolution kernel $K$, as:

$$K(z, \tilde{z}) = 2 \sum_{k=1}^{\infty} \frac{q^k}{1 - q^k} \Big( (z/\tilde{z})^k + (z/\tilde{z})^{-k} \Big) \tag{III.65}$$

and the associated convolution operator $X$:

$$X[f](z) = \oint_S \frac{d\tilde{z}}{i 2\pi \tilde{z}} f(\tilde{z}) K(z, \tilde{z}) \tag{III.66}$$

where $S = \{0, q^k a, q^k \tilde{a}, q^k b, q^k \tilde{b}\}$ for $k$ in $\mathbb{N}$. Note that there is an extra factor 2 in $K$, and that the integration domain of $X$ has changed.

If we now define the function $W(z)$ such that:

$$\boxed{W(z) = -\frac{1}{2} \ln\Big(1 - BF(z) e^{X[W](z)}\Big)} \tag{III.67}$$

we find that:

$$\boxed{\mu = -\oint_S \frac{dz}{i2\pi z} W(z) = -\sum_{k=1}^\infty C_k \frac{B^k}{k}} \qquad (\text{III.68})$$

and

$$\boxed{E(\mu) = -(1-q)\oint_S \frac{dz}{i2\pi(1+z)^2} W(z) = -(1-q)\sum_{k=1}^\infty D_k \frac{B^k}{k}.} \qquad (\text{III.69})$$

The form of this result is exactly the same as what we found for the periodic ASEP in section II.2.2. The only differences are a factor 2 in $X$, a factor $\frac{1}{2}$ in front of the logarithm in the definition of $W(z)$, and $h(z)$ being replaced by the function $F(z)$ that we used in II.2.1 for the mean current (with all the contour integrals being taken over the appropriate integration domain instead of the unit circle). As before, the coefficients $C_k$ and $D_k$ can be expressed in terms of combinations of $F$ and $K$, such as, for instance:

$$C_2 = \frac{1}{2}\oint_S \frac{dz}{i2\pi z} F(z)^2 \;+\; \frac{1}{2}\oint_S \frac{dz_1}{i2\pi z_1}\oint_S \frac{dz_2}{i2\pi z_2} F(z_1)F(z_2)K(z_1,z_2) \qquad (\text{III.70})$$

and a few more explicit examples can be found in section IV.1. The cumulants of the current can then be obtained by inverting eq.(III.68) order by order and injecting the result in eq.(III.69).

Let us note that if we choose the boundary parameters to be $a=1$, $\tilde{a}=-q$, $b=\sqrt{q}$ and $\tilde{b}=-\sqrt{q}$, which is to say $\alpha=\frac{1}{2}$, $\gamma=\frac{q}{2}$, $\beta=1$ and $\delta=q$, we find that $(1+z)(az,\tilde{a}z,bz,\tilde{b}z)_\infty = (z^2)_\infty$ and $(1+z^{-1})(a/z,\tilde{a}/z,b/z,\tilde{b}/z)_\infty = (z^{-2})_\infty$, so that $F(z)$ reduces to $(1+z)^{L+1}(1+z^{-1})^{L+1}$. This is the same as the function $h$ for the periodic ASEP with $2L+2$ sites and $L+1$ particles. Because of the extra factors 2 and $\frac{1}{2}$, the generating function of the cumulants of the current is half that which we found in the periodic case, taken at $2\mu$. This also works if we exchange $a$ with $b$ and $\tilde{a}$ with $\tilde{b}$. Those two special points correspond to $\rho_a=\frac{1}{2}$ and $1-\rho_b=\frac{1}{1+q}$, or the opposite, and are on the transition lines between the MC phase and the LD or HD phase. We will come back to this remark in section IV.4.

In the simpler case of the TASEP, we have $K=0$, and we find:

$$C_k = \frac{1}{2}\oint_{\{0,a,b\}} \frac{dz}{i2\pi z} F^k(z) \qquad (\text{III.71})$$

and

$$D_k = \frac{1}{2}\oint_{\{0,a,b\}} \frac{dz}{i2\pi(1+z)^2} F^k(z) \qquad (\text{III.72})$$

where $F(z)$ reduces to:

$$F(z) = \frac{(1+z)^L(1+z^{-1})^L(1-z^2)(1-z^{-2})}{(1-az)(1-\frac{a}{z})(1-bz)(1-\frac{b}{z})} \qquad (\text{III.73})$$

with $a=\frac{1-\alpha}{\alpha}$ and $b=\frac{1-\beta}{\beta}$.

In the even simpler case where $\alpha = \beta = 1$, so that $a = b = 0$, we can do the integrals explicitly, and find:

$$\mu = -\sum_{k=1}^{\infty} \frac{(2k)!}{k!} \frac{[2k(L+1)]!}{[k(L+1)]! \, [k(L+2)]!} \frac{B^k}{2k}, \tag{III.74}$$

$$E(\mu) = -\sum_{k=1}^{\infty} \frac{(2k)!}{k!} \frac{[2k(L+1)-2]!}{[k(L+1)-1]! \, [k(L+2)-1]!} \frac{B^k}{2k}. \tag{III.75}$$

In chapter IV, we take the large $L$ asymptotics of these formulae to see what we can learn on the behaviour of the system. But first, let us see how we can infer their expression from a few low-order calculations.

## III.4 Appendix - low-order calculations and conjecture

In this section, we make use of formula (III.33) to calculate a few of the first cumulants of the current, first for the TASEP, then for the ASEP. We then use these calculations, along with what we know of the corresponding results in the periodic case, to guess the general formula for the generating function of the cumulants of the current. Finally, we validate that formula using exact numerical calculations on systems of small size, and DMRG calculations provided by M. Gorissen and C. Vanderzande.

Our starting point is equation (III.33):

$$E(\mu) = \frac{\langle 1|M_\mu (U_\mu T_\mu)^k|P^\star\rangle}{\langle 1|(U_\mu T_\mu)^k|P^\star\rangle} + \mathcal{O}\left(\mu^{k+2}\right) = \sum_{l=1}^{k+1} E_l \frac{\mu^l}{l!} + \mathcal{O}\left(\mu^{k+2}\right). \tag{III.76}$$

Our goal is to extract, from this formula, an explicit expression for the cumulants of the current. In principle, this is easy enough: one simply has to take equation (III.76) for every $k$ and keep only the coefficient of order $k+1$ in $\mu$. Unfortunately, the ratio of two matrix products in that equation makes this calculation extremely difficult: it's not enough to calculate just one coefficient in the numerator and one the denominator to get a certain order for the ratio. It so happens that the simplest coefficient to calculate for any expression of the form $\langle 1|(U_\mu T_\mu)^k|P^\star\rangle$ is the one of order $k$ in $\mu$, and the simplest in $\langle 1|M_\mu (U_\mu T_\mu)^k|P^\star\rangle$ is that of order $k+1$. As we said, getting just those will give us only partial information on the corresponding cumulant, but we will see that it will actually be enough to make a guess, which we can then check against numerical data.

More precisely, what we will do now is the following: for the case of the periodic TASEP, we will calculate the coefficient of order $k+1$ in the numerator of eq.(III.76) (with $U_\mu T_\mu$ replaced by $T_\mu^{per}$) and that of order $k$ in the numerator, for small values of $k$, and compare what we get with the exact expressions (II.110) and (II.111), which we

recall here:

$$E(\mu) = -\sum_{k=1}^{\infty} D_k \frac{B^k}{k}, \tag{III.77}$$

$$\mu = -\sum_{k=1}^{\infty} C_k \frac{B^k}{k}, \tag{III.78}$$

with

$$C_k = \oint_{c_1} \frac{dz}{\imath 2\pi z} \frac{(1+z)^{kL}}{z^{kN}} = \binom{kL}{kN}, \tag{III.79}$$

$$D_k = \oint_{c_1} \frac{dz}{\imath 2\pi (1+z)^2} \frac{(1+z)^{kL}}{z^{kN}} = \binom{kL-2}{kN-1}, \tag{III.80}$$

where $c_1$ is the unit circle.

We will find that a part of the denominator matches coefficient $C_k$, and that a part of the numerator matches $D_k$. We will then calculate the same coefficients from eq.(III.76) for the open TASEP, and find good candidates for $C_k$ and $D_k$, assuming that the structure of the answer is a double series in a parameter $B$ as well. Note that, here, this structure is entirely assumed, but we will see in chapter V that it naturally arises from the Bethe Ansatz, just as it did in the periodic case.

### III.4.1 Totally asymmetric case

We first focus our attention on the TASEP.

Let us recall the expressions of the objects we will be calculating in terms of matrix products. The denominators are given, in the open case, by:

$$\langle 1|(U_\mu T_\mu)^k|P^\star\rangle = \langle\!\langle W^{(k)}\| A_\mu^{(k)} (D^{(k)} + E^{(k)})^L \|V^{(k)}\rangle\!\rangle \tag{III.81}$$

and, in the periodic case, by

$$\langle 1|(T_\mu^{per})^k|P^\star\rangle = \oint_{c_1} \frac{dz_0}{\imath 2\pi z_0} \mathrm{Tr}[A_\mu^{(k)}(z_0 D^{(k)} + E^{(k)})^L] z_0^{-N} \tag{III.82}$$

and the numerators, in the open case, by

$$\langle 1|M_\mu(U_\mu T_\mu)^k|P^\star\rangle = (\mathrm{e}^\mu - 1)\langle\!\langle W^{(k)}\| A_\mu^{(k)} E^{(k)} (D^{(k)} + E^{(k)})^{L-1} \|V^{(k)}\rangle\!\rangle \tag{III.83}$$

and, in the periodic case, by

$$\langle 1|M_\mu(T_\mu^{per})^k|P^\star\rangle = (\mathrm{e}^\mu - 1)\oint_{c_1} \frac{dz_0}{\imath 2\pi z_0} \mathrm{Tr}[D^{(k)} A_\mu^{(k)} E^{(k)} (z_0 D^{(k)} + E^{(k)})^{L-2}] z_0^{-N+1}. \tag{III.84}$$

To obtain those last two expressions, one has to write $M_\mu = M + (\mathrm{e}^\mu - 1)M^+$, where $M^+$ is the off-diagonal part of the local matrix acting on the bond over which we count the current, and remember that $\langle 1|M = 0$.

In the expressions concerning the periodic case, the contour integral over $z_0$ serves to select only the configurations with $N$ particles.

In what follows, we will be writing the term of order $\mu^k$ in $\langle 1|(U_\mu T_\mu)^k|P^\star\rangle$ as $X_k$, and the terms of order $\mu^{k+1}$ in $\langle 1|M_\mu(U_\mu T_\mu)^k|P^\star\rangle$ as $Y_{k+1}$ (one order in $\mu$ will be coming from the factor $(e^\mu - 1)$).

For the TASEP, the algebra satisfied by $d$ and $e$ reduces to $de = 1$. The representation that we will be using for $d$ and $e$ is given by the matrices:

$$d = \sum_{n=1}^{\infty} \|n-1\rangle\!\rangle\langle\!\langle n\| = \begin{bmatrix} 0 & 1 & 0 & 0 & \cdots \\ 0 & 0 & 1 & 0 & \\ 0 & 0 & 0 & 1 & \\ 0 & 0 & 0 & 0 & \\ \vdots & & & & \ddots \end{bmatrix} \tag{III.85}$$

and

$$e = \sum_{n=0}^{\infty} \|n+1\rangle\!\rangle\langle\!\langle n\| = \begin{bmatrix} 0 & 0 & 0 & 0 & \cdots \\ 1 & 0 & 0 & 0 & \\ 0 & 1 & 0 & 0 & \\ 0 & 0 & 1 & 0 & \\ \vdots & & & & \ddots \end{bmatrix} \tag{III.86}$$

which can be considered as jump matrices for a random walk on a one-dimensional lattice with a wall at site $-1$.

The matrix $A_\mu$ can be written as:

$$A_\mu = (1 - e^{-\mu}) \sum_{n=0}^{\infty} e^{-n\mu} \|n\rangle\!\rangle\langle\!\langle n\| = (1 - e^{-\mu}) \begin{bmatrix} 1 & 0 & 0 & 0 & \cdots \\ 0 & e^{-\mu} & 0 & 0 & \\ 0 & 0 & e^{-2\mu} & 0 & \\ 0 & 0 & 0 & e^{-3\mu} & \\ \vdots & & & & \ddots \end{bmatrix} \tag{III.87}$$

where the factor $(1 - e^{-\mu})$ is there so that the trace of $A_\mu$ has a finite limit for $\mu = 0$.

Let us also recall that, in the periodic case, $D^{(k)}$ and $E^{(k)}$ are tensor products of order $k$, whereas in the open case they are of order $2k + 1$.

We will now try to determine $X_k$ and $Y_k$ order by order, first for the periodic case, and then for the open case.

**Periodic case - first order**

This one is easy: for the periodic case, $D^{(0)} = E^{(0)} = 1$, so that we find immediately:

$$\boxed{X_0 = \oint_{C_1} \frac{dz_0}{i2\pi z_0} \frac{(1+z_0)^L}{z_0^N}} \tag{III.88}$$

and

$$\boxed{Y_1 = \oint_{C_1} \frac{dz_0}{i2\pi z_0} \frac{(1+z_0)^{L-2}}{z_0^{N-1}}} \tag{III.89}$$

(the one order in $\mu$ corresponding to $Y_1$ comes from the prefactor $(e^\mu - 1)$ in (III.84)).

We recognise $C_1$ and $D_1$, but it's too early to be drawing conclusions.

**Periodic case - second order**

For the second order, we have only one row of matrices $d$ and $e$:

$$X_1 = \oint_{c_1} \frac{dz_0}{\imath 2\pi z_0} \mathrm{Tr}[A_\mu (z_0 D + E)^L] z_0^{-N} \bigg|_{\mu^1}. \tag{III.90}$$

Just as in section II.2.1, we need to look for eigenvectors of $(z_0 D + E)$. Let us define the plane wave vector

$$\|z\rangle\!\rangle = \sum_{n=0}^{\infty} z^n \|n\rangle\!\rangle. \tag{III.91}$$

The action of $d$ and $e$ on $\|z\rangle\!\rangle$ is as follows:

$$d\|z\rangle\!\rangle = z\|z\rangle\!\rangle \quad , \quad e\|z\rangle\!\rangle = \bar{z}(\|z\rangle\!\rangle - \|0\rangle\!\rangle) \tag{III.92}$$

where, since $z$ will always be on the unit circle, we write $\bar{z}$ for the inverse of $z$, which is more compact than $z^{-1}$ or $\frac{1}{z}$.

Because of that $\|0\rangle\!\rangle$ in the right side of this last equation, which is due to the wall in the random walk that $d$ and $e$ represent, we need to use the method of images to build the vector we seek. We ultimately find that:

$$(z_0 d + e)\Big(z\overline{z_0}\|z\overline{z_0}\rangle\!\rangle - \bar{z}\|\bar{z}\rangle\!\rangle\Big) = (z + z_0 \bar{z})\Big(z\overline{z_0}\|z\overline{z_0}\rangle\!\rangle - \bar{z}\|\bar{z}\rangle\!\rangle\Big) \tag{III.93}$$

so that:

$$(z_0 D + E)\Big(z\overline{z_0}\|z\overline{z_0}\rangle\!\rangle - \bar{z}\|\bar{z}\rangle\!\rangle\Big) = (1+z)(1+z_0 \bar{z})\Big(z\overline{z_0}\|z\overline{z_0}\rangle\!\rangle - \bar{z}\|\bar{z}\rangle\!\rangle\Big). \tag{III.94}$$

The closure identity for these combinations of plane waves is:

$$1 = \frac{1}{2}\oint_{c_1} \frac{dz}{\imath 2\pi z}\Big(z\overline{z_0}\|z\overline{z_0}\rangle\!\rangle - \bar{z}\|\bar{z}\rangle\!\rangle\Big)\Big(\bar{z}z_0\langle\!\langle\bar{z}z_0\| - z\langle\!\langle z\|\Big) \tag{III.95}$$

so that, if we inject it at the left of matrix $A_\mu$ in (III.90), we get:

$$X_1 = \frac{1}{2}\oint_{c_1}\!\!\oint \frac{dz_0}{\imath 2\pi z_0}\frac{dz}{\imath 2\pi z}\frac{(1+z)^L(1+z_0\bar{z})^L}{z_0^N}\Big(\bar{z}z_0\langle\!\langle\bar{z}z_0\|-z\langle\!\langle z\|\Big)A_\mu\Big(z\overline{z_0}\|z\overline{z_0}\rangle\!\rangle-\bar{z}\|\bar{z}\rangle\!\rangle\Big)\bigg|_{\mu^1}. \tag{III.96}$$

The right part of the argument of the integral is easily calculated:

$$\Big(\bar{z}z_0\langle\!\langle\bar{z}z_0\|-z\langle\!\langle z\|\Big)A_\mu\Big(z\overline{z_0}\|z\overline{z_0}\rangle\!\rangle-\bar{z}\|\bar{z}\rangle\!\rangle\Big) = (1-e^{-\mu})\Big(\sum_{n=0}^{\infty}(2-(z_0\bar{z}^2)^{n+1}-(\overline{z_0}z^2)^{n+1})e^{-n\mu}\Big) \tag{III.97}$$

and we need to take $\mu$ to order 1 in this expression.

This is where we start throwing terms away for no apparent reason. The part of this last expression which doesn't depend on $z$ (i.e. the factor 2) will produce a contribution to $X_1$ which is equal to $X_0^2$. We are not interested in this part. In what is left, $(1-e^{-\mu})$ is taken at order 1 in $\mu$, so that everything else is taken at $\mu = 0$. This gives us, up to a minus sign: $\sum_{k=-\infty}^{\infty}(z_0\bar{z}^2)^k$.

This series, in a contour integral over the unit circle, is a delta function:

$$\boxed{\sum_{k=-\infty}^{\infty} x^k = \delta(1-x)} \tag{III.98}$$

because, if applied to a function $f(x) = \sum_{k=-\infty}^{\infty} a_k x^k$, it gives:

$$\oint_{c_1} \frac{dx}{i2\pi x} \sum_{k=-\infty}^{\infty} a_k x^k \sum_{l=-\infty}^{\infty} x^l = \sum_{k=-\infty}^{\infty} a_k = f(1). \tag{III.99}$$

This tells us that we must take $z_0 = z^2$ in the left part of (III.96). In the end, we get:

$$\boxed{X_1 \sim \frac{1}{2} \oint_{c_1} \frac{dz}{i2\pi z} \frac{(1+z)^{2L}}{z^{2N}}} \tag{III.100}$$

where the $\sim$ symbol means nothing rigorous.

The reason we kept only the terms that we did is that they produce a contribution to $X_1$ equal to $C_2$, which is what we are looking for. If we're lucky, we might be able to find $C_k$ in every $X_{k-1}$ and $D_k$ in every $Y_k$.

Let's continue with $Y_2$:

$$Y_2 = (e^\mu - 1) \oint_{c_1} \frac{dz_0}{i2\pi z_0} \text{Tr}[DA_\mu E(z_0 D + E)^{L-2}] z_0^{-N+1} \bigg|_{\mu^2}. \tag{III.101}$$

Using the same closure identity as before, we get:

$$Y_2 = \frac{(e^\mu - 1)}{2} \oint \oint_{c_1} \frac{dz_0}{i2\pi z_0} \frac{dz}{i2\pi z} \frac{(1+z)^{L-2}(1+z_0\bar{z})^{L-2}}{z_0^{N-1}}$$
$$\left( \bar{z}z_0 \langle\!\langle \bar{z}z_0 \| - z \langle\!\langle z \| \right) DA_\mu E \left( z\overline{z_0} \| z\overline{z_0} \rangle\!\rangle - \bar{z} \| \bar{z} \rangle\!\rangle \right) \bigg|_{\mu^2} \tag{III.102}$$

and, as before, we need to deal with the right part.

We calculate:

$$DA_\mu E = A_\mu + dA_\mu + A_\mu e + dA_\mu e = A_\mu + e^\mu A_\mu d + e^\mu e A_\mu + e^\mu A_\mu \tag{III.103}$$

so that:

$$\left( \bar{z}z_0 \langle\!\langle \bar{z}z_0 \| - z \langle\!\langle z \| \right) DA_\mu E \left( z\overline{z_0} \| z\overline{z_0} \rangle\!\rangle - \bar{z} \| \bar{z} \rangle\!\rangle \right) =$$
$$(1 - e^{-\mu})\Big( \sum_{n=0}^{\infty} \big((1 + e^{-\mu} + e^{-\mu}\bar{z}z_0 + e^{-\mu}z\overline{z_0})$$
$$- (1 + e^{-\mu} + e^{-\mu}\bar{z}z_0 + e^{-\mu}\bar{z})(z_0\bar{z}^2)^{n+1}$$
$$- (1 + e^{-\mu} + e^{-\mu}z + e^{-\mu}z\overline{z_0})(\overline{z_0}z^2)^{n+1}$$
$$+ (1 + e^{-\mu} + e^{-\mu}z + e^{-\mu}\bar{z})\big)e^{-n\mu} \Big). \tag{III.104}$$

Just as before, we only keep the terms that are part of a delta function, and take $\mu$ to order 2, one from the prefactor $(e^\mu - 1)$ in $Y_2$ and one from the factor $(1 - e^{-\mu})$ in this last equation, the rest being taken at $\mu = 0$. We obtain:

$$\sum_{n=0}^{\infty}(2 + \bar{z}z_0 + \bar{z})(z_0\bar{z}^2)^{n+1} + (2 + z + z\overline{z_0})(\overline{z_0}z^2)^{n+1} \sim (2 + z + \bar{z})\delta(1 - z_0\bar{z}^2). \quad \text{(III.105)}$$

Putting this back into $Y_2$ and taking $z_0 = z^2$, we finally obtain:

$$\boxed{Y_2 \sim \oint_{c_1} \frac{dz}{i2\pi z} \frac{(1+z)^{2L-2}}{z^{2N-1}}} \quad \text{(III.106)}$$

in which we recognise $D_2$.

Let us do one more order, to see how we can deal with the tensor products in $D^{(2)}$ and $E^{(2)}$.

**Periodic case - third order**

We consider:

$$X_2 = \oint_{c_1} \frac{dz_0}{i2\pi z_0} \text{Tr}[A^{(2)}_\mu (z_0 D^{(2)} + E^{(2)})^L] z_0^{-N} \Big|_{\mu^2} \quad \text{(III.107)}$$

where we recall that

$$D^{(2)} = \begin{smallmatrix}1\\1\end{smallmatrix} + \begin{smallmatrix}d\\1\end{smallmatrix} + \begin{smallmatrix}1\\d\end{smallmatrix} + \begin{smallmatrix}d\\e\end{smallmatrix} \ , \quad E^{(2)} = \begin{smallmatrix}1\\1\end{smallmatrix} + \begin{smallmatrix}e\\1\end{smallmatrix} + \begin{smallmatrix}1\\e\end{smallmatrix} + \begin{smallmatrix}e\\d\end{smallmatrix} \ , \quad A^{(2)}_\mu = \begin{smallmatrix}A_\mu\\A_\mu\end{smallmatrix} \quad \text{(III.108)}$$

(we write the tensor product in columns, in a way similar to the diagram we drew earlier in fig.-III.2, for compactness).

The action of $(z_0 D^{(2)} + E^{(2)})$ on a product of plane waves is, if we forget the extra $\|0\rangle\!\rangle$ terms due to the action of $e$:

$$\left(z_0 D^{(2)} + E^{(2)}\right)\begin{smallmatrix}\|\overline{z_1}\rangle\!\rangle\\\|\overline{z_2}\rangle\!\rangle\end{smallmatrix} \sim \left(z_0(1 + \overline{z_1} + \overline{z_2} + z_2\overline{z_1}) + (1 + z_1 + z_2 + z_1\overline{z_2})\right)\begin{smallmatrix}\|\overline{z_1}\rangle\!\rangle\\\|\overline{z_2}\rangle\!\rangle\end{smallmatrix}$$

$$= (1 + z_0\overline{z_1})(1 + z_1\overline{z_2})(1 + z_2)\begin{smallmatrix}\|\overline{z_1}\rangle\!\rangle\\\|\overline{z_2}\rangle\!\rangle\end{smallmatrix}. \quad \text{(III.109)}$$

We will therefore try to build the true eigenvectors of $(z_0 D^{(2)} + E^{(2)})$ using a method of images. Their eigenvalues will be of the form $(1 + z_0\overline{z_1})(1 + z_1\overline{z_2})(1 + z_2)$. The way to go is to find all the transformations that leave that eigenvalue invariant, and combine the corresponding plane waves to force a coefficient 0 where the walls of the random walks are (i.e. on the basis vectors '$\|-1\rangle\!\rangle$').

Let us change variables, and define:

$$\phi_0 = z_0\overline{z_1} \ , \quad \phi_1 = z_1\overline{z_2} \ , \quad \phi_2 = z_2, \quad \text{(III.110)}$$

$$z_2 = \phi_2 \ , \quad z_1 = \phi_1\phi_2 \ , \quad z_0 = \phi_0\phi_1\phi_2. \quad \text{(III.111)}$$

The vectors we seek are obtained by exchanging the $\phi_i$'s, as a determinant:

$$\|\overline{\phi}^{(2)}\rangle\!\rangle = \sum_\sigma \epsilon(\sigma)\, \overline{\phi_{\sigma(1)}} \left(\overline{\phi_{\sigma(2)}}\right)^2 \begin{smallmatrix}\|\overline{\phi_{\sigma(1)}}\ \overline{\phi_{\sigma(2)}}\rangle\!\rangle\\\|\overline{\phi_{\sigma(2)}}\rangle\!\rangle\end{smallmatrix} \quad \text{(III.112)}$$

where $\sigma$ is a permutation on $\{0, 1, 2\}$ and $\epsilon(\sigma)$ is its signature. These vectors are such that:
$$\left(z_0 D^{(2)} + E^{(2)}\right) \|\overline{\phi}^{(2)}\rangle\!\rangle = (1+\phi_0)(1+\phi_1)(1+\phi_2) \|\overline{\phi}^{(2)}\rangle\!\rangle \tag{III.113}$$
and the relevant closure identity is:
$$1 = \frac{1}{3!} \oint_{c_1} \frac{\prod d\phi_i}{\prod (\imath 2\pi \phi_i)} \|\overline{\phi}^{(2)}\rangle\!\rangle \langle\!\langle \phi^{(2)}\|. \tag{III.114}$$

Injecting this into (III.107), we get:
$$X_2 = \frac{1}{3!} \oint_{c_1} \frac{\prod d\phi_i}{\prod (\imath 2\pi \phi_i)} \frac{(1+\phi_0)^L}{\phi_0^N} \frac{(1+\phi_1)^L}{\phi_1^N} \frac{(1+\phi_2)^L}{\phi_2^N} \langle\!\langle \phi^{(2)} \| A_\mu^{(2)} \| \overline{\phi}^{(2)}\rangle\!\rangle \bigg|_{\mu^2} \tag{III.115}$$
where
$$\langle\!\langle \phi^{(2)} \| A_\mu^{(2)} \| \overline{\phi}^{(2)}\rangle\!\rangle = (1-e^{-\mu})^2 \sum_{n,m=0}^{\infty} e^{-(n+m)\mu} \sum_{\sigma, \sigma'} (\phi_{\sigma(1)} \overline{\phi_{\sigma'(1)}})^{n+1} (\phi_{\sigma(2)} \overline{\phi_{\sigma'(2)}})^{n+m+2} \tag{III.116}$$

This is where we need to keep certain terms and throw the others away. The sum over the permutations, in the previous equation, contains 36 terms. We will only keep the 12 terms for which $\sigma' \circ \sigma^{-1}$ is an irreducible permutation (i.e. with only one cycle of size 3). These terms are:

$$\left(\frac{\phi_1}{\phi_2}\right)^{n+1}\left(\frac{\phi_2}{\phi_0}\right)^{n+m+2} + \left(\frac{\phi_0}{\phi_2}\right)^{n+1}\left(\frac{\phi_2}{\phi_1}\right)^{n+m+2} + \left(\frac{\phi_1}{\phi_0}\right)^{n+1}\left(\frac{\phi_0}{\phi_2}\right)^{n+m+2}$$
$$+ \left(\frac{\phi_2}{\phi_1}\right)^{n+1}\left(\frac{\phi_0}{\phi_2}\right)^{n+m+2} + \left(\frac{\phi_2}{\phi_0}\right)^{n+1}\left(\frac{\phi_1}{\phi_2}\right)^{n+m+2} + \left(\frac{\phi_0}{\phi_1}\right)^{n+1}\left(\frac{\phi_2}{\phi_0}\right)^{n+m+2}$$
$$+ \left(\frac{\phi_2}{\phi_1}\right)^{n+1}\left(\frac{\phi_1}{\phi_0}\right)^{n+m+2} + \left(\frac{\phi_2}{\phi_0}\right)^{n+1}\left(\frac{\phi_0}{\phi_1}\right)^{n+m+2} + \left(\frac{\phi_0}{\phi_1}\right)^{n+1}\left(\frac{\phi_1}{\phi_2}\right)^{n+m+2}$$
$$+ \left(\frac{\phi_1}{\phi_2}\right)^{n+1}\left(\frac{\phi_0}{\phi_1}\right)^{n+m+2} + \left(\frac{\phi_0}{\phi_2}\right)^{n+1}\left(\frac{\phi_1}{\phi_0}\right)^{n+m+2} + \left(\frac{\phi_1}{\phi_0}\right)^{n+1}\left(\frac{\phi_2}{\phi_1}\right)^{n+m+2}. \tag{III.117}$$

The first six are the same as the last six with $\phi_1 \leftrightarrow \phi_2$, so that we can keep only the first lot and multiply $X_2$ by 2.

All those terms can be written using only $\psi_1 = \phi_1/\phi_0$ and $\psi_2 = \phi_2/\phi_0$. The exponents of those two variables in each of the first six terms are:

$$1 : \{n+1, m+1\} \qquad 2 : \{-n-m-2, m+1\} \qquad 3 : \{n+1, -n-m-2\}$$
$$4 : \{-n-1, -m-1\} \qquad 5 : \{n+m+2, -m-1\} \qquad 6 : \{-n-1, n+m+2\} \tag{III.118}$$

for $n$ and $m$ in $\mathbb{N}$. If we draw these regions on $\mathbb{Z}^2$ (fig.-III.3), we see that all the powers of $\psi_1$ and $\psi_2$ are there except for a few lines.

If we add those, the sum of all these terms becomes:
$$\sum_{n=-\infty}^{\infty} \sum_{m=-\infty}^{\infty} \psi_1^n \psi_2^m = \delta(1 - \phi_1/\phi_0)\delta(1 - \phi_2/\phi_0) \tag{III.119}$$

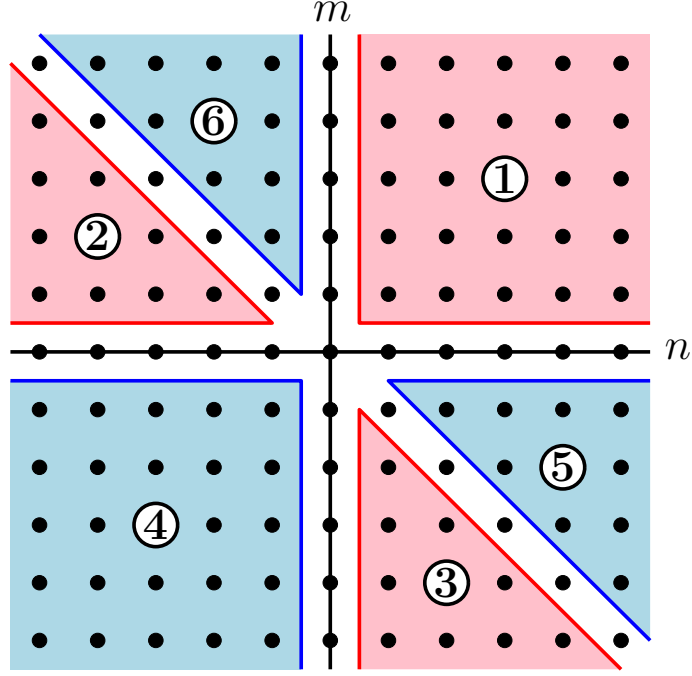

Figure III.3: The six regions listed above for the exponents of $\psi_1$ and $\psi_2$ in $X_2$.

and by putting these deltas back in $X_2$, we finally get:

$$X_2 \sim \frac{1}{3} \oint_{c_1} \frac{dz}{\imath 2\pi z} \frac{(1+z)^{3L}}{z^{3N}}. \tag{III.120}$$

A similar calculation on $Y_3$ gives:

$$Y_3 \sim \frac{1}{3} \oint_{c_1} \frac{dz}{\imath 2\pi z} \frac{(1+z)^{3L-2}}{z^{3N-1}}. \tag{III.121}$$

We recognise those two terms to be $C_3$ and $D_3$, and we have all we need to go to an arbitrary order.

**Periodic case - any order**

We can finally consider $X_k$ for any $k$:

$$X_k = \oint_{c_1} \frac{dz_0}{\imath 2\pi z_0} \text{Tr}[A_\mu^{(k)}(z_0 D^{(k)} + E^{(k)})^L] z_0^{-N} \bigg|_{\mu^k}. \tag{III.122}$$

As in the previous case, the action of $(z_0 D^{(k)} + E^{(k)})$ on a plane wave gives, up to some boundary terms:

$$\left(z_0 D^{(k)} + E^{(k)}\right) \| \{\overline{z_i}\} \rangle\rangle \sim \prod_{i=0}^{k} (1 + z_i \overline{z_{i+1}}) \| \{\overline{z_i}\} \rangle\rangle. \tag{III.123}$$

We define new variables $\phi_i$, under the exchange of which the eigenvalues of $(z_0 D^{(k)} + E^{(k)})$ are invariant, as:

$$\phi_i = z_i \overline{z_{i+1}} \quad , \quad z_i = \prod_{l=i}^{k} \phi_l \tag{III.124}$$

and express the true eigenvectors of $(z_0 D^{(k)} + E^{(k)})$ as determinants of plane waves:

$$\|\overline{\phi}^{(k)}\rangle\rangle = \sum_{\sigma} \epsilon(\sigma) \prod_{i=1}^{k} \left(\overline{\phi_{\sigma(i)}}\right)^i \Big\| \{\prod_{l=i}^{k} \overline{\phi_{\sigma(l)}}\}\Big\rangle\rangle \tag{III.125}$$

such that:

$$\left(z_0 D^{(k)} + E^{(k)}\right)\|\overline{\phi}^{(k)}\rangle\rangle = \prod_{i=0}^{k}(1+\phi_i)\|\overline{\phi}^{(k)}\rangle\rangle. \tag{III.126}$$

The closure identity for those vectors is given by:

$$1 = \frac{1}{(k+1)!} \oint_{c_1} \frac{\prod d\phi_i}{\prod(\imath 2\pi \phi_i)} \|\overline{\phi}^{(k)}\rangle\rangle\langle\langle\phi^{(k)}\| \tag{III.127}$$

which gives, once introduced into eq.(III.122):

$$X_k = \frac{1}{(k+1)!} \oint_{c_1} \frac{\prod d\phi_i}{\prod(\imath 2\pi \phi_i)} \prod_{i=0}^{k} \frac{(1+\phi_i)^L}{\phi_i^N} \langle\langle\phi^{(k)}\|A_\mu^{(k)}\|\overline{\phi}^{(k)}\rangle\rangle\Big|_{\mu^k}. \tag{III.128}$$

The last part of this equation gives:

$$\langle\langle\phi^{(k)}\|A_\mu^{(k)}\|\overline{\phi}^{(k)}\rangle\rangle = (1 - e^{-\mu})^k \sum_{n_i=0}^{\infty} e^{-(\sum n_i)\mu} \sum_{\sigma,\sigma'} \prod_{i=1}^{k} (\phi_{\sigma(i)} \overline{\phi_{\sigma'(i)}})^{\sum_{l=1}^{i}(n_l+1)}. \tag{III.129}$$

From the sum on $\sigma$ and $\sigma'$, which contains $(k+1)!^2$ terms, we keep only those such that $\sigma' \circ \sigma^{-1}$ is an irreducible permutation. This leaves us with $k!(k+1)!$ terms. We can separate them into $k!$ groups, related to one another through the permutation of the $\phi_i$'s for $i \neq 0$. Keeping only one of these groups, and multiplying $X_k$ by $k!$, we can write each term using only the $k$ variables $\psi_i = \phi_i/\phi_0$. The exponents of these variables in each of the terms we kept determine regions in $\mathbb{Z}^k$, arranged as the vertices of a permutohedron, or as the faces of its dual (fig.-III.4), which fill the whole lattice except for a few hyperplanes.

If we add those hyperplanes by hand, we get a product of delta functions. Putting those back into $X_k$, we finally get:

$$\boxed{X_{k-1} \sim \frac{1}{k} \oint_{c_1} \frac{dz}{\imath 2\pi z} \frac{(1+z)^{kL}}{z^{kN}}} \tag{III.130}$$

and

$$\boxed{Y_k \sim \frac{1}{k} \oint_{c_1} \frac{dz}{\imath 2\pi z} \frac{(1+z)^{kL-2}}{z^{kN-1}}} \tag{III.131}$$

where we recognise $C_k$ and $D_k$, multiplied by the factor $\frac{1}{k}$ which is also present in (II.110) and (II.111).

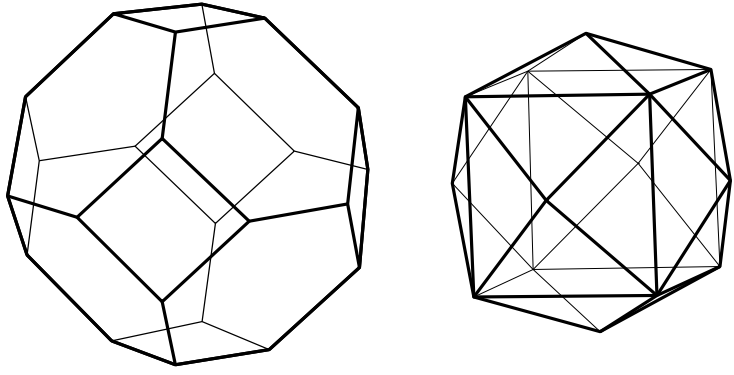

Figure III.4: The permutohedron of order 3 (truncated octahedron) and its dual (tetrakis hexahedron).

Note that all the terms we threw away in calculating these coefficients would have produced combinations of lower order $C_l$'s and $D_l$'s (with $l < k$), which is precisely why we threw them away, to isolate the part that produced $C_k$ and $D_k$.

What we intend to do next is to make the same calculations in the open case, and identify the terms which look like the ones we kept for the periodic case. This should give us a good candidate for the coefficients $C_k$ and $D_k$ for the open ASEP.

**Open case - first order**

As is the periodic case, the first order is easy to deal with. We have $X_0 = Z_L$ and $Y_1 = Z_{L-1}$, as we saw in section II.2.1. Nevertheless, we will redo the calculation here, in the case of the TASEP, just to refresh our memory.

We have:
$$X_0 = \langle\!\langle W \| (D+E)^L \| V \rangle\!\rangle \tag{III.132}$$

with

$$\langle\!\langle W \| = \sum_{n=0}^{\infty} a^n \langle\!\langle n \| = \langle\!\langle a \| \quad , \quad a = \frac{1-\alpha}{\alpha}, \tag{III.133}$$

$$\| V \rangle\!\rangle = \sum_{n=0}^{\infty} b^n \| n \rangle\!\rangle = \| b \rangle\!\rangle \quad , \quad b = \frac{1-\beta}{\beta}. \tag{III.134}$$

For the eigenvectors of $(D+E)$, and, later, of $(D^{(k)} + E^{(k)})$, we can use the ones we found earlier, with $z_0$ set to 1.

This gives us:
$$(D+E)\Big(z\|z\rangle\!\rangle - \overline{z}\|\overline{z}\rangle\!\rangle\Big) = (1+z)(1+\overline{z})\Big(z\|z\rangle\!\rangle - \overline{z}\|\overline{z}\rangle\!\rangle\Big) \tag{III.135}$$

and

$$1 = \frac{1}{2} \oint_{c_1} \frac{dz}{i2\pi z}\Big(z\|z\rangle\!\rangle - \overline{z}\|\overline{z}\rangle\!\rangle\Big)\Big(\overline{z}\langle\!\langle\overline{z}\| - z\langle\!\langle z\|\Big) \tag{III.136}$$

so that

$$X_0 = \frac{1}{2} \oint_{c_1} \frac{dz}{i2\pi z} (1+z)^L (1+\bar{z})^L \langle\!\langle a \| \big(z\|z\rangle\!\rangle - \bar{z}\|\bar{z}\rangle\!\rangle\big)\big(\bar{z}\langle\!\langle \bar{z}\| - z\langle\!\langle z\|\big) \|b\rangle\!\rangle. \tag{III.137}$$

We know that, for $a < 1$ and $b < 1$:

$$\langle\!\langle a\|z\rangle\!\rangle = \frac{1}{1-az} \quad, \quad \langle\!\langle z\|b\rangle\!\rangle = \frac{1}{1-bz}. \tag{III.138}$$

We will stick to that case for now, and generalise to any $a$ and $b$ at the end. All the contour integrals are done on the unit circle for now.

We get:

$$\langle\!\langle W\| \big(z\|z\rangle\!\rangle - \bar{z}\|\bar{z}\rangle\!\rangle\big)\big(\bar{z}\langle\!\langle \bar{z}\| - z\langle\!\langle z\|\big) \|V\rangle\!\rangle = \left(\frac{z}{1-az} - \frac{\bar{z}}{1-a\bar{z}}\right)\left(\frac{\bar{z}}{1-b\bar{z}} - \frac{z}{1-bz}\right)$$
$$= \frac{(z-\bar{z})(\bar{z}-z)}{(1-az)(1-a\bar{z})(1-bz)(1-b\bar{z})} \tag{III.139}$$

so that

$$\boxed{X_0 = \frac{1}{2} \oint_{c_1} \frac{dz}{i2\pi z} \frac{(1+z)^L(1+\bar{z})^L(1-z^2)(1-\bar{z}^2)}{(1-az)(1-a\bar{z})(1-bz)(1-b\bar{z})}} \tag{III.140}$$

and

$$\boxed{Y_1 = \frac{1}{2} \oint_{c_1} \frac{dz}{i2\pi z} \frac{(1+z)^{L-1}(1+\bar{z})^{L-1}(1-z^2)(1-\bar{z}^2)}{(1-az)(1-a\bar{z})(1-bz)(1-b\bar{z})}.} \tag{III.141}$$

By rewriting $\bar{z}$ as $z^{-1}$, and adapting the contour integral to cases where $a$ and $b$ may be larger than 1, we recognise the expressions of $C_1$ and $D_1$ from (III.71) and (III.72).

**Open case - second order**

For the next order, things get tougher: we now have to deal with tensor products of order 3. $X_1$ is expressed as:

$$X_1 = \langle\!\langle W^{(1)}\| A_\mu^{(1)} (D^{(1)} + E^{(1)})^L \|V^{(1)}\rangle\!\rangle \Big|_\mu \tag{III.142}$$

with

$$D^{(1)} = {}^1_1 {}^1_1 + {}^1_d {}^1_1 + {}^1_1 {}^1_d + {}^1_d {}^1_d + {}^d_1 {}^d_1 + {}^d_e {}^d_1 + {}^d_1 {}^d_e + {}^d_e {}^d_e \quad, \quad E^{(1)} = {}^e_1 {}^e_1 + {}^e_1 {}^e_d + {}^e_d {}^e_1 + {}^e_d {}^e_e + {}^1_1 {}^1_e + {}^1_1 {}^1_e + {}^1_e {}^1_1 + {}^1_e {}^1_d \tag{III.143}$$

and

$$\langle\!\langle W^{(1)}\| = \begin{matrix}\langle\!\langle a\| \\ \langle\!\langle 1\| \\ \langle\!\langle a\|\end{matrix} \quad, \quad \|V^{(1)}\rangle\!\rangle = \begin{matrix}\|b\rangle\!\rangle \\ \|1\rangle\!\rangle \\ \|b\rangle\!\rangle\end{matrix} \quad, \quad A_\mu^{(1)} = \begin{matrix}\tilde{A}_\mu \\ A_\mu \\ \tilde{A}_\mu\end{matrix} \tag{III.144}$$

where $\langle\!\langle 1\|$ is not the basis vector, but the plane wave with argument 1, and

$$\tilde{A}_\mu = \sum_{n=0}^\infty e^{-n\mu} \|n\rangle\!\rangle \langle\!\langle n\| = (1-e^{-\mu})^{-1} A_\mu. \tag{III.145}$$

Notice that we have replaced $A_\mu$ with $\tilde{A}_\mu$ for the first and third rows. It is exactly the same matrix, without the normalisation factor $(1 - e^{-\mu})$. The reason for this is that, on those rows, the matrix product is set between vectors $\langle\!\langle a \|$ and $\| b \rangle\!\rangle$ (which are the $q = 0$ limit of vectors $\langle\!\langle W \|$ and $\| V \rangle\!\rangle$ that we found in section II.2.1), of which any scalar product converges if $a < 1$ and $b < 1$ (which we assume for now). On the secod row, however, those boundary vectors are replaced by $\langle\!\langle 1 \|$ and $\| 1 \rangle\!\rangle$ (which can easily be found to be the solutions for vectors $\langle\!\langle \tilde{W} \|$ and $\| \tilde{V} \rangle\!\rangle$ for $q = 0$), and the normalisation becomes necessary.

We can make the same change of variables (III.124) as in the periodic case, but with $z_0 = 1$, i.e.

$$z_0 = \prod_{i=0}^{2k+1} \phi_i = 1 \tag{III.146}$$

We inforce this last condition through a delta function in $X_k$. We then insert the closure identity of order $2k + 1$ next to the right boundary in $X_k$, and get:

$$X_k = \frac{1}{(2k+2)!} \oint_{c_1} \frac{\prod d\phi_i}{\prod (i 2\pi \phi_i)} \delta\left(1 - \prod \phi_i\right) \prod_{i=0}^{k}(1+\phi_i)^L \langle\!\langle W^{(k)} \| A_\mu^{(k)} \| \overline{\phi}^{(2k+1)} \rangle\!\rangle \langle\!\langle \phi^{(2k+1)} \| V^{(k)} \rangle\!\rangle \bigg|_{\mu^k}. \tag{III.147}$$

Contrary to what we had in the periodic case, the left and right vectors of the closure identity don't project onto one another (which they did because of the trace), but onto the boundary vectors, separately. We therefore have two objects to calculate independently.

We first look at the simpler case where $a = b = 0$.
For $k = 1$, we need to calculate:

$$\langle\!\langle \phi^{(3)} \| V^{(1)} \rangle\!\rangle = \sum_\sigma \epsilon(\sigma) \left(\phi_{\sigma(1)} \phi_{\sigma(2)}^2 \phi_{\sigma(3)}^3\right) \sum_{n=0}^\infty (\phi_{\sigma(2)} \phi_{\sigma(3)})^n$$

$$= \frac{1}{4} \sum_\sigma \epsilon(\sigma) \Big((\phi_{\sigma(1)} - \phi_{\sigma(0)})(\phi_{\sigma(3)} - \phi_{\sigma(2)})\Big) \sum_{n=0}^\infty (\phi_{\sigma(2)} \phi_{\sigma(3)})^{n+2} \tag{III.148}$$

$$= \frac{1}{8} \sum_\sigma \epsilon(\sigma) \Big((\phi_{\sigma(1)} - \phi_{\sigma(0)})(\phi_{\sigma(3)} - \phi_{\sigma(2)})\Big) \sum_{n=0}^\infty \Big((\phi_{\sigma(2)} \phi_{\sigma(3)})^{n+2} + (\phi_{\sigma(0)} \phi_{\sigma(1)})^{n+2}\Big).$$

To get from the first to the second line, we collected the terms with $\sigma(0) \leftrightarrow \sigma(1)$ and/or $\sigma(2) \leftrightarrow \sigma(3)$ (for which the prefactor is different but the sum over $n$ is the same). To go from the second to the third, we collected those with $\sigma(0) \leftrightarrow \sigma(2)$ and $\sigma(1) \leftrightarrow \sigma(3)$ (for which the prefactor is the same but the sum over $n$ is different).

We now need to use the delta function imposing $\phi_0 \phi_1 \phi_2 \phi_3 = 1$, to write $\phi_{\sigma(0)} \phi_{\sigma(1)} = (\phi_{\sigma(2)} \phi_{\sigma(3)})^{-1}$ in that last expression. Because of that, the sum over $n$ is only three terms short of a delta function (those terms being the ones for $n = -1$, $n = 0$ and $n = 1$). These missing terms acually compensate between one permutation and another, so that

we finally get:

$$\langle\!\langle \phi^{(3)} \| V^{(1)} \rangle\!\rangle = \Big((\phi_1 - \phi_0)(\phi_3 - \phi_2)\Big)\delta(1 - \phi_2\phi_3)$$
$$+ \Big((\phi_2 - \phi_0)(\phi_3 - \phi_1)\Big)\delta(1 - \phi_1\phi_3)$$
$$+ \Big((\phi_3 - \phi_0)(\phi_2 - \phi_1)\Big)\delta(1 - \phi_1\phi_2). \tag{III.149}$$

For the term from (III.147) involving the left boundary, things are a bit different due to the presence of $A_\mu$. We can make the same transformations except for the very last one, and get:

$$\langle\!\langle W^{(1)} \| A_\mu^{(1)} \| \phi^{(3)} \rangle\!\rangle = (1 - e^{-\mu})\Big((\phi_1 - \phi_0)(\phi_3 - \phi_2)\Big) \sum_{n=0}^{\infty} \Big((\phi_2\phi_3)^{n+2} + (\phi_2\phi_3)^{-n-2}\Big)e^{-n\mu}$$
$$+ (1 - e^{-\mu})\Big((\phi_2 - \phi_0)(\phi_3 - \phi_1)\Big) \sum_{n=0}^{\infty} \Big((\phi_1\phi_3)^{n+2} + (\phi_1\phi_3)^{-n-2}\Big)e^{-n\mu}$$
$$+ (1 - e^{-\mu})\Big((\phi_3 - \phi_0)(\phi_2 - \phi_1)\Big) \sum_{n=0}^{\infty} \Big((\phi_1\phi_2)^{n+2} + (\phi_1\phi_2)^{-n-2}\Big)e^{-n\mu}. \tag{III.150}$$

For the open case, this is the point where we throw a few terms away. In the product between the two previous equations, the first delta from (III.149) transforms each term is the first sum from (III.150) into 1s, and produces contribution to $X_1$ equal to $X_0^2$. We won't keep this term, nor will we keep the two we get by multiplying the second delta with the second sum, or the third delta with the third sum. In all the other combinations, we can take the prefactors $(1 - e^{-\mu})$ to order 1, and the rest at $\mu = 0$. The sums in (III.150) can be completed into delta functions, and we get:

$$X_1 \sim \frac{1}{4!} \oint_{c_1} \frac{\prod d\phi_i}{\prod(i2\pi\phi_i)} \prod_{i=0}^{k}(1 + \phi_i)^L \Big((\phi_1 - \phi_0)(\phi_3 - \phi_2)(\overline{\phi_2} - \overline{\phi_0})(\overline{\phi_3} - \overline{\phi_1})\Big)$$
$$\delta\Big(1 - \prod \phi_i\Big)\delta(1 - \phi_2\phi_3)\delta(1 - \phi_1\phi_3)$$
$$+ \text{permutations} \tag{III.151}$$

which gives us, in the end:

$$\boxed{X_1 \sim \frac{1}{4} \oint_{c_1} \frac{dz}{i2\pi z} \Big((1+z)^L(1+\overline{z})^L(1-z^2)(1-\overline{z}^2)\Big)^2.} \tag{III.152}$$

**Open case - third order**

For the next order, the calculations are pretty much the same, with two more variables. The factors we get from $\langle\!\langle \phi^{(5)} \| V^{(2)} \rangle\!\rangle$ are of the form:

$$\Big((\phi_1 - \phi_0)(\phi_3 - \phi_2)(\phi_5 - \phi_4)\Big)\delta(1 - \phi_2\phi_3)\delta(1 - \phi_4\phi_5) \tag{III.153}$$

plus permutations of the indices.

If we set $\psi_i = \phi_{2i}\phi_{2i+1}$, with $\psi_0 = 1/\psi_1\psi_2$, the terms that we get from $\langle\!\langle W^{(2)} \| A_\mu^{(2)} \| \overline{\phi}^{(5)} \rangle\!\rangle$ are, up to a prefactor $(\phi_1 - \phi_0)(\phi_3 - \phi_2)(\phi_5 - \phi_4)$, of the form:

$$(\psi_1)^{n+2}(\psi_2)^{n+m+4} + (\psi_0)^{n+2}(\psi_2)^{n+m+4} + (\psi_1)^{n+2}(\psi_0)^{n+m+4}$$
$$+(\psi_2)^{n+2}(\psi_1)^{n+m+4} + (\psi_2)^{n+2}(\psi_0)^{n+m+4} + (\psi_0)^{n+2}(\psi_1)^{n+m+4}. \qquad (\text{III}.154)$$

As in the periodic case, if we map those terms onto $\mathbb{Z}^2$ (fig.-III.5), we have six regions:

$1 : \{n+2, n+m+4\} \quad 2 : \{-n-2, m+2\} \quad 3 : \{-n-2, -n-m-4\}$
$4 : \{n+m+4, n+2\} \quad 5 : \{-n-m-4, -n-2\} \quad 6 : \{m+2, -n-2\} \qquad (\text{III}.155)$

and we see that we are only a few lines short of a product of delta functions.

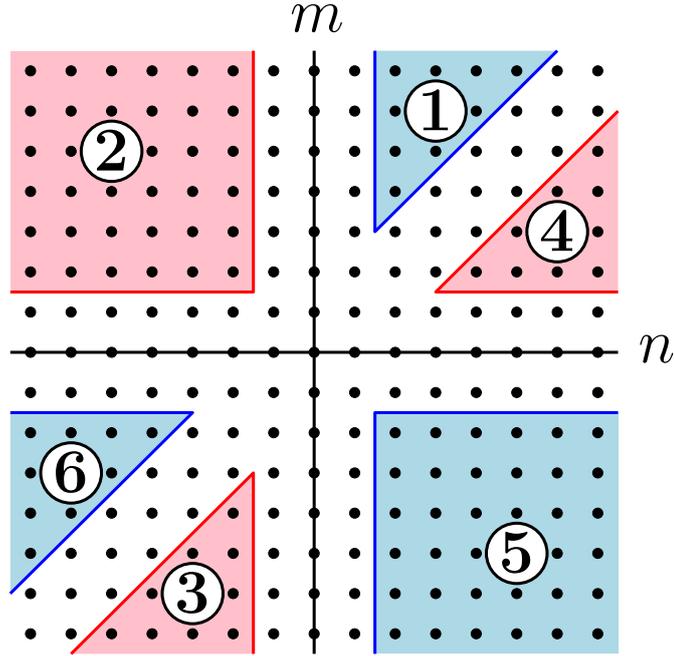

Figure III.5: The six regions listed above for the exponents of $\psi_1$ and $\psi_2$ in $X_2$.

Completing them by hand, we finally get:

$$X_2 \sim \frac{1}{6} \oint_{c_1} \frac{dz}{\imath 2\pi z} \left((1+z)^L(1+\overline{z})^L(1-z^2)(1-\overline{z}^2)\right)^3. \qquad (\text{III}.156)$$

**Open case - any order**

For an arbitrary order $k$, the exact same reasoning gives us a sum of terms corresponding to regions of $\mathbb{Z}^2$ organised as the faces of a permutohedron, or as the vertices of its dual, (fig.-III.4), which is the reverse of the periodic case (that was not clear for $k=2$ because the dual of a hexagon is another hexagon). We can then complete that sum to get a product of delta functions, and obtain, in general:

$$X_{k-1} \sim \frac{1}{2k} \oint_{c_1} \frac{dz}{\imath 2\pi z} \left((1+z)^L(1+\overline{z})^L(1-z^2)(1-\overline{z}^2)\right)^k. \qquad (\text{III}.157)$$

For $Y_k$, we might as well assume that what we've seen in all case until now (i.e. dividing the argument of $X_{k-1}$ by $(1+z)(1+\overline{z})$) still works, which is to say:

$$Y_k \sim \frac{1}{2k} \oint_{c_1} \frac{dz}{i2\pi z} \frac{\left((1+z)^L(1+\overline{z})^L(1-z^2)(1-\overline{z}^2)\right)^k}{(1+z)(1+z^{-1})}. \tag{III.158}$$

To generalise to any $a$ and $b$, we simply replace $(1+z)^L(1+\overline{z})^L(1-z^2)(1-\overline{z}^2)$ by the full function $F(z)$ (which we know from the calculation of the mean current in section II.2.1), rewriting $\overline{z}$ as $z^{-1}$, and we replace the unit circle by the appropriate integration contour:

$$\boxed{X_{k-1} \sim \frac{1}{2k} \oint_{\{0,a,b\}} \frac{dz}{i2\pi z} F(z)^k} \tag{III.159}$$

and

$$\boxed{Y_k \sim \frac{1}{2k} \oint_{\{0,a,b\}} \frac{dz}{i2\pi z} \frac{F(z)^k}{(1+z)(1+z^{-1})}} \tag{III.160}$$

with

$$F(z) = \frac{(1+z)^L(1+z^{-1})^L(1-z^2)(1-z^{-2})}{(1-az)(1-a/z)(1-bz)(1-b/z)}. \tag{III.161}$$

We also remark that the factor $\frac{1}{k}$ from the periodic case has turned into $\frac{1}{2k}$, which accounts for the factor $\frac{1}{2}$ in (III.67). Multiplying $X_{k-1}$ and $Y_k$ by $k$, we get a conjecture for $C_k$ and $D_k$ for the open TASEP.

**Numerical checks**

Naturally, we now need to validate our conjecture.

As we saw at the end of section II.2.2, using the expressions we found for $C_k$ and $D_k$, we can write the first few cumulants of the current, by inverting eq.(III.68) and injecting the result into eq.(III.69). We find:

$$\begin{aligned}
E_1 &= J = \frac{D_1}{C_1}, \\
E_2 &= \frac{D_1 C_2 - D_2 C_1}{C_1^3}, \\
E_3 &= \frac{3D_1 C_2^2 - 2D_1 C_1 C_3 - 3D_2 C_1 C_2 + 2D_3 C_1^2}{C_1^5}, \\
E_4 &= \frac{15D_1 C_2^3 - 20D_1 C_1 C_2 C_3 + 6D_1 C_1^2 C_4 - 15D_2 C_1 C_2^2 + 8D_2 C_1^2 C_3 + 12D_3 C_1^2 C_2 - 6D_4 C_1^3}{C_1^7},
\end{aligned} \tag{III.162}$$

and so on.

We checked our formulae against the first six cumulants for systems of size up to 10, for rational values of $\alpha$ and $\beta$ (which produce rational values of the cumulants), and found them to match in every case. More detail on this can be found in [1].

We were also able to compare our results to numerical calculations provided by M. Gorissen and C. Vanderzande, obtained by a DMRG-like algorithm [137], for system sizes of up to 100. Those are shown in the following plots, and show excellent agreement with our formulae.

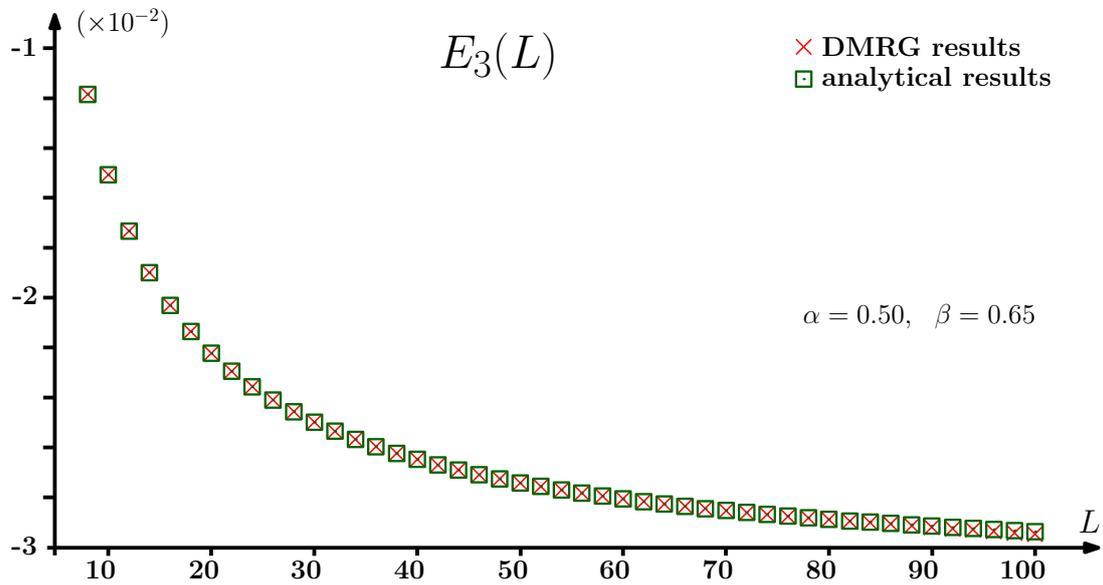

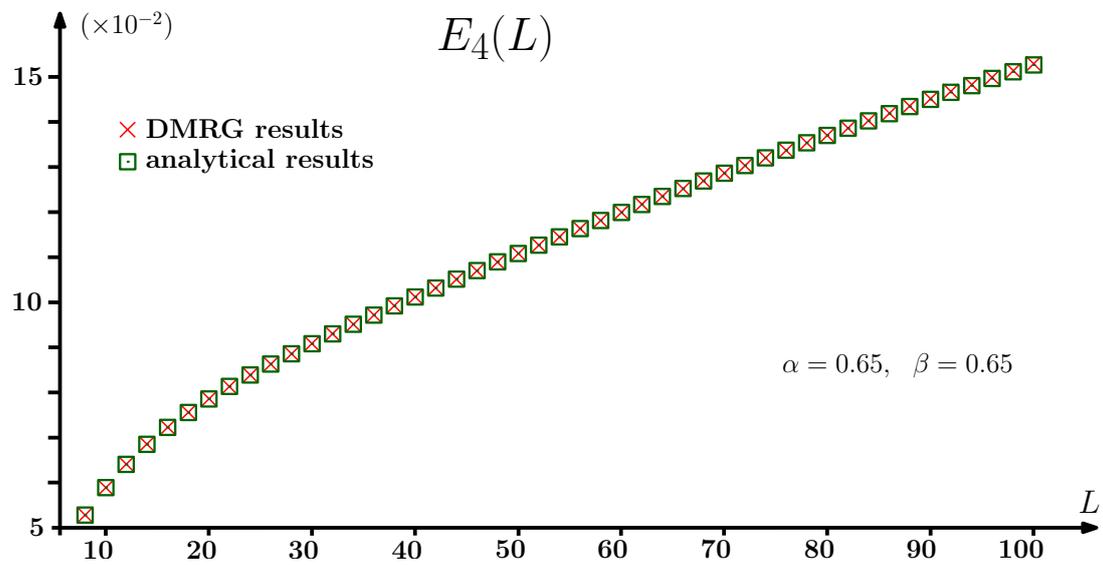

## III.4.2 Partially asymmetric case

We now examine the general case where $q \neq 0$. We expect the function $F(z)$ from the totally asymmetric case to be replaced by its full expression (III.64). We also expect to find expressions involving convolutions with a kernel $K$, as in the periodic case, but it might not be the same one, so we need to find how $K$ appears in the periodic ASEP, and by what it is replaced in the open case.

**Periodic case**

For this, and considering eq.(III.70), we only need to look at $X_1$ in both cases. We start with the periodic ASEP.

We have:
$$X_1 = \oint_{c_1} \frac{dz_0}{i2\pi z_0} \text{Tr}[A_\mu(z_0 D + E)^L] z_0^{-N} \Big|_{\mu^1}. \tag{III.163}$$

As we saw in section II.2.1, the vectors that replace the plane waves we have used for the TASEP involve q-deformed Hermite polynomials. Let's define:
$$\|x,y\rangle\!\rangle = \sum_{n=0}^{\infty} \frac{H_n(x,y)}{(q)_n} \|n\rangle\!\rangle \quad , \quad \langle\!\langle x,y\| = \sum_{n=0}^{\infty} H_n(x,y) \langle\!\langle n\|. \tag{III.164}$$

Those vectors are such that:
$$(z_0 d + e)\|z, z_0 \bar{z}\rangle\!\rangle = (z + z_0 \bar{z})\|z, z_0 \bar{z}\rangle\!\rangle \tag{III.165}$$

and, considering that $H_n(xy, x/y) = x^n H_n(y, 1/y)$, the closure identity (II.75) becomes:
$$1 = \frac{(q)_\infty}{2} \oint_{c_1} \frac{dz}{i2\pi z} (z^2/z_0, z_0/z^2)_\infty \|z, z_0\bar{z}\rangle\!\rangle \langle\!\langle \bar{z}, z\overline{z_0}\|. \tag{III.166}$$

We insert this identity into $X_1$ next to $A_\mu$, and we get:
$$X_1 = \frac{(q)_\infty}{2} \oint_{c_1} \frac{dz_0}{i2\pi z_0} \frac{dz}{i2\pi z} \frac{(1+z)^L(1+z_0\bar{z})^L}{z_0^N} (z^2/z_0, z_0/z^2)_\infty \langle\!\langle \bar{z}, z\overline{z_0}\|A_\mu\|z, z_0\bar{z}\rangle\!\rangle \Big|_{\mu^1}. \tag{III.167}$$

Using the q-Mehler formula (II.71) on the last scalar product, we obtain:
$$(z^2/z_0, z_0/z^2)_\infty \langle\!\langle \bar{z}, z\overline{z_0}\|A_\mu\|z, z_0\bar{z}\rangle\!\rangle = \frac{(1-e^{-\mu})(z^2/z_0, z_0/z^2, e^{-2\mu})_\infty}{(z^2/z_0 e^{-\mu}, z_0/z^2 e^{-\mu}, e^{-\mu}, e^{-\mu})_\infty}. \tag{III.168}$$

We need to separate, in that last equation, the terms that don't contain a $q$ (i.e. the first terms of every q-Pochhammer symbol), from the rest. Those become, when truncated to order 1 in $\mu$:

$$\frac{(1-e^{-\mu})(1-z^2/z_0)(1-z_0/z^2)(1-e^{-2\mu})}{(1-e^{-\mu})^2(1-z^2/z_0 e^{-\mu})(1-z_0/z^2 e^{-\mu})} \sim 1 - \mu\left(1 + 2\sum_{k=1}^{\infty}(z^2/z_0)^k + (z_0/z^2)^k\right)$$
$$= 1 - \mu\big(2\delta(1-z^2/z_0) - 1\big) \tag{III.169}$$

and the rest gives:

$$\frac{(q, qz^2/z_0, qz_0/z^2, qe^{-2\mu})_\infty}{(qz^2/z_0 e^{-\mu}, qz_0/z^2 e^{-\mu}, qe^{-\mu}, qe^{-\mu})_\infty} \sim 1 - \mu \left( \sum_{k=1}^{\infty} \frac{q^k z^2/z}{1 - q^k z^2/z} + \frac{q^k z_0/z^2}{1 - q^k z_0/z^2} \right)$$

$$= 1 - \mu \sum_{k=1}^{\infty} \frac{q^k}{1 - q^k} \left( (z^2/z)^k + (z_0/z^2)^k \right) \quad \text{(III.170)}$$

where we have expanded and re-summed every term between the first and second line.

We now keep only the terms of order $\mu$ in the product of those two terms. The first part, coming from (III.169), is a delta function (we throw the $-1$ away), and gives, as it did for the TASEP:

$$X_1^{(1)} \sim \oint_{c_1} \frac{dz}{i 2\pi z} \frac{(1+z)^{2L}}{z^{2N}}. \quad \text{(III.171)}$$

The second part, from (III.170), is what we are interested in. If we define a new variable $\tilde{z}$ as $\tilde{z} = z_0/z$, we find:

$$\boxed{X_1^{(2)} \sim \frac{1}{2} \oint_{c_1} \frac{dz}{i 2\pi z} \frac{d\tilde{z}}{i 2\pi \tilde{z}} \frac{(1+z)^L}{z^N} \frac{(1+\tilde{z})^L}{\tilde{z}^N} K(z, \tilde{z})} \quad \text{(III.172)}$$

with

$$\boxed{K(z, \tilde{z}) = \sum_{k=1}^{\infty} \frac{q^k}{1 - q^k} \left( (z/\tilde{z})^k + (\tilde{z}/z)^k \right)} \quad \text{(III.173)}$$

which is what we were expecting.

We now only need to do the same calculation for the open ASEP.

**Open case**

To make our lives simpler, we will consider the case where $\tilde{a} = \tilde{b} = 0$, which is to say $\gamma = \delta = 0$.

Unfortunately, we have no idea what the vectors $\|\overline{\phi}^{(3)}\rangle\rangle$ and $\langle\langle\phi^{(3)}\|$ turn into for $q \neq 0$, which forces us to change our method entirely. However, seeing that, for the periodic case, the vectors had changed but not their associated eigenvalues, we will be looking for eigenvectors of $D^{(3)} + E^{(3)}$ with eigenvalues of the form $(1 + \phi_1)(1 + \phi_2)(1 + \phi_3)(1 + \phi_4)$ (using the same notations as for the TASEP), with the constraint that $\phi_1 \phi_2 \phi_3 \phi_4 = 1$.

Let us separate $D^{(3)} + E^{(3)}$ into groups of matrices, defining:

$$S_1 = {}^d_1 {}^1_1 + {}^e_d {}^1_1 + {}^1_e {}^1_d + {}^1_1 {}^1_e, \quad \text{(III.174)}$$

$$S_2 = {}^1_e {}^1_1 + {}^1_d {}^1_1 + {}^e_1 {}^d_e + {}^d_1 {}^e_d + {}^e_e {}^d_1 + {}^d_d {}^e_1, \quad \text{(III.175)}$$

$$S_3 = {}^e_1 {}^1_1 + {}^d_e {}^1_1 + {}^1_1 {}^d_e + {}^1_1 {}^1_d, \quad \text{(III.176)}$$

such that $D^{(3)} + E^{(3)} = 1 + S_1 + S_2 + S_3 + 1$.

Using $de - q\, ed = (1 - q)$, we can easily check that $[S_i, S_j] = 0$ for any $i$ and $j$. By examining those matrices acting on $\|\overline{\phi}^{(3)}\rangle\!\rangle$ for $q = 0$, we find their eigenvalues to be symmetric functions of the $\phi_i$'s:

$$S_1 \to \sigma_1 = \phi_1 + \phi_2 + \phi_3 + \phi_4, \tag{III.177}$$
$$S_2 \to \sigma_2 = \phi_1\phi_2 + \phi_1\phi_3 + \phi_1\phi_4 + \phi_2\phi_3 + \phi_2\phi_4 + \phi_3\phi_4, \tag{III.178}$$
$$S_3 \to \sigma_3 = \phi_1\phi_2\phi_3 + \phi_1\phi_2\phi_4 + \phi_1\phi_3\phi_4 + \phi_2\phi_3\phi_4. \tag{III.179}$$

Now, instead of writing the eigenvectors of $D^{(3)} + E^{(3)}$ on the basis we used before, we will write them as generating functions: for a vector $\sum a_n \|n\rangle\!\rangle$, we define $f(z) = \sum a_n z^n$, which is the projection of this vector onto a plane wave with argument $z$. The action of $d$ and $e$, in this formalism, is as follows:

$$e : f(z) \to z f(z) \quad , \quad d : f(z) \to \frac{1}{z}\big(f(z) - f(qz)\big). \tag{III.180}$$

We are looking for vectors with three integer indices, so we define $f(x, y, z)$, where $x$ bears the first of those indices, $y$ the second, and $z$ the third.

The action of $S_1$, $S_2$ and $S_3$ on this function is:

$$\sigma_1 f(x,y,z) = \frac{1}{x}\big(f(x,y,z) - f(qx,y,z)\big) + \frac{x}{y}\big(f(x,y,z) - f(x,qy,z)\big)$$
$$+ \frac{y}{z}\big(f(x,y,z) - f(x,y,qz)\big) + z f(x,y,z), \tag{III.181}$$

$$\sigma_2 f(x,y,z) = y f(xyz) + \frac{1}{y}\big(f(x,y,z) - f(x,qy,z)\big) + \frac{x}{z}\big(f(x,y,z) - f(x,y,qz)\big)$$
$$+ \frac{z}{x}\big(f(x,y,z) - f(qx,y,z)\big) + \frac{xz}{y}\big(f(x,y,z) - f(x,qy,z)\big)$$
$$+ \frac{y}{xz}\big(f(x,y,z) - f(qx,y,z) - f(x,y,qz) + f(qx,y,qz)\big), \tag{III.182}$$

$$\sigma_3 f(x,y,z) = x f(x,y,z) + \frac{y}{x}\big(f(x,y,z) - f(qx,y,z)\big)$$
$$+ \frac{z}{y}\big(f(x,y,z) - f(x,qy,z)\big) + \frac{1}{z}\big(f(x,y,z) - f(x,y,qz)\big). \tag{III.183}$$

By combining those, we can find simpler equations. For instance, we have:

$$(1-x\sigma_1+x^2\sigma_2-x^3\sigma_3+x^4)f(x,y,z) = (1-xz)(1-\frac{xy}{z})f(qx,y,z) + \frac{xy}{z}f(qx,y,qz) \tag{III.184}$$

$$(1-z\sigma_3+z^2\sigma_2-z^3\sigma_1+z^4)f(x,y,z) = (1-xz)(1-\frac{zy}{x})f(x,y,qz) + \frac{zy}{x}f(qx,y,qz) \tag{III.185}$$

Considering the expressions we have for the $\sigma_i$'s, the terms on the left factorise into:

$$(1 - x\sigma_1 + x^2\sigma_2 - x^3\sigma_3 + x^4) = (1 - x\phi_1)(1 - x\phi_2)(1 - x\phi_3)(1 - x\phi_4), \tag{III.186}$$

$$(1 - z\sigma_3 + z^2\sigma_2 - z^3\sigma_1 + z^4) = (1 - \frac{z}{\phi_1})(1 - \frac{z}{\phi_2})(1 - \frac{z}{\phi_3})(1 - \frac{z}{\phi_4}). \tag{III.187}$$

For $y = 0$, (III.184) and (III.185) take the simpler form:

$$(1 - x\phi_1)(1 - x\phi_2)(1 - x\phi_3)(1 - x\phi_4) f(x, 0, z) = (1 - xz) f(qx, 0, z), \tag{III.188}$$

$$(1 - \frac{z}{\phi_1})(1 - \frac{z}{\phi_2})(1 - \frac{z}{\phi_3})(1 - \frac{z}{\phi_4}) f(x, 0, z) = (1 - xz) f(x, 0, qz), \tag{III.189}$$

which give us, through iteration:

$$f(x, 0, z) = \frac{(xz)_\infty}{(x\phi_1, x\phi_2, x\phi_3, x\phi_4)_\infty} g(z), \qquad (\text{III.190})$$

$$f(x, 0, z) = \frac{(xz)_\infty}{(\frac{z}{\phi_1}, \frac{z}{\phi_2}, \frac{z}{\phi_3}, \frac{z}{\phi_4})_\infty} h(x), \qquad (\text{III.191})$$

where $g$ and $h$ are functions of only one variable.

Combining those two, we finally get:

$$f(x, 0, z) = \frac{(xz)_\infty}{(x\phi_1, x\phi_2, x\phi_3, x\phi_4, \frac{z}{\phi_1}, \frac{z}{\phi_2}, \frac{z}{\phi_3}, \frac{z}{\phi_4})_\infty}. \qquad (\text{III.192})$$

Considering the nice symmetry of those relations, we want to do the same with $y$. A good candidate to be applied to $f$ instead of (III.186) or (III.187) is:

$$(1 - y\phi_1\phi_2)(1 - y\phi_1\phi_3)(1 - y\phi_1\phi_4)(1 - y\phi_2\phi_3)(1 - y\phi_2\phi_4)(1 - y\phi_3\phi_4) \qquad (\text{III.193})$$

which we find to be equal, in terms of the $\sigma_i$'s, to:

$$(1 - y^2)^2(1 - \sigma_2 y + y^2) + y^2(\sigma_1 - \sigma_3 y)(\sigma_3 - \sigma_1 y) \qquad (\text{III.194})$$

and, applying this to $f(x, y, z)$, with the help of (III.181), (III.182) and (III.183), we get:

$$(1 - y\phi_1\phi_2)(1 - y\phi_1\phi_3)(1 - y\phi_1\phi_4)(1 - y\phi_2\phi_3)(1 - y\phi_2\phi_4)(1 - y\phi_3\phi_4) f(x, y, z)$$

$$= \frac{(1 - qy^2)}{(1 - q^2 y^2)} \left( (1 - y^2)(1 - q^2 y^2)(1 + \frac{xz}{q}) - y((1 + qy^2)(\sigma_1 z + \sigma_3 x), \right.$$

$$\left. - y(1 - q)(\sigma_1 x + \sigma_3 z) \right) f(x, qy, z),$$

$$- \frac{xz}{q} \frac{(1 - y^2)(1 - q\frac{xy}{z})(1 - q\frac{zy}{x})}{(1 - q^2 y^2)} f(x, q^2 y, z), \qquad (\text{III.195})$$

which is not terribly useful as is. However, if we take $x = z = 0$, the right hand side of the equation reduces to $(1 - y^2)(1 - qy^2) f(0, qy, 0)$, and we find that:

$$f(0, y, 0) = \frac{(y^2)_\infty}{(y\phi_1\phi_2, y\phi_1\phi_3, y\phi_1\phi_4, y\phi_2\phi_3, y\phi_2\phi_4, y\phi_3\phi_4)_\infty}. \qquad (\text{III.196})$$

It is now time to use the preliminary calculations we have just performed to analyse the factor $\langle\!\langle W^{(1)} \| A_\mu^{(1)} \| \overline{\phi}^{(3)} \rangle\!\rangle \langle\!\langle \phi^{(3)} \| V^{(1)} \rangle\!\rangle$ which appears in $X_1$. First of all, we recall from (III.144) that $\langle\!\langle W^{(1)} \| A_\mu^{(1)}$ is, for $\tilde{a} = 0$, the product of three plane waves of arguments $ae^{-\mu}$, $e^{-\mu}$ and $ae^{-\mu}$, up to a factor $(1 - e^{-\mu})$, so that we have, according to the definition of $f$:

$$\langle\!\langle W^{(1)} \| A_\mu^{(1)} \| \overline{\phi}^{(3)} \rangle\!\rangle = (1 - e^{-\mu}) f(ae^{-\mu}, e^{-\mu}, ae^{-\mu}). \qquad (\text{III.197})$$

On the other side, with an appropriate definition of $f$, we find:

$$\langle\!\langle \phi^{(3)} \| V^{(1)} \rangle\!\rangle = f(b, 1, b). \qquad (\text{III.198})$$

Since we didn't find an expression for $f$ with all three variables, we will get rid of the first and the third by noticing that eq.(III.192), when replacing $x$ and $z$ by $a$ or $b$, and after equating the different $\phi_i$'s using the delta functions that will undoubtedly emerge at some point from the next calculation, gives a contribution to $X_1$ which is simply the denominator in $F(z)$, which we expected. This is not what we are after, and we can take $a = b = 0$ for now.

We now only have to consider eq.(III.196). Expanding the terms without a $q$, as we did for the periodic case, and then taking $y$ to 1, gives us the delta functions we were expecting. Taking, for instance, $\delta(1 - \phi_1\phi_2)$ from the right boundary term, and applying it to the left boundary, gives:

$$\frac{(1-y)(y^2, \phi_1\phi_3, \phi_1/\phi_3, \phi_3/\phi_1, 1/\phi_1\phi_3)_\infty}{(y, y\phi_1\phi_3, y\phi_1/\phi_3, y\phi_3/\phi_1, y/\phi_1\phi_3, y)_\infty} \tag{III.199}$$

with $y = \mathrm{e}^{-\mu}$ (the numerator comes from the closure identity). If we then expand the part of that is left after we take the first term of each q-Pochhammer symbol away, we get what should take the place of the convolution kernel in the open case. We find:

$$\boxed{K(z, \tilde{z}) = \sum_{k=1}^{\infty} \frac{q^k}{1-q^k}\left( \left(z/\tilde{z}\right)^k + \left(\tilde{z}/z\right)^k + \left(z\tilde{z}\right)^k + \left(1/z\tilde{z}\right)^k \right)} \tag{III.200}$$

where we replaced $\phi_1$ and $\phi_3$ by $z$ and $\tilde{z}$. Because of the $z \leftrightarrow 1/z$ symmetry of $F(z)$, this actually gives us twice the kernel we had for the periodic case.

**Numerical checks**

This time, we cannot check our conjecture on small systems, because the extra parameter $q$ makes it difficult to get exact results even for small sizes. What's more, the contour integrals we found have infinitely many poles, and getting an exact value for those is impossible. What we can do, however, is to check numerical evaluations of our results against numerical calculations from DMRG, as we did before. The plots below give a few examples of those comparisons. They are in nearly perfect agreement.

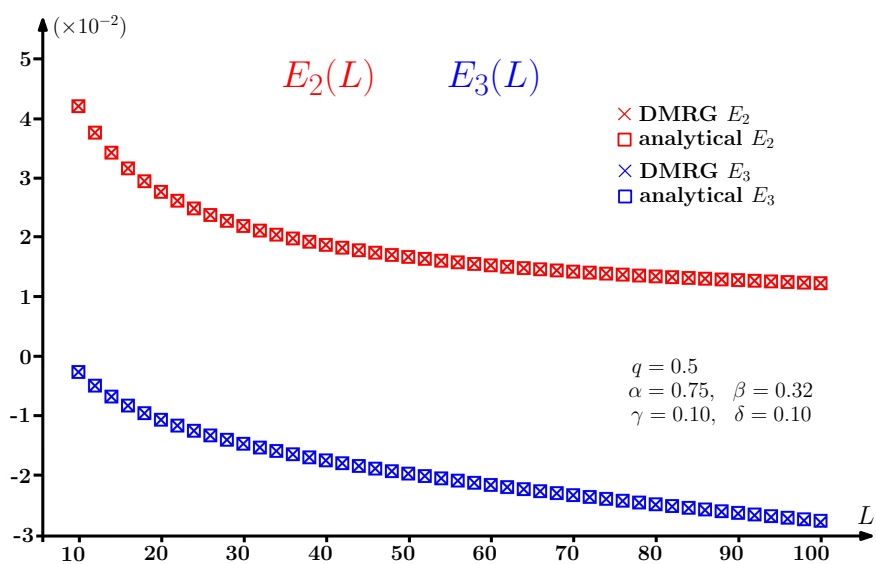

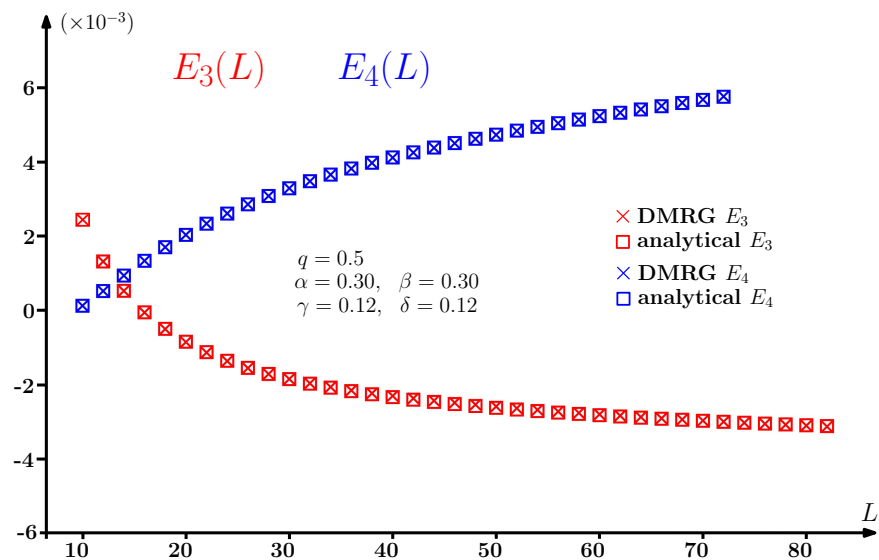

# CHAPTER IV

## Large Deviations of the Current in the S-Ensemble

Calculating the exact generating function for the cumulants of the current is nice, but using it to learn something about the physical behaviour of the system is even better. In section III.3, we presented the expression we found for the exponential generating function $E(\mu)$ of the cumulants of the current in the open ASEP, with generic boundary parameters and asymmetry (where $\mu$ is the variable conjugate to the current). In this chapter, we see how, combining calculations from large size asymptotics of those cumulants, macroscopic fluctuation theory (MFT), and direct diagonalisation in particular limits, we can describe the phase diagram of the system in the s-ensemble (which is to say, with the current as a tunable parameter). We will uncover two new phases, on top of the low density, high density and maximal current ones: a shock phase, which continues the shock line for positive fluctuations of the current, and an anti-shock phase, which can be reached from the maximal current phase, through negative fluctuations of the current. We will also be able to obtain the large deviation function for the current, and the optimal profiles, in all phases but one (the maximal current phase, for which we will have some asymptotic results nevertheless). Those phases will be, incidentally, exactly those where the MFT gives correct results.

In the first section of this chapter, we take the large size limit in the cumulants of the current that we have just obtained, and see that the phase diagram for all the cumulants is the same as that for the current. We get a particularly simple result in the high and low density phases, where $E(\mu)$ takes a closed form in terms of $\mu$. We also note that considering the ASEP instead of the simpler TASEP makes no difference apart from a global $(1-q)$ factor in each cumulant.

In the second section, we try to see what we can learn, from that large size limit, about the large deviations of the current, by taking the Legendre transform of the results from the first section. Except in the high and low density phases, these calculations give us only the first non-trivial order in $j$ of the large deviation functions.

In the third section, we start back from the deformed Markov matrix itself, and take three extreme limits, in which we can calculate what we need directly. The first limit we take is that for an extremely low current (which is to say $\mu \to -\infty$ for the TASEP), in which $M_\mu$ is almost diagonal. The second is for extremely low entry and exit rates (which also implies extremely low current). In those two limits, we see how an effective description of the system in a reduced phase space allows to easily recover the dominant eigenvectors and eigenvalues of $M_\mu$. The third limit we take is that of an extremely large current. In that limit, we will see that the deformed Markov matrix can be related to the Hamiltonian of the open XX spin chain with anti-diagonal boundaries, which can then be solved exactly using free fermions techniques. The distribution of the steady state, in that limit, can be related to a discrete Coulomb gas (also called Dyson-Gaudin gas) with anti-symmetric boundary conditions.

In the fourth and last section, we see how a simple trick, which consists in using the MFT on the weakly asymmetric simple exclusion process (WASEP) and then taking the weak asymmetry to infinity, allows us to conjecture the form of the full phase diagram for the open ASEP in the s-ensemble, as well as an expression for the large deviation function of the current and a description of the typical density profiles in four of the five phases we find. We check that the expressions we find for the large deviation function are consistent with all the calculations from the previous three sections.

Note that the first three sections are mostly calculation details, and can be skipped except for the final results in each subsection (which are boxed). Also note that all the results in this chapter are new (to our knowledge), apart from those in section IV.4.1, where the appropriate reference is given.

## IV.1 Large size asymptotics of the cumulants

Our starting point here is the expression we found in section III.3 for the generating function of the cumulants of the current $E(\mu)$ in the open ASEP. We recall that expression, where both $E(\mu)$ and $\mu$ are written as series in a parameter $B$:

$$\mu = -\oint_S \frac{dz}{\imath 2\pi z} W(z) = -\sum_{k=1}^{\infty} C_k \frac{B^k}{k}, \tag{IV.1}$$

$$E(\mu) = -(1-q)\oint_S \frac{dz}{\imath 2\pi (1+z)^2} W(z) = -(1-q)\sum_{k=1}^{\infty} D_k \frac{B^k}{k}, \tag{IV.2}$$

with $W(z)$ defined as:

$$W(z) = -\frac{1}{2}\ln\Big(1 - BF(z)e^{X[W](z)}\Big). \tag{IV.3}$$

The function $F(z)$ is given by:

$$F(z) = \frac{(1+z)^L(1+z^{-1})^L(z^2, z^{-2})_\infty}{(az, a/z, \tilde{a}z, \tilde{a}/z, bz, b/z, \tilde{b}z, \tilde{b}/z)_\infty} \tag{IV.4}$$

which is symmetric in $z \leftrightarrow z^{-1}$ (this will be useful later), and the convolution operator $X$ by:

$$X[f](z) = \oint_S \frac{d\tilde{z}}{\imath 2\pi \tilde{z}} f(\tilde{z}) K(z, \tilde{z}) \tag{IV.5}$$

with a kernel
$$K(z, \tilde{z}) = 2 \sum_{k=1}^{\infty} \frac{q^k}{1-q^k} \left( (z/\tilde{z})^k + (z/\tilde{z})^{-k} \right). \tag{IV.6}$$

All the contour integrals in $\mu$, $E(\mu)$ and $X$ are taken around a set of points given by $S = \{0, q^k a, q^k \tilde{a}, q^k b, q^k \tilde{b}\}$ for $k$ in $\mathbb{N}$.

From these expressions, each coefficient $C_k$ and $D_k$ can be obtained by expanding $W(z)$ in terms of $B$. The first few of these are given by:

$$C_1 = \frac{1}{2} \oint_S \frac{dz}{i 2\pi z} F(z), \tag{IV.7}$$

$$D_1 = \frac{1}{2} \oint_S \frac{dz}{i 2\pi (1+z)^2} F(z), \tag{IV.8}$$

which are fairly simple, then

$$C_2 = \frac{1}{2} \oint_S \frac{dz}{i 2\pi z} F(z)^2 + \frac{1}{2} \oint_S \frac{dz_1}{i 2\pi z_1} \frac{dz_2}{i 2\pi z_2} F(z_1) F(z_2) K(z_1, z_2), \tag{IV.9}$$

$$D_2 = \frac{1}{2} \oint_S \frac{dz}{i 2\pi (1+z)^2} F(z)^2 + \frac{1}{2} \oint_S \frac{dz_1}{i 2\pi (1+z_1)^2} \frac{dz_2}{i 2\pi z_2} F(z_1) F(z_2) K(z_1, z_2), \tag{IV.10}$$

in which a first convolution appears, then

$$\begin{aligned} C_3 =& \frac{1}{2} \oint_S \frac{dz}{i 2\pi z} F(z)^3 \\ &+ \frac{1}{2} \oint_S \frac{dz_1}{i 2\pi z_1} \frac{dz_2}{i 2\pi z_2} \left( \frac{3}{2} F(z_1)^2 F(z_2) + \frac{3}{4} F(z_1) F(z_2)^2 \right) K(z_1, z_2) \\ &+ \frac{1}{2} \oint_S \frac{dz_1}{i 2\pi z_1} \frac{dz_2}{i 2\pi z_2} \frac{dz_3}{i 2\pi z_3} F(z_1) F(z_2) F(z_3) \Big( \frac{3}{4} K(z_1, z_2) K(z_2, z_3) \\ &\hspace{6cm} + \frac{3}{8} K(z_1, z_2) K(z_1, z_3) \Big), \end{aligned} \tag{IV.11}$$

$$\begin{aligned} D_3 =& \frac{1}{2} \oint_S \frac{dz}{i 2\pi (1+z)^2} F(z)^3 \\ &+ \frac{1}{2} \oint_S \frac{dz_1}{i 2\pi (1+z_1)^2} \frac{dz_2}{i 2\pi z_2} \left( \frac{3}{2} F(z_1)^2 F(z_2) + \frac{3}{4} F(z_1) F(z_2)^2 \right) K(z_1, z_2) \\ &+ \frac{1}{2} \oint_S \frac{dz_1}{i 2\pi (1+z_1)^2} \frac{dz_2}{i 2\pi z_2} \frac{dz_3}{i 2\pi z_3} F(z_1) F(z_2) F(z_3) \Big( \frac{3}{4} K(z_1, z_2) K(z_2, z_3) \\ &\hspace{6cm} + \frac{3}{8} K(z_1, z_2) K(z_1, z_3) \Big), \end{aligned} \tag{IV.12}$$

and so on and so forth. These expressions, in terms of combinations of $F$ and $K$, are equivalent to the 'tree expansion' which can be found in [27] for the case of the periodic ASEP.

Once we have the coefficients $C_k$ and $D_k$, all we have to do to find the cumulants of the current is to invert (IV.1) and inject it in (IV.2), to get the coefficients $E_k$ of $E(\mu)$ expanded as an exponential series in $\mu$, as we did at the end of section II.2.2:

$$\sum_{k=1}^{\infty} E_k \frac{\mu^k}{k!} = E(\mu) = E\bigl(B(\mu)\bigr) \tag{IV.13}$$

The first few of these are given by:

$$\begin{aligned}
E_1 &= J = \frac{D_1}{C_1}, \\
E_2 &= \frac{D_1 C_2 - D_2 C_1}{C_1^3}, \\
E_3 &= \frac{3 D_1 C_2^2 - 2 D_1 C_1 C_3 - 3 D_2 C_1 C_2 + 2 D_3 C_1^2}{C_1^5}, \\
E_4 &= \frac{15 D_1 C_2^3 - 20 D_1 C_1 C_2 C_3 + 6 D_1 C_1^2 C_4 - 15 D_2 C_1 C_2^2 + 8 D_2 C_1^2 C_3 + 12 D_3 C_1^2 C_2 - 6 D_4 C_1^3}{C_1^7},
\end{aligned} \tag{IV.14}$$

etcetera.

In this section, we will find what happens to those cumulants when we take the size of the system to infinity. As we recall, from the end of section II.2.1, the behaviour of the contour integrals in $C_1$ and $D_1$ in that limit depend on the position of $a$ and $b$ with respect to the unit circle. From that, we can read the phase diagram of the system (fig.-IV.1): in the MC phase, both $a$ and $b$ are inside the circle, in the LD phase $a$ is outside and larger than $b$, in the HD phase $b$ is outside and larger than $a$, and on the SL $a$ and $b$ are equal and outside of the circle.

We will see that this remains valid for the other cumulants, so that we can simply treat all the phases an transition lines one by one, and do only one calculation for each to get all the cumulants at once. In every case, we start by looking at the TASEP, and then we justify that considering the general ASEP only contributes a global factor $(1-q)$ to each cumulant.

### IV.1.1 High/Low density phases

We start with the high and low density phases (which are symmetric to one another through $a \leftrightarrow b$). In this case, a few of the poles in $S$ are out of the unit circle. The first thing we need to do is to deform the integration contour around $S$ as shown on fig.IV.2.

We start from an infinite collection of small contours around each point in $S$ (first image, in red). We then deform all the contours inside of the unit circle to a single contour on the circle, removing the poles that are inside of the circle but not in $S$ (which are the inverses of the poles that are in $S$ but not inside of the circle) by integrating around them clockwise (second image, in blue). Thanks to the $z \leftrightarrow z^{-1}$ symmetry of $F(z)$, we can consider that a clockwise integral over a pole at $z_0$ is the same as a counter-clockwise integral around $z_0^{-1}$, and transfer those blue contours to the other side of the circle, changing their direction and adding them to the contours that are already there (third image). In the end, we get one integral around the unit circle, and twice around each pole in $S$ that is out of the circle. There is a finite number of those, and the one furthest away from 1 is $a$ in the low density phase, and $b$ in the high density phase.

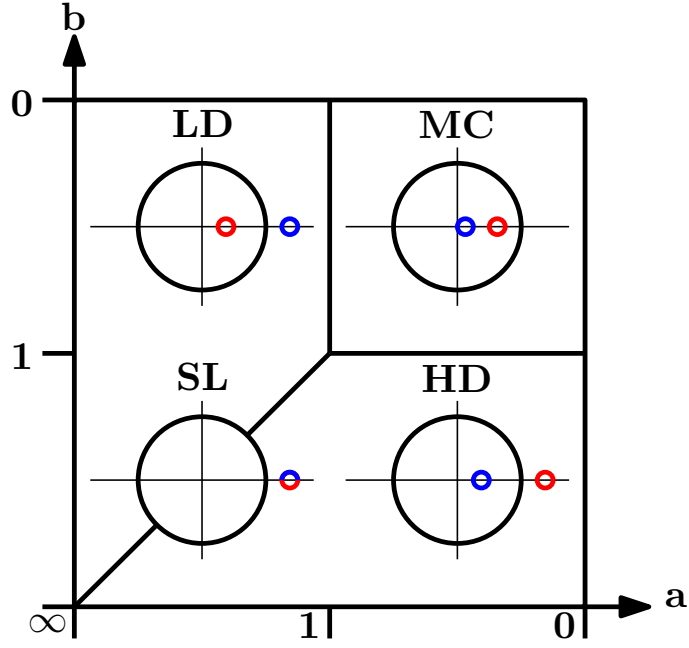

Figure IV.1: Positions of $a$ (blue circles) and $b$ (red circles) with respect to the unit circle (black circles) in each phase of the open ASEP.

Considering the TASEP for the moment, we only keep the first term in each $C_k$ and $D_k$, containing $F(z)^k$ (the other terms vanishing because $K = 0$). The set $S$ is reduced to $\{0, a, b\}$. Let us assume that $a > b$. Since the part of $F(z)$ that depends on $L$ is $(1+z)^L(1+z^{-1})^L$, which is minimal, on the real axis, at $z = 1$, all the contour integrals are dominated by the pole which is the furthest away from 1, which is $a$. We can therefore write:

$$C_k = \frac{1}{2}\oint_S \frac{dz}{2i\pi z} F(z)^k \sim \oint_{\{a\}} \frac{dz}{2i\pi z} \frac{\phi(z)^k}{(z-a)^k} = \frac{1}{(k-1)!} \frac{d^{k-1}}{dz^{k-1}} \left\{\frac{\phi^k(z)}{z}\right\}\bigg|_{z=a} \quad \text{(IV.15)}$$

and

$$D_k = \frac{1}{2}\oint_S \frac{dz}{2i\pi(1+z)^2} F(z)^k \sim \oint_{\{a\}} \frac{dz}{2i\pi(1+z)^2} \frac{\phi(z)^k}{(z-a)^k} = \frac{1}{(k-1)!} \frac{d^{k-1}}{dz^{k-1}} \left\{\frac{\phi^k(z)}{(1+z)^2}\right\}\bigg|_{z=a} \quad \text{(IV.16)}$$

with

$$\phi(z) = (z-a)F(z) = z\frac{(1+z)^L(1+z^{-1})^L(1-z^2)(1-z^{-2})}{(1-az)(1-bz)(1-b/z)}. \quad \text{(IV.17)}$$

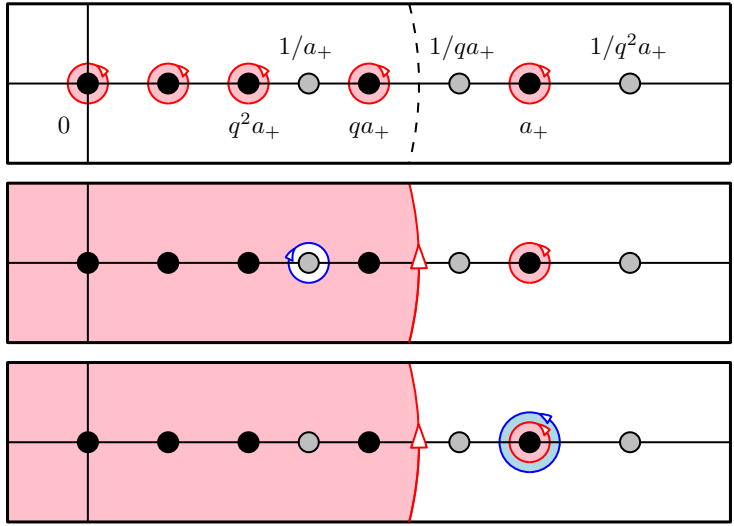

Figure IV.2: Only the pole at 0 and those related to $a_+$ are represented ; the dashed line in the first figure is part of the unit circle ; the three sets of contours are equivalent.

Putting this back in $\mu$ and $E(\mu)$, we get:

$$\mu = -\sum_{k=1}^{\infty} \frac{B^k}{k!} \frac{d^{k-1}}{dz^{k-1}} \left\{ \frac{\phi^k(z)}{z} \right\} \bigg|_{z=a}, \tag{IV.18}$$

$$E(\mu) = -\sum_{k=1}^{\infty} \frac{B^k}{k!} \frac{d^{k-1}}{dz^{k-1}} \left\{ \frac{\phi^k(z)}{(1+z)^2} \right\} \bigg|_{z=a}. \tag{IV.19}$$

This calls for the Lagrange inversion formula [154]. Considering two variables $w$ and $z$ related by:

$$w = z + B\phi(w) \tag{IV.20}$$

we can express a function $f$ taken at $w$ by expanding it around $z$, as:

$$f(w) = f(z) + \sum_{k=1}^{\infty} \frac{B^k}{k!} \frac{d^{k-1}}{dz^{k-1}} \left( \phi^k(z) f'(z) \right). \tag{IV.21}$$

This last sum is exactly what we have, for $f'(z) = -1/z$ in $\mu$ and $1/(1+z)$ in $E(\mu)$. This allows us to write:

$$\mu = -\log(w) + \log(a) \tag{IV.22}$$

and

$$E(\mu) = \frac{1}{w+1} - \frac{1}{a+1} \tag{IV.23}$$

where $w$ is as defined in (IV.20). We can now combine those two last equations and get:

$$\boxed{E(\mu) = \frac{a}{a+1} \frac{e^\mu - 1}{e^\mu + a}.} \tag{IV.24}$$

The same goes for the high density phase, with $b$ replacing $a$.

Using this expression, the cumulants of the current can be expressed as:

$$E_k = \frac{a(-1)^{(k+1)}}{(1+a)^{(k+1)}} A_k(-a) \tag{IV.25}$$

where $A_k$ it the Eulerian polynomial of order $k$, defined by:

$$\sum_{n=0}^{\infty} A_n(t) \frac{x^n}{n!} = \frac{t-1}{t - e^{(t-1)x}}. \tag{IV.26}$$

We now examine the case of the ASEP.
The convolution kernel $K$ can be rewritten as:

$$\begin{aligned} K(z, \tilde{z}) &= \sum_{k=1}^{\infty} \frac{q^k}{1-q^k} \left( (z/\tilde{z})^k + (z/\tilde{z})^{-k} \right) \\ &= \sum_{k,l=1}^{\infty} q^{kl} \left( (z/\tilde{z})^k + (z/\tilde{z})^{-k} \right) \\ &= \sum_{l=1}^{\infty} \left( \frac{q^l z}{\tilde{z} - q^l z} + \frac{q^l \tilde{z}}{z - q^l \tilde{z}} \right). \end{aligned} \tag{IV.27}$$

Now, remembering that the first term in $C_2$, for instance, is of order:

$$\oint_S \frac{dz}{i 2\pi z} F(z)^2 \sim \left( (1+a)(1+1/a) \right)^{2L} \tag{IV.28}$$

(which is the value of $F(z)$ at $z = a$, where we have kept only the part depending on $L$), we must find the behaviour, with respect to $L$, of the other term in $C_2$, which is:

$$\oint_S \frac{dz_1}{i 2\pi z_1} \oint_S \frac{dz_2}{i 2\pi z_2} F(z_1) F(z_2) K(z_1, z_2). \tag{IV.29}$$

The only way we can get a term of order $\left( (1+a)(1+1/a) \right)^{2L}$ out of this is to take both residues at $a$, and since $K$ has no pole at $(a, a)$, it only contributes a global factor $K(a, a)$ rather than change the order of one of the poles (which it would have around $(a, qa)$, for instance). This goes for any $C_k$ and $D_k$: the only contribution not negligible with respect to the first term is the one where all the residues are taken at $a$ for all integrals.

This tells us that, in the large $L$ limit, we can approximate $W(z)$ with:

$$W(z) = -\frac{1}{2} \ln\left( 1 - BF(z) e^{K(a,a) \oint_{\{a\}} \frac{d\tilde{z}}{i 2\pi \tilde{z}} W(\tilde{z})} \right) \tag{IV.30}$$

and, redefining $\phi(z)$ as:

$$\phi(z) = (z-a) F(z) e^{K(a,a) \oint_{\{a\}} \frac{d\tilde{z}}{i 2\pi \tilde{z}} W(\tilde{z})} \tag{IV.31}$$

we can repeat the calculations we did for the TASEP exactly. Since the result doesn't depend on $\phi$, the only difference from before is the factor $(1-q)$ in $E(\mu)$, so that, in the end:

$$\boxed{E(\mu) = (1-q)\frac{a}{a+1}\frac{e^\mu - 1}{e^\mu + a}.} \quad \text{(IV.32)}$$

This result was conjectured in [118] using numerical techniques related to the Bethe Ansatz, and can also be derived from the corresponding large deviation function which was found in [85] using MFT. We will come back to this in section IV.4. Note that this is the only case where we get a closed expression for $E(\mu)$ around $\mu = 0$. In all other cases, we will not be able to get rid of the parametric series in $B$. It is also the only case where the result doesn't depend on the size of the system.

### IV.1.2 HD-LD transition line

We now consider the shock line, i.e. $a = b > 1$. We start with the TASEP.

As before, the residue at $a$ dominates all the contour integrals, so we can write $C_k$ and $D_k$ as:

$$C_k \sim \oint_{\{a\}} \frac{dz}{2i\pi} \frac{(1+z)^{kL}(1+z^{-1})^{kL}\phi(z)^k}{z(z-a)^{2k}} = \frac{1}{(2k-1)!}\frac{d^{2k-1}}{dz^{2k-1}}\left\{(1+z)^{kL}(1+z^{-1})^{kL}\frac{\phi^k(z)}{z}\right\}\bigg|_{z=a} \quad \text{(IV.33)}$$

and

$$D_k \sim \oint_{\{a\}} \frac{dz}{2i\pi} \frac{(1+z)^{kL}(1+z^{-1})^{kL}\phi(z)^k}{(1+z)^2(z-a)^{2k}} = \frac{1}{(2k-1)!}\frac{d^{2k-1}}{dz^{2k-1}}\left\{(1+z)^{kL}(1+z^{-1})^{kL}\frac{\phi^k(z)}{(1+z)^2}\right\}\bigg|_{z=a} \quad \text{(IV.34)}$$

with

$$\phi(z) = z^2 \frac{(1-z^2)(1-z^{-2})}{(1-az)^2}. \quad \text{(IV.35)}$$

Unfortunately, we have here only half of the derivatives we need for the Lagrange inversion formula (the odd ones), so the same method will not work here. Instead, we consider that, for some large $N$:

$$\frac{d^n}{dz^n}\left\{f(z)^N g(z)\right\} \sim N^n f'(z)^n f(z)^{N-n} g(z) \quad \text{(IV.36)}$$

(where we have only differentiated $f^N$ $n$ times, since all the other terms in the derivative are of lower order in $N$), which gives us, for $N = kL$,

$$C_k \sim \frac{1}{(2k-1)!}(kL)^{2k-1}(1-a^{-2})^{2k-1}\Big((1+a)(1+1/a)\Big)^{k(L-2)+1}\frac{\phi^k(a)}{a}$$

$$\sim \frac{2}{L}\frac{a+1}{a-1}\frac{k^{2k}}{(2k)!}\left[L^2(1-a^{-2})\Big((1+a)(1+1/a)\Big)^{L-2}\phi^k(a)\right]^k \quad \text{(IV.37)}$$

and, similarly,

$$D_k \sim C_k \frac{a}{(1+a)^2}. \quad \text{(IV.38)}$$

This tells us that
$$E(\mu) \sim J\mu \quad \text{with} \quad J = \frac{a}{(1+a)^2} \tag{IV.39}$$

which is to say that the current converges to $J$ for $L \to \infty$ (which we already knew). To get the rest of the cumulants, we have to look at the next order in every $D_k$, which we do by subtracting $JC_k$ from $D_k$. Considering that

$$\frac{z}{(1+z)^2} - \frac{a}{(1+a)^2} = \frac{(1-az)(z-a)}{(1+z)^2(1+a)^2} \tag{IV.40}$$

we get

$$
\begin{aligned}
D_k - JC_k &\sim \oint_{\{a\}} \frac{dz}{2i\pi} \frac{(1+z)^{kL}(1+z^{-1})^{kL}(1-az)\phi(z)^k}{z(1+z)^2(1+a)^2(z-a)^{2k-1}} \\
&= \frac{1}{(2k-2)!} \frac{d^{2k-2}}{dz^{2k-2}} \left\{ (1+z)^{kL}(1+z^{-1})^{kL} \frac{(1-az)\phi^k(z)}{z(1+z)^2(1+a)^2} \right\} \Big|_{z=a} \\
&\sim (1-2k)\frac{2}{L^2} \frac{a}{a^2-1} \frac{k^{2k-1}}{(2k)!} \left[ L^2(1-a^{-2})\big((1+a)(1+1/a)\big)^{L-2} \phi^k(a) \right]^k \tag{IV.41}
\end{aligned}
$$

We notice that, in this last result, expanding the first monomial $(1-2k)$, the second term gives:

$$(-2k)\frac{2}{L^2}\frac{a}{a^2-1}\frac{k^{2k-1}}{(2k)!}\left[L^2(1-a^{-2})\big((1+a)(1+1/a)\big)^{L-2}\phi^k(a)\right]^k \sim -\frac{2}{L}JC_k \tag{IV.42}$$

which only contributes for a vanishing term in $J$, so that we can throw it away. What's more, the factor in brackets, which is in both $C_k$ and $D_k - JC_k$, can be gotten rid of by redefining $\left[L^2(1-a^{-2})\big((1+a)(1+1/a)\big)^{L-2}\phi^k(a)\right]B$ as $B$. Since it is the only place where $\phi$ appeared, we see that the final result doesn't depend on $\phi$ at all.

We also note that, as in the previous section, the only difference for the ASEP is a different function $\phi$ (on which the result doesn't depend), for the same reasons as before, and a factor $(1-q)$ in $E(\mu)$.

Putting all those calculations together, we finally get:

$$\mu = -\frac{2}{L}\frac{a+1}{a-1}\sum_{k=1}^{\infty}\frac{k^{2k-1}}{(2k)!}B^k, \tag{IV.43}$$

$$E(\mu) - (1-q)\frac{a}{(1+a)^2}\mu = -(1-q)\frac{2}{L^2}\frac{a}{a^2-1}\sum_{k=1}^{\infty}\frac{k^{2k-2}}{(2k)!}B^k. \tag{IV.44}$$

Considering this and (IV.13), we find that the cumulants behave as:

$$E_k \sim (1-q)\frac{a}{a^2-1}\Big(\frac{a-1}{a+1}\Big)^k L^{k-2} \tag{IV.45}$$

for $k \geq 2$, up to a numerical factor. This means that every cumulant higher than the second diverges for $L \to \infty$, which has an effect on the power law behaviour of $g(j)$ around $j = J$ (as we will see further down).

We can also note that the series in $B$ in $\mu$ and $E(\mu)$, which are pure mathematical functions (all the dependence in $a$ and $L$ has factored out), can be expressed in terms of the Lambert $\mathcal{W}$ function [155], defined as the solution to $x = \mathcal{W}_{\mathcal{L}} e^{\mathcal{W}_{\mathcal{L}}}$. The series expansion of $\mathcal{W}_{\mathcal{L}}(x)$ around 0 is:

$$\mathcal{W}_{\mathcal{L}}(x) = -\sum_{n=1}^{\infty} \frac{n^{n-1}}{n!}(-x)^n \tag{IV.46}$$

so that the sum in $\mu$ can be seen as the even part of $\mathcal{W}_{\mathcal{L}}$:

$$-\sum_{k=1}^{\infty} \frac{k^{2k-1}}{(2k)!} B^k = \mathcal{W}_{\mathcal{L}}(\sqrt{B}/2) + \mathcal{W}_{\mathcal{L}}(-\sqrt{B}/2) \tag{IV.47}$$

and the sum in $E(\mu)$ can be found to be, through integration of the previous one:

$$-\sum_{k=1}^{\infty} \frac{k^{2k-2}}{(2k)!} B^k = 2\left(\mathcal{W}_{\mathcal{L}}(\sqrt{B}/2) + \mathcal{W}_{\mathcal{L}}(-\sqrt{B}/2)\right) + \mathcal{W}_{\mathcal{L}}(\sqrt{B}/2)^2 + \mathcal{W}_{\mathcal{L}}(-\sqrt{B}/2)^2. \tag{IV.48}$$

This will come in handy in section IV.2.2.

### IV.1.3 Maximal current phase

We now consider the maximal current phase, where both $a$ and $b$ are inside of the unit circle. All the poles of $F(z)$ that are in $S$ are inside of that circle, and all those that aren't are outside, so that the contour integrals can be taken around the unit circle instead of $S$. On that circle, because of the factor $(1+z)^L(1+z^{-1})^L$, $F(z)$ has a saddle point at $z = 1$. We will be doing an approximation around that point.

In fact, it is more convenient, for the calculations we need to do, to consider angular integrals. Defining $z = e^{i\theta}$, with $\frac{dz}{iz} = d\theta$, we find:

$$(1+z)(1+z^{-1}) = 2(1+\cos(\theta)) \sim 4(1-\theta^2/4), \tag{IV.49}$$
$$(1-z^2)(1-z^{-2}) = 2(1-\cos(2\theta)) \sim 4\theta^2, \tag{IV.50}$$

so that, around $z = 1$, i.e. $\theta = 0$, we get

$$F(z) \sim 4^{L+1}\theta^2 e^{-L\frac{\theta^2}{4}}. \tag{IV.51}$$

This gives us, quite simply:

$$C_k \sim \frac{1}{2}\int \frac{d\theta}{2\pi} 4^{k(L+1)}\theta^{2k} e^{-kL\frac{\theta^2}{4}} = \frac{L^{-1/2}}{2\sqrt{\pi}} \frac{(2k)!}{k!k^{(k+1/2)}} \frac{4^{k(L+1)}}{L^k} \tag{IV.52}$$

As before, we find that

$$D_k \sim JC_k \quad \text{with} \quad J = \frac{1}{4} \tag{IV.53}$$

Considering that

$$\frac{z}{(1+z)^2} - \frac{1}{4} = -\frac{(1-z)^2}{4(1+z)^2} \sim \frac{\theta^2}{16} \tag{IV.54}$$

we get

$$D_k - \frac{C_k}{4} \sim \frac{1}{2}\int \frac{d\theta}{2\pi} 4^{k(L+1)-2}\theta^{2(k+1)} e^{-kL\frac{\theta^2}{4}} = (2k+1)\frac{L^{-3/2}}{16\sqrt{\pi}} \frac{(2k)!}{k! k^{(k+3/2)}} \frac{4^{k(L+1)}}{L^k}. \quad (IV.55)$$

Expanding the first monomial $(2k+1)$, the first term gives

$$(2k)\frac{L^{-3/2}}{16\sqrt{\pi}} \frac{(2k)!}{k! k^{(k+3/2)}} \sim \frac{1}{L} J C_k \quad (IV.56)$$

which we can throw away, as before. We can also redefine $B$ to get rid of the factor $\frac{4^{k(L+1)}}{L^k}$ in each coefficient, and finally get:

$$\mu = -\frac{L^{-1/2}}{2\sqrt{\pi}} \sum_{k=1}^{\infty} \frac{(2k)!}{k! k^{(k+3/2)}} B^k, \quad (IV.57)$$

$$E(\mu) - \frac{1}{4}\mu = -\frac{L^{-3/2}}{16\sqrt{\pi}} \sum_{k=1}^{\infty} \frac{(2k)!}{k! k^{(k+5/2)}} B^k. \quad (IV.58)$$

In the case of the ASEP, let us look at the two terms in $C_2$. The first gives:

$$\oint_s \frac{dz}{i2\pi z} F(z)^2 \sim \int \frac{d\theta}{2\pi} \theta^4 e^{-2L\frac{\theta^2}{4}} \sim L^{-5/2} \quad (IV.59)$$

and the second, with both variables expanded around $z = 1$, gives:

$$\oint_S \frac{dz_1}{i2\pi z_1} \frac{dz_2}{i2\pi z_2} F(z_1)F(z_2)K(z_1,z_2) \sim \int \frac{d\theta_1}{2\pi} \frac{d\theta_2}{2\pi} \theta_1^2 e^{-L\frac{\theta_1^2}{4}} \theta_2^2 e^{-L\frac{\theta_2^2}{4}} K(1,1) \sim L^{-3}. \quad (IV.60)$$

We see that the extra $d\theta$ causes the expression to go down by half a power in $L$. This goes for any terms with convolutions in all the $C_k$'s and $D_k$'s, so that the only terms worth keeping are those without convolutions, which are the same as for the TASEP. Because of the global $(1-q)$ factor in $E(\mu)$, we finally get:

$$\mu = -\frac{L^{-1/2}}{2\sqrt{\pi}} \sum_{k=1}^{\infty} \frac{(2k)!}{k! k^{(k+3/2)}} B^k \quad (IV.61)$$

$$E(\mu) - (1-q)\frac{1}{4}\mu = -(1-q)\frac{L^{-3/2}}{16\sqrt{\pi}} \sum_{k=1}^{\infty} \frac{(2k)!}{k! k^{(k+5/2)}} B^k. \quad (IV.62)$$

In this case, the cumulants behave as:

$$E_k \sim (1-q)\pi(\pi L)^{(k-3)/2} \quad (IV.63)$$

for $k \geq 2$. They are divergent with respect to $L$ for $k > 3$, which, as previously, has an impact on the power law behaviour of $g(j)$ around $j = J$.

## IV.1.4 HD-MC transition line

We now take $a < 1$ and $b = 1$. The factors $(1 - bz)(1 - b/z)$ from the denominator compensate part of $(1 - z^2)(1 - z^{-2})$ from the numerator, and we get (for the TASEP):

$$F(z) = \frac{(1+z)^{L+1}(1+z^{-1})^{L+1}}{(1-az)(1-a/z)} \sim 4^{L+1} e^{-L\frac{\theta^2}{4}} \tag{IV.64}$$

using the same saddle-point approximation as in the previous section.

We find:

$$C_k \sim \frac{1}{2} \int \frac{d\theta}{2\pi} \, 4^{k(L+1)} e^{-kL\frac{\theta^2}{4}} = \frac{L^{-1/2}}{2\sqrt{\pi}} \frac{1}{k^{1/2}} 4^{k(L+1)} \tag{IV.65}$$

and

$$D_k \sim JC_k \quad \text{with} \quad J = \frac{1}{4} \tag{IV.66}$$

and, furthermore,

$$D_k - \frac{C_k}{4} \sim \frac{1}{2} \int \frac{d\theta}{2\pi} \, 4^{k(L+1)-2} \theta^2 e^{-kL\frac{\theta^2}{4}} = \frac{L^{-3/2}}{16\sqrt{\pi}} \frac{1}{k^{3/2}} 4^{k(L+1)}. \tag{IV.67}$$

In the case of the ASEP, we find for the first term in $C_2$:

$$\oint_s \frac{dz}{i2\pi z} F(z)^2 \sim \int \frac{d\theta}{2\pi} \, e^{-2L\frac{\theta^2}{4}} \sim L^{-1/2} \tag{IV.68}$$

and for the second:

$$\oint_S \frac{dz_1}{i2\pi z_1} \frac{dz_2}{i2\pi z_2} F(z_1) F(z_2) K(z_1, z_2) \sim \int \frac{d\theta_1}{2\pi} \frac{d\theta_2}{2\pi} \, e^{-L\frac{\theta_1^2}{4}} e^{-L\frac{\theta_2^2}{4}} K(1,1) \sim L^{-1} \tag{IV.69}$$

so that, for the same reasons as in the previous section, all the convolutions can be thrown away. We finally get:

$$\boxed{\mu = -\frac{L^{-1/2}}{2\sqrt{\pi}} \sum_{k=1}^{\infty} \frac{B^k}{k^{3/2}},} \tag{IV.70}$$

$$\boxed{E(\mu) - (1-q)\frac{1}{4}\mu = -(1-q)\frac{L^{-3/2}}{16\sqrt{\pi}} \sum_{k=1}^{\infty} \frac{B^k}{k^{5/2}},} \tag{IV.71}$$

and

$$E_k \sim (1-q)\pi(\pi L)^{(k-3)/2} \tag{IV.72}$$

for $k \geq 2$.

We mentioned, in section III.3, that for a special choice of the boundary parameters, namely $a = q^{1/2}, \tilde{a} = -q^{1/2}, b = 1, \tilde{b} = -q$, which is to say $\alpha = 1, \beta = 1/2, \gamma = q, \delta = q/2$, and considering that

$$(z)_\infty (-qz)_\infty (q^{1/2} z)_\infty (-q^{1/2} z)_\infty = \frac{(z^2)_\infty}{(1+z)} \tag{IV.73}$$

we find that $F(z)$ is equal to:

$$F(z) = (1+z)^{L+1}(1+z^{-1})^{L+1} \qquad \text{(IV.74)}$$

which is the same function as for the periodic ASEP with $2L+2$ sites, and at half filling. It is therefore not surprising to see that the large size asymptotics of the cumulants on the HD-MC (or the LD-MC) transition line, on which that choice of parameters sits, are the same (up to a factor $\frac{1}{2}$) as those found in [94] for the periodic TASEP. This fact will be useful to us in section IV.2.3.

### IV.1.5 Triple point

The last region of the phase diagram we must consider is the triple point $a = b = 1$. This one is particularly tricky, as we shall see, and cannot set $a$ and $b$ to 1 right away. Instead, we need to approach the point in the most convenient way that we can find. We will therefore take $b = 1$, with the same simplifications as in the previous section, and consider $a \to 1^-$. We will also need to define $s = -L\log(a) \to 0$.

The function $F$ becomes:

$$F(z) = \frac{(1+z)^{L+1}(1+z^{-1})^{L+1}}{(1-az)(1-a/z)} = \frac{\left(2\cos(\frac{\theta}{2})\right)^{L+1}}{(1+a)^2(1-X\cos(\frac{\theta}{2}))} \sim \frac{4^{L+1}}{(1+a)^2} \frac{e^{-(L+1)\frac{\theta^2}{4}}}{1-Xe^{-\frac{\theta^2}{4}}} \qquad \text{(IV.75)}$$

where $X = \frac{4a}{(1+a)^2} \sim a = e^{-s/L} < 1$.

We will have to consider $F(z)^k$, and use this relation:

$$\frac{y^N}{(1-Xy)^k} = \frac{1}{(k-1)!}\left(\frac{d}{dX}\right)^{k-1}\frac{y^{N-k+1}}{1-Xy} \qquad \text{(IV.76)}$$

with $\frac{d}{dX} \sim \frac{d}{da} = -\frac{L}{a}\frac{d}{ds}$, so that $\left(\frac{d}{dX}\right)^{k-1} \sim (-1)^{k-1}\left(\frac{L}{a}\right)^{k-1}\left(\frac{d}{ds}\right)^{k-1}$ (these are not the same approximation: from the first one to the second one, we need not to differentiate any factor $1/a$ with respect to $s$, which would lower the exponent of $L$).

Putting this into $C_k$, we get:

$$\begin{aligned}
C_k &\sim \frac{1}{2}\frac{1}{(k-1)!}\left(\frac{d}{dX}\right)^{k-1}\int\frac{d\theta}{2\pi}\frac{e^{-(kL+1)\frac{\theta^2}{4}}}{1-Xe^{-\frac{\theta^2}{4}}}\\
&= \frac{1}{2\sqrt{\pi}}\frac{1}{(k-1)!}\left(\frac{d}{dX}\right)^{k-1}\sum_{n=0}^{\infty}\frac{X^n}{\sqrt{kL+1+n}}\\
&\sim \frac{(-1)^{k-1}}{2\sqrt{\pi}}\frac{1}{(k-1)!}\left(\frac{L}{a}\right)^{k-1}\left(\frac{d}{ds}\right)^{k-1}\sum_{n=0}^{\infty}\frac{e^{-sn/L}}{\sqrt{kL+1+n}}
\end{aligned} \qquad \text{(IV.77)}$$

where we get from the first to the second line by expanding the integrand in powers of $X$, and from the second to the third using the approximation we just mentioned.

Now, considering that

$$\sum_{n=0}^{\infty}\frac{e^{-sn/L}}{\sqrt{kL+1+n}} \sim \int_0^{\infty}\frac{e^{-sx}}{\sqrt{k+x}}\sqrt{L/k}\,dx \qquad \text{(IV.78)}$$

we get

$$C_k \sim \frac{1}{2\sqrt{\pi}} \frac{L^{k-1/2}}{(k-1)!} \int_0^\infty \frac{x^{k-1}e^{-sx}}{(k+x)^{1/2}} dx. \tag{IV.79}$$

Once more, for $D_k$, at first order, we find

$$D_k \sim JC_k \quad \text{with} \quad J = \frac{1}{4} \tag{IV.80}$$

and then, through the same calculations as for $C_k$, we get:

$$D_k - \frac{C_k}{4} \sim \frac{1}{16\sqrt{\pi}} \frac{L^{k-3/2}}{(k-1)!} \int_0^\infty \frac{x^{k-1}e^{-sx}}{(k+x)^{3/2}} dx. \tag{IV.81}$$

Assuming that the only difference for the ASEP is, as in all the previous cases, a factor $(1-q)$ in $E(\mu)$, we can finally write:

$$\mu = -\frac{L^{-1/2}}{2\sqrt{\pi}} \sum_{k=1}^\infty \int_0^\infty \frac{x^{k-1}e^{-sx}}{(k+x)^{1/2}} dx \frac{B^k}{k!}, \tag{IV.82}$$

$$E(\mu) - (1-q)\frac{1}{4}\mu = -(1-q)\frac{L^{-3/2}}{16\sqrt{\pi}} \sum_{k=1}^\infty \int_0^\infty \frac{x^{k-1}e^{-sx}}{(k+x)^{3/2}} dx \frac{B^k}{k!}. \tag{IV.83}$$

Notice that although we need to take $s \to 0$, we have kept it until now, otherwise all of the coefficients $C_k$ and $D_k$ would have diverged. Keeping only the dominant term in $1/s$ in each expression is not an option, because their contributions compensate in the cumulants. What we must do, instead, is to use expressions (IV.14) to get the cumulants, and then take $s$ to 0, which should produce a finite value for each of the $E_k$.

Also note that we can replace the integrals in (IV.82) and (IV.83) by hypergeometric series:

$$\mu = \frac{L^{-1/2}}{2\sqrt{\pi}} \sum_{k=1}^\infty \left( s^{1/2-k} \Gamma(k-1/2) \,_1F_1(1/2; 3/2-k; ks) \right.$$
$$\left. + \frac{k^{k-1/2}}{\sqrt{\pi}} \Gamma(1/2-k)\Gamma(k) \,_1F_1(k; k+1/2; ks) \right) \frac{B^k}{k!} \tag{IV.84}$$

$$E(\mu) - \frac{(1-q)}{4}\mu = \frac{(1-q)L^{-3/2}}{16\sqrt{\pi}} \sum_{k=1}^\infty \left( s^{3/2-k} \Gamma(k-3/2) \,_1F_1(3/2; 5/2-k; ks) \right.$$
$$\left. + 2\frac{k^{k-3/2}}{\sqrt{\pi}} \Gamma(3/2-k)\Gamma(k) \,_1F_1(k; k-1/2; ks) \right) \frac{B^k}{k!} \tag{IV.85}$$

which is slightly less compact, but, arguably, slightly more explicit than the previous expressions.

# IV.2 Large deviations of the current

Now that we have the generating function for the cumulants of the current, we can try to get at its large deviation function, of which it is easier to make physical sense. We recall that this function $g(j)$ is given by the Legendre transform of $E(\mu)$:

$$g(j) = \mu j - E(\mu) \quad \text{with} \quad \frac{d}{d\mu}E(\mu) = j. \tag{IV.86}$$

We will try to get whatever information we can on $g(j)$ in each of the cases we considered in the previous section, except for the last one, which is too pathological (meaning that we didn't manage to get a result). Note that, since we threw a lot of terms away in obtaining the expressions from the last section, we will here only get the behaviour of $g(j)$ for small fluctuations of $j$. We will see how to get the rest in section IV.4, and show that the two are consistent.

## IV.2.1 High/Low density phases

We first look at the low density phase. We recall that:

$$E(\mu) = (1-q)\frac{a}{a+1}\frac{e^\mu - 1}{e^\mu + a}. \tag{IV.87}$$

In this case, we can simply do the Legendre transform right away. The second part of equation (IV.86) gives:

$$e^\mu = a\frac{(1-q-2j) - \sqrt{(1-q)(1-q-4j)}}{2j} = \frac{1-\rho_a}{\rho_a}\frac{r}{1-r} \tag{IV.88}$$

where $r$ is such that $j = (1-q)r(1-r)$ and we recall that $\rho_a = \frac{1}{(1+a)}$.

Putting this in the first part of (IV.86), we get:

$$\boxed{g(j) = (1-q)\left[\rho_a - r + r(1-r)\log\left(\frac{1-\rho_a}{\rho_a}\frac{r}{1-r}\right)\right].} \tag{IV.89}$$

There is a very simple physical interpretation of this function, which we will present in section IV.4.

In all the other cases we must now consider, things won't be as easy. We will first have to obtain an asymptotic expression for $E(\mu)$ for large $\mu$, which will be different for positive and negative fluctuations, and only then take a Legendre transform.

## IV.2.2 HD-LD transition line

For $a = b > 1$, we recall that:

$$\mu = \frac{2}{L}\frac{a+1}{a-1}\left[\mathcal{W}_\mathcal{L}(\sqrt{B}/2) + \mathcal{W}_\mathcal{L}(-\sqrt{B}/2)\right] \tag{IV.90}$$

$$E(\mu) - (1-q)\frac{a}{(1+a)^2}\mu = (1-q)\frac{2}{L^2}\frac{a}{a^2-1}\Big[2\big(\mathcal{W}_\mathcal{L}(\sqrt{B}/2) + \mathcal{W}_\mathcal{L}(-\sqrt{B}/2)\big)$$
$$+ \mathcal{W}_\mathcal{L}(\sqrt{B}/2)^2 + \mathcal{W}_\mathcal{L}(-\sqrt{B}/2)^2\Big] \tag{IV.91}$$

where $\mathcal{W}_\mathcal{L}(x)$ is the Lambert $W$ function [155], solution to $x = \mathcal{W}_\mathcal{L} e^{\mathcal{W}_\mathcal{L}}$.

Luckily for us, many things are known about this function, such as its asymptotic behaviour, which is what we need. For the main branch of the function, called $W_0$, which is defined everywhere except on $]-\infty, -1/e]$, it is known that $\mathcal{W}_\mathcal{L}(x)$ behaves as $\log(x)$ (even for $x$ complex, in which case the angular part of $x$ can be neglected and we get $\log(|x|)$). This will be useful for $B < 0$. For $B > 0$, however, the functions $\mathcal{W}_\mathcal{L}(-\sqrt{B}/2)$ in our expressions have to be continued analytically to the second branch $\mathcal{W}_{-1}$, on which $x$ goes back from $-1/e$ to $0$ and behaves as $\log(-x)$ (fig.-IV.3).

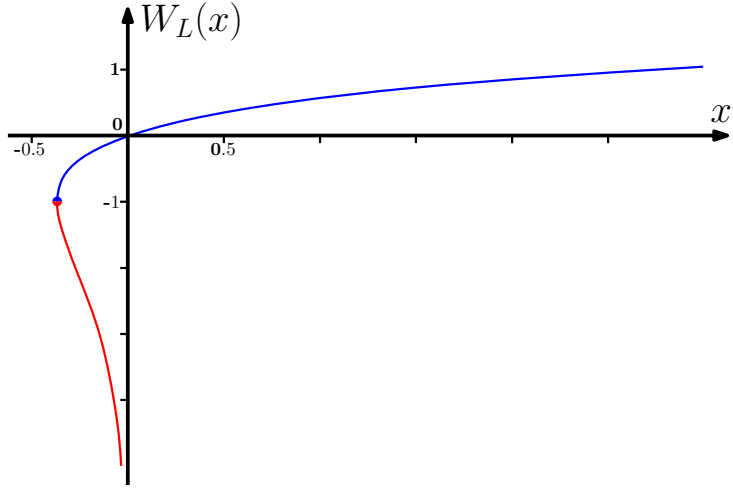

Figure IV.3: Plot of the Lambert $\mathcal{W}$ function. The principal branch (blue) behaves as $\log(x)$ for $x \to \infty$. The second branch (red) behaves as $\log(-x)$ for $x \to 0^-$.

We can now put those remarks to good use. For $B \to -\infty$, we have $\mathcal{W}_\mathcal{L}(\pm\sqrt{B}/2) \sim \log(|B|)/2$, so that:

$$\mu \sim \frac{2}{L}\frac{a+1}{a-1}\log(|B|) \tag{IV.92}$$

$$E(\mu) - (1-q)\frac{a}{(1+a)^2}\mu \sim (1-q)\frac{2}{L^2}\frac{a}{a^2-1}\log(|B|)^2/2 \tag{IV.93}$$

meaning that, for $\mu > 0$ but not too large, we have

$$\boxed{E_+(\mu) - (1-q)\frac{a}{(1+a)^2}\mu \sim (1-q)\frac{a(a-1)}{4(a+1)^3}\mu^2.} \tag{IV.94}$$

We can now take the Legendre transform of this and get, for $j > J = (1-q)\rho_a(1-\rho_a)$:

$$\boxed{\boxed{g_+(j) \sim \frac{(j-J)^2}{J(1-2\rho_a)}}} \tag{IV.95}$$

(where $J = (1-q)\rho_a(1-\rho_a)$ is the mean current).

To do the same for $\mu < 0$, we need to get on the second branch of $\mathcal{W}_\mathcal{L}$, for which we have $B \to 0^+$. In that case, we have $\mathcal{W}_\mathcal{L}(-\sqrt{B}/2) \sim \log(B)/2$, but $\mathcal{W}_\mathcal{L}(\sqrt{B}/2) \sim 0$ (because that part is still on the main branch). This gives us:

$$\mu \sim \frac{2}{L}\frac{a+1}{a-1}\log(|B|)/2 \qquad (\text{IV.96})$$

$$E(\mu) - (1-q)\frac{a}{(1+a)^2}\mu \sim (1-q)\frac{2}{L^2}\frac{a}{a^2-1}\log(|B|)^2/4 \qquad (\text{IV.97})$$

so that, for $\mu < 0$ but not too large, we have

$$\boxed{E_-(\mu) - (1-q)\frac{a}{(1+a)^2}\mu \sim (1-q)\frac{a(a-1)}{2(a+1)^3}\mu^2.} \qquad (\text{IV.98})$$

We can now take the Legendre transform of this and get, for $j < J$:

$$\boxed{g_-(j) \sim \frac{(j-J)^2}{2J(1-2\rho_a)}.} \qquad (\text{IV.99})$$

There are a number of things to be said about these formulae. Notice that the dependence in $L$ has vanished from both cases, so that even though all the cumulants of the current depend on $L$ at $\mu = 0$, none of them do for a finite $\mu$. Notice also that $g_-$ differs from $g_+$ by a factor $\frac{1}{2}$, which comes from the fact that for $\mu > 0$, both functions $\mathcal{W}_\mathcal{L}$ in $\mu$ and $\mathcal{W}_\mathcal{L}^2$ in $E(\mu)$ contribute to the limit, whereas for $\mu < 0$, only one of each does. This results in $g(j)$ not being analytic at $j = J$ (fig.IV.4), which is the signature of a non-equilibrium phase transition (see sections IV.4.3 and IV.4.4 for a description of the phase on each side of the transition). We will see many more of these in this chapter.

### IV.2.3 Maximal current zone

We finally focus on the maximal current zone (i.e. the MC phase and the MC-LD or MC-HD transition lines, which we bundle together because the same method applies to both).

**LD-MC line**

We start with one of the transition lines bordering the MC phase, where we recall that:

$$E(\mu) - \frac{1-q}{4}\mu = (1-q)\frac{L^{-3/2}}{16\sqrt{\pi}}H(B) \qquad (\text{IV.100})$$

$$\mu = \frac{L^{-1/2}}{2\sqrt{\pi}}BH'(B) \qquad (\text{IV.101})$$

with

$$H(B) = -\sum_{k=1}^{\infty}\frac{B^k}{k^{5/2}} \sim \frac{2}{\sqrt{\pi}}\int_{-\infty}^{+\infty}d\theta\,\theta^2\log\left[1-Be^{-\theta^2}\right] \qquad (\text{IV.102})$$

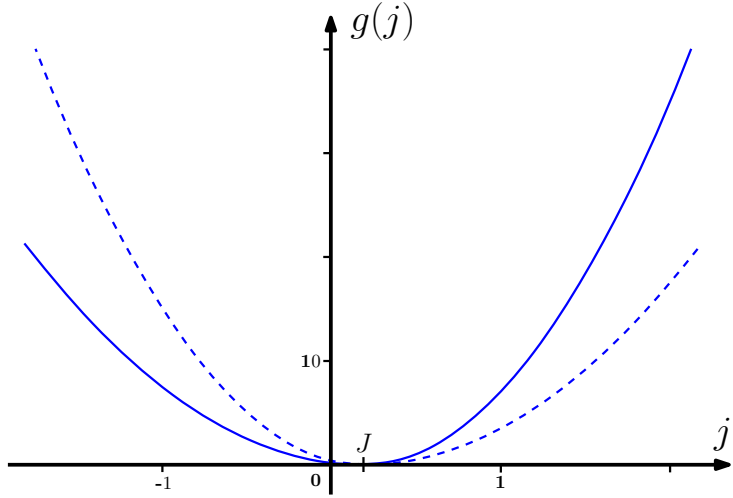

Figure IV.4: Large deviation function of the current around the shock line, for $\rho_a = \frac{1}{4}$. The function is made of two hyperbolic parts with different widths, meeting at $j = J$. The complements of each parabola are represented in dashed lines.

which is a polylogarithm: $H(B) = -Li_{5/2}(B)$.

As in the previous section, the cases $\mu < 0$ and $\mu > 0$ require different approaches.

For $\mu > 0$, we need to take $B \to -\infty$. In this case, the integrand in $H(B)$ can be approximated by:

$$\log\left[1 - Be^{-\theta^2}\right] \sim \log\left[|B|\,e^{-\theta^2}\right] \mathbb{I}\left[\theta^2 < \log(|B|)\right] \tag{IV.103}$$

where $\mathbb{I}[X]$ is the indicator of $X$, equal to 1 if $X$ is true, and 0 if $X$ is false.

We can then estimate:

$$H(B) \sim \frac{4}{\sqrt{\pi}} \int_0^{\log(|B|)^{1/2}} d\theta\, \theta^2 \left(\log(|B|) - \theta^2\right) = \frac{8}{15\sqrt{\pi}} \log(|B|)^{5/2} \tag{IV.104}$$

and

$$BH'(B) \sim \frac{4}{3\sqrt{\pi}} \log(|B|)^{3/2} \tag{IV.105}$$

so that

$$\boxed{E_+(\mu) - \frac{1-q}{4}\mu \sim (1-q)\frac{1}{20}\left(\frac{3}{2}\right)^{2/3} L^{-2/3} \pi^{2/3} \mu^{5/3}.} \tag{IV.106}$$

We can then take the Legendre transform of this result. We find, for $j > J = \frac{1-q}{4}$:

$$\boxed{g_+(j) \sim (j-J)^{5/2} \frac{32\sqrt{3}L}{5\pi(1-q)^{3/2}}} \tag{IV.107}$$

and notice that, for once, it depends on $L$. More on that in section IV.4.

For $\mu < 0$, things get trickier. We have to take into account the fact that the polylogarithm $Li_{5/2}(x)$ has a branch cut at $x = 1$ and is not defined for $x > 1$. Luckily, these calculations have already been done in [95] for the periodic TASEP (which is equivalent to the open one at the transition we are considering), and we only need to reproduce them. What's more, they will serve as a guide for the next case.

The reasoning goes roughly like this: the expressions we have of $E(\mu)$ and $\mu$ in terms of $B$, for all eigenvalues of $M_\mu$, are obtained by taking contour integrals over Bethe roots, and depend on which roots are included in the integral. In the case of the steady state, for $B$ small enough, all the roots we consider are inside of the unit circle. However, it can be shown that, as $B$ gets closer to 1, one of the roots $z_0$ goes to 1, and touches its counterpart $z_0^{-1}$ from outside of the unit circle (fig.-IV.5). Now, since we know, from the Perron-Frobenius theorem, that $E(\mu)$ never crosses any other eigenvalue of $M_\mu$, the choice of roots that consists in taking $z_0^{-1}$ instead of $z_0$ must correspond to $E(\mu)$ as well (because they coincide for $B = 1$). We can therefore find the correct analytic continuations for $\mu$ and $E(\mu)$ in terms of $B$ by finding $z_0$, replacing its contribution in those series by that of $z_0^{-1}$, and taking $B$ back from 1 to 0. This procedure is explained in more detail in [156].

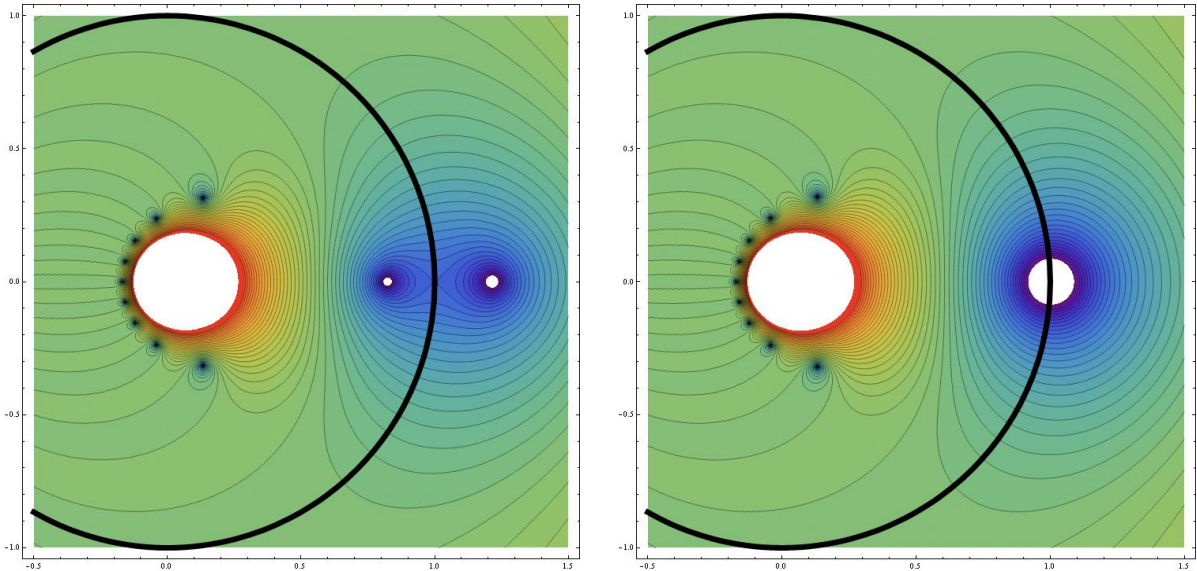

Figure IV.5: Bethe roots for a periodic system with 20 sites and 10 particles. The roots are at the centres of the white discs. The unit circle is represented in black. On the left, where $B < 1$, a pair of roots can be seen to approach the unit circle on the real axis. On the right, where $B = 1$, those roots have merged.

We can find those two roots using equation (IV.102). An integration by parts turns it into:
$$H(B) \sim \frac{2}{3\sqrt{\pi}} \int_{-\infty}^{+\infty} d\theta \; \theta^3 \frac{2\theta B e^{-u}}{B e^{-u} - 1}. \tag{IV.108}$$

For $0 < B < 1$, the poles in this expression are at $\theta_\pm = \pm i\sqrt{-\log(B)}$. The corresponding residues are $i\frac{4\sqrt{\pi}}{3}\theta_\pm^3$ (as explained in [156]), and we must subtract the one corresponding to $\theta_-$ and add the other one to $H$. We get, for $B \to 0$:

$$H(B) = \frac{8}{3}\sqrt{\pi}\bigl[-\log(B)\bigr]^{3/2} - \sum_{k=1}^{\infty} \frac{B^k}{k^{5/2}} \sim \frac{8}{3}\sqrt{\pi}\bigl[-\log(B)\bigr]^{3/2} \qquad \text{(IV.109)}$$

and

$$BH'(B) = -4\sqrt{\pi}\bigl[-\log(B)\bigr]^{1/2} - \sum_{k=1}^{\infty} \frac{B^k}{k^{3/2}} \sim -4\sqrt{\pi}\bigl[-\log(B)\bigr]^{1/2}. \qquad \text{(IV.110)}$$

Putting those together, we find, for $\mu < 0$:

$$\boxed{E_-(\mu) - \frac{1-q}{4}\mu \sim -(1-q)\frac{1}{48}\mu^3} \qquad \text{(IV.111)}$$

and, for $j < J = \frac{1-q}{4}$,

$$\boxed{g_-(j) \sim (J-j)^{3/2}\frac{8}{3(1-q)^{1/2}}.} \qquad \text{(IV.112)}$$

Once more, we find a result that is independent of $L$. We also see that the phase transition that takes place here at $\mu = 0$ is of a different nature than the one on the shock line: the behaviour of $g(j)$ with respect to $L$ changes from one side to the other. This will be explained in section IV.4.

**MC phase**

We will now try to repeat all this for the inside of the maximal current phase. Here, $H(B)$ is replaced by:

$$H(B) = -\sum_{k=1}^{\infty}(2k+1)\frac{(2k)!}{k!k^{(k+5/2)}}\left(\frac{B}{4}\right)^k \sim \frac{2}{\sqrt{\pi}}\int_{-\infty}^{+\infty} d\theta\, \theta^2 \log\bigl[1 - B\theta^2 e^{-\theta^2}\bigr]. \qquad \text{(IV.113)}$$

For $\mu > 0$, i.e. $B \to -\infty$, we have:

$$\log\bigl[1 - B\theta^2 e^{-\theta^2}\bigr] \sim \log\bigl[|B|\theta^2 e^{-\theta^2}\bigr]\, \mathbb{I}\bigl[|B|\theta^2 e^{-\theta^2} > 1\bigr]. \qquad \text{(IV.114)}$$

The upper bound of the integral can therefore be set at $\theta_B$ such that $|B|\theta_B^2 e^{-\theta_B^2} = 1$, in which we recognise the square root of the Lambert $W$ function: $\theta_B = \sqrt{-W_{-1}(-1/B)}$. For large $B$, it behaves as $\log(|B|)^{1/2}$, just as in the previous case.

A few more calculations show that the term $\theta^2$ inside of the log in (IV.113) makes in fact no difference at all for $B \to -\infty$, so that we get the same results:

$$\boxed{E_+(\mu) - \frac{1-q}{4}\mu \sim (1-q)\frac{1}{20}\left(\frac{3}{2}\right)^{2/3}L^{-2/3}\pi^{2/3}\mu^{5/3}} \qquad \text{(IV.115)}$$

and, for $j > J = \frac{1-q}{4}$,

$$\boxed{g_+(j) \sim (j-J)^{5/2}\frac{32\sqrt{3}L}{5\pi(1-q)^{3/2}}.} \qquad \text{(IV.116)}$$

For $\mu < 0$, the story is slightly different. The same scenario as we saw before applies, but this time, there are two pairs of roots crossing the unit circle instead of one (fig.-IV.6), all close to the real axis. Using the same procedure as before, we find them to have exactly the same behaviour with respect to $B$, so that the only difference is that we have twice as many residues as we had then.

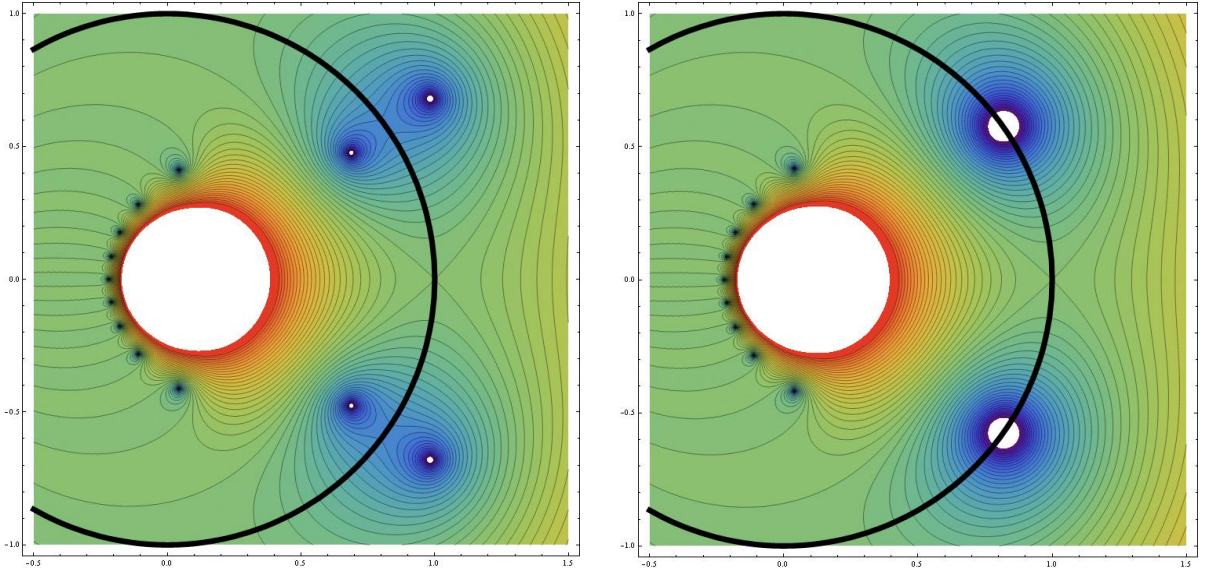

Figure IV.6: Bethe roots for an open system with 9 sites. The unit circle is represented in black. This time, there are two pairs of roots that merge for a critical value of $B$. Those roots get closer to the real axis as $L$ becomes large.

This gives us:
$$H(B) \sim \frac{16}{3}\sqrt{\pi}\bigl[-\log(B)\bigr]^{3/2} \tag{IV.117}$$

and
$$BH'(B) \sim -8\sqrt{\pi}\bigl[-\log(B)\bigr]^{1/2} \tag{IV.118}$$

so that, for $\mu < 0$,
$$\boxed{E_-(\mu) - \frac{1-q}{4}\mu \sim -(1-q)\frac{1}{192}\mu^3} \tag{IV.119}$$

and, for $j < J = \frac{1-q}{4}$,
$$\boxed{g_-(j) \sim (J-j)^{3/2}\frac{16}{3(1-q)^{1/2}}.} \tag{IV.120}$$

The fact that we do not find the same behaviour for negative fluctuations of the current inside of the MC phase and on its boundaries is a sign that there might be a phase transition between those two regions for $\mu < 0$. Since the behaviour is the same for positive fluctuations, there probably isn't one for $\mu > 0$. We will confirm this in section IV.4.

# IV.3 Asymptotics for extreme currents

In this section, we go back to the definition of $M_\mu$ itself, and try to find its largest eigenvalue, and possibly the associated eigenvectors, using direct diagonalisation, in some simplifying limits. The first limit, which we will take in all the cases we will consider, is that of the TASEP (which, from what we've seen before, makes no difference apart from the absence of a factor $(1-q)$ in $E(\mu)$).

As we recall, $M_\mu$ is defined by:

$$M_\mu = m_0(\mu_0) + \sum_{i=1}^{L-1} M_i(\mu_i) + m_L(\mu_l) \qquad \text{(IV.121)}$$

with

$$m_0(\mu_0) = \begin{bmatrix} -\alpha & 0 \\ \alpha e^{\mu_0} & 0 \end{bmatrix} \;,\; M_i(\mu_i) = \begin{bmatrix} 0 & 0 & 0 & 0 \\ 0 & 0 & e^{\mu_i} & 0 \\ 0 & 0 & -1 & 0 \\ 0 & 0 & 0 & 0 \end{bmatrix} \;,\; m_L(\mu_l) = \begin{bmatrix} 0 & \beta e^{\mu_L} \\ 0 & -\beta \end{bmatrix} \qquad \text{(IV.122)}$$

and $\mu = \sum \mu_i$. We have kept the fugacities on all bonds, because we will need to distribute them differently depending on the situation.

The three limits we will consider are that of a very low current ($\mu \to -\infty$), that of very low boundary rates in the shock phase ($\alpha = \beta \to 0$), and that of a very high current ($\mu \to \infty$).

## IV.3.1 Low current limit

We start with the $\mu \to -\infty$ limit. We choose $\mu_i = \frac{\mu}{L+1}$ for every bond, and note $\varepsilon = e^{\mu/(L+1)} \to 0$.

We can then write $M_\mu$ as:
$$M_\mu = M_d + \varepsilon M_j \qquad \text{(IV.123)}$$
where $M_d$ is the matrix containing the diagonal (escape) rates, and $M_j$ is the matrix containing the non-diagonal (jumping) rates. We see that $M_\mu$ is almost diagonal, and we are going to treat it perturbatively in $\varepsilon$.

The entries of $M_d$ are given by:

$$M_d(\mathcal{C}, \mathcal{C}) = -(1-n_1)\alpha - \sum_{i=1}^{L-1} n_i(1-n_{i+1}) - n_L \beta \qquad \text{(IV.124)}$$

where $n_i$ is the occupancy of site $i$ in $\mathcal{C}$. At lowest order in $\varepsilon$, those are the eigenvalues of $M_\mu$.

Since we are looking for the highest eigenvalue of $M_\mu$, we see that there are four possible situations (assuming that $\alpha$ and $\beta$ are limited to $[0,1]$): if $\alpha < \beta$, then the best configuration is empty ($n_i = 0$ for all $i$'s), with an eigenvalue of $E = -\alpha$. If $\beta < \alpha$, we have the same in reverse: the best configuration is full ($n_i = 1$ for all $i$'s) and $E = -\beta$ (those two first cases are symmetric to one another, and we will only be considering the first one). If $\alpha = \beta < 1$, then $E = -\alpha$, and we have two competing configurations: empty or full. Finally, if $\alpha = \beta = 1$, then any configuration with a block of 1's followed by a block of 0's has an eigenvalue of $E = -1$, which is the highest, and there are $L+1$ of those. We therefore have two phases, one line, and one point, to investigate (fig.-IV.7).

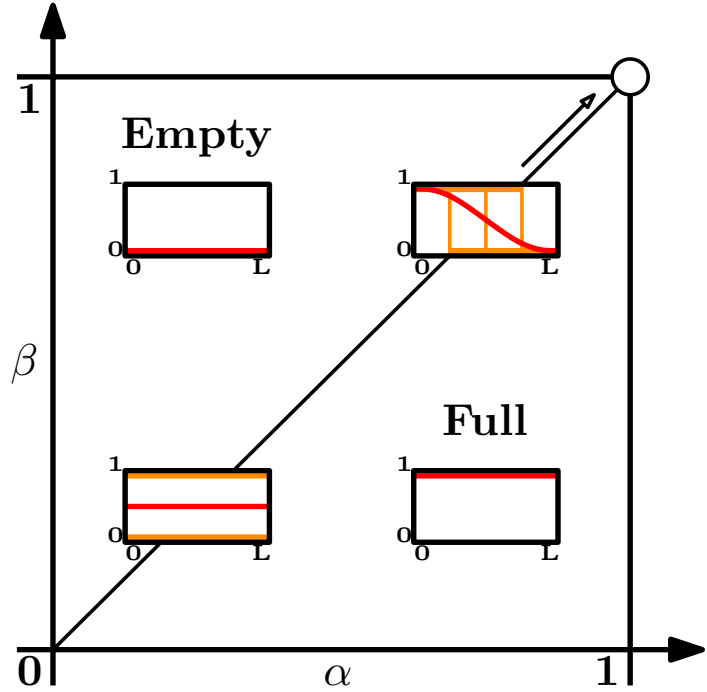

Figure IV.7: Phase diagram of the open ASEP for very low current. The mean density profiles are represented in red the insets. The profiles in orange are the individual configurations which compose the steady state.

**Empty/full phases**

We first consider the case where $\alpha < \beta$. We know that the dominant eigenvalue of $M_\mu$ in this case is equal to $-\alpha$ at leading order in $\varepsilon$, but since we need to differentiate it with respect to $\mu$ to get the current and its large deviation function, we will have to find the next to leading order as well.

We expand that eigenvalue and its corresponding eigenvector as a series in $\varepsilon$:

$$E(\mu) = E_0 + \sum E_k \varepsilon^k \quad , \quad |P_\mu\rangle = |P_0\rangle + \sum \varepsilon^k |P_k\rangle \tag{IV.125}$$

where

$$E(\mu)|P_\mu\rangle = M_\mu |P_\mu\rangle. \tag{IV.126}$$

We already know that

$$|P_0\rangle = |00\ldots 00\rangle \quad , \quad E_0 = -\alpha. \tag{IV.127}$$

The first order in $\varepsilon$ in (IV.126) gives:

$$E_1|P_0\rangle + E_0|P_1\rangle = M_j|P_0\rangle + M_d|P_1\rangle. \tag{IV.128}$$

Since the only state that can be reached from $|P_0\rangle$ through $M_j$ is $|10\ldots00\rangle$, which has no overlap with $|P_0\rangle$, we get:

$$|P_1\rangle = \frac{1}{E_0 - M_d} M_j |P_0\rangle \sim |10\ldots00\rangle \quad , \quad E_1 = 0 \tag{IV.129}$$

so that the first correction to $E(\mu)$ is 0. The second order in $\varepsilon$ in (IV.126) gives:

$$E_2 |P_0\rangle + E_0 |P_2\rangle = M_j |P_1\rangle + M_d |P_2\rangle. \tag{IV.130}$$

Again, the only state that can be reached from $|P_1\rangle$ through $M_j$ is $|010\ldots00\rangle$, which has no overlap with $|P_0\rangle$:

$$|P_2\rangle = \Big(\frac{1}{E_0 - M_d} M_j\Big)^2 |P_0\rangle \sim |010\ldots00\rangle \quad , \quad E_2 = 0 \tag{IV.131}$$

and once more, the correction to $E(\mu)$ is 0.

The first possible non-zero correction to $E(\mu)$ that we might get is when we find a $|P_k\rangle$ which has an overlap with $|P_0\rangle$. The shortest way to go back to $|P_0\rangle$ through jumps is to have one particle enter the system (the first step being $|P_1\rangle$), and then travel all the way to the other end, and exit the system. This can be done in $L+1$ steps, so that $E_k = 0$ for $k : 1..L$, and

$$E_{L+1} |P_0\rangle + E_0 |P_{L+1}\rangle = M_j |P_L\rangle + M_d |P_{L+1}\rangle. \tag{IV.132}$$

Putting those $L+1$ first equations together, we get

$$|P_{L+1}\rangle = M_j \Big(\frac{1}{E_0 - M_d} M_j\Big)^L |P_0\rangle \sim |P_0\rangle + \ldots \tag{IV.133}$$

and

$$E_{L+1} = \langle P_0 | P_{L+1}\rangle = \frac{\alpha}{1 - \alpha}. \tag{IV.134}$$

Putting this back into $E(\mu)$, we get:

$$\boxed{E(\mu) \sim -\alpha + e^\mu \frac{\alpha}{1 - \alpha}} \tag{IV.135}$$

and

$$\boxed{\boxed{g(j) = \alpha + j \log(j) - j\big(\log(\alpha/(1-\alpha)) + 1\big).}} \tag{IV.136}$$

We may note that taking the limit $\mu \to -\infty$ in (IV.32) gives the same result:

$$E(\mu) = \frac{a}{a+1} \frac{e^\mu - 1}{e^\mu + a} \sim -\frac{1}{1+a} + \frac{1}{a} e^\mu = -\alpha + e^\mu \frac{\alpha}{1-\alpha} \tag{IV.137}$$

which indicates that this expression for $E(\mu)$ remains valid for all $\mu < 0$. We will be more specific in section IV.4.

Finally, note that in this case, the second largest eigenvalue is $-\beta$ for $\varepsilon \to 0$ (and corresponds to a completely full system), so that the gap between the first two eigenvalues of $M_\mu$ is finite and equal, at leading order, to $\Delta E = (\beta - \alpha)$.

The corresponding results for $\beta < \alpha$ (the 'full' phase) can be obtained by exchanging $\alpha$ with $\beta$.

**Coexistence line**

We now consider the slightly more complex case where $\alpha = \beta < 1$. This time, there are two states with equal eigenvalues for $\mu = -\infty$:

$$|P_0\rangle = |0\rangle = |00\ldots00\rangle \quad \text{with} \quad E_0 = -\alpha \qquad (IV.138)$$

and

$$|\tilde{P}_0\rangle = |1\rangle = |11\ldots11\rangle \quad \text{with} \quad \tilde{E}_0 = -\alpha. \qquad (IV.139)$$

As in the previous case, the first corrections to those eigenvalues are the rates with which we can go from these configurations back to themselves, but since they are degenerate, we must also consider how we can go from one to the other. As before, it takes the same $L+1$ steps to go from $|0\rangle$ to itself, or from $|1\rangle$ to itself, so that the first correction to both $E$ and $\tilde{E}$ is $e^{\mu}\frac{\alpha}{1-\alpha}$. At this stage, those states are still degenerate. To lift the degeneracy, we have to consider the shortest way to go from $|0\rangle$ to $|1\rangle$, or the opposite. This means completely filling or emptying the system, and can be done in $L(L+1)/2$ steps. This tells us that the difference between the two highest eigenvalues is of order $\varepsilon^{L(L+1)/2} = e^{\frac{L}{2}\mu}$. For symmetry reasons, the main eigenvector is then $\frac{1}{2}(|0\rangle + |1\rangle)$, and the second one is $\frac{1}{2}(|0\rangle - |1\rangle)$.

In conclusion, we have, as in the previous case,

$$\boxed{E(\mu) \sim -\alpha + e^{\mu}\frac{\alpha}{1-\alpha}} \qquad (IV.140)$$

and

$$\boxed{\boxed{g(j) = \alpha + j\log(j) - j\bigl(\log(\alpha/(1-\alpha)) + 1\bigr)}} \qquad (IV.141)$$

but this time, the gap goes like $\Delta E \sim e^{\frac{L}{2}\mu}$.

There is a systematic way to perform the calculations we just did, called the 'resolvent formalism' [157]. We present it here for two reasons: it will be useful to us in the next section, and it allows, in the case of a perturbation around degenerate states, to rigorously define an effective interaction matrix between those states. This gives us, in essence, a reduced dynamics for the system in the subset of phase space which contains only the dominant configurations.

This formalism can be stated as follows: for a general matrix $M$ with eigenvalues $E_i$ and eigenvectors $|P_i\rangle$ and $\langle P_i|$, we have:

$$\oint_C \frac{dz}{i2\pi}\frac{1}{z-M} = \sum_{E_i \in C} |P_i\rangle\langle P_i| \qquad (IV.142)$$

where the sum is over the eigenvalues of $M$ which lie inside of the contour $C$. What's more, we have

$$\oint_C \frac{dz}{i2\pi}\frac{z}{z-M} = \sum_{E_i \in C} E_i |P_i\rangle\langle P_i|. \qquad (IV.143)$$

Now, considering $M_\mu$ with $\alpha = \beta$, a good way to isolate the two dominant eigenvectors is to consider that same contour integral, with a contour close enough to $-\alpha$ so that the two highest eigenvalues are inside it, but not any of the others. The simplest way to do that is to consider a small circle around $-\alpha$. This gives us an effective matrix $M_{eff}$ such that:

$$M_{eff} = -\alpha + \oint_C \frac{dz}{i2\pi} \frac{z}{z - \alpha - M_d - \varepsilon M_j} \tag{IV.144}$$

where $C$ is a small circle centred at 0.

We can now expand this expression in terms of $\varepsilon$:

$$M_{eff} = -\alpha + \oint_C \frac{dz}{i2\pi} \sum_{k=0}^{\infty} \frac{z}{z - (M_d + \alpha)} \left(M_j \frac{1}{z - (M_d + \alpha)}\right)^k \varepsilon^k \tag{IV.145}$$

which is a sum over paths of length $k$, with transitions given by $M_j$ and a 'potential' given by $(z - M_d - \alpha)^{-1}$. We see that the only terms which contribute to the integral (i.e. that give first order poles which yield non-zero residues) are those for which $M_d$ is taken at $-\alpha$ exactly twice, which is to say the paths that go through $|0\rangle$ or $|1\rangle$ twice.

It is now fairly easy to find the amplitudes of $M_{eff}$ between $|0\rangle$ and $|1\rangle$: we only have to project that expression between those states, and since $M_d = -\alpha$ in both of those, we only have to consider all the paths going from one of those states to another without going through them at any other point. Since we are doing a perturbative expansion in $\varepsilon$, we only need the term with the lowest number of steps. Between $|0\rangle$ and itself, or $|1\rangle$ and itself, there is only one path of the lowest length, which is $L+1$, and the amplitude for that path is $\frac{\alpha}{1-\alpha}$. Between $|0\rangle$ and $|1\rangle$, or the opposite, the shortest length is $L(L+1)/2$. There are many suitable paths for that transition, and the total amplitude is some $X$ which we don't need anyway, since that factor doesn't appear in the dominant term in $E(\mu)$ (it would be of order $e^{\frac{L}{2}\mu}$).

All in all, we have an effective matrix given by:

$$M_{eff} = \begin{bmatrix} -\alpha + e^{\mu} \frac{\alpha}{1-\alpha} & X e^{\frac{L}{2}\mu} \\ X e^{\frac{L}{2}\mu} & -\alpha + e^{\mu} \frac{\alpha}{1-\alpha} \end{bmatrix} \tag{IV.146}$$

which is easy to diagonalise, and we can retrieve the results we found earlier.

**Equal rates point**

For the last case, where all the jumping rates are equal ($\alpha = \beta = 1$), we find $L+1$ states with an eigenvalue equal to $-1$ for $\mu = -\infty$. Those states are given by $|k\rangle = |\{1\}_k \{0\}_{L-k}\rangle$, i.e. by configurations made of a block of 1's followed by a block of 0's. Those are called 'anti-shocks', being symmetric to the usual shocks which have a low density region followed by a high density one.

Using the resolvent formalism, we find:

$$\langle k | M_{eff} | k \rangle \sim -1 + \varepsilon^{L+1}, \tag{IV.147}$$
$$\langle k+1 | M_{eff} | k \rangle \sim \varepsilon^{k+1}, \tag{IV.148}$$
$$\langle k-1 | M_{eff} | k \rangle \sim \varepsilon^{L-k+1}, \tag{IV.149}$$

as well as terms of the type

$$\langle k+2|M_{eff}|k\rangle \sim X\varepsilon^{2k+3}, \tag{IV.150}$$

$$\langle k-2|M_{eff}|k\rangle \sim Y\varepsilon^{2L-2k+3}, \tag{IV.151}$$

$$\langle k+3|M_{eff}|k\rangle \sim Z\varepsilon^{3k+6}, \tag{IV.152}$$

and so on. We checked those last terms to be of sub-leading order in the end, so we will save ourselves the trouble of dealing with them and neglect them right away.

We now have:

$$M_{eff} = -1 + \varepsilon^{L+1} + \sum_{k=1}^{L} \varepsilon^k |k\rangle\langle k-1| + \varepsilon^{L-k+1}|k-1\rangle\langle k|. \tag{IV.153}$$

We can transform it through a matrix similarity to have all the non-diagonal coefficients be equal to $\varepsilon^{(L+1)/2}$, which yields:

$$M_{eff} = -1 + \varepsilon^{L+1} + \varepsilon^{(L+1)/2} \sum_{k=1}^{L} |k\rangle\langle k-1| + |k-1\rangle\langle k|. \tag{IV.154}$$

This is a well known tridiagonal matrix, used to model the electronic interactions in conjugated dienes through the Hückel method [158], among other things. It is easily diagonalised. Its eigenvalues are:

$$E^{(k)} = -1 + 2\varepsilon^{(L+1)/2} \cos(k\pi/(L+2)) \tag{IV.155}$$

for $k \in [\![1, L+1]\!]$. The highest one is

$$E^{(1)} = -1 + 2\varepsilon^{(L+1)/2} \cos(\pi/(L+2)) \tag{IV.156}$$

and the gap to the second one is

$$\Delta E = 2\varepsilon^{(L+1)/2}\Big(\cos(\pi/(L+2)) - \cos(2\pi/(L+2))\Big) \sim \frac{3\pi^2}{L^2}e^{\mu/2}. \tag{IV.157}$$

In the end, we find that

$$\boxed{E(\mu) \sim -1 + 2e^{\mu/2}} \tag{IV.158}$$

and

$$\boxed{g(j) = 1 + 2j\log(j) - 2j.} \tag{IV.159}$$

Moreover, knowing that the eigenvector associated to that first eigenvalue is distributed according to a sine function, which is to say that the probability of $|k\rangle$ is:

$$P(k) \sim \sin\left(\frac{\pi k}{L+1}\right)^2 \tag{IV.160}$$

we find that the mean density $\rho_n$ at site $n$ is of the form:

$$\boxed{\rho_n = 1 - \frac{n}{L+1} + \frac{1}{2\pi}\sin\left(\frac{2\pi n}{L+1}\right).} \tag{IV.161}$$

## IV.3.2 Low boundary rates limit

We now look at the case where $\alpha = \beta \to 0$. We choose $\mu_0 = \mu_L = \frac{\mu}{L+1} - \frac{L-1}{L+1}\log(\alpha)$ and $\mu_i = \frac{\mu}{L+1} + \frac{2}{L+1}\log(\alpha)$ for $i \in [\![1, L]\!]$, so that $\alpha e^{\mu_0} = \beta e^{\mu_L} = e^{\mu_i}$, which we shall note as $\varepsilon$. Note that for $\alpha$ fixed, $\varepsilon$ is not small if $\mu$ is too large, and that for $\mu$ finite, it is larger than $\alpha$.

In this case, we can write $M_\mu$ as:

$$M_\mu = M_d + \varepsilon M_j - \alpha(n_L + 1 - n_1) \tag{IV.162}$$

with $M_j$ having all rates equal to 1, and

$$M_d(\mathcal{C}, \mathcal{C}) = \sum_{i=1}^{L-1} n_i(1 - n_{i+1}). \tag{IV.163}$$

This time, because $\alpha$ is close to 0, the configurations with the highest eigenvalues at first order are the ones with the lowest numbers of 10's in them (each of those lowers the eigenvalue by 1), which is to say the shock configurations $|k\rangle = |\{0\}_k\{1\}_{L-k}\rangle$. Using the resolvent formalism again, we find:

$$\langle 0|M_{eff}|0\rangle \sim -\alpha + \varepsilon^{L+1}, \tag{IV.164}$$

$$\langle L+1|M_{eff}|L+1\rangle \sim -\alpha + \varepsilon^{L+1}, \tag{IV.165}$$

$$\langle k|M_{eff}|k\rangle \sim -2\alpha + \varepsilon^{L+1}, \tag{IV.166}$$

$$\langle k+1|M_{eff}|k\rangle \sim \varepsilon^{k+1}, \tag{IV.167}$$

$$\langle k-1|M_{eff}|k\rangle \sim \varepsilon^{L-k+1}, \tag{IV.168}$$

and other rates which we will be neglecting, as in the previous section.

After a matrix similarity, we find that $M_{eff}$ takes the simple form:

$$M_{eff} = -2\alpha + \varepsilon^{L+1} + \alpha\big(|0\rangle\langle 0| + |L+1\rangle\langle L+1|\big) + \varepsilon^{(L+1)/2} \sum_{k=1}^{L} |k\rangle\langle k-1| + |k-1\rangle\langle k|. \tag{IV.169}$$

We were not able to diagonalise it exactly as for (IV.154), but could determine numerically that for $L \to \infty$, the largest eigenvalue behaves as:

$$E = -2\alpha + \varepsilon^{L+1} + \alpha + \varepsilon^{L+1}/\alpha \quad \text{if} \quad \varepsilon^{(L+1)/2} < \alpha \tag{IV.170}$$

$$E = -2\alpha + \varepsilon^{L+1} + 2\varepsilon^{(L+1)/2} \quad \text{if} \quad \varepsilon^{(L+1)/2} > \alpha \tag{IV.171}$$

Replacing $\varepsilon$ by its definition, we find that $\varepsilon^{(L+1)/2} = \alpha e^{\mu/2}$, so that the two cases above become $\mu < 0$ and $\mu > 0$ respectively. We then have, for $\mu < 0$:

$$\boxed{E_-(\mu) = -\alpha + e^\mu \alpha} \tag{IV.172}$$

and

$$\boxed{\boxed{g_-(j) = \alpha + j\log(j) - j(\log(\alpha) + 1)}} \tag{IV.173}$$

and, for $\mu < 0$:

$$\boxed{E_+(\mu) = -2\alpha + 2e^{\mu/2}\alpha} \tag{IV.174}$$

and

$$\boxed{\boxed{g_+(j) = 2\alpha + 2j\log(j) - 2j(\log(\alpha) + 1).}} \tag{IV.175}$$

### IV.3.3 Large current limit

The last limit we consider is that where $\mu \to \infty$.

We first consider $\mu_i = \frac{\mu}{L+1}$ for all $i$'s. In this case, it's the diagonal part of $M_\mu$ that is negligible. We can write:
$$M_\mu \sim e^{\mu/(L+1)} M_j \tag{IV.176}$$

with
$$M_j = \alpha S_1^+ + \sum_{n=1}^{L-1} S_n^- S_{n+1}^+ + \beta S_L^- \tag{IV.177}$$

where $S_n^\pm$ are the operators for the creation or annihilation of a particle at site $n$.

The bulk part of that matrix looks like half that of the Hamiltonian of the XX spin chain [159]. If we could somehow combine this matrix with its transpose, we could rebuild that Hamiltonian, and then diagonalise it using free fermions techniques.

Let us examine the commutator between $M_j$ and ${}^tM_j$ (where all the signs of the operators are inverted). Since all $S_n^\pm$ commute for different sites, we only have to consider a few terms:

$$[\alpha S_1^+, \alpha S_1^-] = \alpha^2 (2n_1 - 1), \tag{IV.178}$$
$$[S_n^- S_{n+1}^+, S_n^+ S_{n+1}^-] = n_{n+1} - n_n, \tag{IV.179}$$
$$[\beta S_L^-, \beta S_L^+] = \beta^2 (1 - 2n_L). \tag{IV.180}$$

We see that if (and only if) $\alpha = \beta = 1/\sqrt{2}$, those terms cancel one another and we get $[M_j, {}^tM_j] = 0$. Let us therefore change our fugacities to:

$$\mu_0 = \frac{\mu}{L+1} + \frac{1}{L+1} \log(2\alpha\beta) - \log(\sqrt{2}\alpha), \tag{IV.181}$$
$$\mu_i = \frac{\mu}{L+1} + \frac{1}{L+1} \log(2\alpha\beta), \tag{IV.182}$$
$$\mu_L = \frac{\mu}{L+1} + \frac{1}{L+1} \log(2\alpha\beta) - \log(\sqrt{2}\beta), \tag{IV.183}$$

for which we have:
$$M_\mu \sim A\, M_j \tag{IV.184}$$

with $A = (2\alpha\beta e^\mu)^{\frac{1}{L+1}}$ and
$$M_j = \frac{1}{\sqrt{2}} S_1^+ + \sum_{n=1}^{L-1} S_n^- S_{n+1}^+ + \frac{1}{\sqrt{2}} S_L^-. \tag{IV.185}$$

Note that, since $A$ is the only thing that depends on $\alpha$ and $\beta$, and that for $L \to \infty$ we have $A \sim e^{\mu/L}$, all the results in this section will be entirely independent of the boundary parameters.

We know, from the Perron-Frobenius theorem, that the highest eigenvalue of that matrix is real and non-degenerate. It is therefore also the highest eigenvalue of its transpose, with the same eigenvectors (because they commute). This allows us to define $H = \frac{1}{2}(M_j + {}^t M_j)$, which has the same highest eigenvalue and the same eigenvectors as $M_j$. $H$ is given by:

$$H = \frac{1}{\sqrt{8}} S_1^x + \frac{1}{2} \sum_{n=1}^{L-1} (S_n^- S_{n+1}^+ + S_n^+ S_{n+1}^-) + \frac{1}{\sqrt{8}} S_L^x \qquad \text{(IV.186)}$$

which is the Hamiltonian for the open XX chain with spin $1/2$ and extra boundary terms $S_1^x$ and $S_L^x$ (with $S^x = S^+ + S^-$). This spin chain was studied for general boundary conditions in [159]. We will present here a simpler version of their calculations, only applicable to our situation but much less intricate than their general solution. Note that applying their solution to our simpler case turned out to be more difficult than solving it ourselves directly, which is what we did in the end.

The first step in diagonalising this Hamiltonian is to consider two extra sites, one at $0$ and one at $L+1$, which we couple with our system by defining an new Hamiltonian:

$$\tilde{H} = \frac{1}{\sqrt{8}} S_0^x S_1^x + \frac{1}{2} \sum_{n=1}^{L-1} (S_n^- S_{n+1}^+ + S_n^+ S_{n+1}^-) + \frac{1}{\sqrt{8}} S_L^x S_{L+1}^x. \qquad \text{(IV.187)}$$

Since $[\tilde{H}, S_0^x] = [\tilde{H}, S_{L+1}^x] = 0$, this modified Hamiltonian has four sectors, corresponding to the eigenspaces of $S_0^x$ and $S_{L+1}^x$. Since each of those has two eigenvalues $1$ and $-1$, we can recover $H$ by projecting $\tilde{H}$ onto the eigenspaces of $S_0^x$ and $S_{L+1}^x$ where both eigenvalues are $1$:

$$H = \frac{1}{4} \big( \langle 0_0 | + \langle 1_0 | \big) \otimes \big( \langle 0_{L+1} | + \langle 1_{L+1} | \big) \tilde{H} \big( |0_0\rangle + |1_0\rangle \big) \otimes \big( |0_{L+1}\rangle + |1_{L+1}\rangle \big). \qquad \text{(IV.188)}$$

We are now left with diagonalising $\tilde{H}$, which is better than $H$ in that it is quadratic.

The next step in solving that problem is to do a Jordan-Wigner transformation on the operators $S_n^\pm$:

$$c_n = \left( \prod_{m=0}^{n-1} (-1)^{n_m} \right) S_n^- \quad , \quad c_n^\dagger = \left( \prod_{m=0}^{n-1} (-1)^{n_m} \right) S_n^+, \qquad \text{(IV.189)}$$

$$S_n^- = \left( \prod_{m=0}^{n-1} (-1)^{c_m^\dagger c_m} \right) c_n \quad , \quad S_n^+ = \left( \prod_{m=0}^{n-1} (-1)^{c_m^\dagger c_m} \right) c_n^\dagger, \qquad \text{(IV.190)}$$

which yields fermionic operators:

$$\{c_n^\dagger, c_m\} = \delta_{n,m}, \qquad \text{(IV.191)}$$
$$\{c_n^\dagger, c_m^\dagger\} = 0, \qquad \text{(IV.192)}$$
$$\{c_n, c_m\} = 0. \qquad \text{(IV.193)}$$

The elements of $\tilde{H}$ become:

$$S_n^+ S_{n+1}^- = c_n^\dagger c_{n+1}, \qquad \text{(IV.194)}$$
$$S_n^- S_{n+1}^+ = c_{n+1}^\dagger c_n, \qquad \text{(IV.195)}$$
$$S_0^x S_1^x = (c_0^\dagger - c_0)(c_1^\dagger + c_1), \qquad \text{(IV.196)}$$
$$S_L^x S_{L+1}^x = (c_L^\dagger - c_L)(c_{L+1}^\dagger + c_{L+1}), \qquad \text{(IV.197)}$$

so that

$$\tilde{H} = \frac{1}{\sqrt{8}}(c_0^\dagger - c_0)(c_1^\dagger + c_1) + \frac{1}{2}\sum_{n=1}^{L-1}(c_n^\dagger c_{n+1} + c_{n+1}^\dagger c_n) + \frac{1}{\sqrt{8}}(c_L^\dagger - c_L)(c_{L+1}^\dagger + c_{L+1}). \quad \text{(IV.198)}$$

We now perform a Bogoliubov transformation on $\tilde{H}$, writing it as

$$\tilde{H} = \mathcal{E}_0 + \sum_k \mathcal{E}_k d_k^\dagger d_k \quad \text{(IV.199)}$$

with all the $\mathcal{E}_k > 0$, and the $d_k$'s to be determined.

We want the $d_k$'s to be fermionic, so that $[\tilde{H}, d_k^\dagger] = \mathcal{E}_k d_k^\dagger$, which is the equation we will now try to solve. We have two trivial solutions with energy 0 (called zero-modes): $(c_0^\dagger + c_0)$ and $(c_{L+1}^\dagger - c_{L+1})$. For the other solutions, we write:

$$d_k^\dagger = \frac{X^{(k)}}{\sqrt{2}}(c_0^\dagger - c_0) + \sum_{n=1}^{L} A_n^{(k)} c_n^\dagger + B_n^{(k)} c_n + \frac{Y^{(k)}}{\sqrt{2}}(c_{L+1}^\dagger + c_{L+1}) \quad \text{(IV.200)}$$

and $[\tilde{H}, d_k^\dagger] = \mathcal{E}_k d_k^\dagger$ becomes:

$$A_{n+1}^{(k)} + A_{n-1}^{(k)} = 2\mathcal{E}_k A_n^{(k)} \quad \text{for} \quad n \in [\![2, L-1]\!], \quad \text{(IV.201)}$$

$$-B_{n+1}^{(k)} - B_{n-1}^{(k)} = 2\mathcal{E}_k B_n^{(k)} \quad \text{for} \quad n \in [\![2, L-1]\!], \quad \text{(IV.202)}$$

$$X^{(k)} + A_2^{(k)} = 2\mathcal{E}_k A_1^{(k)}, \quad \text{(IV.203)}$$

$$X^{(k)} - B_2^{(k)} = 2\mathcal{E}_k B_1^{(k)}, \quad \text{(IV.204)}$$

$$A_1^{(k)} + B_1^{(k)} = 2\mathcal{E}_k X^{(k)}, \quad \text{(IV.205)}$$

$$Y^{(k)} + A_{L-1}^{(k)} = 2\mathcal{E}_k A_L^{(k)}, \quad \text{(IV.206)}$$

$$-Y^{(k)} - B_{L-1}^{(k)} = 2\mathcal{E}_k B_L^{(k)}, \quad \text{(IV.207)}$$

$$A_L^{(k)} - B_L^{(k)} = 2\mathcal{E}_k Y^{(k)}. \quad \text{(IV.208)}$$

All those equations can be written in a more compact form by defining:

$$A_{L+1}^{(k)} = Y^{(k)}, \quad \text{(IV.209)}$$

$$A_{L+1+n}^{(k)} = (-1)^n B_{L+1-n}^{(k)}, \quad \text{(IV.210)}$$

$$A_0^{(k)} = X^{(k)}, \quad \text{(IV.211)}$$

for which they become:

$$A_{n+1}^{(k)} + A_{n-1}^{(k)} = 2\mathcal{E}_k A_n^{(k)} \quad \text{for} \quad n \in [\![1, 2L]\!], \quad \text{(IV.212)}$$

$$A_{2L}^{(k)} + (-1)^L A_0^{(k)} = 2\mathcal{E}_k A_{2L+1}^{(k)}, \quad \text{(IV.213)}$$

$$(-1)^L A_{2L+1}^{(k)} + A_1^{(k)} = 2\mathcal{E}_k A_0^{(k)}. \quad \text{(IV.214)}$$

These are the same equations which we would have found for a periodic $XX$ spin chain with $2L + 2$ sites, the only (but important) difference being that $d_k^\dagger$ mixes $c_k$'s and $c_k^\dagger$'s, so that the total spin is not conserved.

We look for plane wave solutions of the form $A_n = r^n$, with $2\mathcal{E} = r + \frac{1}{r}$. This automatically solves eq.(IV.212). The other two equations become:

$$r^{2L} + (-1)^L = r^{2L} + r^{2L+2}, \tag{IV.215}$$

$$(-1)^L r^{2L+1} + r = r + \frac{1}{r} \tag{IV.216}$$

and both simplify into

$$r^{2L+2} = (-1)^L. \tag{IV.217}$$

We have $2L+2$ solutions to this equation, given by $r = \omega_k = e^{\frac{i\pi(L-2k+2)}{2L+2}}$ for $k \in [\![1, 2L+2]\!]$, so that $A_n^{(k)} = \omega_k^n$, and the energies are given by:

$$\boxed{\mathcal{E}_k = \cos\left(\frac{(L-2k+2)\pi}{2L+2}\right) = \sin\left(\frac{(2k-1)\pi}{2L+2}\right) \quad \text{for} \quad k \in [\![1, 2L+2]\!].} \tag{IV.218}$$

We can then write the $d_k^\dagger$'s as:

$$d_k^\dagger = \frac{1}{\sqrt{2L+2}} \left( \frac{1}{\sqrt{2}}(c_0^\dagger - c_0) + \sum_{n=1}^{L} \omega_k^n c_n^\dagger - (-\omega_k)^{-n} c_n + \frac{\omega_k^{L+1}}{\sqrt{2}}(c_{L+1}^\dagger + c_{L+1}) \right) \tag{IV.219}$$

and the inverse relations as:

$$\frac{1}{\sqrt{2}}(c_0^\dagger - c_0) = \frac{1}{\sqrt{2L+2}} \sum_{k=1}^{2L+2} d_k^\dagger, \tag{IV.220}$$

$$c_n^\dagger = \frac{1}{\sqrt{2L+2}} \sum_{k=1}^{2L+2} \omega_k^{-n} d_k^\dagger, \tag{IV.221}$$

$$c_n = \frac{1}{\sqrt{2L+2}} \sum_{k=1}^{2L+2} -(-\omega_k)^n d_k^\dagger, \tag{IV.222}$$

$$\frac{1}{\sqrt{2}}(c_{L+1}^\dagger + c_{L+1}) = \frac{1}{\sqrt{2L+2}} \sum_{k=1}^{2L+2} \omega_k^{-L-1} d_k^\dagger. \tag{IV.223}$$

Note that the $d_k^\dagger$'s are fermions, but, because there are $2L+2$ of them, and only $L+2$ $c_k^\dagger$'s, they are not all independent: we have $\omega_{L+1+k} = -\omega_k$, so that $d_{L+1+k}^\dagger = -d_k$.

We now need to determine the constant term $\mathcal{E}_0$ in $\tilde{H}$. Considering only the scalar terms in $\sum_k \mathcal{E}_k d_k^\dagger d_k$, we find:

$$\sum_{k=1}^{L+1} \mathcal{E}_k \frac{1}{2L+2} \left( \frac{1}{2}(c_0^\dagger - c_0)(c_0 - c_0^\dagger) + \sum_{n=1}^{L} c_n^\dagger c_n + c_n c_n^\dagger + \frac{1}{2}(c_{L+1}^\dagger + c_{L+1})(c_{L+1} + c_{L+1}^\dagger) \right)$$

$$= \frac{1}{2} \sum_{k=1}^{L+1} \mathcal{E}_k. \tag{IV.224}$$

Since there is no scalar part in $\tilde{H}$, we must therefore have:

$$\mathcal{E}_0 = -\frac{1}{2} \sum_{k=1}^{L+1} \mathcal{E}_k. \tag{IV.225}$$

The highest eigenvalue can then be obtained by considering the state with all the energy levels occupied:

$$\boxed{E = \mathcal{E}_0 + \sum_{k=1}^{L+1} \mathcal{E}_k = \frac{1}{2} \sin\left(\frac{\pi}{2L+2}\right)^{-1} \sim \frac{L}{\pi}} \tag{IV.226}$$

for the eigenstate defined by

$$\boxed{|\psi\rangle = \prod_{k=1}^{L+1} d_k^\dagger |\Omega\rangle} \tag{IV.227}$$

which is such that $d_k^\dagger |\psi\rangle = 0$ for $k \in [\![1, L+1]\!]$. The vector $|\Omega\rangle$ is arbitrary (provided that it is not in the kernel of any of the $d_k^\dagger$'s that we apply to it).

Remembering the factor $A$ which we took out of $M_\mu$ at the beginning of this section, we finally get:

$$\boxed{E(\mu) \sim \frac{L}{\pi} e^{\mu/L}} \tag{IV.228}$$

and

$$\boxed{g(j) \sim Lj \log(j) - Lj(1 - \log(\pi))} \tag{IV.229}$$

which is proportional to $L$, consistently with eq.(IV.116) which was obtained for positive fluctuations of the current in the MC phase.

We now have what we wanted, but while we're at it, we might as well analyse the eigenstate (IV.227) too.

The first and easiest calculation that we can do here is that of the two-point correlations in $|\psi\rangle$. The connected correlation between the occupancies of sites $n$ and $m$ is given by:

$$\begin{aligned} C_{nm} &= \langle\psi|c_n^\dagger c_n c_m^\dagger c_m|\psi\rangle - \langle\psi|c_n^\dagger c_n|\psi\rangle\langle\psi|c_m^\dagger c_m|\psi\rangle \\ &= -\langle\psi|c_n^\dagger c_m^\dagger|\psi\rangle\langle\psi|c_n c_m|\psi\rangle + \langle\psi|c_n^\dagger c_m|\psi\rangle\langle\psi|c_n c_m^\dagger|\psi\rangle \end{aligned} \tag{IV.230}$$

(using Wick's theorem).

We find that:

$$\langle\psi|c_n^\dagger c_m|\psi\rangle = \frac{1}{2L+2}\sum_{k=1}^{L+1}\omega_k^{m-n}, \tag{IV.231}$$

$$\langle\psi|c_n c_m^\dagger|\psi\rangle = \frac{1}{2L+2}\sum_{k=1}^{L+1}(-\omega_k)^{n-m}, \tag{IV.232}$$

$$\langle\psi|c_n^\dagger c_m^\dagger|\psi\rangle = -\frac{1}{2L+2}\sum_{k=1}^{L+1}\omega_k^{-n}(-\omega_k)^{-m}, \tag{IV.233}$$

$$\langle\psi|c_n c_m|\psi\rangle = -\frac{1}{2L+2}\sum_{k=1}^{L+1}(-\omega_k)^n \omega_k^m, \tag{IV.234}$$

If $n$ and $m$ have same parity, each of those terms sum to 0 (unless $n=m$ in the first two sums, but here we consider two different sites). If not, we get:

$$\langle\psi|c_n^\dagger c_m|\psi\rangle = \frac{1}{L+1}\frac{e^{i\pi(L)(m-n)/(2L+2)}}{1-e^{-i\pi(m-n)/(L+1)}}, \tag{IV.235}$$

$$\langle\psi|c_n^\dagger c_m^\dagger|\psi\rangle = -\frac{1}{L+1}(-1)^m\frac{e^{i\pi(L)(-m-n)/(2L+2)}}{1-e^{i\pi(m+n)/(L+1)}}, \tag{IV.236}$$

$$\langle\psi|c_n c_m|\psi\rangle = -\frac{1}{L+1}(-1)^n\frac{e^{i\pi(L)(m+n)/(2L+2)}}{1-e^{-i\pi(m+n)/(L+1)}}, \tag{IV.237}$$

so that

$$\boxed{C_{mn} = \frac{1}{4(L+1)^2\sin^2\left(\frac{\pi(m+n)}{(2L+2)}\right)} - \frac{1}{4(L+1)^2\sin^2\left(\frac{\pi(m-n)}{(2L+2)}\right)}.} \tag{IV.238}$$

The correlations are therefore exactly 0 for sites which are an even number of bonds apart (this was also the case in [100] for a half-filled periodic chain), and behave as

$$C_{mn} \sim -\frac{1}{\pi^2(m-n)^2} \tag{IV.239}$$

otherwise, if the two sites are far away enough from the boundaries. Note that those correlations do not vanish with the size of the system, in contrast with the steady state of the ASEP at $\mu=0$, where they behaved as $L^{-1}$ in the maximal current phase and vanished exponentially in the high and low density phases.

We can now look at the (un-normalised) probability of any given configuration. This can be expressed as

$$P(\{h_n, p_n\}) = \langle\psi|\prod_{n=0}^{N} c_{h_n} c_{h_n}^\dagger \frac{(c_{L+1}^\dagger + c_{L+1})^2}{2} \prod_{n=N+1}^{L} c_{p_n}^\dagger c_{p_n}|\psi\rangle \tag{IV.240}$$

which is to say that we select only the configuration that has holes at positions $\{h_n\}$ and particles at positions $\{p_n\}$ in $|\psi\rangle$, and project it onto its Hermitian conjugate. Note that the term $\frac{(c_{L+1}^\dagger + c_{L+1})^2}{2}$ makes no difference (since it is equal to a constant factor $\frac{1}{2}$), but is there to have effectively $L+1$ sites instead of $L$, which will be useful shortly. From the expression of this term in (IV.223), we see that it corresponds to having a hole (or, in fact, a particle) at site $L+1$. Note also that the terms in (IV.240) can be reordered as long as any pair $\{c_n, c_n^\dagger\}$ is kept in the same order, so that we can regroup all the $c$'s from the first product to the left, for instance.

We can now use Wick's theorem [160] on this expression, and write it as the Pfaffian of an anti-symmetric matrix $\mathcal{A}$ whose upper triangle consists of all the mean values $\langle\psi|ab|\psi\rangle$ where $a$ and $b$ are two terms from the product in (IV.240), taken in the same order. We can write it as a block matrix:

$$\mathcal{A} = \begin{bmatrix} A_1 & \langle\psi|c_{h_n}c_{h_m}^\dagger|\psi\rangle & \langle\psi|c_{h_n}c_{p_m}^\dagger|\psi\rangle & \langle\psi|c_{h_n}c_{p_m}|\psi\rangle \\ -\langle\psi|c_{h_m}c_{h_n}^\dagger|\psi\rangle & A_2 & \langle\psi|c_{h_n}^\dagger c_{p_m}^\dagger|\psi\rangle & \langle\psi|c_{h_n}^\dagger c_{p_m}|\psi\rangle \\ -\langle\psi|c_{h_m}c_{p_n}^\dagger|\psi\rangle & -\langle\psi|c_{h_m}^\dagger c_{p_n}^\dagger|\psi\rangle & A_3 & \langle\psi|c_{p_n}^\dagger c_{p_m}|\psi\rangle \\ -\langle\psi|c_{h_m}c_{p_n}|\psi\rangle & -\langle\psi|c_{h_m}^\dagger c_{p_n}|\psi\rangle & -\langle\psi|c_{p_m}^\dagger c_{p_n}|\psi\rangle & A_4 \end{bmatrix} \quad \text{(IV.241)}$$

where $A_1|_{n,m} = \langle\psi|c_{h_n}c_{h_m}|\psi\rangle$ if $n<m$ and $A_1|_{n,m} = -\langle\psi|c_{h_m}c_{h_n}|\psi\rangle$ if $n>m$. The same goes for $A_2$ with $c_{h_n}^\dagger$, $A_3$ with $c_{p_n}$ and $A_4$ with $c_{p_n}^\dagger$.

Looking at the expression given in (IV.231) to (IV.234), we see that $-\langle\psi|c_m c_n|\psi\rangle = \langle\psi|c_n c_m|\psi\rangle$, $-\langle\psi|c_m^\dagger c_n^\dagger|\psi\rangle = \langle\psi|c_n^\dagger c_m^\dagger|\psi\rangle$ and $-\langle\psi|c_m c_n^\dagger|\psi\rangle = \langle\psi|c_n^\dagger c_m|\psi\rangle - \delta_{n,m}$. We also note that all those block matrices can be factorised in a simple way: if we define

$$X_{n,k}^- = -(-w_k)^{h_n}, \qquad X_{n,k}^+ = w_k^{-h_n}, \quad \text{(IV.242)}$$

$$Y_{n,k}^- = -(-w_k)^{p_n}, \qquad Y_{n,k}^+ = w_k^{-p_n}, \quad \text{(IV.243)}$$

then $\mathcal{A}$ can be rewritten as

$$\mathcal{A} = \begin{bmatrix} X^-(X^+)^\dagger & X^-(X^-)^\dagger & X^-(Y^-)^\dagger & X^-(Y^+)^\dagger \\ X^+(X^+)^\dagger - 1_h & X^+(X^-)^\dagger & X^+(Y^-)^\dagger & X^+(Y^+)^\dagger \\ Y^+(X^+)^\dagger & Y^+(X^-)^\dagger & Y^+(Y^-)^\dagger & Y^+(Y^+)^\dagger \\ Y^-(X^+)^\dagger & Y^-(X^-)^\dagger & Y^-(Y^-)^\dagger - 1_p & Y^-(Y^+)^\dagger \end{bmatrix} \quad \text{(IV.244)}$$

where $1_h$ and $1_p$ are identity matrices whose respective sizes are the numbers of holes and particles (one of which is occupying site $L+1$, so that the sum of the two numbers is $L+1$).

In order to calculate $P(\{h_n, p_n\}) = \text{Pf}[\mathcal{A}]$, we need to first consider its square $P(\{h_n, p_n\})^2 = \text{Det}[\mathcal{A}]$. After exchanging a few lines and columns in that determinant, we can write it in a factorised form:

$$\text{Det}[\mathcal{A}] = \text{Det}\begin{bmatrix} \begin{smallmatrix} X^+ \\ Y^- \\ X^- \\ Y^+ \end{smallmatrix} & -1_{L+1} \\ & 0 \end{bmatrix} \cdot \begin{bmatrix} (X^+)^\dagger \ (Y^-)^\dagger & (X^-)^\dagger \ (Y^+)^\dagger \\ 1_{L+1} & 0 \end{bmatrix}. \quad \text{(IV.245)}$$

Each block in this last expression is of a square matrix of size $L+1$ (which is where the fact that we included site $L+1$ becomes useful). That determinant then reduces to:

$$P(\{h_n, p_n\})^2 = \text{Det}\begin{bmatrix} X^- \\ Y^+ \end{bmatrix} \text{Det}[(X^-)^\dagger \ (Y^+)^\dagger] = \left|\text{Det}\begin{bmatrix} X^- \\ Y^+ \end{bmatrix}\right|^2. \quad \text{(IV.246)}$$

After a few final simplifications, we find

$$\boxed{P(\{h_n, p_n\}) = \text{Det}[\omega^{h_n k}, \omega^{-p_n k}]} \quad \text{(IV.247)}$$

where $\omega = e^{\frac{i\pi}{L+1}}$. We recognise this to be Vandermonde determinant which gives, for any configuration:

$$P(\{n_n\}) = \prod_{n_n = n_m} [\sin(r_m - r_n)] \prod_{n_n \neq n_m} [\sin(r_m + r_n)] \quad \text{(IV.248)}$$

where $n_n$ is the occupancy of site $n$, and $r_n = n\pi/(2L+2)$. Note that all these probabilities are still un-normalised.

This distribution is exactly that of a Dyson-Gaudin gas [161], which is a discrete version of the Coulomb gas, on a periodic lattice of size $2L+2$, with two defect sites (at $0$ and $L+1$) that have no occupancy, and a reflection anti-symmetry between one side of the system and the other (fig.-IV.8). The first (upper) part of the gas is given by the configuration we are considering, and the second (lower) is deduced by anti-symmetry. The interaction potential between two particles at positions $r_n$ and $r_m$ is then given by:

$$V(r_n, r_m) = -\log\big(\sin(r_m - r_n)\big). \quad \text{(IV.249)}$$

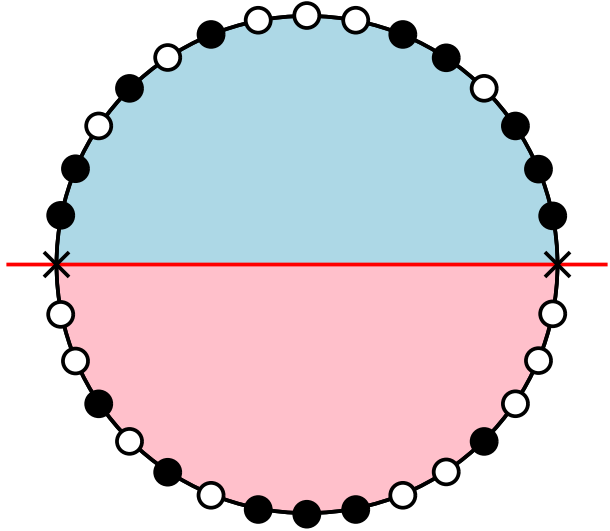

Figure IV.8: Dyson-Gaudin gas equivalent of the configuration (110101000110111) for the open ASEP conditioned on a large current. The lower part of the system is deduced from the upper part by an axial anti-symmetry.

It was noticed in [100] that the large current limit of the steady state of the periodic ASEP of size $L$ also converges to a Dyson-Gaudin gas (this time, the standard periodic case, without defects or symmetry).

## IV.4 Phase diagram of the open ASEP in the s-ensemble

In this section, we put all the information we have gathered in the previous ones together, and attempt to describe the full phase diagram of the open ASEP in the s-ensemble (i.e.

with $j$ or $\mu$ as a variable).

We first show how to retrieve the large deviation function of the current in the low and high density phases, given in eq.(IV.89), using a simple hydrodynamic description of the system, called the macroscopic fluctuation theory (MFT), and a method first devised in [85]. We then examine the domain of validity of that expression, and, assuming that it is large enough, we see how this same method gives us access to the large deviation function of the current in three other domains: a shock phase, which continues the shock line for positive fluctuations of $j$, a coexistence plane, which separates the high and low density phases for negative fluctuations of $j$, and an anti-shock phase, which sits on top of the maximal current phase for negative fluctuations of $j$. In each of those cases, we also obtain an expression for $E(\mu)$, and we check that everything is consistent with the limits we have found before. Finally, in the extended maximal current phase, where we cannot find an expression for $g(j)$ and $E(\mu)$, we review whatever information we do have on the behaviour of the system.

### IV.4.1 Macroscopic fluctuations for the WASEP

We will now present a method, which was pointed out to us by B. Derrida, and can be found in [85], with which we can retrieve the large deviation function of the current in the low and high density phases through a hydrodynamic description of the system. This relies on a not entirely rigorous trick which consists in starting from the weakly asymmetric simple exclusion process (WASEP), in which the asymmetry scales as $L^{-1}$, and in which the MFT is applicable, and then take that weak asymmetry to be finite again.

Let us therefore define $\nu$ such that $(1-q) = \frac{\nu}{L}$. We will first consider $\nu$ to be finite, and, when it suits us, replace it by $L(1-q)$ and see what happens. In all the following, we will consider the continuous limit of the system, rescaled to have a size 1, and recall that the mean field description of the system is governed by eq.(II.21):

$$j_L = LJ = \nu\rho(1-\rho) - \lambda\nabla\rho \qquad (\text{IV.250})$$

with boundary conditions $\rho_a$ and $\rho_b$, and where $j_L$ is the space-integrated current and $\lambda = \frac{1+q}{2}$.

The macroscopic fluctuation theory, as stated, for instance, in [75], consists in assuming that the fluctuations of the density profile are Gaussian around the mean-field profile given by (IV.250). The large deviation functional of a history $\rho(t)$ with a current $j(x,t)$ is written as:

$$g_t(j,\rho) = \frac{1}{t}\int_0^t d\tau \int_0^1 \frac{\left[j - \nu\rho(1-\rho) + \lambda\nabla\rho\right]^2}{2\rho(1-\rho)} dx \qquad (\text{IV.251})$$

which is to say that the difference between a profile and the mean field solution is a Gaussian white noise with a variance $\rho(1-\rho)$.

The contraction principle which we presented in eq.(I.9) allows us to get the large deviation function for $j$ alone by taking the optimal value of $\rho$ in the previous equation. What's more, as we saw at the end of section I.2.2, the joint probability distribution of the current and of the configurations (therefore of the density) is independent of time in the large time limit, so that we can get rid of the integral on $\tau$, and obtain:

$$\begin{aligned}g(j) &= \min_\rho \int_0^1 \frac{\left[j - \nu\rho(1-\rho) + \lambda\nabla\rho\right]^2}{2\rho(1-\rho)} dx \\ &= \min_\rho \int_0^1 \frac{\left[j - \nu\rho(1-\rho)\right]^2 + \left[\lambda\nabla\rho\right]^2}{2\rho(1-\rho)} dx + 2\lambda \int_{\rho_a}^{\rho_b} \frac{\left[j - \nu\rho(1-\rho)\right]}{2\rho(1-\rho)} d\rho\end{aligned} \quad \text{(IV.252)}$$

(where we just expanded the square in the first line, and find that the cross product is conservative in $\rho$). Note that, if $t$ is not taken to be large, the time integral cannot be removed, because of transient regimes at the beginning and end of the evolution, which depend on the initial and final conditions, and make the profile dependent on time.

If we now define $X(\rho) = \frac{[j-\nu\rho(1-\rho)]^2}{2\rho(1-\rho)}$, we find that the profile that extremises the first integral satisfies the Euler-Lagrange equation:

$$X'(\rho) - \frac{\lambda^2 \Delta\rho}{\rho(1-\rho)} + (\lambda\nabla\rho)^2 \frac{1-2\rho}{2\rho(1-\rho)} = 0. \quad \text{(IV.253)}$$

Multiplying that by $\nabla\rho$, we get

$$\nabla\big[X(\rho)\big] - \nabla\Big[\frac{(\lambda\nabla\rho)^2}{2\rho(1-\rho)}\Big] = 0 \quad \text{(IV.254)}$$

which is to say

$$\big[j - \nu\rho(1-\rho)\big]^2 = \big[\lambda\nabla\rho\big]^2 + K\ 2\rho(1-\rho) \quad \text{(IV.255)}$$

where $K$ is a constant, which can be found to be 0 for $\nu \to \infty$ [85] (since in a region where the profile is constant, i.e. where $\nabla\rho = 0$, the current is given by $j = \nu\rho(1-\rho)$).

This tells us that we must have $\lambda\nabla\rho = \pm(j - \nu\rho(1-\rho))$. The sign is determined by the position of $\rho$ with respect to the two densities $r$ and $(1-r)$ that produce a current $j = \nu r(1-r)$, compared to the sign of $\nabla\rho$.

Putting this back into (IV.252), we get 0 if $\rho$ is such that $\lambda\nabla\rho = -(j - \nu\rho(1-\rho))$ (which is to say if $\rho$ satisfies the mean field equation (IV.250)), and otherwise:

$$\frac{\big[j - \nu\rho(1-\rho) + \lambda\nabla\rho\big]^2}{2\rho(1-\rho)} dx = \frac{\big[2(j - \nu\rho(1-\rho))\big]\big[2\lambda\nabla\rho\big]}{2\rho(1-\rho)} dx = 2\lambda \frac{\big[j - \nu\rho(1-\rho)\big]}{\rho(1-\rho)} d\rho$$

which is to say that the only parts of a density profile that contribute to $g(j)$ are those where the sign of $\nabla\rho$ is inconsistent with the mean field equation (fig.IV.9).

For instance, if the imposed current $j = \nu r(1-r)$ (with $r < \frac{1}{2}$) and the boundary densities $\rho_a$ and $\rho_b$ are such that $\rho_a < 1-r$, $\rho_b < 1-r$ and $\rho_a < 1-\rho_b$, the only part of the optimal profile that doesn't satisfy the mean field equation is the first boundary layer going from $\rho_a$ to $r$, so that:

$$g(j) = 2\lambda \int_{\rho_a}^r \frac{\big[j - \nu\rho(1-\rho)\big]}{\rho(1-\rho)} d\rho = 2\lambda j \log\Big(\frac{1-\rho_a}{\rho_a}\frac{r}{1-r}\Big) - 2\lambda\nu(r - \rho_a) \quad \text{(IV.256)}$$

Taking the limit $q = 0$ in (IV.256), which is to say $\lambda = \frac{1}{2}$ and $\nu = L$, and going from the space-integrated current to the local current (i.e. taking $j \to Lj$), we find:

$$g(j) = j \log\Big(\frac{1-\rho_a}{\rho_a}\frac{r}{1-r}\Big) - (r - \rho_a) \quad \text{(IV.257)}$$

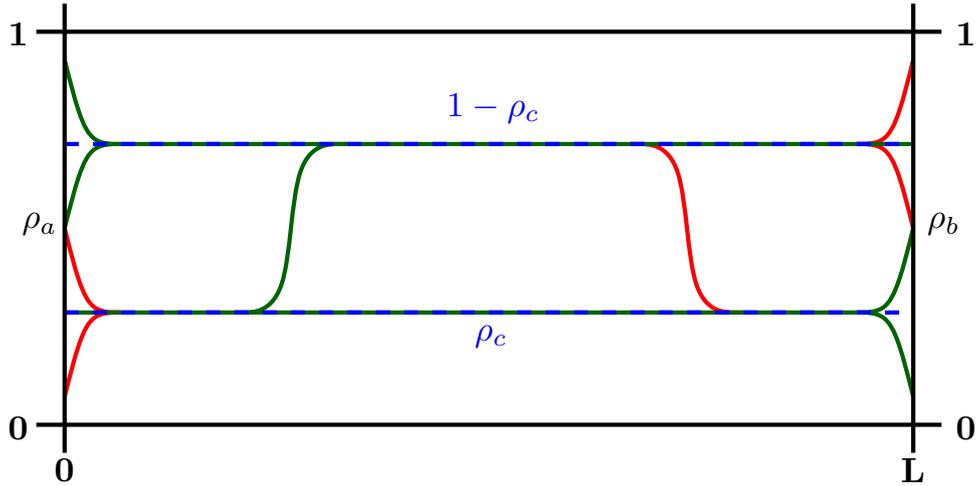

Figure IV.9: Various optimal profiles for a fixed $\rho_c$. The parts in green satisfy the mean field equation, and do not contribute to $g(j)$, whereas the portions in red do.

which is the same as eq.(IV.89).

In other cases, the contribution to $g(j)$ of each portion $[r_1, r_2]$ for which the variations of $\rho$ contradicts the mean field equation is given by:

$$f(j; r1, r2) = \int_{r_1}^{r_2} \frac{[j - \rho(1-\rho)]}{\rho(1-\rho)} d\rho = j \log\left(\frac{1-r_1}{r_1} \frac{r_2}{1-r_2}\right) + r_1 - r_2 \quad \text{(IV.258)}$$

where $j$ has to be equal to either $r_1(1-r_1)$ or $r_2(1-r_2)$. This is the same as the function $F^{\text{res}}$ from [85].

What we are now going to do is to find all the possible configurations for $j = r(1-r)$, $\rho_a$ and $\rho_b$ that lead to different forms of $g(j)$ (i.e. with different combinations of the function $f$), and determine, in each of those phases, the expressions of $g(j)$, $E(\mu)$, $j(\mu)$, and the boundaries of the phase. We will also compare the asymptotic behaviours of those results with everything we found in sections IV.2 and IV.3, to confirm their validity. We will then summarise all we know about the maximal current phase, which is not accessible by this method (and which is defined by $j > \frac{1}{4}$). Finally, we will put all this together in order to draw the phase diagram of the open ASEP with respect to $\rho_a$, $\rho_b$ and $\mu$.

In all cases, we will be noting $r$ the density for which $j = r(1-r)$ which is below $\frac{1}{2}$. Also note that we will do all the calculations for the TASEP, knowing that the same for the ASEP can be obtained merely by multiplying $E(\mu)$ by $(1-q)$. We also recall that we have defined two other boundary parameters $a = \frac{1-\rho_a}{\rho_a}$ and $b = \frac{\rho_b}{1-\rho_b}$, which we will use in certain formulae to make them more compact. Finally, we will, for the same reason, sometimes use, instead of $\mu$, the variable $u$ defined by:

$$u = \frac{1}{1 + e^\mu}. \quad \text{(IV.259)}$$

The non-perturbed case is given by $u = \frac{1}{2}$, $u = 0$ corresponds to an infinite current, and $u = 1$ to zero current.

## IV.4.2 Low/high density phases

We start with the low density phase, from which we can deduce the high density phase through $\rho_a \leftrightarrow 1 - \rho_b$.

This phase is defined by $\rho_a < 1 - \rho_b$, $\rho_a < 1 - r$ and $\rho_b < 1 - r$. The optimal profile is almost always on $\rho = r$, with possible boundary layers at both ends, with only the one at the left boundary contributing to $g(j)$ (fig.-IV.10).

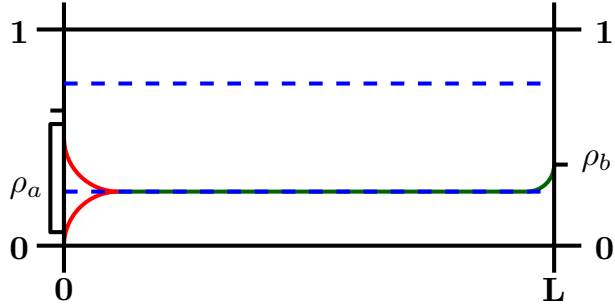

Figure IV.10: Optimal profiles for a fixed $\rho_c$ in the low density phase. Only the portion in red contribute to $g(j)$.

As we saw earlier, the large deviation function of the current is in this case:

$$g(j) = f(j; \rho_a, r) = j \log\left(\frac{1-\rho_a}{\rho_a}\frac{r}{1-r}\right) + \rho_a - r. \tag{IV.260}$$

The generating function of the cumulants of the current is

$$E(\mu) = \frac{a}{a+1}\frac{e^\mu - 1}{e^\mu + a} = \frac{e^\mu}{e^\mu + a} - \frac{1}{1+a} \tag{IV.261}$$

and the current is, in terms of $\mu$:

$$j(\mu) = \frac{a\,e^\mu}{(e^\mu + a)^2} = \rho_a(1-\rho_a)\frac{u(1-u)}{(\rho_a + u - 2u\rho_a)^2}. \tag{IV.262}$$

The boundaries of the phase are given by:

$$\rho_a < 1 - \rho_b \tag{IV.263}$$

$$u > \frac{\rho_a^2}{1 - 2\rho_a + 2\rho_a^2} \qquad \text{with} \quad \rho_a > \frac{1}{2}, \tag{IV.264}$$

$$u > \frac{\rho_a\rho_b}{1 - \rho_b - \rho_a + 2\rho_a\rho_b} \qquad \text{with} \quad \rho_b > \frac{1}{2}, \tag{IV.265}$$

$$u > \rho_a \qquad \text{with} \quad \rho_a < \frac{1}{2}\ ,\ \rho_b < \frac{1}{2}, \tag{IV.266}$$

where this last condition corresponds to $j < \frac{1}{4}$, which is the boundary with the MC phase.

According to this, the LD phase goes all the way up to $u = 1$. We have already checked in section IV.3.1 that this expression of $E(\mu)$ is consistent with what we found for $\mu \to -\infty$, i.e. $u \to 1$.

We may also note that, on the line $\rho_a = \frac{1}{2}$, which corresponds to the LD-MC transition line for $\mu = 0$, we find:

$$E(\mu) = \frac{1}{2} \frac{e^\mu - 1}{e^\mu + 1} \tag{IV.267}$$

which is consistent with the expression found in [95] for the half-filled periodic TASEP (we recall that the open system with $\rho_a = \frac{1}{2}$ and $\rho_b < 1/2$ is equivalent to a half-filled periodic system of twice the size). The $\mu \to 0^-$ limit gives:

$$E(\mu) \sim \frac{\mu}{4} - \frac{\mu^3}{48} \tag{IV.268}$$

which is the same as eq.(IV.111).

### IV.4.3 LD-HD coexistence plane

We now look at the case where $\rho_a = 1 - \rho_b$ and $r < \rho_a < 1 - r$. There are two possible profiles in this case: one around $\rho = r$ with a left boundary layer contributing to $g(j)$, and one around $\rho = 1 - r$ with a right boundary layer contributing to $g(j)$ for the same amount (fig.-IV.11).

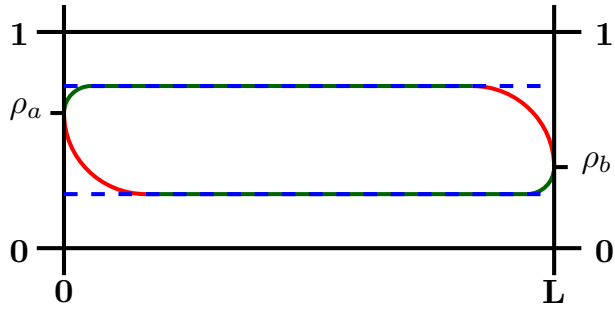

Figure IV.11: The two optimal profiles for a fixed $\rho_c$ in the LD-HD coexistence plane. Only the portions in red contribute to $g(j)$.

Since the large deviation function of any of those two profiles is the same, and equal to that of the LD phase (or of the HD phase with $b = a$), the expressions for $g(j)$, $E(\mu)$ and $j(\mu)$ are exactly the same as in the previous case. The boundaries of this plane are given by:

$$\rho_a = 1 - \rho_b, \tag{IV.269}$$

$$u > \frac{\rho_a^2}{1 - 2\rho_a + 2\rho_a^2} \quad \text{with} \quad \rho_a > \frac{1}{2}, \tag{IV.270}$$

$$u > \frac{1}{2} \quad \text{with} \quad \rho_a < \frac{1}{2}. \tag{IV.271}$$

We have three asymptotic results to check in this case. The one for $\mu \to -\infty$, from section IV.3.1, is the same as for the LD phase. We also looked at the case $\rho_a \to 0$ in section IV.3.2. The corresponding limit for $E(\mu)$ given by eq.(IV.261) is:

$$E(\mu) \sim -\rho_a + e^\mu \rho_a \tag{IV.272}$$

which is consistent with eq.(IV.172).

Finally, for $\rho_a < \frac{1}{2}$ and $\mu \to 0^-$, we have:

$$E(\mu) \sim \frac{a}{(1+a)^2}\mu + \frac{a(a-1)}{2(a+1)^3}\mu^2 \tag{IV.273}$$

which is consistent with eq.(IV.98) from section IV.2.2.

### IV.4.4  Shock phase

We consider the case where $\rho_a < r$ and $\rho_b > 1-r$. Here, there is a number of optimal profiles of order $L$. Each of them has a boundary layer around each boundary, both of them contributing to $g(j)$, and two constant regions, where $\rho = r$ near the left boundary and $\rho = 1-r$ near the right boundary, separated by a shock that can be placed anywhere in the system (fig.-IV.12).

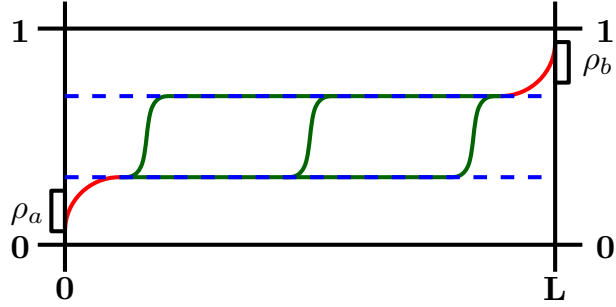

Figure IV.12: A few optimal profiles for a fixed $\rho_c$ in the shock phase. Only the portions in red contribute to $g(j)$.

The large deviation function of the current is given by:

$$g(j) = f(j;\rho_a,r) + f(j;1-r,\rho_b) = j\log\left(\frac{(1-\rho_a)\rho_b}{\rho_a(1-\rho_b)}\frac{r^2}{(1-r)^2}\right) + \rho_a - \rho_b + 1 - 2r. \tag{IV.274}$$

The generating function of the cumulants of the current is

$$E(\mu) = \frac{2e^{\mu/2}}{e^{\mu/2} + \sqrt{ab}} - \frac{1}{1+a} - \frac{1}{1+b} \tag{IV.275}$$

and the current is:

$$j(\mu) = \frac{\sqrt{ab}\, e^{\mu/2}}{(e^{\mu/2} + \sqrt{ab})^2} = \sqrt{\frac{(1-\rho_a)\rho_b}{\rho_a(1-\rho_b)}\frac{(1-u)}{u}}\left(\sqrt{\frac{(1-\rho_a)\rho_b}{\rho_a(1-\rho_b)}} + \sqrt{\frac{(1-u)}{u}}\right)^{-2}. \tag{IV.276}$$

The boundaries of the phase are given by:

$$u < \frac{\rho_a\rho_b}{1-\rho_b-\rho_a+2\rho_a\rho_b} \quad \text{with} \quad \rho_a < \frac{1}{2}\ ,\ \rho_b > \frac{1}{2}, \tag{IV.277}$$

$$u < \frac{(1-\rho_a)(1-\rho_b)}{1-\rho_b-\rho_a+2\rho_a\rho_b} \quad \text{with} \quad \rho_a < \frac{1}{2}\ ,\ \rho_b > \frac{1}{2}, \tag{IV.278}$$

$$u > \frac{\rho_a(1-\rho_b)}{\rho_b+\rho_a-2\rho_a\rho_b} \quad \text{with} \quad \rho_a < \frac{1}{2}\ ,\ \rho_b > \frac{1}{2}, \tag{IV.279}$$

where the last condition corresponds to $j < \frac{1}{4}$. Strangely enough, the volume that is defined by these boundaries is symmetric under any permutation of $\rho_a$, $1 - \rho_b$ and $u$.

The shock phase concerns two of the asymptotic results we found before. For $\mu \to 0^+$, which imposes $\rho_a = 1 - \rho_b$, we find:

$$E(\mu) \sim \frac{a}{(1+a)^2}\mu + \frac{a(a-1)}{4(a+1)^3}\mu^2 \qquad (IV.280)$$

which is what we found in eq.(IV.94). For $\rho_a \to 0$, which also imposes $\rho_a = 1 - \rho_b$, we find:

$$E(\mu) \sim -2\rho_a + 2e^{\mu/2}\rho_a \qquad (IV.281)$$

which is identical to eq.(IV.174).

### IV.4.5 Anti-shock phase

The last phase we can access through the MFT is for $\rho_a > (1-r)$ and $\rho_b < r$. In this case, there also is a number of possible profiles of order $L$: the first go down from $\rho_a$ to $(1-r)$, then down from $(1-r)$ to $r$ through an anti-shock that can be placed anywhere, and that contributes to $g(j)$, and finally down from $r$ to $\rho_b$ (fig.-IV.13).

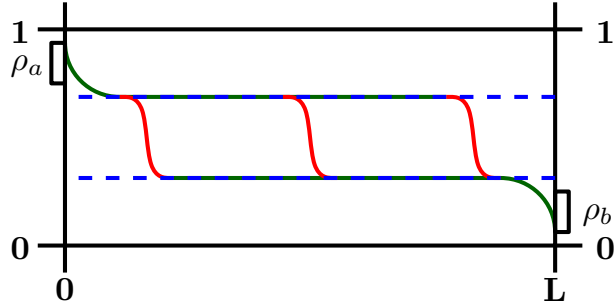

Figure IV.13: A few optimal profiles for a fixed $\rho_c$ in the Anti-shock phase. Only the portions in red contribute to $g(j)$.

The large deviation function of the current is given by:

$$g(j) = f(j; 1-r, r) = 2j \log\left(\frac{r}{1-r}\right) + 1 - 2r. \qquad (IV.282)$$

The generating function of the cumulants of the current is

$$E(\mu) = \frac{2e^{\mu/2}}{e^{\mu/2}+1} - 1 = \tanh(\mu/4) \qquad (IV.283)$$

and the current is:

$$j(\mu) = \frac{1-\tanh(\mu/4)}{4} = \frac{-2(u-u^2)+\sqrt{u-u^2}}{1-4(u-u^2)}. \qquad (IV.284)$$

The boundaries of the phase are given by:

$$u < \frac{\rho_a^2}{1 - 2\rho_a + 2\rho_a^2} \quad \text{with} \quad \rho_a > \frac{1}{2} \ , \ \rho_a > 1 - \rho_b, \tag{IV.285}$$

$$u < \frac{(1-\rho_b)^2}{1 - 2\rho_b + 2\rho_b^2} \quad \text{with} \quad \rho_b < \frac{1}{2} \ , \ \rho_a < 1 - \rho_b, \tag{IV.286}$$

$$u > \frac{1}{2}, \tag{IV.287}$$

where the last condition corresponds to $j < \frac{1}{4}$.

We note that this phase corresponds to one of the examples that can be found in [85]. The expression for $E(\mu)$ also comes up as a side note in [118].

The limit $\mu \to 0^-$ gives:

$$E(\mu) \sim \frac{\mu}{4} - \frac{\mu^3}{192} \tag{IV.288}$$

which is consistent with eq.(IV.120). The limit $\mu \to -\infty$, which implies $\rho_a = 1 - \rho_b = 1$, gives:

$$E(\mu) \sim -1 + 2\mathrm{e}^{\mu/2} \tag{IV.289}$$

which is the same as equation (IV.158), and this is the last asymptotic limit that we had to check.

### IV.4.6 Maximal current phase

There is one phase left for us to examine, to a much lesser extent than all the the others because the MFT breaks down in this case: the maximal current phase. Once we take out the phases we have already considered, we are left with a volume, in the three-dimensional phase space with variables $\rho_a$, $\rho_b$ and $u$, defined by:

$$u < \frac{1}{2} \quad \text{with} \quad \rho_a > \frac{1}{2} \ , \ \rho_b < \frac{1}{2}, \tag{IV.290}$$

$$u < \rho_a \quad \text{with} \quad \rho_a > \frac{1}{2} \ , \ \rho_b > \frac{1}{2}, \tag{IV.291}$$

$$u < 1 - \rho_b \quad \text{with} \quad \rho_a < \frac{1}{2} \ , \ \rho_b < \frac{1}{2}, \tag{IV.292}$$

$$u < \frac{\rho_a(1-\rho_b)}{\rho_b + \rho_a - 2\rho_a\rho_b} \quad \text{with} \quad \rho_a < \frac{1}{2} \ , \ \rho_b > \frac{1}{2}. \tag{IV.293}$$

We know that, asymptotically:

$$g(j) \sim (j - J)^{5/2} \frac{32\sqrt{3}L}{5\pi(1-q)^{3/2}} \tag{IV.294}$$

for $\mu \to 0^+$ with $\rho_a > \frac{1}{2}$ and $\rho_b < \frac{1}{2}$ (i.e. right next to the MC phase for the steady state), which we found in eq.(IV.107), and that:

$$g(j) \sim Lj \log(j) - Lj(1 - \log(\pi)) \tag{IV.295}$$

for $\mu \to \infty$, as we saw in eq.(IV.229).

Since this last result is valid independently of the boundary parameters, we know that all the plane $u=0$ belongs to the same phase. We have, however, no way to be certain that *all* of the volume we have described above is just one phase. That being said, we know that the whole region corresponds to a mean current higher than $\frac{1}{4}$. There is no way for the system to produce such a current through a hydrodynamic profile, for which the maximal possible current is $\frac{1}{4}$ if $\rho = \frac{1}{2}$, so to go higher than that, the system must produce correlations, which is why the MFT breaks down. Those correlations must be negative for neighbouring sites (if the particles are next to holes, they will jump more easily and produce more current), which is consistent with what we found in the large current limit in eq.(IV.239). What's more, those correlations must be created everywhere in the system, because a single non-correlated zone would cause a blockage and bring the current back down to $\frac{1}{4}$. We can also argue that the mean density should be around $\frac{1}{2}$, because it is easier to get to a large current starting from $\frac{1}{4}$ than from anything lower. In all these remarks, the boundaries play little part: independently of them, the system must be around $\rho = \frac{1}{2}$, and anti-correlated at every point. We therefore don't expect any sub-phases in this region.

### IV.4.7 Phase diagram

Now that we have considered all the possible combinations of $\rho_a$, $\rho_b$ and $u$, we can draw the phase diagram of the ASEP in the s-ensemble (fig.-IV.14). Each phase is represented using a different colour: blue for the low density phase, green for the high density phase, orange for the shock phase, purple for the anti-shock phase, and red/pink for the maximal current phase. The full diagram can be seen in the centre of the figure, with black lines marking the corners of the phases, and an exploded view of the LD, MC, shock (S) and anti-shock (AS) phases is also shown, with coloured lines representing slices for regularly spaced values of $u$. The HD phase can be deduced from the LD phase through the symmetry $\rho_a \leftrightarrow 1 - \rho_b$. The top and bottom parts of the figure contain slices of the diagram for specific values of $u$ (top) and $\rho_a$ (bottom), with a few iso-current lines drawn in all phases except the MC phase. Note that those iso-current lines do not represent evenly spaced values of the current ($j$ varies, in fact, more slowly as one approaches the MC phase).

We will conclude this section with a few remarks.

First of all, in this section, we focused on the TASEP for the sake of simplicity, but, as we remarked throughout section IV.1, the only difference for the ASEP (with $q<1$) is an overall factor $(1-q)$ in $E(\mu)$, which translates into the fact that $\frac{g((1-q)\tilde{j})}{(1-q)}$, where $\tilde{j}$ is the current rescaled by $(1-q)$, is independent of $q$. In other terms, the probability of producing a current $(1-q)\tilde{j}$ for a given $\tilde{j}$ during a time $\frac{t}{(1-q)}$ depends neither on $q$ nor on $t$ (that is, for a given value of the boundary densities, not of the boundary rates). This can be understood through a simple argument: considering that $(1-q)$ is effectively the driving field that is applied to the bulk of the system, any event such as a particle entering the system from the left, going through it under the action of the field, and exiting it at the right, is done at a speed (i.e. current) proportional to the driving field $(1-q)$, and takes a time which is therefore inversely proportional to it. In yet other terms: the only time scale in the system (in the continuous limit, where the boundary rates have become boundary densities) is the instantaneous speed of the particles, which is proportional to the field $(1-q)$. We can therefore rescale time by that factor so that everything becomes independent of $q$.

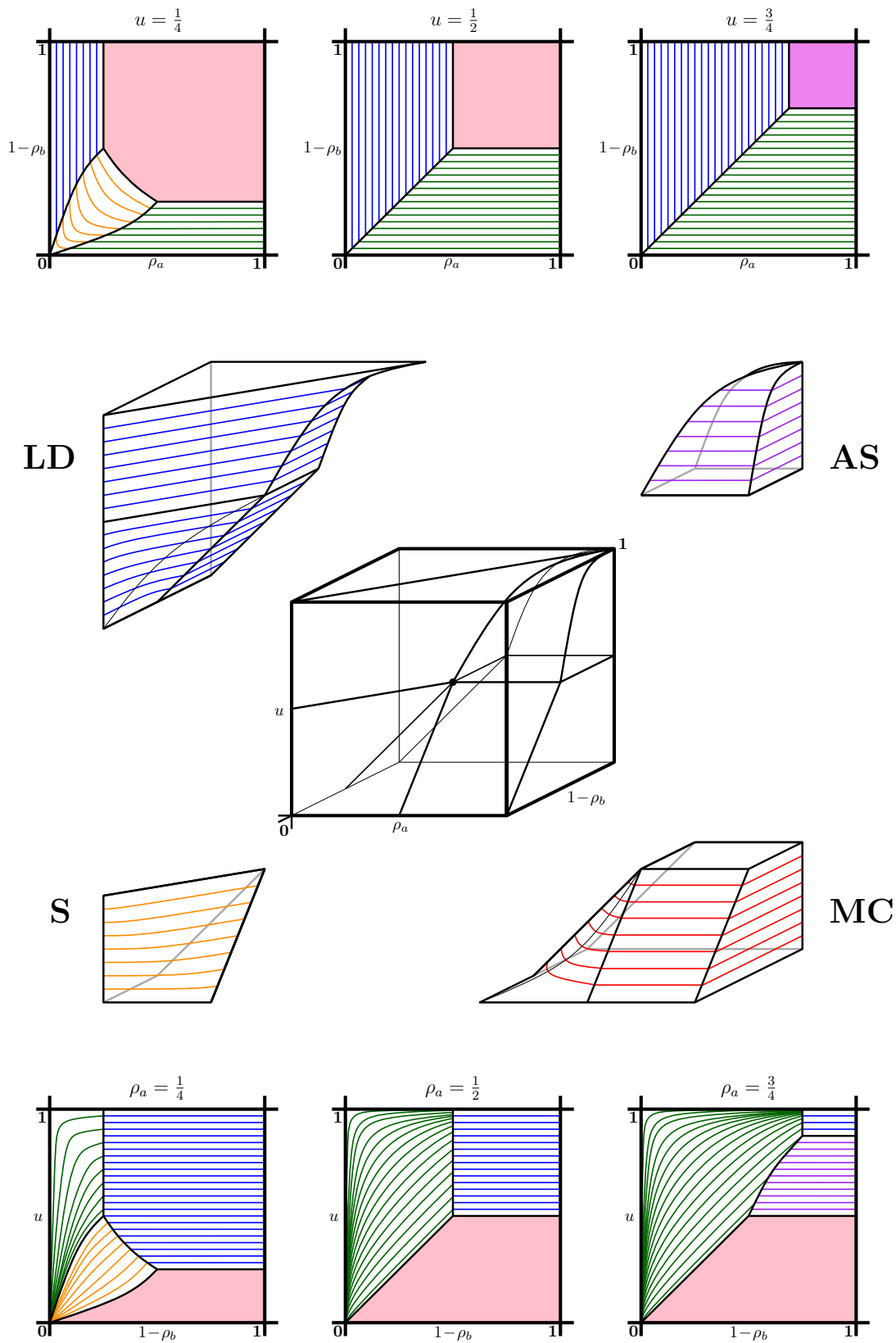

Figure IV.14: Phase diagram of the open ASEP in the s-ensemble. Top: diagrams at fixed $u$. Centre: complete diagram with phase boundaries and exploded view. Bottom: diagrams at fixed $\rho_a$.

As we noticed before, the boundary of the shock phase, which looks somewhat like a triangular dipyramid with the three middle edges smoothed out, is symmetric through $\rho_a \leftrightarrow 1-\rho_b \leftrightarrow u$. The first part of that symmetry comes from the global particle↔hole and left↔right symmetry of the system, but we do not know whether the second part can be understood in any meaningful way.

From the expressions we gave of $E(\mu)$ and $j$ in each phase, one can see that they are both continuous throughout the system, so that there are no first order transitions with respect to the current. This does not contradict the numerical results that were found in [138], where the total current is considered instead of the local current, so that the scale of $u$ is dramatically changed, and the whole diagram is flattened vertically into a plane, with an apparent first order between zero current and an infinite current everywhere.

As for the density profiles, they are discontinuous for all transitions except the ones involving the MC phase.

We have established, on the line $\rho_a = \frac{1}{2}$ with $\rho_b < \frac{1}{2}$, a correspondence between the current in the open ASEP and that in periodic system of twice the size, at half filing. This indicates that the transition between the HD or LD phases and the MC phase is the same as that which was found on the periodic ASEP, between a flat density profile and a modulated profile [90–93] which may be time-dependent (if the density is not $\frac{1}{2}$). Unfortunately, we have not yet found a precise correspondence between the density profiles in those two cases.

Finally, by examining the expressions we found for $g(j)$ in all phases, we notice that it is independent on the size $L$ of the system everywhere except in the MC phase, where it is proportional to $L$. This can be understood by considering that, in all cases where the MFT is valid, only a finite number of particles need to be controlled (i.e. slowed down or accelerated, depending on the sign of the fluctuation of $j$) for the current to be modified, the rest of the system being governed by its normal hydrodynamics. For instance, in the LD phase, in order to obtain a smaller current, it is enough to slow down the first few bonds of the system, in order to create an effective reservoir with a density lower than $\rho_a$, and the rest of the system can then obey its unmodified dynamics. This shows in the expression of the large deviation function $g(j)$, where only a small portion of the profile has a contribution, near the left boundary. This reasoning can be applied everywhere except in the MC phase, where no hydrodynamic profile can create the desired current $j > \frac{1}{4}$, so that no small modification of the density or jumping rates may bring us into that phase. Instead, as we saw, the system has to create anti-correlations between the particles in order to counter the exclusion constraint and increase the current. These correlations have to be created everywhere in the system, because a single mean field region would be enough to create a blockage and bring the current back down to $\frac{1}{4}$, so that the cost of that operation, in terms of large deviations, must be proportional to $L$.

# CHAPTER V

## Bethe Ansatz and Q-Operator for the Open ASEP

We have seen, in section II.2.2, how the periodic ASEP with current-counting deformations could be treated through the coordinate Bethe Ansatz. For the open ASEP, unfortunately, the presence of reservoirs makes it impossible to number the particles, which makes that method inapplicable. There is however another version of the Bethe Ansatz, called the 'algebraic Bethe Ansatz' [39], which can in principle be used instead. Its formulation comes from the fact that the Markov matrix of the ASEP (or the Hamiltonian of the XXZ spin chain), can be expressed in relation to the row-by-row transfer matrix of an ice-type two-dimensional equilibrium statistical system called the 'six-vertex model' [40]. That transfer matrix commutes with the Markov matrix, much like the one we constructed in section III.2, but unlike ours, depends on a free parameter, called the 'spectral parameter'. It has the form of a product of local tensors, called 'Lax matrices', traced over an 'auxiliary space' which is usually of dimension 2, like the physical space of a single site, although any dimension can be chosen (in contrast, the auxiliary space of our transfer matrix is of infinite dimension). That product of Lax matrices, if the trace on the auxiliary space is not taken, is called the 'monodromy matrix' of the system, and creation and annihilation operators for the particles can be extracted from it, that depend on their own spectral parameters. The algebraic Bethe Ansatz then consists in finding a trivial eigenvector of the transfer matrix (a 'vacuum state'), on which the creation operator is then applied to obtain the other eigenvectors. Commutation relations between the transfer matrix and the creation operators, depending on their respective spectral parameters, produce the same Bethe equations as for the coordinate Bethe Ansatz, where the role of the Bethe roots is played by those spectral parameters.

In the case of a periodic system, with a fixed number of particles (or a fixed magnetisation sector), that vacuum state can be chosen as either completely empty, or completely full, and the rest of the resolution is rather straightforward. In the case of an open system, there are two extra difficulties. Firstly, the transfer matrix is more complicated, and involves two rows of the vertex model instead of one, with certain 'reflection operators' at the boundaries [162, 163] (this is also the case for our own transfer matrix: in the open case, we need two rows of tensors, with special vectors at the boundaries). This makes things harder, but not intractable. The second difficulty, however, does: for an open system, with no occupancy sectors and no trivial eigenvectors, we do not know, in general, how to find a suitable vacuum state. Such states have been found for certain special boundary conditions, such as triangular boundary matrices [164, 165], or full matrices with constraints on their coefficients [41, 42, 44, 45], for which some pseudo-particles are conserved and the full construction can be performed. Recent progress has also been made for a semi-infinite chain [166, 167], which has only one boundary, through a method deriving from the Bethe Ansatz. The case of completely general boundary conditions, however, remains unsolved.

Luckily, there is a way to obtain the eigenvalues of the Markov matrix (or Hamiltonian, or transfer matrix) we are interested in without having to deal with the eigenvectors at all. This can be done through Baxter's so-called 'Q-operator'. It was first used as an alternative method to solve the six-vertex model [40], but was later discovered to be, in fact, a limit of the algebraic Bethe Ansatz with an infinite-dimensional auxiliary space [168, 169]. Certain algebraic relations between the Q-operator and the Bethe transfer matrix, called 'T-Q' relations, allow to obtain the functional Bethe Ansatz equations for the eigenvalues directly, without need of the eigenvectors [170–174]. However, even with that method, the open case was solved only for certain constraints on the boundary parameters [174–176] (in the case of the ASEP deformed to count the current, those constraints involve all four boundary parameters, the current-counting fugacity, and the size of the system). In [174], it was even argued that, in all probability, those constraints are necessary in order to construct the appropriate Q-operator.

In this chapter, we show that it is in fact possible to solve the most general case, by constructing explicitly the Q-operator (or rather, the PQ-operator, since what we construct is the product of two different operators, the second of which we will call P), for the open ASEP with any boundary parameters and current-counting deformation, and obtaining the functional Bethe equations for the eigenvalues of the deformed Markov matrix. Our PQ-operator has two spectral parameters instead of one (the second of which is what is usually called the 'representation parameter' of the $U_q[SU(2)]$ algebra [146], and often fixed to a specific value, but we will see that it is essential to us to treat it as a free parameter), and can be constructed as a natural generalisation of the transfer matrix presented in section III.2, with the boundary vectors playing the part of the reflection operators we mentioned. We will also see how, taking special values of these two spectral parameters, we can recover Bethe transfer matrices with any auxiliary space dimension, and the T-Q relations for all of those matrices. We will conclude by connecting this approach to that of the functional Bethe Ansatz presented in section II.2.2, noticing that the polynomials $P$ and $Q$ that we constructed then are the eigenvalues of the operators P and Q that we consider here, and we will then proceed to use the same method as in II.2.2 to obtain the expression of the generating function of the cumulants of the current in the open ASEP that we presented in section III.3.

This work was done in collaboration with V. Pasquier.

## V.1 Periodic ASEP

In this first section, we treat the periodic case, for which we know what to expect. By generalising the tensors $X$ that we defined in chapter III, as well as the algebraic relations their elements satisfy, we construct a transfer matrix with two free parameters, which commutes with the deformed Markov matrix of the periodic ASEP for any values of those parameters. We then show that, for certain special values, the transfer matrix decomposes into two independent blocks, one of which is the Bethe transfer matrix for some dimension of the auxiliary space. We also show that our transfer matrix is in fact the product of two one-parameter operators $P$ and $Q$. Putting these results together, we are able to recover the functional Bethe equations for the periodic ASEP, which we saw in section II.2.2.

We recall that the Markov matrix of the periodic ASEP is given by:

$$M = \sum_{i=1}^{L} M_i \tag{V.1}$$

with

$$M_i = \begin{bmatrix} 0 & 0 & 0 & 0 \\ 0 & -q & 1 & 0 \\ 0 & q & -1 & 0 \\ 0 & 0 & 0 & 0 \end{bmatrix} \tag{V.2}$$

and where $m_L$ connects site $L$ with site 1.

## V.1.1 Bulk algebra and commutation relations

The starting point for everything we are about to do in this chapter is to realise that the definition of matrices $d$ and $e$ that we used until now, with the algebraic relation that they satisfy, $de - q\,ed = (1-q)$, is in fact a special representation of the $\mathrm{U}_q[\mathrm{SU}(2)]$ algebra (up to a simple gauge transformation that we present in section V.2.5). Knowing this, it seems natural to wonder whether a more general representation might be used, and produce different, and perhaps better, results.

Let us therefore redefine:

$$X(x,y) = \begin{bmatrix} n_0 & e \\ d & n_1 \end{bmatrix}, \quad \hat{X} = \begin{bmatrix} \hat{n}_0 & \hat{e} \\ \hat{d} & \hat{n}_1 \end{bmatrix} \tag{V.3}$$

with

$$n_0 = 1 + xA, \tag{V.4}$$

$$n_1 = 1 + yA, \tag{V.5}$$

$$\hat{n}_0 = \frac{(1-q)}{2}(1 - xA), \tag{V.6}$$

$$\hat{n}_1 = \frac{(1-q)}{2}(-1 + yA), \tag{V.7}$$

$$\hat{e} = \frac{(1-q)}{2}e, \tag{V.8}$$

$$\hat{d} = -\frac{(1-q)}{2}d, \tag{V.9}$$

where matrices $A$, $d$ and $e$ satisfy:

$$de - q\ ed = (1-q)(1-xyA^2), \tag{V.10}$$
$$Ae = q\ eA, \tag{V.11}$$
$$dA = q\ Ad. \tag{V.12}$$

The solution to these equations which we will use is a generalisation of the one we used in chapter III:

$$A = \sum_{n=0}^{\infty} q^n \|n\rangle\!\rangle \langle\!\langle n\| \tag{V.13}$$

$$d = \sum_{n=1}^{\infty} (1-q^n) \|n-1\rangle\!\rangle \langle\!\langle n\| = S^-(1-A) \tag{V.14}$$

and

$$e = \sum_{n=0}^{\infty} (1-xyq^n) \|n+1\rangle\!\rangle \langle\!\langle n\| = S^+(1-xyA) \tag{V.15}$$

where $S^+$ and $S^-$ are simply operators increasing or decreasing $n$ by 1 (not to be confused with the spin operators that we saw in section IV.3.3). We recover the simpler versions of these matrices simply by taking $x=y=0$.

A few remarks need to be made here. First of all, the matrix $A$ that we have just defined plays an important role in building the matrix Ansatz for the multispecies periodic ASEP [121]. Secondly, we could have chosen for $d$ and $e$ their contragredient representation $\bar{e} = {}^t d$ and $\bar{d} = {}^t e$, which is equivalent to a gauge transformation on $d$ and $e$. We will be using this fact abundantly in the rest of the chapter. Finally, we can actually define $A$, $S^+$ and $S^-$ over $\mathbb{Z}$ rather than $\mathbb{N}$, so that $S^+$ and $S^-$ are the inverse of one another: $S^+S^- = 1$ (which wouldn't work on $\mathbb{N}$ because of the cut at $-1$). Because of the term $(1-q^n)$ in $d$, which is 0 between states $\|0\rangle\!\rangle$ and $\|-1\rangle\!\rangle$, we are assured that, if starting from a state $\|n\rangle\!\rangle$ with $n \geq 0$, we can never go to one with $n < 0$ through any combination of $d$, $e$ and $A$. We just need to make sure that those expressions are always applied to vectors that have non-zero coefficients only for $n \geq 0$, which is enforced by the matrix $A_\mu$ that we keep as defined in chapter III, on $\mathbb{N}$ alone. This fact will make many future calculations much easier.

All our matrices are now combinations of only $A$ and $S^+$, which satisfy a simple algebra:

$$AS^+ = q\ S^+A, \tag{V.16}$$

with $S^- = (S^+)^{-1}$.

The first thing that we need to show is, as in chapter III, that the transfer matrix

$$T_\mu^{per}(x,y) = \mathrm{Tr}[A_\mu \prod_{i=1}^{L} X^{(i)}] \tag{V.17}$$

commutes with the deformed Markov matrix $M_\mu$. Note that, as in chapter III, the product symbol refers to a matrix product in the auxiliary space and a tensor product in configuration space, and that the trace is taken only on the auxiliary space.

This can be shown through a calculation almost identical to that for the simpler case (which can be found in the appendixes of [3]). The parts of the transfer matrix and of the Markov matrix involving sites $i$ and $i+1$ are:

$$X^{(i)}X^{(i+1)} = \begin{bmatrix} n_0n_0 & n_0e & en_0 & ee \\ n_0d & n_0n_1 & ed & en_1 \\ dn_0 & de & n_1n_0 & n_1e \\ dd & dn_1 & n_1d & n_1n_1 \end{bmatrix} \quad , \quad M_i = \begin{bmatrix} 0 & 0 & 0 & 0 \\ 0 & -q & 1 & 0 \\ 0 & q & -1 & 0 \\ 0 & 0 & 0 & 0 \end{bmatrix}. \quad \text{(V.18)}$$

The commutator of those two gives:

$$[M_i, X^{(i)}X^{(i+1)}] = \begin{bmatrix} 0 & 0 & 0 & 0 \\ dn_0 - q\, n_0d & de - q\, n_0n_1 & n_1n_0 - q\, ed & n_1e - q\, en_1 \\ q\, n_0d - dn_0 & q\, n_0n_1 - de & q\, ed - n_1n_0 & q\, en_1 - n_1e \\ 0 & 0 & 0 & 0 \end{bmatrix}$$

$$- \begin{bmatrix} 0 & q(en_0 - n_0e) & n_0e - en_0 & 0 \\ 0 & q(ed - n_0n_1) & n_0n_1 - ed & 0 \\ 0 & q(n_1n_0 - de) & de - n_1n_0 & 0 \\ 0 & q(n_1d - dn_1) & dn_1 - n_1d & 0 \end{bmatrix}$$

$$= \begin{bmatrix} 0 & q(en_0 - n_0e) & n_0e - en_0 & 0 \\ dn_0 - q\, n_0d & de - q\, ed & (1-q)ed & n_1e - q\, en_1 \\ q\, n_0d - dn_0 & (q-1)de & q\, ed - de & q\, en_1 - n_1e \\ 0 & q(n_1d - dn_1) & dn_1 - n_1d & 0 \end{bmatrix}$$

$$= \begin{bmatrix} 0 & q(en_0 - n_0e) & n_0e - en_0 & 0 \\ dn_0 - q\, n_0d & de - q\, ed & (1-q)ed & n_1e - q\, en_1 \\ q\, n_0d - dn_0 & (q-1)de & q\, ed - de & q\, en_1 - n_1e \\ 0 & q(n_1d - dn_1) & dn_1 - n_1d & 0 \end{bmatrix}$$

$$= \boxed{\hat{X}^{(i)}X^{(i+1)} - X^{(i)}\hat{X}^{(i+1)}}. \quad \text{(V.19)}$$

where we got from the second to the third line using:

$$\begin{aligned} de - qed &= (1-q)(1-xyA^2) = \hat{n}_0 n_1 - n_0 \hat{n}_1, \\ q(n_0e - en_0) &= (q-1)Ae = \hat{n}_0 e - n_0 \hat{e}, \\ en_0 - n_0e &= (1-q)eA = \hat{e}n_0 - e\hat{n}_0, \\ n_1e - q\, en_1 &= (1-q)e = \hat{e}n_1 - e\hat{n}_1, \\ q(dn_1 - n_1d) &= (q-1)dA = \hat{d}n_1 - d\hat{n}_1, \\ n_1d - dn_1 &= (1-q)Ad = \hat{n}_1 d - n_1 \hat{d}, \\ dn_0 - q\, n_0d &= (1-q)d = \hat{n}_0 d - n_0 \hat{d}. \end{aligned} \quad \text{(V.20)}$$

The equivalent commutation relation for the part of the matrix bearing the current counter $A_\mu$ (i.e. between sites 0 and $L$) is proven through the exact same calculation, up to a few terms $e^{\pm\mu}$ here and there, and is left as an exercise to the reader. Note that this relation is in fact the infinitesimal equivalent of the so called 'RLL equation' for the commutation of matrices $X$ with different parameters, where $\hat{X}$ is the derivative of $X$ for a well chosen variable.

We can now recover $M_\mu$ in its entirety by summing over $i$ in (V.19). The hat matrices cancel out from one term to the next, and we are left with 0, so that:

$$\boxed{[M_\mu, T_\mu^{per}(x,y)] = 0.} \tag{V.21}$$

This is the same result as before, but we now have two non-trivial parameters to play with. Which we will do right away.

Note that the results from the end of section III.2.1 hold, namely the fact that for a general set of fugacities $\{\mu_i\}$, the corresponding transfer matrix is the same as the one we defined here, with matrices $A_{\mu_i}$ inserted at their appropriate place in the matrix product:

$$T^{per}_{\{\mu_i\}}(x,y) = \text{Tr}[A_{\mu_0} \prod_{i=1}^{L} X^{(i)} A_{\mu_i}]. \tag{V.22}$$

## V.1.2 Decomposition of the transfer matrix

Considering the representation we chose for matrix $e$ in (V.15), namely $S^+(1-xyA)$, there is a good chance that something might happen for $xy = q^{-k+1}$ with $k \in \mathbb{N}^\star$ (which sets one coefficient to 0 in $e$).

Let us therefore impose $y = 1/q^{k-1}x$. The four matrices in $X$ become:

$$d = \sum_{n=1}^{\infty}(1-q^n)\|n-1\rangle\!\rangle\langle\!\langle n\| \quad , \quad e = \sum_{n=0}^{\infty}(1-q^{n-k+1})\|n+1\rangle\!\rangle\langle\!\langle n\| \tag{V.23}$$

$$n_0 = \sum_{n=0}^{\infty}(1+q^n x)\|n\rangle\!\rangle\langle\!\langle n\| \quad , \quad n_1 = \sum_{n=0}^{\infty}(1+q^{n-k+1}/x)\|n\rangle\!\rangle\langle\!\langle n\| \tag{V.24}$$

and the coefficient of $\|k\rangle\!\rangle\langle\!\langle k-1\|$ in $e$ vanishes. This makes all these matrices lower block-triangular ($n_0$ and $n_1$ obviously are, since they are diagonal, and $d$ already was, but not $e$ in general) with a block of size $k$ (for $n$ from 0 to $k-1$ in the four series above) and one of infinite size (for $n$ from $k$ to $\infty$).

The coefficients of that second block happen to be the same as the coefficients of the whole matrix for $x \to q^k x$ and $y \to q/x$ and in the dual representation of $d$ and $e$ (i.e. exchanging and transposing them):

$$^t e = \sum_{n=0}^{\infty}(1-q^{n+k})\|n-1\rangle\!\rangle\langle\!\langle n\| \quad , \quad ^t d = \sum_{n=1}^{\infty}(1-q^{n+1})\|n+1\rangle\!\rangle\langle\!\langle n\|, \tag{V.25}$$

$$n_0 = \sum_{n=0}^{\infty}(1+q^{n+k}x)\|n\rangle\!\rangle\langle\!\langle n\| \quad , \quad n_1 = \sum_{n=0}^{\infty}(1+q^{n+1}/x)\|n\rangle\!\rangle\langle\!\langle n\|. \tag{V.26}$$

Indeed, taking $n \to n+k$ in (V.23) and (V.24), and removing the $k$ negative indices, we get exactly (V.25) and (V.26).

Since the trace of a product of block-diagonal matrices is the sum of the traces of the products of the blocks, this gives us an equation for $T_\mu^{per}$, which is one of the two results essential to our derivation of the functional Bethe Ansatz:

$$T_\mu^{per}(x, 1/q^{k-1}x) = (1 - e^{-\mu})t^{(k)}(x) + e^{-k\mu}T_\mu^{per}(q^k x, q/x) \qquad (V.27)$$

where the factor $(1 - e^{-\mu})$ comes from the normalisation of $A_\mu$, and the factor $e^{-k\mu}$ is the first coefficient of $A_\mu$ on the second block. The transfer matrix $t^{(k)}$ is the contribution coming from the first block, which can be written as:

$$(1 - e^{-\mu})t^{(k)}(x) = \text{Tr}[A_\mu \prod_{i=1}^{L} X_k^{(i)}(x)] \qquad (V.28)$$

where $X_k(x)$ contains the same entries as $X(x, 1/q^{k-1}x)$, but truncated at $n = k - 1$ (so that the auxiliary space is $k$-dimensional).

We saw that $T_\mu^{per}$ commutes with $M_\mu$ for any values of $x$ and $y$, so it also commutes with another $T_\mu^{per}$ at different values of the parameters (this is in fact not certain, because any one of those matrices could have a degenerate eigenspace, but we will assume that it is true, for now; there is a better way to show that two matrices $T_\mu^{per}$ at different values of $x$ and $y$ commute, and we will come back to it in the next section). This tells us that those matrices also commute with $t^{(k)}(x)$ for any $k$, and that the $t^{(k)}(x)$'s with different $k$'s commute together. This matrix equation therefore implies the same relation for the eigenvalues $\Lambda_i(x,y)$ of $T_\mu^{per}(x,y)$ and the eigenvalues $\Lambda_i^{(k)}(x)$ of $t^{(k)}(x)$.

To go further, we need to examine the first two cases in this last equation.
For $k = 1$, the first block of $X$ is given by:

$$X_1(x) = \left[\begin{array}{c|c} 1+x & 0 \\ \hline 0 & 1+\frac{1}{x} \end{array}\right] \qquad (V.29)$$

(where we separated the blocks from $n_0$, $e$, $n_1$ and $d$). The matrix $t^{(1)}(x)$, which is scalar inside of a given occupancy sector, is then given by:

$$t^{(1)}(x) = (1+x)^{L-N}(1+x^{-1})^N = h(x). \qquad (V.30)$$

This is the same as the function $h(x)$ that we defined in section II.2.2, and is usually called the 'quantum determinant'.

For $k = 2$, the $2 \times 2$ blocks from $n_0$, $e$, $n_1$ and $d$ are:

$$X_2(x) = \left[\begin{array}{cc|cc} 1+x & 0 & 0 & 0 \\ 0 & 1+qx & 1-\frac{1}{q} & 0 \\ \hline 0 & 1-q & 1+\frac{1}{qx} & 0 \\ 0 & 0 & 0 & 1+\frac{1}{x} \end{array}\right] \qquad (V.31)$$

and

$$t^{(2)}(x) = \text{Tr}[A_\mu^{(2)} \prod_{i=1}^{L} X_2^{(i)}(x)] \qquad (V.32)$$

with $A_\mu^{(2)} = \begin{bmatrix} 1 & 0 \\ 0 & e^{-\mu} \end{bmatrix}$.

This matrix is, in fact, the standard Bethe transfer matrix for the periodic XXZ spin chain, with a two-dimensional auxiliary space. To write it in its usual form, we need to make a few transformations and change variables. To that effect, let us consider:

$$\frac{1}{1+x} \begin{bmatrix} 1 & 0 \\ 0 & x \end{bmatrix} \cdot X_2(x) = \begin{bmatrix} 1 & 0 & 0 & 0 \\ 0 & \frac{1+qx}{1+x} & \frac{q-1}{q(1+x)} & 0 \\ 0 & \frac{x(1-q)}{1+x} & \frac{1+qx}{q(1+x)} & 0 \\ 0 & 0 & 0 & 1 \end{bmatrix} = \begin{bmatrix} 1 & 0 & 0 & 0 \\ 0 & q\lambda & 1-\lambda & 0 \\ 0 & 1-q\lambda & \lambda & 0 \\ 0 & 0 & 0 & 1 \end{bmatrix} \quad (V.33)$$

with $\lambda = \frac{1+qx}{q(1+x)}$, i.e. $x = -\frac{1-q\lambda}{q(1-\lambda)}$. We recall that the symbol $\cdot$ is used to signify a product in configuration space, so that it not be confused with a product in the auxiliary space (for which we use the usual product notation). This is the common form of the Lax matrix for the ASEP:

$$L_i(\lambda) = \begin{bmatrix} 1 & 0 & 0 & 0 \\ 0 & q\lambda & 1-\lambda & 0 \\ 0 & 1-q\lambda & \lambda & 0 \\ 0 & 0 & 0 & 1 \end{bmatrix} = P_i(1+\lambda M_i) \quad (V.34)$$

where $P_i$ is a permutation matrix which has the effect of exchanging the physical space at site $i$ with the auxiliary space (fig.-V.1).

$$L_i(\lambda) = a \boxed{\phantom{XX}}_i^i a + \lambda \ a \boxed{M}_i^i a$$

Figure V.1: Schematic representation of $L_i(\lambda)$. The first box represents $P_i$, which exchanges the occupancies of the auxiliary space $a$ and the physical space $i$ (the matrix is seen as acting from SE to NW). The second box is the same exchange operator, with the local matrix $M$ acting during the exchange.

The matrix we applied to $X_2$ from the left in (V.33) multiplies every entry by $x$ for each occupied site (and since this number is conserved between the left and right entries of the transfer matrix, this operation actually commutes with $t^{(2)}$, so that we haven't modified its eigenvectors). All in all, this operation multiplies $t^{(2)}$ by $x^N/(1+x)^L = 1/h(x)$. We therefore define:

$$\hat{t}^{(2)}(\lambda) = \frac{x^N}{(1+x)^L} t^{(2)}(x) = \text{Tr}[A_\mu^{(2)} \prod_{i=1}^L L_i(\lambda)] \quad (V.35)$$

which is the Bethe transfer matrix for the periodic ASEP with one marked bond.

Since $L_i(0) = P_i$ is a permutation matrix, and its derivative with respect to $\lambda$ at 0 is $\frac{d}{d\lambda}L_i(0) = P_i M_i$, we find that $\hat{t}^{(2)}(0)$ is the matrix that transposes the whole system back one step, and that, for the whole transfer matrix $\hat{t}^{(2)}$ (fig.-V.2):

$$M_\mu = \left(\hat{t}^{(2)}(\lambda)\right)^{-1} \frac{d}{d\lambda} \hat{t}^{(2)}(\lambda)\Big|_{\lambda=0} = \frac{d}{d\lambda} \log\left(\hat{t}^{(2)}(\lambda)\right)\Big|_{\lambda=0} \quad (V.36)$$

(we have not considered how $A_\mu$ comes into play, but we can easily check that it gives the correct term in $M_\mu$).

<center>
[Figure V.2 diagram: schematic representation of $\hat{t}^{(2)}(0)$ as a translation matrix, and $\hat{t}^{(2)-1}\frac{d}{d\lambda}\hat{t}^{(2)}(0)$ with local insertion $M$, yielding $M$ at sites 1,2.]
</center>

Figure V.2: Schematic representation of the value and first logarithmic derivative of $\hat{t}^{(2)}$ at 0. The first is a translation matrix which takes each site $i$ to $i-1$. The second (of which only one part of the sum over sites is represented) gives the local jump matrix $M_i$.

Written in terms of $t^{(2)}(x)$, this becomes:

$$M_\mu = -(L-N) - Nq + (1-1/q)\frac{d}{dx}\log\bigl(t^{(2)}(x)\bigr)\bigr|_{x=-1/q}. \tag{V.37}$$

We recognise eq.(II.98), which we found through the coordinate Bethe Ansatz in section II.2.2. This identifies the eigenvalues of $t^{(2)}(x)$ with the functions $T(x)$ which we introduced back then (we recall that there was a different $T$ for each eigenspace of $M_\mu$).

Doing the same calculations at $\lambda = \infty$ instead of 0, after a few modifications (such as taking the contragredient representation for $X$ and multiplying everything by $h(x)/h(qx)$) would have given us eq.(II.97) instead:

$$M_\mu = -(L-N)q - N + (1-q)\frac{d}{dx}\log\bigl(t^{(2)}(x)\bigr)\bigr|_{x=-1} \tag{V.38}$$

which is also what we would have obtained if we had considered a system with $L-N$ particles, exchanging $Q$ with $P$, and kept $y=1/qx$ as a variable. This identity is usually called the 'crossing symmetry' in the language of quantum spin chains.

### V.1.3 R matrix

The next step is to show that the eigenvalues $\Lambda_i(x,y)$ of $T^{per}_\mu$ are in fact a product of a function of $x$ and a function of $y$, which we will note as $P_i(x)$ and $Q_i(y)$. This factorisation property is a well known fact for the periodic XXX [177, 178] and XXZ [152, 172] spin chains. This result will actually be derived in the next section, as there are a few preliminary calculations which need to be done first, mainly in order to find the R matrix of our system.

One way to go about this is through a method used in [179, 180] which consists in introducing two new Lax matrices, defined by:

$$L_1 = L(a_1, b_1, c_1, d_1) = \begin{bmatrix} a_1 A_1 & b_1 S_1^+ \\ -c_1 S_1^- A_1 & d_1 \end{bmatrix}, \tag{V.39}$$

$$\tilde{L}_2 = \tilde{L}(a_2, b_2, c_2, d_2) = \begin{bmatrix} a_2 A_2 & -c_2 S_2^+ A_2 \\ b_2 S_2^- & d_2 \end{bmatrix}, \tag{V.40}$$

where the operators $A_i$ and $S_j^\pm$ obey (V.16) and (??) for $i = j$ (i.e. if they act on the same space), and commute otherwise.

We then take the product of those matrices (which is a matrix product on the physical two-dimensional space and a tensor product on the infinite-dimensional auxiliary spaces of $L_1$ and $\tilde{L}_2$):

$$L_1 \tilde{L}_2 = \begin{bmatrix} a_1 a_2 A + b_1 b_2 S_1^+ S_2^- & b_1 d_2 S_1^+ - a_1 c_2 S_2^+ A \\ d_1 b_2 S_2^- - c_1 a_2 S_1^- A & c_1 c_2 S_1^+ S_2^- A + d_1 d_2 \end{bmatrix} \tag{V.41}$$

with $A = A_1 A_2$. We will omit the notation $\cdot$ for the product between those new Lax matrices, since the indices are there to signify that their elements act on different spaces.

We now need to consider two special cases for the coefficients of $L_1$ and $\tilde{L}_2$. Let us first set them as follows: $a_1 = x$, $c_2 = y$ and the rest is 1. We write the corresponding matrices as $L_1^+$ and $\tilde{L}_2^-$:

$$L_1^+(x) \tilde{L}_2^-(y) = \begin{bmatrix} xA + S_1^+ S_2^- & S_1^+ - xy S_2^+ A \\ S_2^- - S_1^- A & y S_1^+ S_2^- A + 1 \end{bmatrix} \tag{V.42}$$

and, by projecting each element on $S_1^+ = S_2^+ = S^+$ (i.e. by applying $\sum \|i, j\rangle\!\rangle\langle\!\langle i+j\|$ to the right and its contragredient to the left), we get:

$$\boxed{L_1^+(x) \tilde{L}_2^-(y) = \begin{bmatrix} xA + 1 & S^+(1 - xyA) \\ S^-(1 - A) & yA + 1 \end{bmatrix} = X(x, y).} \tag{V.43}$$

Naturally, we check that $A$ and $S^+$ satisfy the correct relations.

Let us now set $a_2 = x$, $c_1 = y/q$, $c_2 = q$ and the rest to 1. We write the corresponding matrices as $L_1^-$ and $\tilde{L}_2^+$:

$$L_1^-(y) \tilde{L}_2^+(x) = \begin{bmatrix} xA + S_1^+ S_2^- & S_1^+ - q S_2^+ A \\ S_2^- - xy/q S_1^- A & y S_1^+ S_2^- A + 1 \end{bmatrix} \tag{V.44}$$

and, through the same operation as before, we get:

$$\boxed{L_1^-(y) \tilde{L}_2^+(x) = \begin{bmatrix} xA + 1 & (1 - A) S^+ \\ (1 - xyA) S^- & yA + 1 \end{bmatrix} = \overline{X}(x, y).} \tag{V.45}$$

In both of those special cases, one of the non-diagonal elements has a factor $(1 - A)$ which allows us to truncate the representation at state $\|0\rangle\!\rangle$ and avoid some convergence issues. It would not be the case, however, if we were to construct $L_1^+(x) \tilde{L}_2^+(y)$ or $L_1^-(x) \tilde{L}_2^-(y)$, which we will therefore avoid at any cost.

We will now use this formalism in order to find the so-called '$R$ matrix' which is such that:

$$X(x,y) \cdot X(x',y') \ R(x,y;x',y') = R(x,y;x',y') \ X(x',y') \cdot X(x,y) \tag{V.46}$$

where $R$ acts on the two auxiliary spaces of both $X$ matrices. We will comment on the use of such a matrix at the end of this section.

Considering that $X(x,y) \cdot X(x',y') = L_1^+(x)\tilde{L}_2^-(y)L_3^+(x')\tilde{L}_4^-(y')$, we will perform this exchange of parameters in steps, exchanging the parameters of only two $L_i$'s at a time. The first thing we could try is to exchange $y$ and $x'$, but this would transform $\tilde{L}_2^-(y)$ into $\tilde{L}_2^+(x')$, so we would have $L_1^+(x)\tilde{L}_2^+(x')$ on the left, and we can't have this. The solution is then to first exchange $x'$ and $y'$, then $y'$ and $y$, and finally $y$ and $x'$, to obtain $X(x,y') \cdot X(x',y)$, and then do the same once more to exchange $x$ and $x'$.

We first need to find $f_{12}(x,y)$ such that $L_1^+(x)\tilde{L}_2^-(y)f_{12}(x,y) = f_{12}(x,y)L_1^-(y)\tilde{L}_2^+(x)$, which is to say:

$$X(x,y)f_{12}(x,y) = f_{12}(x,y)\overline{X}(x,y). \tag{V.47}$$

That $f_{12}$ may depend on $A$ and $S^+$. In terms of the elements of $X(x,y)$, this writes:

$$[1 + xA, f_{12}] = 0, \tag{V.48}$$
$$[1 + yA, f_{12}] = 0, \tag{V.49}$$
$$S^+(1 - xyA) \ f_{12} = f_{12}(1 - A)S^+, \tag{V.50}$$
$$S^-(1 - A) \ f_{12} = f_{12}(1 - xyA)S^-. \tag{V.51}$$

The first and second equations tell us that $f_{12}$ commutes with $A$ (i.e. it is diagonal), so it should be a function of $A$ alone. The third or fourth equations then give us:

$$S^+(1-xyA)f_{12}[A] = f_{12}[A](1-A)S^+ = S^+(1-qA)f_{12}[qA] \tag{V.52}$$

(where the second equality is due to the commutation of $S^+$ with $A$), which we can rewrite as

$$\frac{f_{12}[A]}{f_{12}[qA]} = \frac{(1-qA)}{(1-xyA)}. \tag{V.53}$$

Iterating this last equation, we finally find:

$$\boxed{f_{12}(xy) = \frac{(qA)_\infty}{(xyA)_\infty}.} \tag{V.54}$$

To exchange the parameters back, one simply has to apply $f_{12}^{-1}(xy)$.

This was for the exchange of parameters between the first and second or third and fourth matrices in $L_1^+(x)\tilde{L}_2^-(y)L_3^+(x')\tilde{L}_4^-(y')$ (i.e. inside of a same $X$ matrix). We now need to exchange parameters between the second and third matrices in that product.

Let us consider $\tilde{L}_2 L_3$ with $c_2 = y$, $c_3 = y'/q$, and the rest set to 1:

$$\tilde{L}_2^-(y)L_3^-(y') = \begin{bmatrix} (1+yy'/qS_2^+S_3^-)A_2A_3 & (S_3^+ - yS_2^+)A_2 \\ (S_2^- - y'/qS_3^-)A_3 & S_2^-S_3^+ + 1 \end{bmatrix}. \tag{V.55}$$

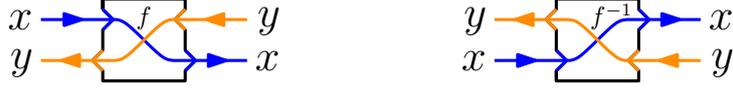

Figure V.3: Schematic representation of $f_{12}$ and $f_{12}^{-1}$. The exponents $+$ and $-$ of the $L$ matrices on one side and the other are represented by, respectively, right and left arrows (the top row being $L_1$, and the second $\tilde{L}_2$), and their arguments are represented by colours (blue for $x$ and orange for $y$).

We need to find $g_{23}^-(y, y')$ such that $\tilde{L}_2^-(y) L_3^-(y') g_{23}^-(y, y') = g_{23}^-(y, y') \tilde{L}_2^-(y') L_3^-(y)$. As before, the commutation for each of the four elements of $\tilde{L}_2^-(y) L_3^-(y')$ is:

$$[(1 + yy'/q S_2^+ S_3^-) A_2 A_3, g_{23}^-] = 0, \tag{V.56}$$
$$[1 + S_2^- S_3^+, g_{23}^-] = 0, \tag{V.57}$$
$$(S_3^+ - y S_2^+) A_2 \, g_{23}^- = g_{23}^- (S_3^+ - y' S_2^+) A_2, \tag{V.58}$$
$$(S_2^- - y'/q S_3^-) A_3 \, g_{23}^- = g_{23}^- (S_2^- - y/q S_3^-) A_3. \tag{V.59}$$

The first and second equations tell us that $g_{23}^-$ commutes with $S_2^+ S_3^-$ and with $A_2 A_3$. the third and fourth suggest that $g_{23}^-$ keeps $A_2$ and $A_3$ separated, so it should only depend on $S_2^+ S_3^-$. The third equation then gives:

$$(S_3^+ - y S_2^+) A_2 \, g_{23}^-[S_2^+ S_3^-] = g_{23}^-[q S_2^+ S_3^-](S_3^+ - y S_2^+) A_2 = g_{23}^-[S_2^+ S_3^-](S_3^+ - y' S_2^+) A_2 \tag{V.60}$$

which can be rewritten as:

$$\frac{g_{23}^-[S_2^+ S_3^-]}{g_{23}^-[q S_2^+ S_3^-]} = \frac{(1 - y S_2^+ S_3^-)}{(1 - y' S_2^+ S_3^-)} \tag{V.61}$$

and produces, through iteration:

$$\boxed{g_{23}^-(y, y') = \frac{(y S_2^+ S_3^-)_\infty}{(y' S_2^+ S_3^-)_\infty}.} \tag{V.62}$$

Finally, we look for $g_{23}^+(x, x')$ such that $\tilde{L}_2^+(x) L_3^+(x') g_{23}^+(x, x') = g_{23}^+(x, x') \tilde{L}_2^+(x') L_3^+(x)$. Setting $a_2 = x$, $c_2 = q$, $a_3 = x'$, and the rest to 1, in $\tilde{L}_2 L_3$, we find:

$$\tilde{L}_2^+(x) L_3^+(x') = \begin{bmatrix} (xx' + q S_2^+ S_3^-) A_2 A_3 & (x S_3^+ - q S_2^+) A_2 \\ (x' S_2^- - S_3^-) A_3 & S_2^- S_3^+ + 1 \end{bmatrix} \tag{V.63}$$

and the exact same operations as before produce:

$$\boxed{g_{23}^+(x, x') = \frac{(x' S_2^- S_3^+)_\infty}{(x S_2^- S_3^+)_\infty}.} \tag{V.64}$$

Let us note that $g^+$ and $g^-$ commute with the projection which we perform on $L_1 \tilde{L}_2$ and $L_3 \tilde{L}_4$ to get the $X$ matrices.

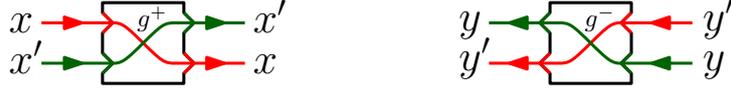

Figure V.4: Schematic representation of $g_{23}^+$ and $g_{23}^-$. The top row represents $\tilde{L}_2$, and the second $L_3$, and their arguments are represented by colours (red for $x$ and $y$ and green for $x'$ and $y'$).

We can finally forget everything about that alternative construction of $X(x,y)$, and simply define those operators $f$ and $g^{\pm}$ as what we found them to be. After re-indexing them so that 1 refers to the auxiliary space of the first $X$ matrix, and 2 to the second, we can write:

$$\boxed{X_1(x,y)\centerdot X_2(x',y')\ R_y(x,y;x',y') = R_y(x,y;x',y')\ X_1(x,y')\centerdot X_2(x',y),} \quad \text{(V.65)}$$
$$\boxed{X_1(x,y)\centerdot X_2(x',y')\ R_x(x,y;x',y') = R_x(x,y;x',y')\ X_1(x',y)\centerdot X_2(x,y'),} \quad \text{(V.66)}$$

with

$$\boxed{R_y(x,y;x',y') = f_2(x'y')g_{12}^-(y,y')f_2^{-1}(x'y),} \quad \text{(V.67)}$$
$$\boxed{R_x(x,y;x',y') = f_1(xy)g_{12}^+(x,x')f_1^{-1}(x'y).} \quad \text{(V.68)}$$

where we relabelled $f_{12}$ as $f_1$, $f_{34}$ as $f_2$, and $g_{23}^{\pm}$ as $g_{12}^{\pm}$, consistently with the indexes of the $X$ matrices.

Applying those one after the other, we find the full $R$ matrix:

$$\boxed{X_1(x,y)\centerdot X_2(x',y')\ R(x,y;x',y') = R(x,y;x',y')\ X_1(x',y')\centerdot X_2(x,y)} \quad \text{(V.69)}$$

with

$$\boxed{\begin{aligned} R(x,y;x',y') &= R_y(x,y;x',y')R_x(x,y';x',y) \\ &= \frac{(qA_2)_\infty}{(x'y'A_2)_\infty}\frac{(yS_1^+S_2^-)_\infty}{(y'S_1^+S_2^-)_\infty}\frac{(x'yA_2)_\infty}{(qA_2)_\infty}\frac{(qA_1)_\infty}{(xy'A_1)_\infty}\frac{(x'S_1^-S_2^+)_\infty}{(xS_1^-S_2^+)_\infty}\frac{(x'y'A_1)_\infty}{(qA_1)_\infty} \end{aligned}} \quad \text{(V.70)}$$

(cf. fig.-V.5), or an equivalent expression if we apply $R_x$ to the left of $R_y$ instead.

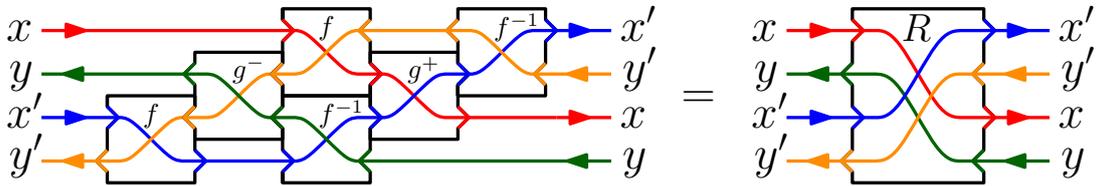

Figure V.5: Schematic representation of the $R$ matrix for the periodic ASEP.

There are many things to be said about that $R$ matrix. First of all, it is the rigorous way to go if we want to prove that $T_\mu^{per}(x,y)$ commutes with $T_\mu^{per}(x',y')$ for any value of those parameters: to do that, we insert $R(x,y;x',y')R^{-1}(x,y;x',y')$ at any point in the matrix product expression of $T_\mu^{per}(x,y)T_\mu^{per}(x',y')$, and make $R(x,y;x',y')$ commute to the left all the way around the trace (fig.-V.6), exchanging parameters between the two rows along the way. When crossing the marked bond, we just need to note that $R$ commutes with $A_\mu \cdot A_\mu$.

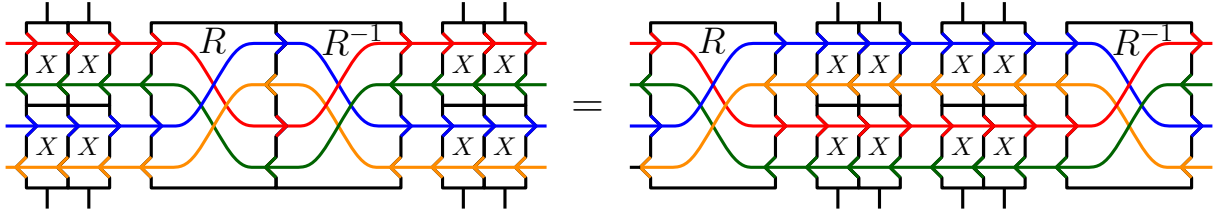

Figure V.6: Schematic representation of the commutation between $R$ matrices and $X$ matrices with different parameters (represented by different colours).

Equations (V.65) and (V.66) are also of interest by themselves: they tell us that $T_\mu^{per}(x,y)$ and $T_\mu^{per}(x',y')$ can actually exchange only one of their parameters and keep the other, instead of commuting altogether (by applying the procedure we just described, but with $R_x$ or $R_y$ instead of $R$). This will be extremely useful to us in the next section.

Moreover, considering the decomposition (V.27) which we obtained in the previous section for special values of the spectral parameters in $T_\mu^{per}(x,y)$, by taking $y = 1/q^{k-1}x$ and $y' = 1/q^{l-1}x'$ in $R(x,y;x',y')$, we should be able to recover the $R$ matrix between auxiliary spins $\frac{k-1}{2}$ and $\frac{l-1}{2}$ as an independent part of the whole matrix. In other words, this $R$ matrix of twice infinite dimension should contain all the smaller $R$ matrices for the XXZ chain. Taking $y = 1/qx$, in particular, should yield the Lax matrix $X$ (possibly up to a permutation). Because of the complexity of eq.(V.70), those facts are yet to be verified.

### V.1.4 Q-operator and Bethe equations

From the previous section, we now know that $T^{per}$ satisfies:

$$T^{per}(x,y)T^{per}(x',y') = T^{per}(x,y')T^{per}(x',y). \qquad (\text{V.71})$$

Fixing $x' = y' = 0$ in that equation (although any other constants would do), we can therefore write:

$$\boxed{T^{per}(x,y) = (1 - e^{-\mu})P(x)Q(y)} \qquad (\text{V.72})$$

with:

$$P(x) = \left[T_\mu^{per}(0,0)\right]^{-1} T_\mu^{per}(x,0) \quad , \quad Q(y) = (1 - e^{-\mu})^{-1} T_\mu^{per}(0,y) \qquad (\text{V.73})$$

(the factor $(1-e^{-\mu})^{-1}$ in $Q$ is there to make our notations consistent with section II.2.2). Since the matrices $P$ and $Q$ are combinations of the matrix $T_\mu^{per}$ taken at various values of its parameters, and considering eq.(V.71), we immediately see that $P(x)$ and $Q(y)$ commute for any values of $x$ and $y$.

This relation is crucial to our reasoning, and, put together with (V.27), will allow us to reach our conclusion in just a few more lines.

Using this, we can now rewrite (V.27) as:

$$\boxed{P(x)Q(1/q^{k-1}x) = t^{(k)}(x) + e^{-k\mu}P(q^kx)Q(q/x).} \tag{V.74}$$

The first and second orders of this equation give:

$$P(x)Q(1/x) = h(x) + e^{-\mu}P(qx)Q(q/x), \tag{V.75}$$
$$P(qx)Q(1/qx) = h(qx) + e^{-\mu}P(q^2x)Q(1/x), \tag{V.76}$$
$$P(x)Q(1/qx) = t^{(2)}(x) + e^{-2\mu}P(q^2x)Q(q/x), \tag{V.77}$$

where we have written the first one twice (once at $x$, once at $qx$). We immediately recognise equation (V.75) to be exactly the same as (II.101) (with $\mu$ instead of $L\mu$, which comes from the fact that in section II.2.2, we had marked every bond instead of just one), written as one single matrix equation rather than a functional equation for each eigenspace of $M_\mu$.

We still need to make sure that the eigenvalues of $Q$ really are the same as the functions we defined in section II.2.2. To do that, we consider a combination of the three previous equations: $Q(1/qx)\times$(V.75)$+e^{-\mu}Q(q/x)\times$(V.76)$-Q(1/x)\times$(V.77), which yields:

$$\boxed{t^{(2)}(x)Q(1/x) = h(x)Q(1/qx) + e^{-\mu}h(qx)Q(q/x).} \tag{V.78}$$

We saw earlier that the eigenvalues of $t^{(2)}$ are the same as the functions $T$ we introduced for the coordinate Bethe Ansatz, so that this equation can be identified with (II.94). This confirms that the matrix $Q$ that we have defined is indeed Baxter's Q-operator. Equation (V.78) is called the T-Q relation, and its eigenvalues give the functional Bethe equations. Note that we could have obtained equation (II.100) instead through a different combination of equations (V.75), (V.76) and (V.77).

From here, we just need to repeat the final steps from section II.2.2 to obtain an expression for $E(\mu)$. We start from (II.101), which is the same as (V.75). Equation (II.95) follows from (V.37) and (V.78). The constant $B$ from (II.104) becomes a matrix:

$$B = -e^{-\mu}\big(Q(0)\big)^{-1}. \tag{V.79}$$

We then only need to show eq.(II.108), which is specific to the steady state, and states that the first eigenvalue of $B$ (which corresponds to the steady state) goes to 0 with $\mu \to 0$ while the others do not.

This can be proven by first showing that $T_\mu^{per}(x,y) \to |1\rangle\langle 1|$ for $\mu \to 0$. Every entry in $T_\mu^{per}(x,y)$ is the trace of $(1-e^{-\mu})A_\mu$ times a product of matrices $n_0$, $n_1$, $d$ and $e$. Each of those matrices is the sum of a term proportional to the identity (or $S^\pm$) and a term proportional to $A$. If we expand the trace on all those sums, there is only one contribution not proportional to $(1-e^{-\mu})$, which is the one where we took the term not containing $A$ in every matrix. This contribution is just the trace of $(1-e^{-\mu})A_\mu$ multiplied by some product of $S^+$ and $S^-$, which is equal to 1 for every entry in $T_\mu^{per}(x,y)$. This proves that $T_\mu^{per}(x,y)$ is a projector on $|1\rangle\langle 1|$ for $\mu \to 0$. All the other eigenvalues of $T_\mu^{per}(x,y)$ are at least of order $\mu$.

Knowing that, and the fact that $P(0) = 1$, we get from (V.72) that:

$$B = -\mathrm{e}^{-\mu}(1 - \mathrm{e}^{-\mu})\Big(T_\mu^{per}(0,0)\Big)^{-1} \qquad (\text{V.80})$$

so that its first eigenvalue goes to 0 for $\mu \to 0$, but not the others.

Let us finally note that using eqs.(V.74), we can obtain all the equations from the so-called 'fusion hierarchy' [173,174], which gives equations on the decomposition of products of matrices $t^{(k)}$, as well as the T-Q equations for any $t^{(k)}$ (which involve products of $k-1$ matrices $Q$).

## V.2  Open ASEP

We will now try to apply the same procedure to the open ASEP. Considering the same generalised $X$ matrix as before, we first need to find out what the boundary vectors become. We then show the PQ factorisation of the transfer matrix (which is much easier to prove this time). Finally, we see how it decomposes into blocks, one of which is the Bethe transfer matrix, according to the same equation as the periodic case but with a different quantum determinant (the boundaries make this much harder to prove than for the previous case). We also show, as an appendix, what this all becomes in the language of the XXZ chain with spin $\frac{1}{2}$.

We recall that the Markov matrix for the open ASEP is given by:

$$M = m_0 + \sum_{i=1}^{L-1} M_i + m_L \qquad (\text{V.81})$$

with

$$m_0 = \begin{bmatrix} -\alpha & \gamma \\ \alpha & -\gamma \end{bmatrix} \;,\; M_i = \begin{bmatrix} 0 & 0 & 0 & 0 \\ 0 & -q & 1 & 0 \\ 0 & q & -1 & 0 \\ 0 & 0 & 0 & 0 \end{bmatrix} \;,\; m_L = \begin{bmatrix} -\delta & \beta \\ \delta & -\beta \end{bmatrix} \qquad (\text{V.82})$$

### V.2.1  Boundary algebra and commutation relations

Let us first find out how the presence of boundaries make this case different from the previous one. We define:

$$U_\mu(x) = \frac{1}{Z_L}\langle\!\langle W \| A_\mu \prod_{i=1}^{L} X^{(i)}(x,x) \| V \rangle\!\rangle, \qquad (\text{V.83})$$

$$T_\mu(y) = \langle\!\langle \tilde{W} \| A_\mu \prod_{i=1}^{L} X^{(i)}(y,y) \| \tilde{V} \rangle\!\rangle, \qquad (\text{V.84})$$

which has the same structure as the transfer matrix from chapter III, but with the $X$ matrices replaced by their generalisation. Note that, since we have two rows of matrices, we have, in principle, four free parameters (two in each row), but we will see in a moment that we must in fact put the same parameter twice in each row (so that $U_\mu$ depends only on $x$, and $T_\mu$ on $y$) if we want to be able to find suitable boundary vectors. To simplify notations, we will therefore rewrite $X$ and $\hat{X}$ as:

$$X(x) = \begin{bmatrix} n_x & e \\ d & n_x \end{bmatrix} \;,\; \hat{X} = \frac{(1-q)}{2} \begin{bmatrix} \tilde{n}_x & e \\ -d & -\tilde{n}_x \end{bmatrix} \tag{V.85}$$

with $n_x = 1 + xA$ and $\tilde{n}_x = 1 - xA$.

Note that for a general set of fugacities, the generalisation (V.22) holds, with the matrices $A_{\mu_i}$ being inserted in both $U$ and $T$.

The conditions that these boundary vectors must satisfy is:

$$[\beta(d + n_x) - \delta(e + n_x) - (1 - q)] \|V\rangle\!\rangle = 0, \tag{V.86}$$
$$\langle\!\langle W\| [\alpha(e + n_x) - \gamma(d + n_x) - (1 - q)] = 0, \tag{V.87}$$
$$[\beta(d - n_y) - \delta(e - n_y) + (1-q)yA] \|\tilde{V}\rangle\!\rangle = 0, \tag{V.88}$$
$$\langle\!\langle \tilde{W}\| [\alpha(e - n_y) - \gamma(d - n_y) + (1-q)yA] = 0, \tag{V.89}$$

where we notice that the first two are the same as in chapter III with $n_x$ replacing 1, and the next two also have an extra term $(1 - q)yA$. For $x = y = 0$, we naturally recover the conditions we had in chapter III. Note that those conditions were found by trial and error, so that they may not be unique.

**Commutation relations**

We now show that these conditions lead to $U_\mu(x)T_\mu(y)$ commuting with $M_\mu$ for any $x$ and $y$. The bulk part of the derivation is the same as for the periodic case, and produces, for the left boundary, a term $\hat{U}_\mu^{(1)}T_\mu + U_\mu \hat{T}_\mu^{(1)}$ (with the same notations as in (III.46)) which must be cancelled by the left boundary matrix. We have:

$$\left[m_0, \langle\!\langle W\| X^{(1)}\right] = \langle\!\langle W\| \begin{bmatrix} \gamma d - \alpha e & -(\alpha - \gamma)e \\ (\alpha - \gamma)d & \alpha e - \gamma d \end{bmatrix}$$
$$= (\alpha - \gamma)\langle\!\langle W\| \begin{bmatrix} n_x & -e \\ d & -n_x \end{bmatrix} + (1-q)\langle\!\langle W\| \begin{bmatrix} -1 & 0 \\ 0 & 1 \end{bmatrix} \tag{V.90}$$

$$\left[m_0, \langle\!\langle \tilde{W}\| X^{(1)}\right] = \langle\!\langle \tilde{W}\| \begin{bmatrix} \gamma d - \alpha e & -(\alpha - \gamma)e \\ (\alpha - \gamma)d & \alpha e - \gamma d \end{bmatrix}$$
$$= (\alpha - \gamma)\langle\!\langle \tilde{W}\| \begin{bmatrix} -n_y & -e \\ d & n_y \end{bmatrix} + (1-q)\langle\!\langle \tilde{W}\| \begin{bmatrix} yA & 0 \\ 0 & -yA \end{bmatrix} \tag{V.91}$$

where we used (V.87) and (V.89). Equivalent relations can be found for the right boundary.

The first terms in each of those two equations cancel one another:

$$\begin{bmatrix} n_x & -e \\ d & -n_x \end{bmatrix} \cdot X(y) + X(x) \cdot \begin{bmatrix} -n_y & -e \\ d & n_y \end{bmatrix} = 0 \tag{V.92}$$

We note that if we had had two different parameters for the two diagonal terms in $X(x)$ or $X(y)$, this relation wouldn't have been possible (or we would have needed two equations on each boundary vector instead of one).

As for the second terms, a straightforward calculation gives:

$$(1-q)\left(\begin{bmatrix} -1 & 0 \\ 0 & 1 \end{bmatrix} \cdot X(y) + X(x) \cdot \begin{bmatrix} yA & 0 \\ 0 & -yA \end{bmatrix}\right) = -\hat{X}(x) \cdot X(y) - X(x) \cdot \hat{X}(y). \quad (V.93)$$

This compensates exactly the contribution of the bulk, coming, as in the periodic case, from eq.(V.19) (these cancellations are identical to those we saw in section III.2.3). The exact same calculations can be done at the right boundary. We have omitted the matrix $A_\mu$ here, because, as in the periodic case, its action is trivial.

We can now conclude that, for any values of $x$ and $y$, we have:

$$\boxed{[M_\mu, U_\mu(x)T_\mu(y)] = 0.} \quad (V.94)$$

**Explicit expressions for the boundary vectors**

We will need to do a few explicit calculations involving the four boundary vectors in a few pages, so we might as well calculate them now.

In terms of the parameters $a$, $\tilde{a}$, $b$ and $\tilde{b}$ that we have defined in section III.3, equations (V.86)-(V.89) become:

$$[d + b\tilde{b}e + (1 + b\tilde{b})xA - (b + \tilde{b})] \, \|V\rangle\!\rangle = 0, \quad (V.95)$$

$$\langle\!\langle W\| \, [e + a\tilde{a}d + (1 + a\tilde{a})xA - (a + \tilde{a})] = 0, \quad (V.96)$$

$$[d + b\tilde{b}e + (b + \tilde{b})yA - (1 + b\tilde{b})] \, \|\tilde{V}\rangle\!\rangle = 0, \quad (V.97)$$

$$\langle\!\langle \tilde{W}\| \, [e + a\tilde{a}d + (a + \tilde{a})yA - (1 + a\tilde{a})] = 0. \quad (V.98)$$

We first focus on the right boundary. We saw that a possible representation for $d$ and $e$ is $e = S_1(1 - x^2 A_1)$ in $U_\mu$ or $e = S_2(1 - y^2 A_2)$ in $T_\mu$, and $d = S_i^{-1}(1 - A_i)$ (where we write $S = S^+$ and $S^{-1} = S^-$, and indices 1 and 2 refer to the auxiliary spaces of $U_\mu$ and $T_\mu$, respectively). Equations (V.95) and (V.96) become:

$$[S_1^{-1}(1 - A_1) + b\tilde{b}S_1(1 - x^2 A_1) + (1 + b\tilde{b})xA_1 - (b + \tilde{b})] \, \|V\rangle\!\rangle = 0, \quad (V.99)$$

$$[S_2^{-1}(1 - A_2) + b\tilde{b}S_2(1 - y^2 A_2) + (b + \tilde{b})yA_2 - (1 + b\tilde{b})] \, \|\tilde{V}\rangle\!\rangle = 0, \quad (V.100)$$

which is to say, multiplying by $S_i$ to the left:

$$[(1 - bS_1)(1 - \tilde{b}S_1 v) - (1 - xS_1)(1 - b\tilde{b}xS_1)A_1] \, \|V\rangle\!\rangle = 0, \quad (V.101)$$

$$[(1 - S_2)(1 - b\tilde{b}S_2) - (1 - byS_2)(1 - \tilde{b}yS_2)A_2] \, \|\tilde{V}\rangle\!\rangle = 0. \quad (V.102)$$

Much as we did in section III.4.2, we can write those vectors as generating functions, which will make it easier for us to manipulate them. Let us then write $\|V\rangle\!\rangle = V(S_1)\|0\rangle\!\rangle =$ $\sum\limits_{k=0}^{\infty} V_k S_1^k \|0\rangle\!\rangle$, which is such that $A_1 \|V\rangle\!\rangle = V(qS_1)\|0\rangle\!\rangle$. Those two last equations now become:

$$\frac{V(S_1)}{V(qS_1)}\|0\rangle\!\rangle = \frac{(1-xS_1)(1-b\tilde{b}xS_1)}{(1-bS_1)(1-\tilde{b}S_1)}\|0\rangle\!\rangle, \tag{V.103}$$

$$\frac{\tilde{V}(S_2)}{\tilde{V}(qS_2)}\|0\rangle\!\rangle = \frac{(1-byS_2)(1-\tilde{b}yS_2)}{(1-S_2)(1-b\tilde{b}S_2)}\|0\rangle\!\rangle, \tag{V.104}$$

which we can iterate to get:

$$\boxed{\|V\rangle\!\rangle = \frac{(xS_1)_\infty (b\tilde{b}xS_1)_\infty}{(bS_1)_\infty (\tilde{b}S_1)_\infty}\|0\rangle\!\rangle,} \tag{V.105}$$

$$\boxed{\|\tilde{V}\rangle\!\rangle = \frac{(byS_2)_\infty (\tilde{b}yS_2)_\infty}{(S_2)_\infty (b\tilde{b}S_2)_\infty}\|0\rangle\!\rangle.} \tag{V.106}$$

As for the left boundary, it is simpler to treat it using the contragredient representation $\overline{X}$ of $X$ on both vectors (which are then the exact symmetric of $\|V\rangle\!\rangle$ and $\|\tilde{V}\rangle\!\rangle$, but with $a$ and $\tilde{a}$ replacing $b$ and $\tilde{b}$), and then apply the operator $f_i$ that we found in section V.1.3 in order to go back to $X$. This gives us:

$$\boxed{\langle\!\langle W\| = \langle\!\langle 0\| \frac{(x/S_1)_\infty (a\tilde{a}x/S_1)_\infty}{(a/S_1)_\infty (\tilde{a}/S_1)_\infty} \frac{(x^2 A_1)_\infty}{(qA_1)_\infty},} \tag{V.107}$$

$$\boxed{\langle\!\langle \tilde{W}\| = \langle\!\langle 0\| \frac{(ay/S_2)_\infty (\tilde{a}y/S_2)_\infty}{(1/S_2)_\infty (a\tilde{a}/S_2)_\infty} \frac{(y^2 A_2)_\infty}{(qA_2)_\infty},} \tag{V.108}$$

to which we could add factors $\frac{(q)_\infty}{(x^2)_\infty}$ and $\frac{(q)_\infty}{(y^2)_\infty}$ for normalisation.

In fact, in most future calculations, we will use $X$ for $U_\mu$ and $\overline{X}$ for $T_\mu$, so that the factor $\frac{(y^2 A_2)_\infty}{(qA_2)_\infty}$ goes to $\|\tilde{V}\rangle\!\rangle$ instead of $\langle\!\langle \tilde{W}\|$.

There are two things that we now need to show: that the transfer matrix is a product of two commuting matrices, one depending on $x$ and one on $y$, and that for special values of the parameters it decomposes into two independent blocks.

### V.2.2 PQ factorisation and R matrix

Unlike what we did for the periodic case, we will first show that $U_\mu(x)T_\mu(y)$ factorises into two commuting matrices $P(x)$ and $Q(y)$. It is much easier to show than the previous time, because of the structure of the transfer matrix, in which $x$ and $y$ are already separated. Notice however that it is not trivial: $U_\mu$ and $T_\mu$ do not commute, so $P$ is not simply equal to $U_\mu$ and $Q$ to $T_\mu$.

To show this, we need to assume that eq.(V.94), which states that $U_\mu(x)T_\mu(y)$ commutes with $M_\mu$ for any $x$ and $y$, implies that $U_\mu(x)T_\mu(y)$ commutes with $U_\mu(x')T_\mu(y')$ for any values of $x$, $y$, $x'$ and $y'$ (which is true unless $M_\mu$ has a degenerate eigenspace which is not degenerate for $U_\mu(x)T_\mu(y)$). We could, in principle, prove this more rigorously later using R matrices, as we did before, but we will see that it is much more complicated to do in the open case. We also need to assume that $T_\mu$ is, generically, invertible.

We start from the commutation of $U_\mu(x)T_\mu(y)$ with $U_\mu(x')T_\mu(y)$:

$$U_\mu(x)T_\mu(y) \ U_\mu(x')T_\mu(y) = U_\mu(x')T_\mu(y) \ U_\mu(x)T_\mu(y), \quad \text{(V.109)}$$
$$U_\mu(x)T_\mu(y) \ U_\mu(x') \qquad\quad = U_\mu(x')T_\mu(y) \ U_\mu(x), \quad \text{(V.110)}$$
$$U_\mu(x)T_\mu(y) \ U_\mu(x')T_\mu(y') = U_\mu(x')T_\mu(y) \ U_\mu(x)T_\mu(y'), \quad \text{(V.111)}$$

where we applied $\big(T_\mu(y)\big)^{-1}$ to the right between the first and second lines, and $T_\mu(y')$ between the second and the third. This tells us that, as in the periodic case, we may exchange just one of the parameters between the two transfer matrices. Taking $x' = y' = 0$, we get:

$$\boxed{U_\mu(x)T_\mu(y) = (1 - e^{-\mu})P(x)Q(y)} \quad \text{(V.112)}$$

with

$$P(x) = U_\mu(x)\Big[U_\mu(0)\Big]^{-1}, \qquad Q(y) = (1 - e^{-\mu})^{-1}U_\mu(0)T_\mu(y). \quad \text{(V.113)}$$

As for the R matrix, from our calculations on the periodic case, we actually already know its expression here. Since we have two matrix rows instead of one, this can now be written as a product of 20 ratios of q-Pochhammer symbols instead of the 6 we had in eq.(V.70). We will not white it here, but we represent it schematically on fig.-V.7.

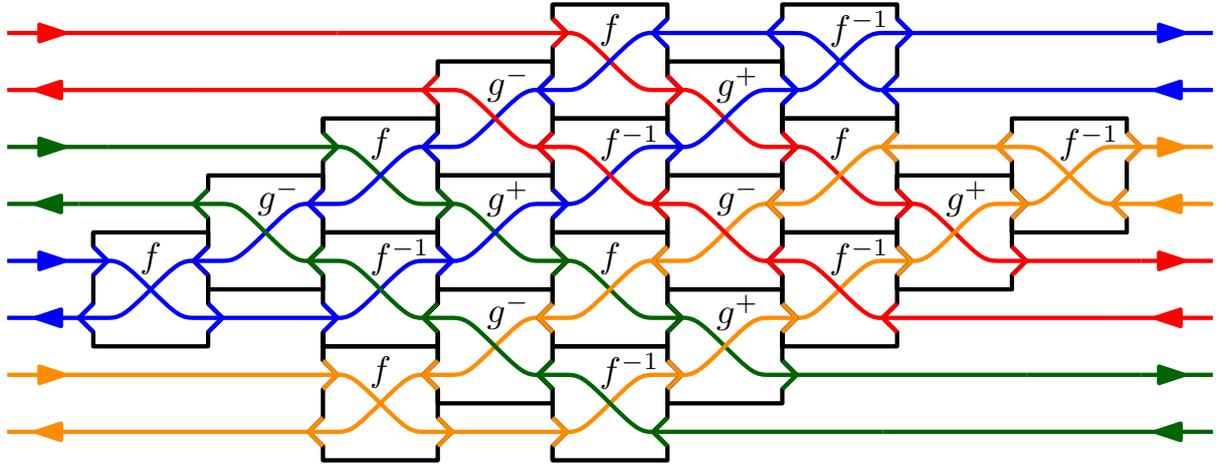

Figure V.7: Schematic representation of the $R$ matrix for the open ASEP.

Unfortunately, we cannot use it as easily as we did before, because of the boundaries: instead of taking $R$ around a trace to exchange the top and bottom matrices, we now need to show that when applied to a boundary, $R$ transfers the spectral parameters from the top vectors to the bottom ones (fig.-V.8). Considering that each boundary is the product of 8 more ratios of q-Pochhammer symbols, we will not try to make that full calculation directly, but search for a clever way to do it as simply as possible. We haven't found it yet.

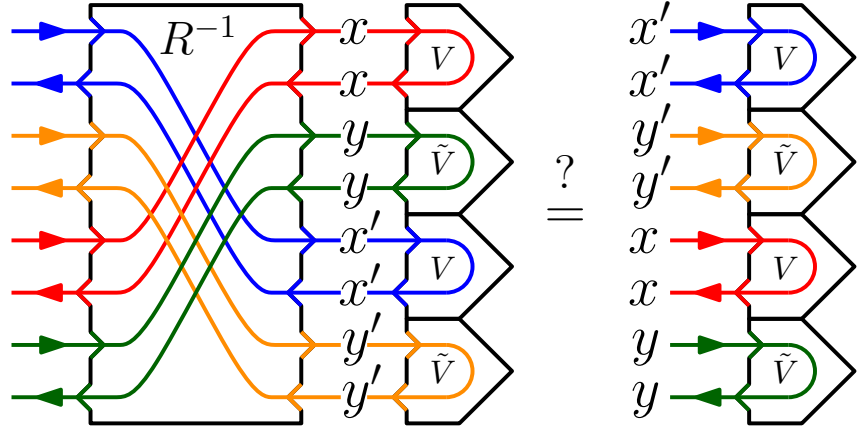

Figure V.8: Schematic representation of the equation involving the $R$ matrix and the right boundary vectors. This has yet to be verified.

### V.2.3 Decomposition of the transfer matrix

The last thing we need to investigate is whether anything happens for those special values of $x$ and $y$ that we considered in section V.1.2. The calculations involved are slightly more complicated that they were then, not only because of the boundaries, but also because $x$ and $y$ are now separated, and the factors $(1-xyA)$ in $e$ have been replaced by $(1-x^2 A_1)$ and $(1-y^2 A_2)$, which don't give anything useful for $xy = q^{1-k}$.

The simple solution to this problem is to put $x$ and $y$ back together, using $g_{12}^-(x,y)$ (as defined in section V.1.3) to go from $X(x,x) \cdot \overline{X}(y,y)$ to $X(x,y) \cdot \overline{X}(y,x)$. We choose to keep the second row of matrices in the contragredient representation to simplify future calculations. This operation makes the boundary vectors more complicated: they are not a product of two simple vectors acting each on one row any more (see fig.-V.9). We now have two complicated operators instead, which we will write as $K^+$ for the right boundary and $K^-$ for the left one:

$$K^- = \langle\!\langle 0_1, 0_2 \| \frac{(x/S_1)_\infty (a\tilde{a}x/S_1)_\infty}{(a/S_1)_\infty (\tilde{a}/S_1)_\infty} \frac{(x^2 A_1)_\infty}{(qA_1)_\infty} \frac{(ay/S_2)_\infty (\tilde{a}y/S_2)_\infty}{(1/S_2)_\infty (a\tilde{a}/S_2)_\infty} \frac{(xS_1/S_2)_\infty}{(yS_1/S_2)_\infty}, \quad (\text{V.114})$$

$$K^+ = \frac{(yS_1/S_2)_\infty}{(xS_1/S_2)_\infty} \frac{(xS_1)_\infty (b\tilde{b}xS_1)_\infty}{(bS_1)_\infty (\tilde{b}S_1)_\infty} \frac{(y^2 A_2)_\infty}{(qA_2)_\infty} \frac{(byS_2)_\infty (\tilde{b}yS_2)_\infty}{(S_2)_\infty (b\tilde{b}S_2)_\infty} \| 0_1, 0_2 \rangle\!\rangle. \quad (\text{V.115})$$

We will denote by $K^+_{i,j}$ the coefficient of $\| i,j \rangle\!\rangle$ in $K^+$ (i.e. the coefficient of $S_1^i S_2^j$ in the expansion of the ratios of q-Pochhammer symbols in $K^+$), and by $K^-_{j,i}$ the coefficient of $\langle\!\langle i,j \|$ in $K^-$.

Before getting on with the calculations, there are a few things to be said about the structure of what we have here. Let us for a moment forget that the upper and lower rows of matrices are connected, except through the boundary operators $K^\pm$. Since the lower row is written in the contragredient representation of $d$ and $e$, it is natural to consider it as acting towards the left rather than towards the right, and 'untwist' the whole lower row by cutting it somewhere in the middle and flipping both ends to their respective side (fig.-V.10). While doing that, we implicitly transpose the elements of $\overline{X}(y,x)$ on the lower row and invert their order. Now, the transposed elements of $\overline{X}(y,x)$ are exactly the same as the elements of $X(x,y)$, with the occupancies of the physical space exchanged (i.e. $n_0$ exchanged with $n_1$ and $d$ with $e$). This can be interpreted as an exchange of particles with holes on the whole lower row of the matrix product. What's more, the boundary operators $K^\pm$ are now matrices acting on one of the auxiliary spaces to the right, and the other to the left, so that those spaces can in fact be considered as one and the same. What we have, in the end, is a chain of size $2L+2$, with periodic boundary conditions, two defects where $K^+$ and $K^-$ are, and an anti-symmetry on the occupancies of one half of the chain. We can write it as:

$$UT_\mu(x,y) = \mathrm{Tr}[A_\mu K^- A_\mu \prod_{i=1}^{L} X^{(i)}(x,y) K^+ \prod_{i=L}^{1} X^{(i)}(x,y)] \qquad \text{(V.116)}$$

which is the usual formalism for the Bethe transfer matrix for an open system [163]. Note that there is an implicit transpose in the physical space on the second product of $X^{(i)}$'s and a trace between the ingoing configurations of the first product and the outgoing configurations of the second product.

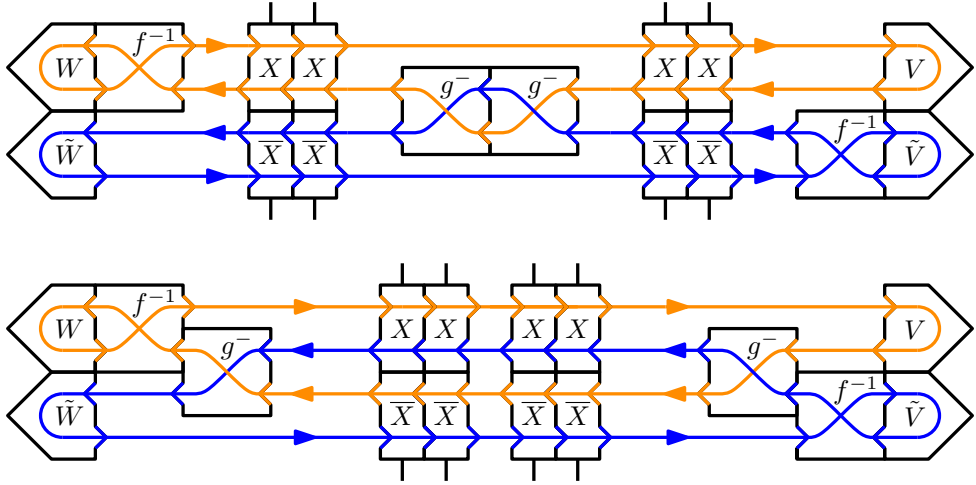

Figure V.9: Two equivalent representations for $U_\mu(x)T_\mu(y)$. In the first one, the two rows of $X$ matrices are independent, and the two spectral parameters $x$ (orange) and $y$ (blue) are separated. In the second, $x$ and $y$ are back together in $X$, but the two rows are now intertwined.

This remark can be related to two results that we obtained previously. One is the fact that, as we saw in section III.3, for certain values of the boundary parameters, the cumulants of the current for the open ASEP are exactly those of a periodic system with $2L+2$ sites, at half filling. The other is that the dominant eigenstate of the open ASEP for an extremely large current, which we saw in section IV.3.3, is that of an anti-periodic Dyson-Gaudin gas on a periodic lattice with $2L+2$, two being defects (without occupancies), where the boundaries should be. These similarities cannot be coincidental, but we have not found any way to make sense of them yet.

Let us now go back to the matter at hand, which is to find whether anything useful happens for $xy = q^{1-k}$ in the open case. Half of the question, we just answered: in the bulk of the system, now that $x$ and $y$ have been reunited, the same decomposition (into block triangular matrices) happens as did in the periodic case. We now only have to see whether the same happens to the boundary matrices $K^\pm$.

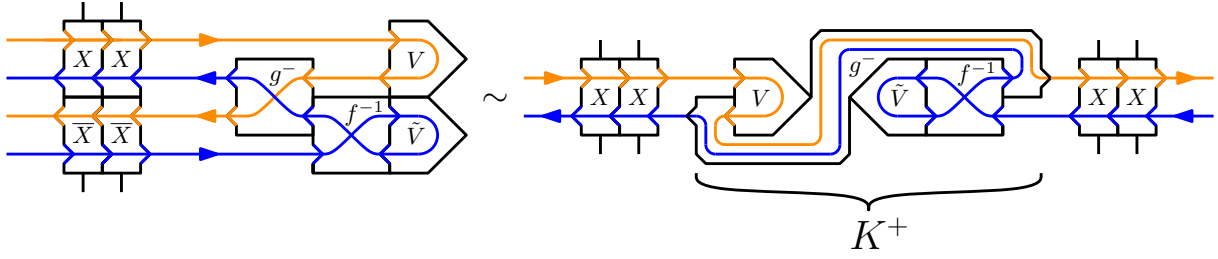

Figure V.10: Opening the loop at the right boundary.

**One-way boundaries**

We first focus on the simpler case where $\tilde{a} = \tilde{b} = 0$, which is to say $\gamma = \delta = 0$. Considering that $K^+$ and $K^-$ are roughly the same if we exchange $a$ and $b$, we will do all the calculations only on $K^+$ and simply give the equivalent results for $K^-$ in passing.

For this simpler case, we have:

$$K^+ = \frac{(xS_1)_\infty}{(bS_1)_\infty} \frac{(yS_1/S_2)_\infty}{(xS_1/S_2)_\infty} \frac{(y^2 A_2)_\infty}{(qA_2)_\infty} \frac{(byS_2)_\infty}{(S_2)_\infty} \frac{(q)_\infty}{(y^2)_\infty} \tag{V.117}$$

where the factor $|0_1, 0_2\rangle\!\rangle$ on which this whole function acts is implicit.

This can be expanded into:

$$K^+ = \frac{(xS_1)_\infty}{(bS_1)_\infty} \sum_{k=0}^\infty \frac{(y/x)_k}{(q)_k} (xS_1/S_2)^k \sum_{n=0}^\infty \frac{(by)_n}{(y^2)_n} (S_2)^n \tag{V.118}$$

where the first sum comes from $\frac{(yS_1/S_2)_\infty}{(xS_1/S_2)_\infty}$, and the second from everything that is to the right of that.

We will shortly be in need of one of Heine's transformation formulae for basic hypergeometric series, which we give now. For a function ${}_2\phi_1(a, b; c; z)$ defined as:

$$ {}_2\phi_1(a, b; c; z) = \sum_{n=0}^\infty \frac{(a)_n (b)_n}{(q)_n (c)_n} z^n \tag{V.119}$$

we have [147]:

$$\boxed{{}_2\phi_1(a, b; c; z) = \frac{(zab/c)_\infty}{(z)_\infty} \, {}_2\phi_1(c/a, c/b; c; zab/c).} \tag{V.120}$$

We will also need this simple identity on q-Pochhammer symbols:

$$\boxed{(x)_{j+k} = (x)_j (xq^j)_k.} \tag{V.121}$$

Coming back to $K^+$, we have:

$$K^+ = \frac{(xS_1)_\infty}{(bS_1)_\infty} \sum_{j=0}^{\infty} \frac{(by)_j}{(y^2)_j}(S_2)^j \sum_{k=0}^{\infty} \frac{(y/x)_k}{(q)_k}(xS_1/S_2)^k \frac{(byq^j)_k}{(y^2q^j)_k}(S_2)^k$$

$$= \sum_{j=0}^{\infty} \frac{(by)_j}{(y^2)_j}(S_2)^j \frac{(xS_1)_\infty}{(bS_1)_\infty} {}_2\phi_1(y/x, byq^j; y^2q^j; xS_1)$$

$$= \sum_{j=0}^{\infty} \frac{(by)_j}{(y^2)_j}(S_2)^j {}_2\phi_1(xyq^j, y/b; y^2q^j, bS_1)$$

$$= \sum_{j=0}^{\infty} \frac{(by)_j}{(y^2)_j}(S_2)^j \sum_{i=0}^{\infty} \frac{(xyq^j)_i(y/b)_i}{(q)_i(y^2q^j)_i}(bS_1)^i \qquad (V.122)$$

where we used (V.121) between the first line and the second, and (V.120) between the second and the third. This finally gives us:

$$\boxed{K^+_{i,j}(x,y) = \frac{(xyq^j)_i(y/b)_i b^i (by)_j}{(q)_i(y^2)_{i+j}}} \qquad (V.123)$$

and, at the left boundary:

$$\boxed{K^-_{j,i}(x,y) = \frac{(xyq^i)_j(x/a)_i a^i (ax)_j}{(q)_j(x^2)_{i+j}}.} \qquad (V.124)$$

We need to compare those values for $(x,y)$ equal to $(1/q^{k-1}y, y)$ or $(q/y, q^k y)$ (which are the same values as before, but keeping $y$ as a variable instead of $x$). We get:

$$K^+_{i,j}(1/q^{k-1}y, y) = \frac{(q^{j-k+1})_i(y/b)_i b^i (by)_j}{(q)_i(y^2)_{i+j}}, \qquad (V.125)$$

$$K^+_{i,j}(q/y, q^k y) = \frac{(q^{j+k+1})_i(yq^k/b)_i b^i (byq^k)_j}{(q)_i(y^2q^{2k})_{i+j}}. \qquad (V.126)$$

Since $(q^{j-k+1})_i = 0$ for $j-k+1 \leq 0$ and $i+j-k+1 \geq 0$, i.e. for $j \leq k-1$ and $i \geq k-1-j$, the first matrix is, as we expected, block triangular.

We now consider, for $j \geq k-1$, the ratio:

$$\frac{K^+_{i+k,j+k}(1/q^{k-1}y, y)}{K^+_{i,j}(q/y, q^k y)} = \frac{(y/b)_k b^k (by)_k}{(y^2)_{2k}} \frac{(q^{j+1})_k}{(q^{i+1})_k}. \qquad (V.127)$$

The term $\frac{(q^{j+1})_k}{(q^{i+1})_k}$ accounts for going to the contragredient representation on both lines (consistently with what happens to the $X$ matrices), and can be transferred to the left boundary, where they compensate similar terms, emerging from the same calculation:

$$\frac{K^-_{j+k,i+k}(x, 1/q^{k-1}x)}{K^-_{j,i}(q^k x, q/x)} = \frac{(x/a)_k a^k (ax)_k}{(x^2)_{2k}} \frac{(q^{i+1})_k}{(q^{j+1})_k}. \qquad (V.128)$$

All the other terms depend only on $k$, and factor out of the matrix product. We will now put them together.

First, we need to make a few transformations on those factors. For $y = 1/q^{k-1}x$, we have:

$$(y/b)_k b^k = (-x)^{-k} q^{-k(k-1)/2} (bx)_k, \tag{V.129}$$

$$(by)_k = (bq^{-k+1}/x)_k, \tag{V.130}$$

$$(x/a)_k a^k = (-x)^k q^{k(k-1)/2} (aq^{-k+1}/x)_k, \tag{V.131}$$

which gives us:

$$\frac{(x/a)_k a^k (ax)_k (y/b)_k b^k (by)_k}{(x^2)_{2k} (y^2)_{2k}} = \frac{(aq^{-k+1}/x)_k (bq^{-k+1}/x)_k (ax)_k (bx)_k}{(x^2)_{2k} (x^{-2} q^{-2k+2})_{2k}}$$

$$= \frac{h_b(q^k x) h_b(q/x)}{h_b(x) h_b(1/q^{k-1}x)} \tag{V.132}$$

with

$$\boxed{h_b(x) = \frac{(x^2)_\infty}{(ax, bx)_\infty}.} \tag{V.133}$$

What we have, then, is that the boundary matrices, just as the bulk matrices, are upper block triangular, with a first block of auxiliary dimension $k$, and a second of infinite auxiliary dimension, which is the same as the full matrix at different values of the spectral parameters, up to a global factor which we have just computed. Put into equations, this becomes:

$$P(x) Q(1/q^{k-1}x) = t^{(k)}(x) + e^{-2k\mu} \frac{h_b(q^k x) h_b(q/x)}{h_b(x) h_b(1/q^{k-1}x)} P(q^k x) Q(q/x) \tag{V.134}$$

where the factor $e^{-2k\mu}$ comes, as before, from the first coefficient of the matrices $A_\mu$ (of which there are now 2) in the second block.

To make things simpler, we can redefine all our matrices as:

$$\tilde{P}(x) = h_b(x) P(x) \;,\; \tilde{Q}(x) = h_b(x) Q(x) \;,\; \tilde{t}^{(k)}(x) = h_b(x) h_b(1/q^{k-1}x) t^{(k)}(x) \quad (V.135)$$

which is equivalent to choosing another normalisation for $U_\mu$ and $T_\mu$. Written in terms of those new transfer matrices, this last equation takes its definitive form:

$$\boxed{\tilde{P}(x) \tilde{Q}(1/q^{k-1}x) = \tilde{t}^{(k)}(x) + e^{-2k\mu} \tilde{P}(q^k x) \tilde{Q}(q/x).} \tag{V.136}$$

This has the exact same form as eq.(V.74), the only difference being a factor $2\mu$ instead of $\mu$ in the right hand side.

The rest of the reasoning is the same as in the periodic case. We first consider the case where $k = 1$. This gives us quite simply: $t^{(1)}(x) = (1+x)^L (1+1/x)^L = h(x)$ (where half of the monomials come from the upper matrix row, and the other half from the lower; the occupancy anti-symmetry between the two rows imply that whatever the configuration, there will be $L$ particles and $L$ holes in total). From this, we get:

$$\tilde{t}^{(1)}(x) = h(x)h_b(x)h_b(x^{-1}) = \frac{(1+x)^L(1+1/x)^L(x^2,x^{-2})_\infty}{(ax,bx,a/x,b/x)_\infty} = F(x) \qquad \text{(V.137)}$$

which is the quantum determinant for the open ASEP with one-way boundaries. This explains how the function $F(x)$ replaces $h(x)$ (from the periodic case) in all the expressions for the cumulants of the current.

Now, for $k=2$, we need to verify that $\tilde{t}^{(2)}(x)$ is related to $M_\mu$ through some derivative. We will do this directly in the general case (with all four boundary rates), in a few pages. For now, we just give the two-dimensional boundary matrices that we find from $K^+$ and $K^-$:

$$K_2^+ = \begin{bmatrix} 1 & \frac{qx(qx-b)}{(q^2x^2-1)} \\ \frac{x(1-qxb)}{(q^2x^2-1)} & 0 \end{bmatrix} \quad , \quad K_2^- = \begin{bmatrix} 1 & \frac{(a-x)}{(1-x^2)} \\ \frac{(ax-1)}{q(1-x^2)} & 0 \end{bmatrix}. \qquad \text{(V.138)}$$

**Two-way boundaries**

In the general case, where both boundaries have two non-zero rates, we were not able to do the first step of calculation leading to the decomposition of each boundary matrix into blocks (i.e. the step equivalent to (V.122)). We were, however, able to find the result of that calculation using a formal analysis software. We find that $K^+$ can be written as:

$$K_{i,j}^+ = \sum_{n=0}^{\min[i,j]} (-1)^n \frac{(q)_j}{(q)_n(q)_{j-n}} \frac{(xyq^n)_{i-n}}{(q)_{i-n}(y^2q^{2n})_{i+j-2n}} A_{i-n,j-n}^+(q^n y) \prod_{l=0}^{n-1} \tau^+(q^l y) \qquad \text{(V.139)}$$

where $A_{i,j}^+$ is a polynomial and

$$\tau^+(y) = \frac{(1-by)(1-\tilde{b}y)(y-b)(y-\tilde{b})}{y(1-y^2)(1-qy^2)}. \qquad \text{(V.140)}$$

As before, we take $y = 1/q^{k-1}x$, so that $(xyq^n)_{i-n} = (q^{n-k+1})_{i-n}$. This is 0 for $i-k+1 \geq 0$ and $n-k+1 \leq 0$ for all $n: 0..j$, i.e. for $i-k+1 \geq 0$ and $j-k+1 \leq 0$, which says precisely that it is upper block triangular with a first block of size $k$.

We then calculate:

$$\frac{K_{i+k,j+k}^+(1/q^{k-1}y,y)}{K_{i,j}^+(q/y,q^k y)} = (-1)^k \prod_{l=0}^{k-1} \tau^+(q^l y) \frac{(q^{j+1})_k}{(q^{i+1})_k} \qquad \text{(V.141)}$$

where

$$\prod_{l=0}^{k-1} \tau^+(q^l y) = q^{-k(k-1)/2} \frac{(by)_k(y/b)_k b^k (\tilde{b}y)_k (y/\tilde{b})_k \tilde{b}^k}{y^k(y^2)_{2k}}. \qquad \text{(V.142)}$$

At the other boundary, we get:

$$K_{j,i}^- = \sum_{n=0}^{\min[i,j]} (-1)^n \frac{(q)_i}{(q)_n(q)_{i-n}} \frac{(xyq^n)_{j-n}}{(q)_{j-n}(x^2q^{2n})_{i+j-2n}} A_{j-n,i-n}^-(q^n x) \prod_{l=0}^{n-1} \tau^-(q^l x) \qquad \text{(V.143)}$$

where $A_{j,i}^-$ a polynomial, and
$$\tau^-(x) = \frac{(1-ax)(1-\tilde{a}x)(x-a)(x-\tilde{a})}{x(1-x^2)(1-qx^2)} \tag{V.144}$$

which gives us
$$\frac{K_{j+k,i+k}^-(x, 1/q^{k-1}x)}{K_{j,i}^-(q^k x, q/x)} = (-1)^k \prod_{l=0}^{k-1} \tau^-(q^l x) \frac{(q^{i+1})_k}{(q^{j+1})_k} \tag{V.145}$$

where
$$\prod_{l=0}^{k-1} \tau^-(q^l x) = q^{-k(k-1)/2} \frac{(ax)_k (x/a)_k a^k (\tilde{a}x)_k (x/\tilde{a})_k \tilde{a}^k}{x^k (x^2)_{2k}}. \tag{V.146}$$

As previously, apart from factors $\frac{(q^{j+1})_k}{(q^{i+1})_k}$ and $\frac{(q^{i+1})_k}{(q^{j+1})_k}$ which account for the change from $X$ to $\overline{X}$ on both rows, everything else depends only on $k$, factors out of the matrix products, and gives, when put together:
$$\prod_{l=0}^{k-1} \tau^-(q^l x)\tau^+(q^{l-k+1}/x) = \frac{h_b(q^k x) h_b(q/x)}{h_b(x) h_b(1/q^{k-1}x)} \tag{V.147}$$

with
$$\boxed{h_b(x) = \frac{(x^2)_\infty}{(ax, \tilde{a}x, bx, \tilde{b}x)_\infty}} \tag{V.148}$$

which replaces the simpler version (V.133).

Using this, we do the exact same operations as in the previous case, and we get the complete quantum determinant for the open ASEP with all non-zero boundary rates:
$$\boxed{F(x) = \frac{(1+x)^L (1+1/x)^L (x^2, x^{-2})_\infty}{(ax, \tilde{a}x, bx, \tilde{b}x, a/x, \tilde{a}/x, b/x, \tilde{b}/x)_\infty}}. \tag{V.149}$$

We finally look at the transfer matrix $\tilde{t}^{(2)}(x)$ and try to relate it to $M_\mu$, as we did in section V.1.2 for the periodic case. By analogy with eq.(V.35), we will try to rewrite $\tilde{t}^{(2)}(h)/F(x)$ as the standard Bethe transfer matrix.

The two-dimensional blocks from $K^+$ and $K^-$ for $y = 1/qx$ are:
$$K_2^+ = \begin{bmatrix} 1 & \frac{qx(qx+qx b\tilde{b}-b-\tilde{b})}{(q^2x^2-1)} \\ \frac{x(1+b\tilde{b}-qxb-qx\tilde{b})}{(q^2x^2-1)} & -b\tilde{b}x \end{bmatrix} , \quad K_2^- = \begin{bmatrix} 1 & \frac{(a+\tilde{a}-x-a\tilde{a}x)}{(1-x^2)} \\ \frac{(ax+\tilde{a}x-1-a\tilde{a})}{q(1-x^2)} & -\frac{a\tilde{a}}{qx} \end{bmatrix} \tag{V.150}$$

and the corresponding blocks from the bulk are:
$$X_2(x) = \left[\begin{array}{cc|cc} 1+x & 0 & 0 & 0 \\ 0 & 1+qx & 1-\frac{1}{q} & 0 \\ \hline 0 & 1-q & 1+\frac{1}{qx} & 0 \\ 0 & 0 & 0 & 1+\frac{1}{x} \end{array}\right], \tag{V.151}$$

$$\overline{X}_2(x) = \left[\begin{array}{cc|cc} 1+\frac{1}{qx} & 0 & 0 & 0 \\ 0 & 1+\frac{1}{x} & 1-q & 0 \\ \hline 0 & 1-\frac{1}{q} & 1+x & 0 \\ 0 & 0 & 0 & 1+qx \end{array}\right]. \tag{V.152}$$

Consider these transformations, with $\lambda = \frac{1+qx}{q(1+x)}$, i.e. $x = -\frac{1-q\lambda}{q(1-\lambda)}$:

$$L_i(\lambda) = \frac{1}{1+x} \left( \begin{bmatrix} 1 & 0 \\ 0 & -q \end{bmatrix} X_2(x) \begin{bmatrix} 1 & 0 \\ 0 & -1/q \end{bmatrix} \right) \cdot \left[ \begin{array}{c|c} 1 & 0 \\ \hline 0 & x \end{array} \right] = \left[ \begin{array}{cc|cc} 1 & 0 & 0 & 0 \\ 0 & q\lambda & 1-q\lambda & 0 \\ 0 & 1-\lambda & \lambda & 0 \\ 0 & 0 & 0 & 1 \end{array} \right] \quad (V.153)$$

and

$$\overline{L}_i(\lambda) = \frac{1}{1+x} \left[ \begin{array}{c|c} x & 0 \\ \hline 0 & 1 \end{array} \right] \cdot \left( \begin{bmatrix} 0 & 1 \\ 1 & 0 \end{bmatrix} \overline{X}_2(x) \begin{bmatrix} 0 & 1 \\ 1 & 0 \end{bmatrix} \right) = \left[ \begin{array}{cc|cc} 1 & 0 & 0 & 1-q\lambda \\ 0 & \lambda & 0 & 0 \\ \hline 0 & 0 & q\lambda & 0 \\ 1-\lambda & 0 & 0 & 1 \end{array} \right] \quad (V.154)$$

where the matrix products inside the parentheses are done in the auxiliary space, on each element of $X_2$ or $\overline{X}_2$, and the third product is done on the physical space at each site. Notice that the inner products cancel out between one site and the next, and that the outer products are done between $X_2$ and $\overline{X}_2$ and amount to a global factor $\frac{x}{(1+x)^2}$ on each site. Taking a product of $L$ matrices $X_2 \cdot \overline{X}_2$, this transformation gives, apart from the inner products at each end of the chain, a global factor $\frac{x^L}{(1+x)^{2L}}$, which accounts for the bulk part $h(x)$ of $F(x)$. Also note that the matrices from $\overline{L}_i$ need to be transposed if multiplied from right to left.

Considering that $\tilde{t}^{(2)}$ has a factor $h_b(x)h_b(1/qx)$, and that

$$h(x)h_b(x)h_b(1/qx) = F(x) \frac{(1-1/q^2x^2)(1-1/qx^2)}{(1-a/qx)(1-\tilde{a}/qx)(1-b/qx)(1-\tilde{b}/qx)} \quad (V.155)$$

the transformations we need to do on the boundary matrices are:

$$\hat{K}_2^+(\lambda) = \frac{(1-1/q^2x^2)}{(1-b/qx)(1-\tilde{b}/qx)} \begin{bmatrix} 1 & 0 \\ 0 & -q \end{bmatrix} K_2^+ \begin{bmatrix} 0 & 1 \\ 1 & 0 \end{bmatrix}, \quad (V.156)$$

$$\hat{K}_2^-(\lambda) = \frac{(1-1/qx^2)}{(1-a/qx)(1-\tilde{a}/qx)} \begin{bmatrix} 0 & 1 \\ 1 & 0 \end{bmatrix} K_2^- \begin{bmatrix} 1 & 0 \\ 0 & -1/q \end{bmatrix}, \quad (V.157)$$

which are too complicated to be written in terms of $\lambda$. Their values and first derivatives at $\lambda = 0$, which is all we will need, are, in terms of the original boundary parameters:

$$\hat{K}_2^+ = \begin{bmatrix} 1 & 0 \\ 0 & 1 \end{bmatrix} \quad, \quad \frac{d}{d\lambda}\hat{K}_2^+ = \begin{bmatrix} -2\delta & 2\beta \\ 2\delta & 1-q-2\beta \end{bmatrix}, \quad (V.158)$$

$$\hat{K}_2^- = \begin{bmatrix} \frac{1-\alpha+\gamma}{1+q} & \frac{\gamma}{q} \\ \alpha & \frac{\alpha-\gamma+q}{1+q} \end{bmatrix} \quad, \quad \frac{d}{d\lambda}\hat{K}_2^- = \begin{bmatrix} -\alpha+\gamma+A-B & \frac{\gamma(2q-\alpha-\gamma)}{q} \\ \alpha(1+q-\alpha-\gamma) & -2\gamma-q-A+B \end{bmatrix}, \quad (V.159)$$

with $A = \frac{2+(\alpha-2)\alpha-\gamma^2}{1+q}$ and $B = \frac{2(1-\alpha+\gamma)}{(1+q)^2}$.

We now put all these matrices together, obtaining $\tilde{t}^{(2)}(h)/F(x)$ as desired, and see what happens. The Lax matrices $L_i$ are not exactly the same as those we had for the periodic case. Their values at $\lambda = 0$ are $L_i(0) = P_i$ and $\overline{L}_i(0) = \overline{P}_i$, where $P_i$ is the permutation matrix exchanging the auxiliary space with the physical space when applied to the right in the matrix product, and $\overline{P}_i$ is the same but when applied to the left. We see that the two boundaries don't play the same role. The right boundary matrix is simply the identity at $\lambda = 0$, and serves to connect the Lax matrices on the last site. The left boundary matrix is traced, at $\lambda = 0$, because of the Lax matrices on the first site, and we see that its trace is 1. All in all, at $\lambda = 0$, the whole transfer matrix is simply the identity (see fig.-V.11-a).

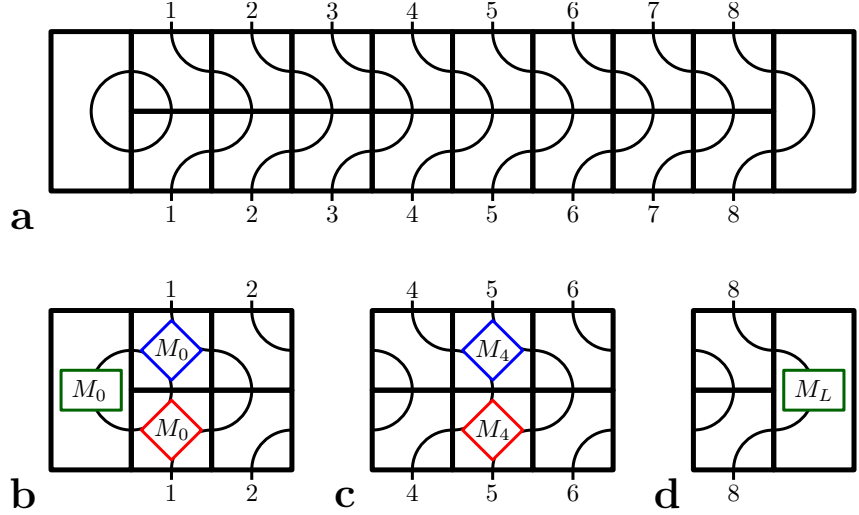

Figure V.11: Schematic representation of the value and first logarithmic derivative of $\hat{t}^{(2)}$ at 0. The first is the identity matrix (a). The second is a sum of terms adding up to $M_\mu$: three terms at the left boundary (b), two for each bond in the bulk (c) and one term at the right boundary (d).

As for its first derivative with respect to $\lambda$, we find that each pair of Lax matrices $L_i$ and $\overline{L}_i$ have a total contribution of $2M_i + (1-q)(n_i - n_{i+1})$. The right boundary matrix, as can be seen in (V.158), gives a term $2m_L + (1-q)n_L$. At the left boundary, we have three terms to consider, one involving the derivative of $K^-$ and the first Lax matrices taken at 0, and two more where we take the derivative of the first Lax matrices, and the boundary matrix at 0. The sum of those terms gives a total contribution of $2m_0 - (1-q)n_1$. If we now sum all the terms that we have found, the parts proportional to $(1-q)$ all cancel out, and we are simply left with $2M_\mu$.

At the end of the day, we find that:

$$M_\mu = \frac{1}{2}\left(1 - \frac{1}{q}\right)\frac{d}{dx}\log\left(\frac{\tilde{t}^{(2)}(x)}{F(x)}\right)\bigg|_{x=-1/q} \quad \text{(V.160)}$$

which is the same as eq.(V.37), with $h(x)$ replaced by $F(x)$, and an extra factor $\frac{1}{2}$. We haven't mentioned the matrices $A_\mu$ in those last calculations, because their behaviour is trivial, and it is left to the reader, as an exercise, to check that adding one between two sites gives the correct deformation for $M_\mu$.

As for the periodic case, we could have done all those calculations around $\lambda = \infty$ instead of 0, which would have given an equivalent result:

$$M_\mu = \frac{1}{2}(1-q)\frac{d}{dx}\log\left(\frac{\tilde{t}^{(2)}(x)}{F(qx)}\right)\bigg|_{x=-1}. \tag{V.161}$$

This is the last equation we needed in order to obtain the results from section III.3. In the next section, we put everything together and give a summary of the whole procedure.

### V.2.4 Summary - Functional Bethe Ansatz for the open ASEP

In this section, we collect all the results we found for the open ASEP, and show how they lead to the expressions for the cumulants of the current from section III.3.

The first step is to construct two transfer matrices $U_\mu(x)$ and $T_\mu(y)$:

$$U_\mu(x) = h_b(x)\frac{1}{Z_L}\langle\!\langle W\|A_\mu\prod_{i=1}^{L}X^{(i)}(x,x)\|V\rangle\!\rangle, \tag{V.162}$$

$$T_\mu(y) = h_b(y)\langle\!\langle \tilde{W}\|A_\mu\prod_{i=1}^{L}X^{(i)}(y,y)\|\tilde{V}\rangle\!\rangle \tag{V.163}$$

(involving the function $h_b$ defined in eq.(V.148)), such that, for any $x$ and $y$, we have:

$$[M_\mu, U_\mu(x)T_\mu(y)] = 0. \tag{V.164}$$

Using these, we can construct two commuting matrices $P(x)$ and $Q(y)$ as:

$$P(x) = U_\mu(x)\left[U_\mu(0)\right]^{-1} \quad , \quad Q(y) = (1-\mathrm{e}^{-\mu})^{-1}U_\mu(0)T_\mu(y) \tag{V.165}$$

such that:
$$U_\mu(x)T_\mu(y) = (1-\mathrm{e}^{-\mu})P(x)Q(y). \tag{V.166}$$

Note that a completely rigorous proof of the commutation of $P$ and $Q$, involving the R-matrix, has not yet been obtained.

We can show that those matrices verify, for any positive integer $k$:

$$P(x)Q(1/q^{k-1}x) = t^{(k)}(x) + \mathrm{e}^{-2k\mu}P(q^k x)Q(q/x) \tag{V.167}$$

(where we omit the tildes, since we have correctly normalised $P$ and $Q$ from the start). The matrix $t^{(k)}(x)$ is the Bethe transfer matrix with a $k$-dimensional auxiliary space.

The first two of those relations write:

$$P(x)Q(1/x) = F(x) + \mathrm{e}^{-2\mu}P(qx)Q(q/x), \tag{V.168}$$
$$P(x)Q(1/qx) = t^{(2)}(x) + \mathrm{e}^{-4\mu}P(q^2 x)Q(q/x), \tag{V.169}$$

where $F(x)$ is the quantum determinant, and $t^{(2)}$ is such that:

$$M_\mu = \frac{1}{2}(1-q)\frac{d}{dx}\log\left(\frac{t^{(2)}(x)}{F(qx)}\right)\bigg|_{x=-1}. \tag{V.170}$$

Using eq.(V.168) at $x$ and $qx$, and eq.(V.169), we can find the T-Q equation:

$$t^{(2)}(x)Q(1/x) = F(x)Q(1/qx) + e^{-2\mu}F(qx)Q(q/x) \tag{V.171}$$

which allows us to express $M_\mu$ in terms of $Q$ instead:

$$M_\mu = \frac{1}{2}(1-q)\frac{d}{dx}\log\left(\frac{Q(q/x)}{Q(1/x)}\right)\bigg|_{x=-1}. \tag{V.172}$$

We now consider:

$$B = -e^{2\mu}\big(Q(0)\big)^{-1} = -e^{2\mu}(1-e^{-\mu})\big(U_\mu(0)T_\mu(0)\big)^{-1}. \tag{V.173}$$

We saw in section III.2 that $U_\mu(x)T_\mu(y) = |P^\star\rangle\langle 1|$ for $\mu = 0$ (in fact, we saw that only for $x = y = 0$, but the derivation also works in this case), which tells us that the dominant eigenvalue of $U_\mu(x)T_\mu(y)$ goes to 1 when $\mu \to 0$, while the others are at least of order $\mu$. From this, we find that the first eigenvalue of $B$ goes to 0 for $\mu \to 0$, and the others remain finite. This also tells us that all the roots of the first eigenvalue of $Q(1/x)$ go to 0 for $\mu \to 0$, while those of $P(x)$ go to $\infty$.

Notice that for $a < 1$ and $b < 1$, $P(x)$ is expandable as a series in $x$ around of the unit circle, while $Q(1/x)$ is expandable as a series in $1/x$. This is because $P(x)$ is a product of $L$ matrices containing only monomials in $x$, two boundary vectors which contain polynomials in $x$, and a function $h_b(x)$ which is expandable in $x$ as long as all the constants in its denominator are inside of the unit circle (hence the constraint on $a$ and $b$). The same can be said of $Q(1/x)$ with respect to $1/x$. This tells us that, for $\mu$ small enough, and in the first eigenspace of $P$ and $Q$, $P$ is holomorphic inside of the unit circle, so that a contour integral over that circle will only pick up contributions from $Q$ and not from $P$. This is the same reasoning as we did in section II.2.2 to justify the contour integral expressions for $\mu$ and $E(\mu)$.

Knowing all this, we can do the same calculations as in section II.2.2, starting from eq.(II.103), with only a few minor differences coming from the factor 2 in front of $\mu$ in (V.168) and the factor $\frac{1}{2}$ in (V.172). We define a function $W(x)$ as:

$$W(x) = -\frac{1}{2}\log\left(\frac{P(x)Q(1/x)}{e^{-2\mu}P(qx)Q(q/x)}\right), \tag{V.174}$$

and a convolution kernel $K$, as:

$$K(z,\tilde{z}) = 2\sum_{k=1}^{\infty}\frac{q^k}{1-q^k}\left((z/\tilde{z})^k + (z/\tilde{z})^{-k}\right) \tag{V.175}$$

along with the associated convolution operator $X$:

$$X[f](z) = \oint_{c_1}\frac{d\tilde{z}}{i2\pi\tilde{z}}f(\tilde{z})K(z,\tilde{z}). \tag{V.176}$$

Using those, one can find that $-\log\big(P(qx)Q(q/x)/Q(0)\big) = X[W](x)$, and we can finally rewrite eq.(V.168) in terms of only one unknown function $W$:

$$\boxed{W(x) = -\frac{1}{2}\ln\left(1 - BF(x)e^{X[W](x)}\right)} \tag{V.177}$$

which is the same as eq.(III.67).

The last step is to take eq.(V.172) in the first eigenspace of $M_\mu$, and eq.(V.174) at $x = 0$, to find:

$$E(\mu) = \frac{1}{2}(1-q)\frac{d}{dx}\log\left(\frac{Q(q/x)}{Q(1/x)}\right)\bigg|_{x=-1} \quad , \quad \mu = -W(0) \tag{V.178}$$

Considering what we said before about $P$ being holomorphic inside of the unit circle, we can replace $\frac{1}{2}\log\left(\frac{Q(q/x)}{Q(1/x)}\right)$ by $-W(x)$ when expressing $E(\mu)$ as a contour integral (since $P$ will not contribute), and obtain:

$$\boxed{\mu = -\oint_{c_1}\frac{dz}{i2\pi z}W(z)} \tag{V.179}$$

and

$$\boxed{E(\mu) = -(1-q)\oint_{c_1}\frac{dz}{i2\pi(1+z)^2}W(z),} \tag{V.180}$$

in which we recognise (III.68) and (III.69) from section III.3. All this is done for $a < 1$ and $b < 1$, but can then be generalised to any $a$ and $b$ through the same reasoning as in section II.2.1 for the mean current, replacing the unit circle $c_1$ by small contours around $S = \{0, q^k a, q^k \tilde{a}, q^k b, q^k \tilde{b}\}$.

## V.2.5 Appendix - XXZ spin chain with general boundary conditions

In this section, we explain how our construction for the open ASEP can be translated for the spin-$\frac{1}{2}$ XXZ chain with non-diagonal boundary conditions [38].

Let us first define the bulk Hamiltonian of the XXZ spin chain of length $L$:

$$H_b = \frac{1}{2}\sum_{k=1}^{L-1} h_i \tag{V.181}$$

with $h_i$ acting as:

$$h_i = \begin{bmatrix} \Delta & 0 & 0 & 0 \\ 0 & -\Delta & 1 & 0 \\ 0 & 1 & -\Delta & 0 \\ 0 & 0 & 0 & \Delta \end{bmatrix} \tag{V.182}$$

on sites $i$ and $i+1$ (in basis $\{00, 01, 10, 11\}$, as usual), and as the identity on the rest of the chain. We define $\Delta$ as $\frac{1}{2}(q^{-1/2} - q^{1/2})$, which is not the usual definition for the XXZ chain (that can be obtained simply by replacing $q$ by $q^2$).

Let us also write the deformed Markov matrix $M_{\{\mu_i\}}$ for the special choice of weights defined by:

$$\{\mu_0 = \frac{1}{2}\log\left(\frac{\gamma}{\alpha}\right) + \nu_0, \quad \mu_i = \frac{1}{2}\log(q), \quad \mu_L = \frac{1}{2}\log\left(\frac{\delta}{\beta}\right) + \nu_L\} \tag{V.183}$$

which is on the line $\mu = \frac{1}{2} \log\left(\frac{\gamma\delta}{\alpha\beta} q^{L-1}\right) + \imath\mathbb{R}$ if $\nu_0$ and $\nu_L$ are imaginary numbers (in which case $M_{\{\mu_i\}}$ is Hermitian). The deformed local matrices become:

$$m_0(\mu_0) = \begin{bmatrix} -\alpha & \sqrt{\alpha\gamma}\,e^{-\nu_0} \\ \sqrt{\alpha\gamma}\,e^{\nu_0} & -\gamma \end{bmatrix}, \tag{V.184}$$

$$M_i(\mu_i) = \begin{bmatrix} 0 & 0 & 0 & 0 \\ 0 & -q & \sqrt{q} & 0 \\ 0 & \sqrt{q} & -1 & 0 \\ 0 & 0 & 0 & 0 \end{bmatrix}, \tag{V.185}$$

$$m_L(\mu_l) = \begin{bmatrix} -\delta & \sqrt{\beta\delta}e^{\nu_L} \\ \sqrt{\beta\delta}e^{-\nu_L} & -\beta \end{bmatrix}. \tag{V.186}$$

It is straightforward to check that in this case, we have $M_{\{\mu_i\}} = \sqrt{q}H + \epsilon$, where $\epsilon$ is a constant, with the boundary matrices being equal to:

$$h_0 = \frac{1}{2\sqrt{q}} \begin{bmatrix} (1-q-\alpha+\gamma) & 2\sqrt{\alpha\gamma}\,e^{-\nu_0} \\ 2\sqrt{\alpha\gamma}\,e^{\nu_0} & (-1+q+\alpha-\gamma) \end{bmatrix}, \tag{V.187}$$

$$h_L = \frac{1}{2\sqrt{q}} \begin{bmatrix} (-1+q+\beta-\delta) & 2\sqrt{\beta\delta}e^{\nu_L} \\ 2\sqrt{\beta\delta}e^{-\nu_L} & (1-q-\beta+\delta) \end{bmatrix}. \tag{V.188}$$

Since we have three nontrivial parameters in each of those matrices, they are completely general: we can write (without restricting ourselves to hermitian matrices)

$$h_0 = a_z \sigma_z + a_+ \sigma^+ + a_- \sigma^- = \begin{bmatrix} a_z & a_- \\ a_+ & -a_z \end{bmatrix}, \tag{V.189}$$

$$h_L = b_z \sigma_z + b_+ \sigma^+ + b_- \sigma^- = \begin{bmatrix} b_z & b_- \\ b_+ & -b_z \end{bmatrix}, \tag{V.190}$$

with

$$a_z = (1-q-\alpha+\gamma)/2\sqrt{q}\ ,\quad a_+ = \sqrt{\alpha\gamma/q}\,e^{\nu_0}\ ,\quad a_- = \sqrt{\alpha\gamma/q}\,e^{-\nu_0}, \tag{V.191}$$

$$b_z = (-1+q+\beta-\delta)/2\sqrt{q}\ ,\quad b_+ = \sqrt{\beta\delta/q}\,e^{-\nu_L}\ ,\quad b_- = \sqrt{\beta\delta/q}\,e^{\nu_L}, \tag{V.192}$$

which is to say

$$\nu_0 = -2\log(a_+/a_-), \tag{V.193}$$

$$\alpha = \sqrt{(\sqrt{q}a_z - (1-q)/2)^2 + a_+ a_-} - \sqrt{q}a_z + (1-q)/2, \tag{V.194}$$

$$\gamma = \sqrt{(\sqrt{q}a_z - (1-q)/2)^2 + a_+ a_-} + \sqrt{q}a_z - (1-q)/2, \tag{V.195}$$

and

$$\nu_L = -2\log(b_-/b_+), \tag{V.196}$$

$$\beta = \sqrt{(\sqrt{q}b_z + (1-q)/2)^2 + b_+ b_-} + \sqrt{q}b_z + (1-q)/2, \tag{V.197}$$

$$\delta = \sqrt{(\sqrt{q}a_z + (1-q)/2)^2 + b_+ b_-} - \sqrt{q}b_z - (1-q)/2. \tag{V.198}$$

Those are well defined for any values of $a_z$, $a_+$, $a_-$, $b_z$, $b_+$ and $b_-$.

Considering expression (V.22), or its equivalent for an open chain, and noting that $A_{\frac{\mu_i}{2}} = A^{-1/4}$, we can rewrite $U(x)$ and $T(y)$ in a way better suited to this situation:

$$U(x) = \frac{1}{Z_L} \langle\!\langle \phi \| \prod_{i=1}^{L} Y^{(i)}(x) \| \psi \rangle\!\rangle, \tag{V.199}$$

$$T(y) = \langle\!\langle \tilde\phi \| \prod_{i=1}^{L} Y^{(i)}(y) \| \tilde\psi \rangle\!\rangle, \tag{V.200}$$

with

$$Y(x) = \begin{bmatrix} N_x & \Sigma_+ \\ \Sigma_- & N_x \end{bmatrix} \tag{V.201}$$

where

$$N_x = A^{-1/2} - x A^{1/2} \tag{V.202}$$
$$\Sigma_+ = A^{-1/4} e A^{-1/4} = q^{-1/4} S^+ (A^{-1/2} - x^2 A^{1/2}), \tag{V.203}$$
$$\Sigma_- = A^{-1/4} d A^{-1/4} = q^{1/4} S^- (A^{-1/2} - A^{1/2}). \tag{V.204}$$

The boundary vectors become:

$$\langle\!\langle \phi \| = \langle\!\langle W \| A_{\mu_0} A^{1/4}, \tag{V.205}$$
$$\| \psi \rangle\!\rangle = A_{\mu_L} A^{1/4} \| V \rangle\!\rangle, \tag{V.206}$$
$$\langle\!\langle \tilde\phi \| = \langle\!\langle \tilde W \| A_{\mu_0} A^{1/4}, \tag{V.207}$$
$$\| \tilde\psi \rangle\!\rangle = A_{\mu_L} A^{1/4} \| \tilde V \rangle\!\rangle. \tag{V.208}$$

Matrices $N_x$, $\Sigma_+$ and $\Sigma_-$ satisfy the $U_q[SU(2)]$ algebra [146]:

$$[\Sigma_+, \Sigma_-] = (q^{-1/2} - q^{1/2})(A^{-1} - x^2 A), \tag{V.209}$$
$$\Sigma_- A = q\, A \Sigma_-, \tag{V.210}$$
$$A \Sigma_+ = q\, \Sigma_+ A, \tag{V.211}$$

and the conditions on the boundary vectors become:

$$\langle\!\langle \phi \| \left[ a_+ \Sigma_+ - a_- \Sigma_- - 2 a_z N_x - (q^{-1/2} - q^{1/2}) x A^{1/2} \right] = 0, \tag{V.212}$$
$$\left[ b_- \Sigma_- - b_+ \Sigma_+ + 2 b_z N_x - (q^{-1/2} - q^{1/2}) x A^{1/2} \right] \| \psi \rangle\!\rangle = 0, \tag{V.213}$$
$$\langle\!\langle \tilde\phi \| \left[ a_+ \Sigma_+ - a_- \Sigma_- - 2 a_z N_y + (q^{-1/2} - q^{1/2}) A^{-1/2} \right] = 0, \tag{V.214}$$
$$\left[ b_- \Sigma_- - b_+ \Sigma_+ + 2 b_z N_y + (q^{-1/2} - q^{1/2}) A^{-1/2} \right] \| \tilde\psi \rangle\!\rangle = 0. \tag{V.215}$$

It was surprising to find that this solution has a structure almost identical to that of the Lindblad master equation found in [181], where $Y$ is noted $\Omega$ and $\hat Y$ is noted $\Xi$. In that case, the algebraic relations satisfied by the boundary vectors $\langle\!\langle \phi \|$ and $\| \psi \rangle\!\rangle$ are different from ours, as there is only one vector per boundary but two equations per vector. It would be interesting to understand the precise relation between those two *a priori* very different situations.

# Conclusion

We have studied, in this thesis, a very simple, and yet very rich model from non-equilibrium statistical physics: the asymmetric simple exclusion process.

In doing so, we have used various methods, from simple mean field calculations to the matrix Ansatz, perturbation theory, fluctuating hydrodynamics, and the Bethe Ansatz. Through those, we have been able to find new and interesting results on the fluctuation of the macroscopic current of particles, which is one of the fundamental characteristics of systems out of equilibrium.

Our main result, which we presented in chapter III, is an exact formula for the cumulants of the current in the steady state of the open ASEP, valid for any size and values of the boundary parameters. We first showed, in that chapter, how we derived that result using a generalisation of the matrix Ansatz and a fair amount of guesswork. Later, in chapter V, we gave a more rigorous proof of that result, using a variant of the algebraic Bethe Ansatz applicable to the most general open ASEP, which was, to our knowledge, not known to be possible. We also saw how the matrix Ansatz could be obtained as a special case of our method.

Starting from our newly found formula for the cumulants of the current, we analysed, in chapter IV, their behaviour in the limit of a large system. Combining it to results from the macroscopic fluctuation theory, and from direct diagonalisation in a few extreme cases, we were able to describe the large deviations of the current and the associated density profiles in four of the five phases that the system might find itself in, and give asymptotic results for the fifth.

All the results that we have obtained are, *a priori*, specific to the ASEP, but this is not the end of the story. The next step is precisely to find out what general insights we can extract from our specific results: what methods could be applied to other models, and what aspects of this system's behaviour might be in fact universal. These are the two most important of the many open questions that remain unanswered for now. Here are a few of those that we might want to tackle in the near future:

- The first one is, expectedly, the gaping hole left in chapter IV by the absence of an expression for the large deviations of the current inside of the maximal current phase. A precise analysis of the local anti-correlations in the system might shed some light on this, as we argued that they are an important feature of that phase.

- We uncovered, in that same chapter, a few non-equilibrium phase transitions, for which the order parameter is $\mu$, the quantity conjugate to the current in the system. However, that parameter does not correspond to any physical variable, and the signature of those transitions on a dynamical level (as would be observed in experiments, for instance) is very unclear. We saw in particular, in section IV.1.5, that the mathematical form of the cumulants of the current was most complex at the point $\rho_a = \rho_b = u = \frac{1}{2}$, which sits at the interface between all five phases of the system. We do not know if that complexity is due to any physically relevant behaviour.

- We might wonder to what extent the results from chapter IV could be generalised to other systems, and in particular to other interacting particle models, to which the macroscopic fluctuation theory is in principle applicable. It would be especially interesting to know precisely which part of our results comes from integrability, and which part might be transposable to other non-equilibrium systems.

- For those results that are not due to integrability, we could ask ourselves whether they could be obtained directly from a macroscopic description of the system rather than from its microscopic details. In particular, we saw that the generating function of the cumulants of the current around a phase transition (i.e. in the steady state MC phase or on the shock line) is best expressed as an implicit function of a parameter $B$. We do not know if and how a similar structure could arise from large scale calculations. We also do not know if these expressions are in any way universal.

- The Bethe Ansatz calculations we performed in chapter V were aimed specifically at deriving rigorously the result that we had guessed in chapter III, but it would certainly be nice to use our method in order to obtain some new results as well. We could, for instance, consider the multispecies ASEP, for which it might be fairly straightforward to generalise the existing matrix Ansatz into the algebraic Bethe Ansatz.

- The few final steps of those Bethe Ansatz calculations were specific to the steady state of the ASEP. It would be useful to generalise them to other eigenstates, which are in principle obtainable through our method, although the calculations involved would probably be much more complex. It would, in particular, give us access to the transient regime, which holds interesting information on the physical behaviour of the system. The first step in this would be to retrieve and extend the results of [43] on the spectral gap in the open ASEP.

- Moreover, we could only obtain an eigenvalue of the deformed Markov matrix, but no expression for the corresponding eigenvectors (other than perturbatively, as in chapter III). Such an expression would allow us to see precisely how the hydrodynamic regime which we observed in chapter IV emerges from the microscopic structure of that state.

- Finally, we saw that the original matrix Ansatz can be retrieved as a special case of the transfer matrix we built in chapter V. It would be interesting to know whether there is anything general about that remark, which is to say if matrix product states, in other systems that exhibit them, are generically a special value of a Bethe-like transfer matrix, or if it is merely a coincidence.

# Appendix : Published articles

[1] **An Exact Formula for the Statistics of the Current in the TASEP with Open Boundaries**, Alexandre Lazarescu and Kirone Mallick, Journal of Physics A: Mathematical and Theoretical 44 (2011) 315001.
*Received the 2012 "Best Paper Prize" of Journal of Physics A.*

[2] **Exact Current Statistics of the ASEP with Open Boundaries**, Mieke Gorissen, Alexandre Lazarescu, Kirone Mallick and Carlo Vanderzande, Physical Review Letters 109, 170601 (2012).
*Selected as a Physical Review Letters editors' suggestion and for a Viewpoint in Physics.*

[3] **Matrix Ansatz for the Fluctuations of the Current in the ASEP with Open Boundaries**, Alexandre Lazarescu , Journal of Physics A: Mathematical and Theoretical 46 (2013) 145003.
*Selected for IOPselect.*

# Alexandre Lazarescu

# Exact Large Deviations of the Current in the Asymmetric Simple Exclusion Process with Open Boundaries

## Abstract
In this thesis, we consider one of the most popular models of non-equilibrium statistical physics: the Asymmetric Simple Exclusion Process, in which particles jump stochastically on a one-dimensional lattice, between two reservoirs at fixed densities, with the constraint that each site can hold at most one particle at a given time. This model has the mathematical property of being integrable, which makes it a good candidate for exact calculations. What interests us in particular is the current of particles that flows through the system (which is a sign of it being out of equilibrium), and how it fluctuates with time. We present a method, based on the 'matrix Ansatz' devised by Derrida, Evans, Hakim and Pasquier, that allows to access the exact cumulants of that current, for any finite size of the system and any value of its parameters. We also analyse the large size asymptotics of our result, and make a conjecture for the phase diagram of the system in the so-called 's-ensemble'. Finally, we show how our method relates to the algebraic Bethe Ansatz, which was thought not to be applicable to this situation.

Keywords: Exclusion process, non-equilibrium, large deviations, Bethe Ansatz

## Résumé
Dans cette thèse, on considère un des modèles les plus étudiés en physique statistique hors équilibre : le processus d'exclusion simple asymétrique, qui décrit des particules se déplaçant stochastiquement sur un réseau unidimensionnel, entre deux réservoirs de densités fixées, avec la contrainte que chaque site ne peut porter qu'une particule à un instant donné. Ce modèle a la propriété mathématique d'être intégrable, ce qui en fait un bon candidat à une résolution exacte. Ce qui nous intéresse, en particulier, est de décrire le courant de particules qui traverse le système (ce qui est une caractéristique des systèmes hors équilibre) et comment ce dernier fluctue avec le temps. Nous présentons une méthode inspirée de l'Ansatz matriciel de Derrida, Evans, Hakim et Pasquier, qui nous permet d'obtenir une expression exacte des cumulants de ce courant, et ce pour une taille finie du système et quelle que soit la valeur de ses paramètres. Nous analysons également le comportement asymptotique de ce résultat à la limite d'un système de grande taille, et émettons une conjecture quant au diagramme de phase du système dans 'l'ensemble-s'. Enfin, nous montrons en quoi notre méthode est reliée à l'Ansatz de Bethe algébrique, que l'on pensait ne pas être appliquable à cette situation.

Mots-clés : Processus d'exclusion, hors-équilibre, grandes déviations, Anastz de Bethe